\documentclass{article}[10pt,letterpaper]
\usepackage{fancyhdr}
\usepackage{supertabular}
\renewcommand{\sectionmark}[1]{\markboth{#1}{}}
\usepackage{color}
\usepackage{amsmath}
\usepackage{pbox}
\usepackage{amssymb}
\usepackage{amsbsy}
\usepackage{scalefnt}
\usepackage{ifpdf}
\usepackage{graphicx}
\usepackage{multicol}
\usepackage[colorlinks=true, linkcolor=blue, linktocpage=true,urlcolor=blue]{hyperref}
\usepackage{longtable}
\usepackage{needspace}
\usepackage{eso-pic}
\usepackage{mathptmx}
\usepackage{multirow}
\usepackage{lscape}
\usepackage{textcomp}
\usepackage{changepage}
\usepackage{xfrac}
\usepackage{txfonts}
\usepackage[absolute]{textpos}
\ifpdf
\fi
\makeatletter
\renewcommand{\l@section}{\@dottedtocline{2}{1em}{0em}}
\renewcommand{\l@subsection}{\@dottedtocline{3}{2.5cm}{0em}}
\renewcommand{\hrulefill}{\leavevmode \leaders \hrule \@height 1pt \hfill \kern\z@}
\makeatother
\renewcommand{\underline}[1]{\begin{tabular}{@{\extracolsep{\fill}}c@{\extracolsep{\fill}}}#1\\[-0.2cm]\hrulefill\end{tabular}}
\renewcommand{\footrulewidth}{1pt}
\renewcommand{\headrulewidth}{1pt}
\newcounter{chappage}
\AddToShipoutPicture{\stepcounter{chappage}}
\fancypagestyle{plain}{\lhead{}\rhead{}}
\fancypagestyle{single}{\lhead{\leftmark \ \\}\rhead{\leftmark \ \\}}
\fancypagestyle{bob}{\lhead{\leftmark -\thechappage \ \\}\rhead{\leftmark -\thechappage \ \\}}
\begin{document}
\fontsize{9}{10}\selectfont
\fontdimen2\font=1.3\fontdimen2\font
%Abstract
\thispagestyle{empty}
\setcounter{page}{1}
\begin{center}

\vspace{0.5cm}
{ \huge Nuclear Data Sheets for {}A=201*}\\
\vspace{1.0cm}
{ \normalsize F.G. KONDEV}\\
\vspace{0.2in}
{ \small \it Physics Division\\
  Argonne National Laboratory\\
  9700 South Cass Avenue\\
  Lemont, Illinois 60439, USA}\\
\vspace{0.2in}
\end{center}

\setlength{\parindent}{-0.5cm}
\addtolength{\leftskip}{2cm}
\addtolength{\rightskip}{2cm}
{\bf Abstract: }
Evaluated nuclear structure and decay data for all nuclei with mass number A=201 (\ensuremath{^{\textnormal{201}}}Os, \ensuremath{^{\textnormal{201}}}Ir, \ensuremath{^{\textnormal{201}}}Pt, \ensuremath{^{\textnormal{201}}}Au, \ensuremath{^{\textnormal{201}}}Hg, \ensuremath{^{\textnormal{201}}}Tl, \ensuremath{^{\textnormal{201}}}Pb, \ensuremath{^{\textnormal{201}}}Bi, \ensuremath{^{\textnormal{201}}}Po, \ensuremath{^{\textnormal{201}}}At, \ensuremath{^{\textnormal{201}}}Rn, \ensuremath{^{\textnormal{201}}}Fr, \ensuremath{^{\textnormal{201}}}Ra) are presented. All available experimental data are compiled and evaluated, and best values for level and \ensuremath{\gamma}-ray energies, quantum numbers, lifetimes, \ensuremath{\gamma}-ray intensities and transition probabilities, as well as other nuclear properties, are recommended. Inconsistencies and discrepancies that exist in the literature are discussed. A number of computer codes (https://www-nds.iaea.org/public/ensdf\_pgm/index.htm) developed by members of the NSDD network were used during the evaluation process. For example, the reported absolute \ensuremath{\gamma}-ray emission probabilities and their uncertainties in various decay data sets were determined using the \textit{GABS} code. The \ensuremath{\gamma}-ray transition probabilities were determined using the \textit{RULER} code and the corresponding uncertainties were determined using a Monte-Carlo approach. This work supersedes the earlier evaluation by F.G. Kondev (\href{https://www.nndc.bnl.gov/nsr/nsrlink.jsp?2007Ko06,B}{2007Ko06}), published in \textit{Nuclear Data Sheets} \textbf{108}, 365 (2007).\\

{\bf Cutoff Date: }
Literature up to July 4, 2021 was consulted. The main source used for bibliographic information is the NSR database (\href{https://www.nndc.bnl.gov/nsr/nsrlink.jsp?2014Pr09,B}{2014Pr09}): www.nndc.bnl.gov/nsr/.\\

\vfill

* This work is supported by the Office of Nuclear Physics, Office of Science, U.S. Department of Energy under contract DE-AC02-06CH11357\\

\setlength{\parindent}{+0.5cm}
\addtolength{\leftskip}{-2cm}
\addtolength{\rightskip}{-2cm}
\newpage
\pagestyle{plain}
\setlength{\columnseprule}{1pt}
\setlength{\columnsep}{1cm}
\begin{center}
\underline{\normalsize Index for A=201}
\end{center}
\hspace{.3cm}\raggedright\underline{Nuclide}\hspace{1cm}\underline{Data Type\mbox{\hspace{2.3cm}}}\hspace{2cm}\underline{Page}\hspace{1cm}
\raggedright\underline{Nuclide}\hspace{1cm}\underline{Data Type\mbox{\hspace{2.3cm}}}\hspace{2cm}\underline{Page}
\begin{adjustwidth}{}{0.05\textwidth}
\begin{multicols}{2}
\setcounter{tocdepth}{3}
\renewcommand{\contentsname}{\protect\vspace{-0.8cm}}
\tableofcontents
\end{multicols}
\end{adjustwidth}
\clearpage
\thispagestyle{empty}
\mbox{}
\subsection[\hspace{2cm}Skeleton Scheme for A=201]{ }
\begin{figure}[h]
\begin{center}
\includegraphics[angle=90]{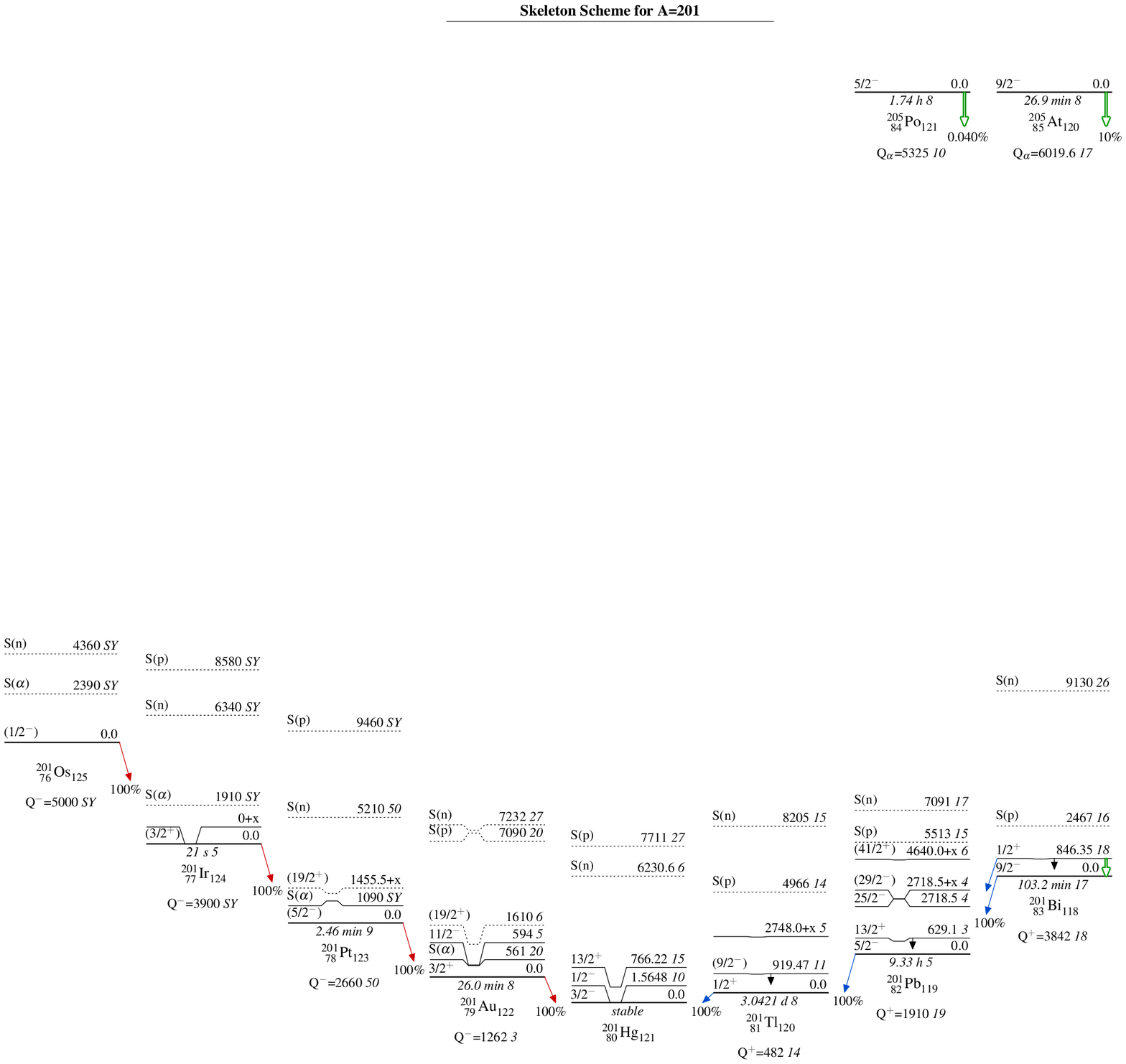}\\
\end{center}
\end{figure}
\clearpage
\begin{figure}[h]
\begin{center}
\includegraphics[angle=90]{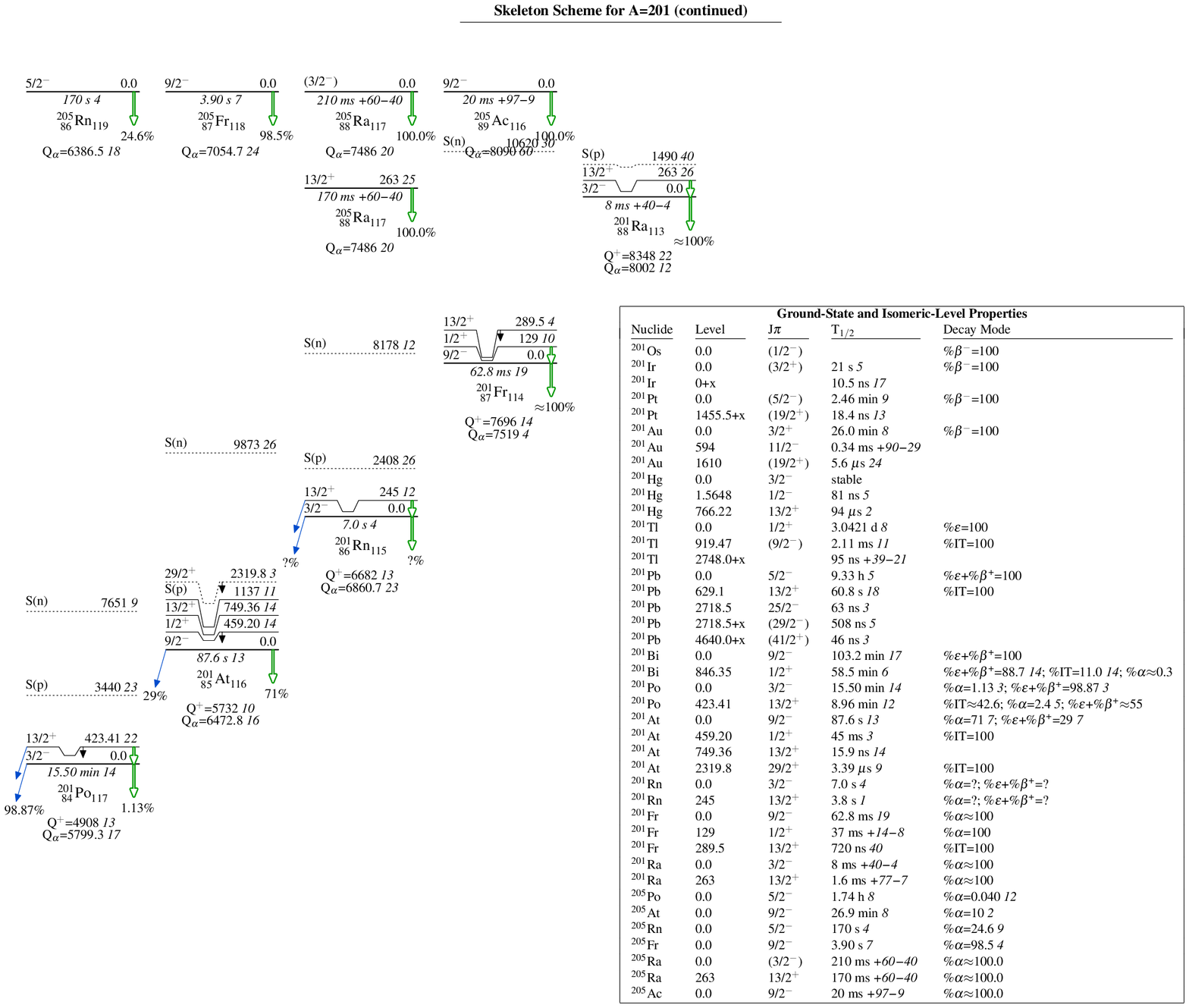}\\
\end{center}
\end{figure}
\clearpage
\pagestyle{bob}
\begin{center}
%ADOPTED LEVELS
\section[\ensuremath{^{201}_{\ 76}}Os\ensuremath{_{125}^{~}}]{ }
\vspace{-30pt}
\setcounter{chappage}{1}
\subsection[\hspace{-0.2cm}Adopted Levels]{ }
\vspace{-20pt}
\vspace{0.3cm}
\hypertarget{OS0}{{\bf \small \underline{Adopted \hyperlink{201OS_LEVEL}{Levels}}}}\\
\vspace{4pt}
\vspace{8pt}
\parbox[b][0.3cm]{17.7cm}{\addtolength{\parindent}{-0.2in}Q(\ensuremath{\beta^-})=5000 {\it SY}; S(n)=4360 {\it SY}; Q(\ensuremath{\alpha})=$-$2390 {\it SY}\hspace{0.2in}\href{https://www.nndc.bnl.gov/nsr/nsrlink.jsp?2021Wa16,B}{2021Wa16}}\\
\parbox[b][0.3cm]{17.7cm}{\addtolength{\parindent}{-0.2in}Estimated \textit{SY} uncertainties: 360 keV for Q(\ensuremath{\beta}\ensuremath{^{-}}), 420 keV for S(n) and 500 keV for Q(\ensuremath{\alpha}) (\href{https://www.nndc.bnl.gov/nsr/nsrlink.jsp?2021Wa16,B}{2021Wa16}).}\\

\parbox[b][0.3cm]{17.7cm}{\addtolength{\parindent}{-0.2in}\href{https://www.nndc.bnl.gov/nsr/nsrlink.jsp?2009St16,B}{2009St16}: \ensuremath{^{\textnormal{201}}}Os produced and identified in \ensuremath{^{\textnormal{9}}}Be(\ensuremath{^{\textnormal{208}}}Pb,x), E=1 GeV/nucleon from the UNILAC and SIS-18 accelerator complex at}\\
\parbox[b][0.3cm]{17.7cm}{GSI. Target=2.5 g/cm\ensuremath{^{\textnormal{2}}} \ensuremath{^{\textnormal{9}}}Be.}\\
\parbox[b][0.3cm]{17.7cm}{\addtolength{\parindent}{-0.2in}\href{https://www.nndc.bnl.gov/nsr/nsrlink.jsp?2012Ku26,B}{2012Ku26}: \ensuremath{^{\textnormal{201}}}Os produced and identified in \ensuremath{^{\textnormal{9}}}Be(\ensuremath{^{\textnormal{238}}}U,x), E=1 GeV/nucleon from the UNILAC and SIS-18 accelerator complex at}\\
\parbox[b][0.3cm]{17.7cm}{GSI. Target=1.6 g/cm\ensuremath{^{\textnormal{2}}} \ensuremath{^{\textnormal{9}}}Be.}\\
\vspace{12pt}
\hypertarget{201OS_LEVEL}{\underline{$^{201}$Os Levels}}\\
\begin{longtable}{cccc@{\extracolsep{\fill}}c}
\multicolumn{2}{c}{E(level)$^{}$}&J$^{\pi}$$^{}$&Comments&\\[-.2cm]
\multicolumn{2}{c}{\hrulefill}&\hrulefill&\hrulefill&
\endfirsthead
\multicolumn{1}{r@{}}{0}&\multicolumn{1}{@{}l}{}&\multicolumn{1}{l}{(1/2\ensuremath{^{-}})}&\parbox[t][0.3cm]{15.235281cm}{\raggedright \%\ensuremath{\beta}\ensuremath{^{-}}=100\vspace{0.1cm}}&\\
&&&\parbox[t][0.3cm]{15.235281cm}{\raggedright J\ensuremath{^{\pi}}: assuming spherical shape and systematics of N=125 isotones in the region; shell model predictions.\vspace{0.1cm}}&\\
&&&\parbox[t][0.3cm]{15.235281cm}{\raggedright T\ensuremath{_{1/2}}: \ensuremath{>}160 ns from from the time-of-flight in \href{https://www.nndc.bnl.gov/nsr/nsrlink.jsp?2012Ku26,B}{2012Ku26}. \ensuremath{\approx} 3 s estimated in \href{https://www.nndc.bnl.gov/nsr/nsrlink.jsp?2021Ko07,B}{2021Ko07}. Predicted T\ensuremath{_{\textnormal{1/2}}}(\ensuremath{\beta}): 56 s\vspace{0.1cm}}&\\
&&&\parbox[t][0.3cm]{15.235281cm}{\raggedright {\ }{\ }{\ }(\href{https://www.nndc.bnl.gov/nsr/nsrlink.jsp?2019Mo01,B}{2019Mo01}, FRDM16), 2.1 s (\href{https://www.nndc.bnl.gov/nsr/nsrlink.jsp?2016Ma12,B}{2016Ma12}, CDFT) and 87 s (\href{https://www.nndc.bnl.gov/nsr/nsrlink.jsp?2020Ne08,B}{2020Ne08}, QRPA with Skyrme EDF).\vspace{0.1cm}}&\\
&&&\parbox[t][0.3cm]{15.235281cm}{\raggedright configuration: \ensuremath{\nu} p\ensuremath{_{\textnormal{1/2}}^{\textnormal{$-$1}}} from systematics of N=125 isotones in the region; shell model predictions.\vspace{0.1cm}}&\\
\end{longtable}
\clearpage
%ADOPTED LEVELS
\section[\ensuremath{^{201}_{\ 77}}Ir\ensuremath{_{124}^{~}}]{ }
\vspace{-30pt}
\setcounter{chappage}{1}
\subsection[\hspace{-0.2cm}Adopted Levels]{ }
\vspace{-20pt}
\vspace{0.3cm}
\hypertarget{IR1}{{\bf \small \underline{Adopted \hyperlink{201IR_LEVEL}{Levels}}}}\\
\vspace{4pt}
\vspace{8pt}
\parbox[b][0.3cm]{17.7cm}{\addtolength{\parindent}{-0.2in}Q(\ensuremath{\beta^-})=3900 {\it SY}; S(n)=6340 {\it SY}; S(p)=8580 {\it SY}; Q(\ensuremath{\alpha})=$-$1910 {\it SY}\hspace{0.2in}\href{https://www.nndc.bnl.gov/nsr/nsrlink.jsp?2021Wa16,B}{2021Wa16}}\\
\parbox[b][0.3cm]{17.7cm}{\addtolength{\parindent}{-0.2in}Estimated \textit{SY} uncertainties: 210 keV for Q(\ensuremath{\beta}\ensuremath{^{-}}), 280 keV for S(n), 360 keV for S(p) and 360 keV for Q(\ensuremath{\alpha}) (\href{https://www.nndc.bnl.gov/nsr/nsrlink.jsp?2021WA16,B}{2021WA16}).}\\

\vspace{12pt}
\hypertarget{201IR_LEVEL}{\underline{$^{201}$Ir Levels}}\\
\begin{longtable}[c]{ll}
\multicolumn{2}{c}{\underline{Cross Reference (XREF) Flags}}\\
 \\
\hyperlink{IR2}{\texttt{A }}& \ensuremath{^{\textnormal{9}}}Be(\ensuremath{^{\textnormal{208}}}Pb,X\ensuremath{\gamma})\\
\end{longtable}
\vspace{-0.5cm}
\begin{longtable}{ccccccc@{\extracolsep{\fill}}c}
\multicolumn{2}{c}{E(level)$^{}$}&J$^{\pi}$$^{}$&\multicolumn{2}{c}{T$_{1/2}$$^{}$}&XREF&Comments&\\[-.2cm]
\multicolumn{2}{c}{\hrulefill}&\hrulefill&\multicolumn{2}{c}{\hrulefill}&\hrulefill&\hrulefill&
\endfirsthead
\multicolumn{1}{r@{}}{0}&\multicolumn{1}{@{}l}{}&\multicolumn{1}{l}{(3/2\ensuremath{^{+}})}&\multicolumn{1}{r@{}}{21}&\multicolumn{1}{@{ }l}{s {\it 5}}&\multicolumn{1}{l}{\texttt{\hyperlink{IR2}{A}} }&\parbox[t][0.3cm]{11.952881cm}{\raggedright \%\ensuremath{\beta}\ensuremath{^{-}}=100\vspace{0.1cm}}&\\
&&&&&&\parbox[t][0.3cm]{11.952881cm}{\raggedright J\ensuremath{^{\pi}}: assuming spherical shape and systematics of neighboring Z=77 nuclei; shell model\vspace{0.1cm}}&\\
&&&&&&\parbox[t][0.3cm]{11.952881cm}{\raggedright {\ }{\ }{\ }predictions.\vspace{0.1cm}}&\\
&&&&&&\parbox[t][0.3cm]{11.952881cm}{\raggedright T\ensuremath{_{1/2}}: from \href{https://www.nndc.bnl.gov/nsr/nsrlink.jsp?2014Mo15,B}{2014Mo15}, using Monte Carlo analysis of ion-\ensuremath{\beta}\ensuremath{\gamma}(time) data.\vspace{0.1cm}}&\\
&&&&&&\parbox[t][0.3cm]{11.952881cm}{\raggedright configuration: \ensuremath{\pi} d\ensuremath{_{\textnormal{3/2}}^{\textnormal{$-$1}}} from systematics of known Z=77 isotopes; shell model\vspace{0.1cm}}&\\
&&&&&&\parbox[t][0.3cm]{11.952881cm}{\raggedright {\ }{\ }{\ }predictions.\vspace{0.1cm}}&\\
\multicolumn{1}{r@{}}{0}&\multicolumn{1}{@{}l}{+x}&&\multicolumn{1}{r@{}}{10}&\multicolumn{1}{@{.}l}{5 ns {\it 17}}&\multicolumn{1}{l}{\texttt{\hyperlink{IR2}{A}} }&\parbox[t][0.3cm]{11.952881cm}{\raggedright E(level): 439.6 keV, 452.0 keV and 680.9 keV \ensuremath{\gamma} rays observed below the isomer\vspace{0.1cm}}&\\
&&&&&&\parbox[t][0.3cm]{11.952881cm}{\raggedright {\ }{\ }{\ }(\href{https://www.nndc.bnl.gov/nsr/nsrlink.jsp?2011St21,B}{2011St21}), but the ordering is unknown. Given the reported \ensuremath{\gamma}-ray intensities in\vspace{0.1cm}}&\\
\end{longtable}
\clearpage
%9BE(208PB,XG)
\subsection[\hspace{-0.2cm}\ensuremath{^{\textnormal{9}}}Be(\ensuremath{^{\textnormal{208}}}Pb,X\ensuremath{\gamma})]{ }
\vspace{-27pt}
\vspace{0.3cm}
\hypertarget{IR2}{{\bf \small \underline{\ensuremath{^{\textnormal{9}}}Be(\ensuremath{^{\textnormal{208}}}Pb,X\ensuremath{\gamma})\hspace{0.2in}\href{https://www.nndc.bnl.gov/nsr/nsrlink.jsp?2011St21,B}{2011St21}}}}\\
\vspace{4pt}
\vspace{8pt}
\parbox[b][0.3cm]{17.7cm}{\addtolength{\parindent}{-0.2in}\href{https://www.nndc.bnl.gov/nsr/nsrlink.jsp?2011St21,B}{2011St21}: \ensuremath{^{\textnormal{201}}}Ir nuclide produced by in-flight fragmentation of 1 GeV/A \ensuremath{^{\textnormal{208}}}Pb beam at the GSI UNILAC and SIS-18 accelerator}\\
\parbox[b][0.3cm]{17.7cm}{complex. Target thickness=2.526 g/cm\ensuremath{^{\textnormal{2}}}, backed by a 0.223 g/cm\ensuremath{^{\textnormal{2}}} thick \ensuremath{^{\textnormal{93}}}Nb foil. Fragments identified by the Fragment Separator}\\
\parbox[b][0.3cm]{17.7cm}{(FRS), based on time of flight, B\ensuremath{\rho} and energy loss. The ions were slowed down in Al degraders and stopped in a plastic catcher.}\\
\parbox[b][0.3cm]{17.7cm}{The stopper was surrounded by the RISING \ensuremath{\gamma}-ray spectrometer. Measured: E\ensuremath{\gamma}, I\ensuremath{\gamma}, delayed \ensuremath{\gamma} rays, isomer lifetime. Others (same}\\
\parbox[b][0.3cm]{17.7cm}{authors): \href{https://www.nndc.bnl.gov/nsr/nsrlink.jsp?2009St16,B}{2009St16}, \href{https://www.nndc.bnl.gov/nsr/nsrlink.jsp?2008StZY,B}{2008StZY}.}\\
\vspace{12pt}
\underline{$^{201}$Ir Levels}\\
\begin{longtable}{cccccc@{\extracolsep{\fill}}c}
\multicolumn{2}{c}{E(level)$^{}$}&J$^{\pi}$$^{{\hyperlink{IR2LEVEL0}{\dagger}}}$&\multicolumn{2}{c}{T$_{1/2}$$^{{\hyperlink{IR2LEVEL0}{\dagger}}}$}&Comments&\\[-.2cm]
\multicolumn{2}{c}{\hrulefill}&\hrulefill&\multicolumn{2}{c}{\hrulefill}&\hrulefill&
\endfirsthead
\multicolumn{1}{r@{}}{0}&\multicolumn{1}{@{}l}{}&\multicolumn{1}{l}{(3/2\ensuremath{^{+}})}&\multicolumn{1}{r@{}}{21}&\multicolumn{1}{@{ }l}{s {\it 5}}&&\\
\multicolumn{1}{r@{}}{0}&\multicolumn{1}{@{}l}{+x}&&\multicolumn{1}{r@{}}{10}&\multicolumn{1}{@{.}l}{5 ns {\it 17}}&\parbox[t][0.3cm]{13.172501cm}{\raggedright T\ensuremath{_{1/2}}: from sum of 440,452,681\ensuremath{\gamma}(t) in \href{https://www.nndc.bnl.gov/nsr/nsrlink.jsp?2011St21,B}{2011St21}.\vspace{0.1cm}}&\\
&&&&&\parbox[t][0.3cm]{13.172501cm}{\raggedright Experimental isomeric state population ratio \ensuremath{\geq}3\% \textit{2} (\href{https://www.nndc.bnl.gov/nsr/nsrlink.jsp?2011St21,B}{2011St21}).\vspace{0.1cm}}&\\
\end{longtable}
\parbox[b][0.3cm]{17.7cm}{\makebox[1ex]{\ensuremath{^{\hypertarget{IR2LEVEL0}{\dagger}}}} From Adopted Levels, unless otherwise stated.}\\
\vspace{0.5cm}
\underline{$\gamma$($^{201}$Ir)}\\
\begin{longtable}{cccccc@{\extracolsep{\fill}}c}
\multicolumn{2}{c}{E\ensuremath{_{\gamma}}\ensuremath{^{\hyperlink{IR2GAMMA0}{\dagger}}}}&\multicolumn{2}{c}{I\ensuremath{_{\gamma}}\ensuremath{^{\hyperlink{IR2GAMMA0}{\dagger}}}}&\multicolumn{2}{c}{E\ensuremath{_{i}}(level)}&\\[-.2cm]
\multicolumn{2}{c}{\hrulefill}&\multicolumn{2}{c}{\hrulefill}&\multicolumn{2}{c}{\hrulefill}&
\endfirsthead
\multicolumn{1}{r@{}}{$^{x}$439}&\multicolumn{1}{@{.}l}{6\ensuremath{^{\hyperlink{IR2GAMMA1}{\ddagger}}}}&\multicolumn{1}{r@{}}{39}&\multicolumn{1}{@{ }l}{{\it 9}}&&&\\
\multicolumn{1}{r@{}}{$^{x}$452}&\multicolumn{1}{@{.}l}{0\ensuremath{^{\hyperlink{IR2GAMMA1}{\ddagger}}}}&\multicolumn{1}{r@{}}{51}&\multicolumn{1}{@{ }l}{{\it 9}}&&&\\
\multicolumn{1}{r@{}}{$^{x}$680}&\multicolumn{1}{@{.}l}{9\ensuremath{^{\hyperlink{IR2GAMMA1}{\ddagger}}}}&\multicolumn{1}{r@{}}{100}&\multicolumn{1}{@{ }l}{{\it 13}}&&&\\
\end{longtable}
\parbox[b][0.3cm]{17.7cm}{\makebox[1ex]{\ensuremath{^{\hypertarget{IR2GAMMA0}{\dagger}}}} From \href{https://www.nndc.bnl.gov/nsr/nsrlink.jsp?2011St21,B}{2011St21}.}\\
\parbox[b][0.3cm]{17.7cm}{\makebox[1ex]{\ensuremath{^{\hypertarget{IR2GAMMA1}{\ddagger}}}} observed below the 10.5-ns isomer, but the ordering is unknown.}\\
\parbox[b][0.3cm]{17.7cm}{\makebox[1ex]{\ensuremath{^{\hypertarget{IR2GAMMA2}{x}}}} \ensuremath{\gamma} ray not placed in level scheme.}\\
\vspace{0.5cm}
\clearpage
%ADOPTED LEVELS, GAMMAS
\section[\ensuremath{^{201}_{\ 78}}Pt\ensuremath{_{123}^{~}}]{ }
\vspace{-30pt}
\setcounter{chappage}{1}
\subsection[\hspace{-0.2cm}Adopted Levels, Gammas]{ }
\vspace{-20pt}
\vspace{0.3cm}
\hypertarget{PT3}{{\bf \small \underline{Adopted \hyperlink{201PT_LEVEL}{Levels}, \hyperlink{201PT_GAMMA}{Gammas}}}}\\
\vspace{4pt}
\vspace{8pt}
\parbox[b][0.3cm]{17.7cm}{\addtolength{\parindent}{-0.2in}Q(\ensuremath{\beta^-})=2660 {\it 50}; S(n)=5210 {\it 50}; S(p)=9460 {\it SY}; Q(\ensuremath{\alpha})=$-$1090 {\it SY}\hspace{0.2in}\href{https://www.nndc.bnl.gov/nsr/nsrlink.jsp?2021Wa16,B}{2021Wa16}}\\
\parbox[b][0.3cm]{17.7cm}{\addtolength{\parindent}{-0.2in}Estimated \textit{SY} uncertainties: 200 keV for S(p) and 210 keV for Q(\ensuremath{\alpha}) (\href{https://www.nndc.bnl.gov/nsr/nsrlink.jsp?2021WA16,B}{2021WA16}).}\\

\vspace{12pt}
\hypertarget{201PT_LEVEL}{\underline{$^{201}$Pt Levels}}\\
\begin{longtable}[c]{ll}
\multicolumn{2}{c}{\underline{Cross Reference (XREF) Flags}}\\
 \\
\hyperlink{IR4}{\texttt{A }}& \ensuremath{^{\textnormal{201}}}Ir \ensuremath{\beta}\ensuremath{^{-}} decay\\
\hyperlink{PT5}{\texttt{B }}& Be(\ensuremath{^{\textnormal{208}}}Pb,X\ensuremath{\gamma})\\
\end{longtable}
\vspace{-0.5cm}
\begin{longtable}{ccccccc@{\extracolsep{\fill}}c}
\multicolumn{2}{c}{E(level)$^{{\hyperlink{PT3LEVEL0}{\dagger}}}$}&J$^{\pi}$$^{{\hyperlink{PT3LEVEL3}{@}}}$&\multicolumn{2}{c}{T$_{1/2}$$^{}$}&XREF&Comments&\\[-.2cm]
\multicolumn{2}{c}{\hrulefill}&\hrulefill&\multicolumn{2}{c}{\hrulefill}&\hrulefill&\hrulefill&
\endfirsthead
\multicolumn{1}{r@{}}{0}&\multicolumn{1}{@{.}l}{0}&\multicolumn{1}{l}{(5/2\ensuremath{^{-}})}&\multicolumn{1}{r@{}}{2}&\multicolumn{1}{@{.}l}{46 min {\it 9}}&\multicolumn{1}{l}{\texttt{\hyperlink{IR4}{A}\hyperlink{PT5}{B}} }&\parbox[t][0.3cm]{11.26808cm}{\raggedright \%\ensuremath{\beta}\ensuremath{^{-}}=100\vspace{0.1cm}}&\\
&&&&&&\parbox[t][0.3cm]{11.26808cm}{\raggedright J\ensuremath{^{\pi}}: From systematics in neighboring N=123 nuclides and shell model predictions.\vspace{0.1cm}}&\\
&&&&&&\parbox[t][0.3cm]{11.26808cm}{\raggedright T\ensuremath{_{1/2}}: Weighted average of 2.5 m \textit{1} (\href{https://www.nndc.bnl.gov/nsr/nsrlink.jsp?1963Go06,B}{1963Go06}) and 2.3 m \textit{2} (\href{https://www.nndc.bnl.gov/nsr/nsrlink.jsp?1962Fa06,B}{1962Fa06}).\vspace{0.1cm}}&\\
&&&&&&\parbox[t][0.3cm]{11.26808cm}{\raggedright configuration: \ensuremath{\nu} f\ensuremath{_{\textnormal{5/2}}^{\textnormal{$-$1}}}.\vspace{0.1cm}}&\\
\multicolumn{1}{r@{}}{332}&\multicolumn{1}{@{.}l}{6\ensuremath{^{{\hyperlink{PT3LEVEL2}{\#}}}} {\it 17}}&&&&\multicolumn{1}{l}{\texttt{\hyperlink{IR4}{A}\ } }&&\\
\multicolumn{1}{r@{}}{373}&\multicolumn{1}{@{.}l}{9\ensuremath{^{{\hyperlink{PT3LEVEL1}{\ddagger}}}} {\it 5}}&\multicolumn{1}{l}{(9/2\ensuremath{^{-}})}&&&\multicolumn{1}{l}{\texttt{\ \hyperlink{PT5}{B}} }&\parbox[t][0.3cm]{11.26808cm}{\raggedright configuration: \ensuremath{\nu} (f\ensuremath{_{\textnormal{5/2}}^{\textnormal{$-$1}}})\ensuremath{\otimes}2\ensuremath{^{\textnormal{+}}}. The assignment is tentative.\vspace{0.1cm}}&\\
\multicolumn{1}{r@{}}{389}&\multicolumn{1}{@{.}l}{4\ensuremath{^{{\hyperlink{PT3LEVEL2}{\#}}}} {\it 13}}&&&&\multicolumn{1}{l}{\texttt{\hyperlink{IR4}{A}\ } }&&\\
\multicolumn{1}{r@{}}{655}&\multicolumn{1}{@{.}l}{0\ensuremath{^{{\hyperlink{PT3LEVEL2}{\#}}}} {\it 20}}&&&&\multicolumn{1}{l}{\texttt{\hyperlink{IR4}{A}\ } }&&\\
\end{longtable}
\parbox[b][0.3cm]{17.7cm}{\makebox[1ex]{\ensuremath{^{\hypertarget{PT3LEVEL0}{\dagger}}}} From a least-squares fit to E\ensuremath{\gamma}.}\\
\parbox[b][0.3cm]{17.7cm}{\makebox[1ex]{\ensuremath{^{\hypertarget{PT3LEVEL1}{\ddagger}}}} Level populated only in Be(\ensuremath{^{\textnormal{208}}}Pb,x\ensuremath{\gamma}).}\\
\parbox[b][0.3cm]{17.7cm}{\makebox[1ex]{\ensuremath{^{\hypertarget{PT3LEVEL2}{\#}}}} Level populated only in \ensuremath{^{\textnormal{201}}}Ir \ensuremath{\beta}\ensuremath{^{-}} decay (\ensuremath{J^{\pi}}=(3/2\ensuremath{^{+}})).}\\
\parbox[b][0.3cm]{17.7cm}{\makebox[1ex]{\ensuremath{^{\hypertarget{PT3LEVEL3}{@}}}} From Be(\ensuremath{^{\textnormal{208}}}Pb,x\ensuremath{\gamma}), unless otherwise stated.}\\
\vspace{0.5cm}
\hypertarget{201PT_GAMMA}{\underline{$\gamma$($^{201}$Pt)}}\\
\begin{longtable}{ccccccccc@{}c@{\extracolsep{\fill}}c}
\multicolumn{2}{c}{E\ensuremath{_{i}}(level)}&J\ensuremath{^{\pi}_{i}}&\multicolumn{2}{c}{E\ensuremath{_{\gamma}}\ensuremath{^{\hyperlink{PT3GAMMA0}{\dagger}}}}&\multicolumn{2}{c}{I\ensuremath{_{\gamma}}\ensuremath{^{\hyperlink{PT3GAMMA0}{\dagger}}}}&\multicolumn{2}{c}{E\ensuremath{_{f}}}&J\ensuremath{^{\pi}_{f}}&\\[-.2cm]
\multicolumn{2}{c}{\hrulefill}&\hrulefill&\multicolumn{2}{c}{\hrulefill}&\multicolumn{2}{c}{\hrulefill}&\multicolumn{2}{c}{\hrulefill}&\hrulefill&
\endfirsthead
\multicolumn{1}{r@{}}{332}&\multicolumn{1}{@{.}l}{6}&&\multicolumn{1}{r@{}}{332}&\multicolumn{1}{@{.}l}{6\ensuremath{^{\hyperlink{PT3GAMMA1}{\ddagger}}} {\it 17}}&\multicolumn{1}{r@{}}{100}&\multicolumn{1}{@{}l}{\ensuremath{^{\hyperlink{PT3GAMMA1}{\ddagger}}}}&\multicolumn{1}{r@{}}{0}&\multicolumn{1}{@{.}l}{0}&\multicolumn{1}{@{}l}{(5/2\ensuremath{^{-}})}&\\
\multicolumn{1}{r@{}}{373}&\multicolumn{1}{@{.}l}{9}&\multicolumn{1}{l}{(9/2\ensuremath{^{-}})}&\multicolumn{1}{r@{}}{373}&\multicolumn{1}{@{.}l}{9 {\it 5}}&\multicolumn{1}{r@{}}{100}&\multicolumn{1}{@{}l}{}&\multicolumn{1}{r@{}}{0}&\multicolumn{1}{@{.}l}{0}&\multicolumn{1}{@{}l}{(5/2\ensuremath{^{-}})}&\\
\multicolumn{1}{r@{}}{389}&\multicolumn{1}{@{.}l}{4}&&\multicolumn{1}{r@{}}{389}&\multicolumn{1}{@{.}l}{4\ensuremath{^{\hyperlink{PT3GAMMA1}{\ddagger}}} {\it 15}}&\multicolumn{1}{r@{}}{100}&\multicolumn{1}{@{}l}{\ensuremath{^{\hyperlink{PT3GAMMA1}{\ddagger}}}}&\multicolumn{1}{r@{}}{0}&\multicolumn{1}{@{.}l}{0}&\multicolumn{1}{@{}l}{(5/2\ensuremath{^{-}})}&\\
\multicolumn{1}{r@{}}{655}&\multicolumn{1}{@{.}l}{0}&&\multicolumn{1}{r@{}}{655}&\multicolumn{1}{@{}l}{\ensuremath{^{\hyperlink{PT3GAMMA1}{\ddagger}}} {\it 2}}&\multicolumn{1}{r@{}}{100}&\multicolumn{1}{@{}l}{\ensuremath{^{\hyperlink{PT3GAMMA1}{\ddagger}}}}&\multicolumn{1}{r@{}}{0}&\multicolumn{1}{@{.}l}{0}&\multicolumn{1}{@{}l}{(5/2\ensuremath{^{-}})}&\\
\multicolumn{1}{r@{}}{710}&\multicolumn{1}{@{.}l}{0}&&\multicolumn{1}{r@{}}{710}&\multicolumn{1}{@{.}l}{1\ensuremath{^{\hyperlink{PT3GAMMA1}{\ddagger}}} {\it 15}}&\multicolumn{1}{r@{}}{100}&\multicolumn{1}{@{}l}{\ensuremath{^{\hyperlink{PT3GAMMA1}{\ddagger}}}}&\multicolumn{1}{r@{}}{0}&\multicolumn{1}{@{.}l}{0}&\multicolumn{1}{@{}l}{(5/2\ensuremath{^{-}})}&\\
\end{longtable}
\begin{textblock}{29}(0,27.3)
Continued on next page (footnotes at end of table)
\end{textblock}
\clearpage
\begin{longtable}{cccccccc@{}c@{\extracolsep{\fill}}c}
\\[-.4cm]
\multicolumn{10}{c}{{\bf \small \underline{Adopted \hyperlink{201PT_LEVEL}{Levels}, \hyperlink{201PT_GAMMA}{Gammas} (continued)}}}\\
\multicolumn{10}{c}{~}\\
\multicolumn{10}{c}{\underline{$\gamma$($^{201}$Pt) (continued)}}\\
\multicolumn{10}{c}{~~~}\\
\multicolumn{2}{c}{E\ensuremath{_{i}}(level)}&\multicolumn{2}{c}{E\ensuremath{_{\gamma}}\ensuremath{^{\hyperlink{PT3GAMMA0}{\dagger}}}}&\multicolumn{2}{c}{I\ensuremath{_{\gamma}}\ensuremath{^{\hyperlink{PT3GAMMA0}{\dagger}}}}&\multicolumn{2}{c}{E\ensuremath{_{f}}}&J\ensuremath{^{\pi}_{f}}&\\[-.2cm]
\multicolumn{2}{c}{\hrulefill}&\multicolumn{2}{c}{\hrulefill}&\multicolumn{2}{c}{\hrulefill}&\multicolumn{2}{c}{\hrulefill}&\hrulefill&
\endhead
\multicolumn{1}{r@{}}{1706}&\multicolumn{1}{@{.}l}{3}&\multicolumn{1}{r@{}}{996}&\multicolumn{1}{@{.}l}{5\ensuremath{^{\hyperlink{PT3GAMMA1}{\ddagger}}} {\it 16}}&\multicolumn{1}{r@{}}{42}&\multicolumn{1}{@{}l}{\ensuremath{^{\hyperlink{PT3GAMMA1}{\ddagger}}} {\it 10}}&\multicolumn{1}{r@{}}{710}&\multicolumn{1}{@{.}l}{0 }&&\\
&&\multicolumn{1}{r@{}}{1317}&\multicolumn{1}{@{}l}{\ensuremath{^{\hyperlink{PT3GAMMA1}{\ddagger}}} {\it 2}}&\multicolumn{1}{r@{}}{92}&\multicolumn{1}{@{}l}{\ensuremath{^{\hyperlink{PT3GAMMA1}{\ddagger}}} {\it 17}}&\multicolumn{1}{r@{}}{389}&\multicolumn{1}{@{.}l}{4 }&&\\
&&\multicolumn{1}{r@{}}{1706}&\multicolumn{1}{@{}l}{\ensuremath{^{\hyperlink{PT3GAMMA1}{\ddagger}}} {\it 2}}&\multicolumn{1}{r@{}}{100}&\multicolumn{1}{@{}l}{\ensuremath{^{\hyperlink{PT3GAMMA1}{\ddagger}}} {\it 17}}&\multicolumn{1}{r@{}}{0}&\multicolumn{1}{@{.}l}{0}&\multicolumn{1}{@{}l}{(5/2\ensuremath{^{-}})}&\\
\end{longtable}
\parbox[b][0.3cm]{17.7cm}{\makebox[1ex]{\ensuremath{^{\hypertarget{PT3GAMMA0}{\dagger}}}} From Be(\ensuremath{^{\textnormal{208}}}Pb,x\ensuremath{\gamma}), unless otherwise stated.}\\
\parbox[b][0.3cm]{17.7cm}{\makebox[1ex]{\ensuremath{^{\hypertarget{PT3GAMMA1}{\ddagger}}}} From \ensuremath{^{\textnormal{201}}}Ir \ensuremath{\beta}\ensuremath{^{-}} decay.}\\
\vspace{0.5cm}
\clearpage
\begin{figure}[h]
\begin{center}
\includegraphics{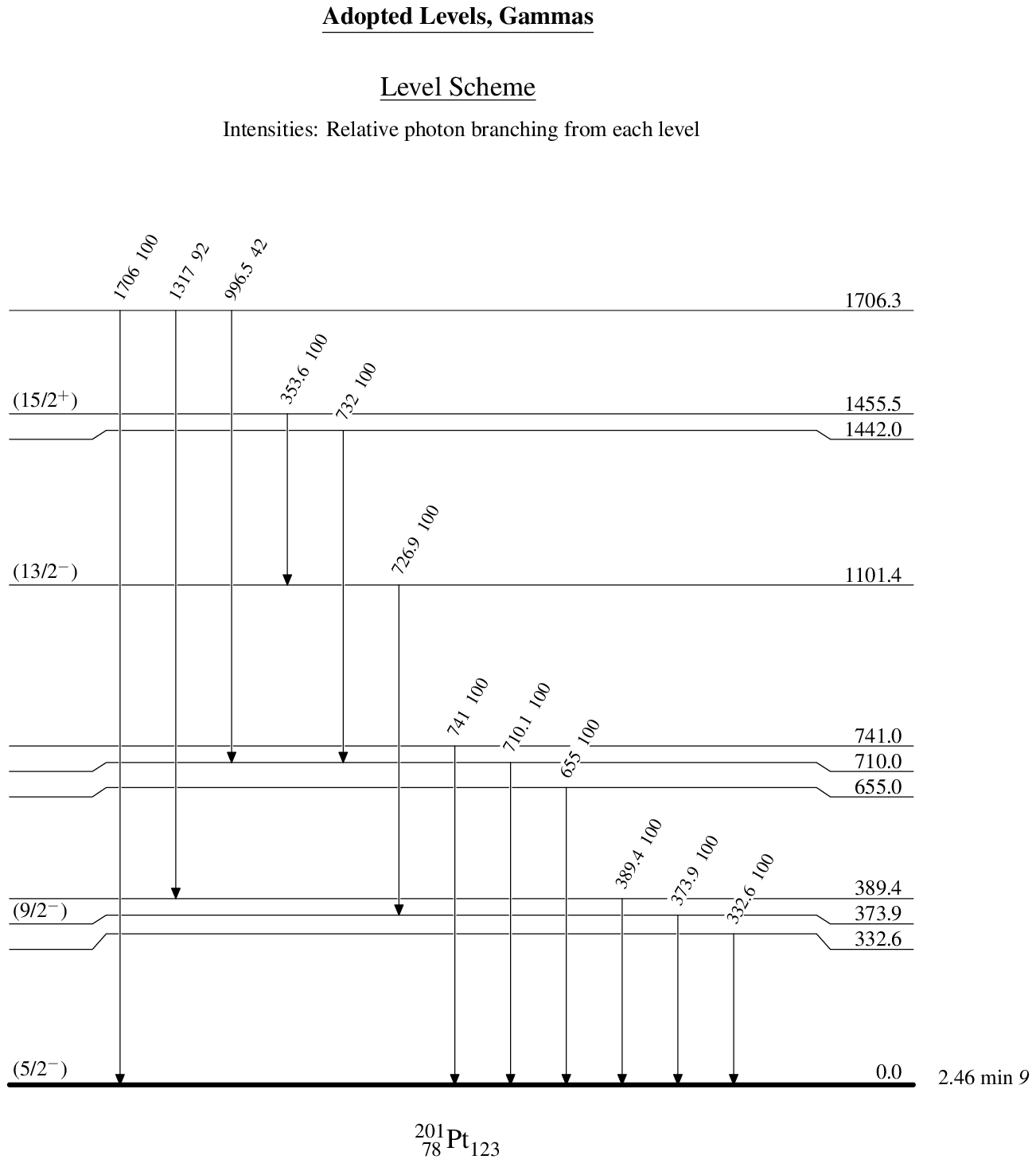}\\
\end{center}
\end{figure}
\clearpage
%201IR B- DECAY
\subsection[\hspace{-0.2cm}\ensuremath{^{\textnormal{201}}}Ir \ensuremath{\beta}\ensuremath{^{-}} decay]{ }
\vspace{-27pt}
\vspace{0.3cm}
\hypertarget{IR4}{{\bf \small \underline{\ensuremath{^{\textnormal{201}}}Ir \ensuremath{\beta}\ensuremath{^{-}} decay\hspace{0.2in}\href{https://www.nndc.bnl.gov/nsr/nsrlink.jsp?2013Mo20,B}{2013Mo20}}}}\\
\vspace{4pt}
\vspace{8pt}
\parbox[b][0.3cm]{17.7cm}{\addtolength{\parindent}{-0.2in}Parent: $^{201}$Ir: E=0.0; J$^{\pi}$=(3/2\ensuremath{^{+}}); T$_{1/2}$=21 s {\it 5}; Q(\ensuremath{\beta}\ensuremath{^{-}})=3900 {\it SY}; \%\ensuremath{\beta}\ensuremath{^{-}} decay=100.0

}\\
\parbox[b][0.3cm]{17.7cm}{\addtolength{\parindent}{-0.2in}\href{https://www.nndc.bnl.gov/nsr/nsrlink.jsp?2013Mo20,B}{2013Mo20}: \ensuremath{^{\textnormal{201}}}Ir produced in cold fragmentation reactions with E=1 GeV/A \ensuremath{^{\textnormal{208}}}Pb beam impinging a 2.5 g/cm\ensuremath{^{\textnormal{2}}} thick Be target.}\\
\parbox[b][0.3cm]{17.7cm}{The beam was provided by SIS-18 synchrotron at GSI facility. Residues were separated using Fragment Separator. Measured E\ensuremath{\gamma},}\\
\parbox[b][0.3cm]{17.7cm}{I\ensuremath{\gamma}, \ensuremath{\gamma}\ensuremath{\gamma}-coin, \ensuremath{\beta}\ensuremath{\gamma}-coin, fragment-\ensuremath{\gamma} correlated event using RISING array of 15 cluster detectors. Others (same authors): \href{https://www.nndc.bnl.gov/nsr/nsrlink.jsp?2011MoZP,B}{2011MoZP}.}\\
\vspace{12pt}
\underline{$^{201}$Pt Levels}\\
\begin{longtable}{cccccc@{\extracolsep{\fill}}c}
\multicolumn{2}{c}{E(level)$^{{\hyperlink{PT4LEVEL0}{\dagger}}}$}&J$^{\pi}$$^{}$&\multicolumn{2}{c}{T$_{1/2}$$^{}$}&Comments&\\[-.2cm]
\multicolumn{2}{c}{\hrulefill}&\hrulefill&\multicolumn{2}{c}{\hrulefill}&\hrulefill&
\endfirsthead
\multicolumn{1}{r@{}}{0}&\multicolumn{1}{@{.}l}{0}&\multicolumn{1}{l}{(5/2\ensuremath{^{-}})}&\multicolumn{1}{r@{}}{2}&\multicolumn{1}{@{.}l}{46 min {\it 9}}&\parbox[t][0.3cm]{13.226cm}{\raggedright \%\ensuremath{\beta}\ensuremath{^{-}}=100\vspace{0.1cm}}&\\
&&&&&\parbox[t][0.3cm]{13.226cm}{\raggedright J\ensuremath{^{\pi}},T\ensuremath{_{1/2}}: From Adopted Levels.\vspace{0.1cm}}&\\
\multicolumn{1}{r@{}}{332}&\multicolumn{1}{@{.}l}{6 {\it 17}}&&&&&\\
\multicolumn{1}{r@{}}{389}&\multicolumn{1}{@{.}l}{4 {\it 13}}&&&&&\\
\multicolumn{1}{r@{}}{655}&\multicolumn{1}{@{.}l}{0 {\it 20}}&&&&&\\
\multicolumn{1}{r@{}}{710}&\multicolumn{1}{@{.}l}{0 {\it 13}}&&&&&\\
\multicolumn{1}{r@{}}{741}&\multicolumn{1}{@{.}l}{0 {\it 20}}&&&&&\\
\multicolumn{1}{r@{}}{1442}&\multicolumn{1}{@{.}l}{0 {\it 24}}&&&&&\\
\multicolumn{1}{r@{}}{1706}&\multicolumn{1}{@{.}l}{3 {\it 13}}&&&&&\\
\end{longtable}
\parbox[b][0.3cm]{17.7cm}{\makebox[1ex]{\ensuremath{^{\hypertarget{PT4LEVEL0}{\dagger}}}} From least-squares fit to E\ensuremath{\gamma}.}\\
\vspace{0.5cm}
\underline{$\gamma$($^{201}$Pt)}\\
\vspace{0.34cm}
\parbox[b][0.3cm]{17.7cm}{\addtolength{\parindent}{-0.254cm}I\ensuremath{\gamma} normalization: Since the ground state to ground state \ensuremath{\beta}-decay feeding is not known, the decay scheme is uncertain and no \ensuremath{\beta}\ensuremath{^{-}}}\\
\parbox[b][0.3cm]{17.7cm}{intensities and log ft values are provided.}\\
\vspace{0.34cm}
\begin{longtable}{cccccccc@{}c@{\extracolsep{\fill}}c}
\multicolumn{2}{c}{E\ensuremath{_{\gamma}}\ensuremath{^{\hyperlink{IR4GAMMA0}{\dagger}}}}&\multicolumn{2}{c}{I\ensuremath{_{\gamma}}\ensuremath{^{\hyperlink{IR4GAMMA0}{\dagger}}}}&\multicolumn{2}{c}{E\ensuremath{_{i}}(level)}&\multicolumn{2}{c}{E\ensuremath{_{f}}}&J\ensuremath{^{\pi}_{f}}&\\[-.2cm]
\multicolumn{2}{c}{\hrulefill}&\multicolumn{2}{c}{\hrulefill}&\multicolumn{2}{c}{\hrulefill}&\multicolumn{2}{c}{\hrulefill}&\hrulefill&
\endfirsthead
\multicolumn{1}{r@{}}{332}&\multicolumn{1}{@{.}l}{6 {\it 17}}&\multicolumn{1}{r@{}}{31}&\multicolumn{1}{@{ }l}{{\it 3}}&\multicolumn{1}{r@{}}{332}&\multicolumn{1}{@{.}l}{6}&\multicolumn{1}{r@{}}{0}&\multicolumn{1}{@{.}l}{0}&\multicolumn{1}{@{}l}{(5/2\ensuremath{^{-}})}&\\
\multicolumn{1}{r@{}}{389}&\multicolumn{1}{@{.}l}{4 {\it 15}}&\multicolumn{1}{r@{}}{26}&\multicolumn{1}{@{ }l}{{\it 3}}&\multicolumn{1}{r@{}}{389}&\multicolumn{1}{@{.}l}{4}&\multicolumn{1}{r@{}}{0}&\multicolumn{1}{@{.}l}{0}&\multicolumn{1}{@{}l}{(5/2\ensuremath{^{-}})}&\\
\multicolumn{1}{r@{}}{655}&\multicolumn{1}{@{ }l}{{\it 2}}&\multicolumn{1}{r@{}}{100}&\multicolumn{1}{@{ }l}{{\it 8}}&\multicolumn{1}{r@{}}{655}&\multicolumn{1}{@{.}l}{0}&\multicolumn{1}{r@{}}{0}&\multicolumn{1}{@{.}l}{0}&\multicolumn{1}{@{}l}{(5/2\ensuremath{^{-}})}&\\
\multicolumn{1}{r@{}}{710}&\multicolumn{1}{@{.}l}{1 {\it 15}}&\multicolumn{1}{r@{}}{46}&\multicolumn{1}{@{ }l}{{\it 4}}&\multicolumn{1}{r@{}}{710}&\multicolumn{1}{@{.}l}{0}&\multicolumn{1}{r@{}}{0}&\multicolumn{1}{@{.}l}{0}&\multicolumn{1}{@{}l}{(5/2\ensuremath{^{-}})}&\\
\multicolumn{1}{r@{}}{732}&\multicolumn{1}{@{ }l}{{\it 2}}&\multicolumn{1}{r@{}}{28}&\multicolumn{1}{@{ }l}{{\it 3}}&\multicolumn{1}{r@{}}{1442}&\multicolumn{1}{@{.}l}{0}&\multicolumn{1}{r@{}}{710}&\multicolumn{1}{@{.}l}{0 }&&\\
\multicolumn{1}{r@{}}{741}&\multicolumn{1}{@{ }l}{{\it 2}}&\multicolumn{1}{r@{}}{22}&\multicolumn{1}{@{ }l}{{\it 3}}&\multicolumn{1}{r@{}}{741}&\multicolumn{1}{@{.}l}{0}&\multicolumn{1}{r@{}}{0}&\multicolumn{1}{@{.}l}{0}&\multicolumn{1}{@{}l}{(5/2\ensuremath{^{-}})}&\\
\multicolumn{1}{r@{}}{996}&\multicolumn{1}{@{.}l}{5 {\it 16}}&\multicolumn{1}{r@{}}{5}&\multicolumn{1}{@{.}l}{0 {\it 12}}&\multicolumn{1}{r@{}}{1706}&\multicolumn{1}{@{.}l}{3}&\multicolumn{1}{r@{}}{710}&\multicolumn{1}{@{.}l}{0 }&&\\
\end{longtable}
\parbox[b][0.3cm]{17.7cm}{\makebox[1ex]{\ensuremath{^{\hypertarget{IR4GAMMA0}{\dagger}}}} From \href{https://www.nndc.bnl.gov/nsr/nsrlink.jsp?2013Mo20,B}{2013Mo20}.}\\
\vspace{0.5cm}
\clearpage
\begin{figure}[h]
\begin{center}
\includegraphics{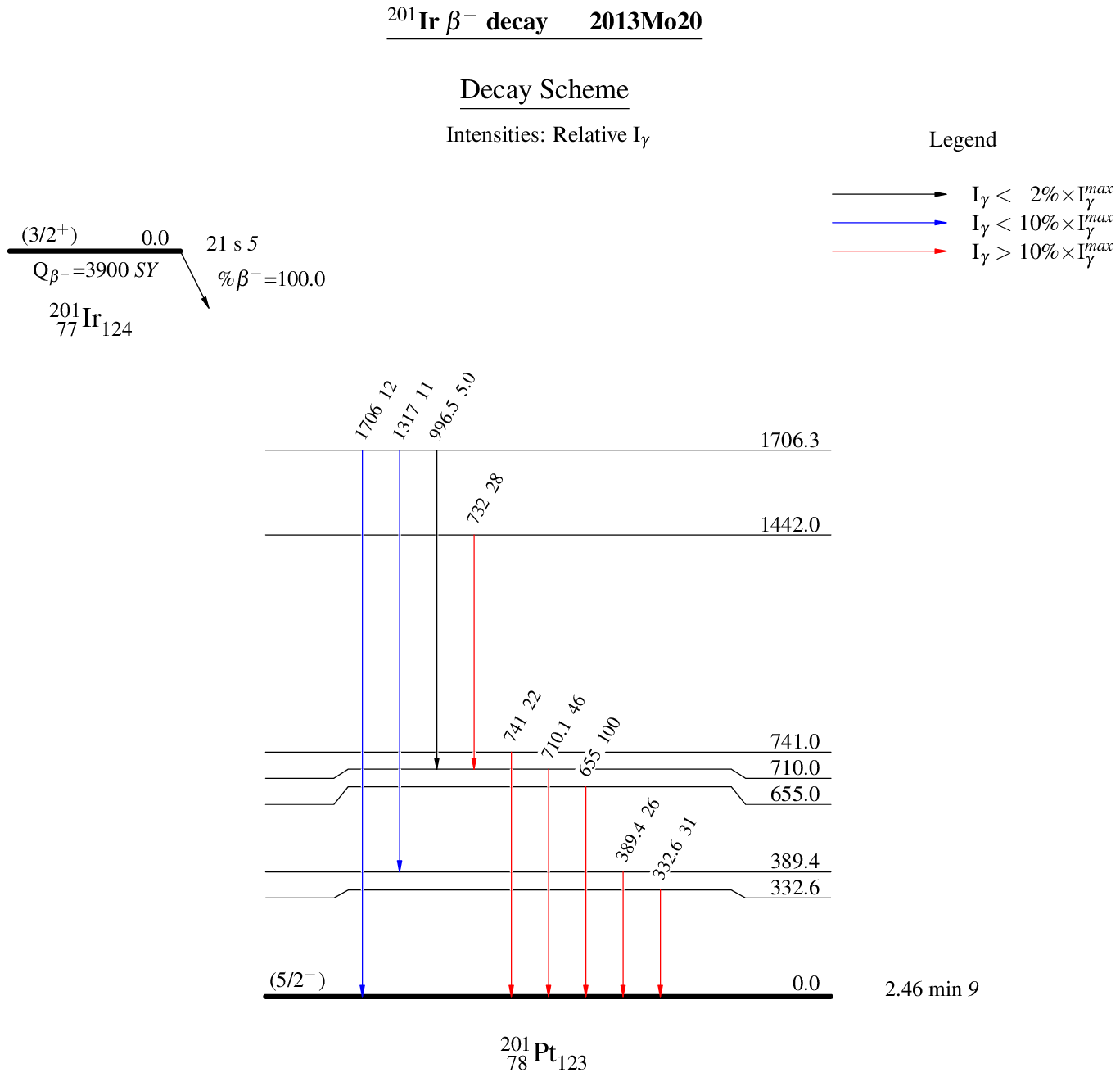}\\
\end{center}
\end{figure}
\clearpage
%BE(208PB,XG)
\subsection[\hspace{-0.2cm}Be(\ensuremath{^{\textnormal{208}}}Pb,X\ensuremath{\gamma})]{ }
\vspace{-27pt}
\vspace{0.3cm}
\hypertarget{PT5}{{\bf \small \underline{Be(\ensuremath{^{\textnormal{208}}}Pb,X\ensuremath{\gamma})\hspace{0.2in}\href{https://www.nndc.bnl.gov/nsr/nsrlink.jsp?2005Ca02,B}{2005Ca02},\href{https://www.nndc.bnl.gov/nsr/nsrlink.jsp?2011St21,B}{2011St21}}}}\\
\vspace{4pt}
\vspace{8pt}
\parbox[b][0.3cm]{17.7cm}{\addtolength{\parindent}{-0.2in}\href{https://www.nndc.bnl.gov/nsr/nsrlink.jsp?2005Ca02,B}{2005Ca02}: Projectile fragmentation of \ensuremath{^{\textnormal{208}}}Pb beam at 1 GeV/A on a 1.6 g/cm\ensuremath{^{\textnormal{2}}} Be target. Fragment Recoil Separator at GSI.}\\
\parbox[b][0.3cm]{17.7cm}{Measured E\ensuremath{\gamma}, I\ensuremath{\gamma}, \ensuremath{\gamma}\ensuremath{\gamma}, \ensuremath{\gamma}\ensuremath{\gamma}(t) using four ``Clover'' type Ge detectors (providing 16 independent Ge crystals). Others (same}\\
\parbox[b][0.3cm]{17.7cm}{collaboration): \href{https://www.nndc.bnl.gov/nsr/nsrlink.jsp?2001Ca13,B}{2001Ca13}, \href{https://www.nndc.bnl.gov/nsr/nsrlink.jsp?2002Po15,B}{2002Po15}, \href{https://www.nndc.bnl.gov/nsr/nsrlink.jsp?2003Po14,B}{2003Po14}, \href{https://www.nndc.bnl.gov/nsr/nsrlink.jsp?2001MaZV,B}{2001MaZV}, \href{https://www.nndc.bnl.gov/nsr/nsrlink.jsp?2000PoZY,B}{2000PoZY}.}\\
\parbox[b][0.3cm]{17.7cm}{\addtolength{\parindent}{-0.2in}\href{https://www.nndc.bnl.gov/nsr/nsrlink.jsp?2011St21,B}{2011St21}: in-flight fragmentation of \ensuremath{^{\textnormal{208}}}Pb beam at 1 GeV/A on a 2.526 g/cm\ensuremath{^{\textnormal{2}}} Be target, backed by 0.223 g/cm\ensuremath{^{\textnormal{2}}}-thick \ensuremath{^{\textnormal{93}}}Nb foil.}\\
\parbox[b][0.3cm]{17.7cm}{Fragment Recoil Separator at GSI. Measured E\ensuremath{\gamma}, I\ensuremath{\gamma}, \ensuremath{\gamma}\ensuremath{\gamma}, \ensuremath{\gamma}\ensuremath{\gamma}(t) using the RISING \ensuremath{\gamma}-ray spectrometer. Other: \href{https://www.nndc.bnl.gov/nsr/nsrlink.jsp?2008StZY,B}{2008StZY}.}\\
\vspace{12pt}
\underline{$^{201}$Pt Levels}\\
\begin{longtable}{cccccc@{\extracolsep{\fill}}c}
\multicolumn{2}{c}{E(level)$^{{\hyperlink{PT5LEVEL0}{\dagger}}}$}&J$^{\pi}$$^{{\hyperlink{PT5LEVEL1}{\ddagger}}}$&\multicolumn{2}{c}{T$_{1/2}$$^{}$}&Comments&\\[-.2cm]
\multicolumn{2}{c}{\hrulefill}&\hrulefill&\multicolumn{2}{c}{\hrulefill}&\hrulefill&
\endfirsthead
\multicolumn{1}{r@{}}{0}&\multicolumn{1}{@{.}l}{0}&\multicolumn{1}{l}{(5/2\ensuremath{^{-}})}&\multicolumn{1}{r@{}}{2}&\multicolumn{1}{@{.}l}{46 min {\it 9}}&\parbox[t][0.3cm]{12.718821cm}{\raggedright J\ensuremath{^{\pi}},T\ensuremath{_{1/2}}: From Adopted Levels.\vspace{0.1cm}}&\\
&&&&&\parbox[t][0.3cm]{12.718821cm}{\raggedright configuration: \ensuremath{\nu} f\ensuremath{_{\textnormal{5/2}}^{\textnormal{$-$1}}}.\vspace{0.1cm}}&\\
\multicolumn{1}{r@{}}{374}&\multicolumn{1}{@{.}l}{2 {\it 10}}&\multicolumn{1}{l}{(9/2\ensuremath{^{-}})}&&&\parbox[t][0.3cm]{12.718821cm}{\raggedright configuration: \ensuremath{\nu} (f\ensuremath{_{\textnormal{5/2}}^{\textnormal{$-$1}}})\ensuremath{\otimes}2\ensuremath{^{\textnormal{+}}}. The assignment is tentative.\vspace{0.1cm}}&\\
\multicolumn{1}{r@{}}{1101}&\multicolumn{1}{@{.}l}{4 {\it 15}}&\multicolumn{1}{l}{(13/2\ensuremath{^{-}})}&&&\parbox[t][0.3cm]{12.718821cm}{\raggedright configuration: \ensuremath{\nu} (f\ensuremath{_{\textnormal{5/2}}^{\textnormal{$-$1}}})\ensuremath{\otimes}4\ensuremath{^{\textnormal{+}}}. The assignment is tentative.\vspace{0.1cm}}&\\
\multicolumn{1}{r@{}}{1455}&\multicolumn{1}{@{.}l}{5 {\it 18}}&\multicolumn{1}{l}{(15/2\ensuremath{^{+}})}&&&&\\
\multicolumn{1}{r@{}}{1455}&\multicolumn{1}{@{.}l}{5+x}&\multicolumn{1}{l}{(19/2\ensuremath{^{+}})}&\multicolumn{1}{r@{}}{18}&\multicolumn{1}{@{.}l}{4 ns {\it 13}}&\parbox[t][0.3cm]{12.718821cm}{\raggedright E(level): x\ensuremath{<}90 keV in both \href{https://www.nndc.bnl.gov/nsr/nsrlink.jsp?2005Ca02,B}{2005Ca02} and \href{https://www.nndc.bnl.gov/nsr/nsrlink.jsp?2011St21,B}{2011St21}. Direct \ensuremath{\gamma}-ray decay to the 1455.5 keV\vspace{0.1cm}}&\\
&&&&&\parbox[t][0.3cm]{12.718821cm}{\raggedright {\ }{\ }{\ }level was not observed.\vspace{0.1cm}}&\\
\end{longtable}
\parbox[b][0.3cm]{17.7cm}{\makebox[1ex]{\ensuremath{^{\hypertarget{PT5LEVEL0}{\dagger}}}} From a least-squares fit to E\ensuremath{\gamma} in \href{https://www.nndc.bnl.gov/nsr/nsrlink.jsp?2011St21,B}{2011St21}.}\\
\parbox[b][0.3cm]{17.7cm}{\makebox[1ex]{\ensuremath{^{\hypertarget{PT5LEVEL1}{\ddagger}}}} From \href{https://www.nndc.bnl.gov/nsr/nsrlink.jsp?2005Ca02,B}{2005Ca02}, based on systematics and shell model predictions. Different \ensuremath{J^{\pi}} values are proposed in \href{https://www.nndc.bnl.gov/nsr/nsrlink.jsp?2011St21,B}{2011St21}, where the}\\
\parbox[b][0.3cm]{17.7cm}{{\ }{\ }observed \ensuremath{\gamma}-ray cascade is placed above an expected, but not yet observed, \ensuremath{J^{\pi}}=13/2\ensuremath{^{+}} state. This alternative was also discussed in}\\
\parbox[b][0.3cm]{17.7cm}{{\ }{\ }\href{https://www.nndc.bnl.gov/nsr/nsrlink.jsp?2005Ca02,B}{2005Ca02}, but was not adopted due to the resulting large measured isomeric ratio, which would exceed the sharp-cutoff model}\\
\parbox[b][0.3cm]{17.7cm}{{\ }{\ }value.}\\
\vspace{0.5cm}
\underline{$\gamma$($^{201}$Pt)}\\
\begin{longtable}{ccccccccc@{}cc@{\extracolsep{\fill}}c}
\multicolumn{2}{c}{E\ensuremath{_{\gamma}}\ensuremath{^{\hyperlink{PT5GAMMA0}{\dagger}}}}&\multicolumn{2}{c}{I\ensuremath{_{\gamma}}\ensuremath{^{\hyperlink{PT5GAMMA0}{\dagger}}}}&\multicolumn{2}{c}{E\ensuremath{_{i}}(level)}&J\ensuremath{^{\pi}_{i}}&\multicolumn{2}{c}{E\ensuremath{_{f}}}&J\ensuremath{^{\pi}_{f}}&Comments&\\[-.2cm]
\multicolumn{2}{c}{\hrulefill}&\multicolumn{2}{c}{\hrulefill}&\multicolumn{2}{c}{\hrulefill}&\hrulefill&\multicolumn{2}{c}{\hrulefill}&\hrulefill&\hrulefill&
\endfirsthead
\multicolumn{1}{r@{}}{353}&\multicolumn{1}{@{.}l}{6 {\it 5}}&\multicolumn{1}{r@{}}{76}&\multicolumn{1}{@{ }l}{{\it 6}}&\multicolumn{1}{r@{}}{1455}&\multicolumn{1}{@{.}l}{5}&\multicolumn{1}{l}{(15/2\ensuremath{^{+}})}&\multicolumn{1}{r@{}}{1101}&\multicolumn{1}{@{.}l}{4 }&\multicolumn{1}{@{}l}{(13/2\ensuremath{^{-}})}&\parbox[t][0.3cm]{9.80624cm}{\raggedright E\ensuremath{_{\gamma}},I\ensuremath{_{\gamma}}: E\ensuremath{\gamma}=354.1 keV \textit{2}, I\ensuremath{\gamma}=95 \textit{6} in \href{https://www.nndc.bnl.gov/nsr/nsrlink.jsp?2005Ca02,B}{2005Ca02}.\vspace{0.1cm}}&\\
\multicolumn{1}{r@{}}{373}&\multicolumn{1}{@{.}l}{9 {\it 5}}&\multicolumn{1}{r@{}}{80}&\multicolumn{1}{@{ }l}{{\it 5}}&\multicolumn{1}{r@{}}{374}&\multicolumn{1}{@{.}l}{2}&\multicolumn{1}{l}{(9/2\ensuremath{^{-}})}&\multicolumn{1}{r@{}}{0}&\multicolumn{1}{@{.}l}{0}&\multicolumn{1}{@{}l}{(5/2\ensuremath{^{-}})}&\parbox[t][0.3cm]{9.80624cm}{\raggedright E\ensuremath{_{\gamma}},I\ensuremath{_{\gamma}}: E\ensuremath{\gamma}=374.4 keV \textit{2}, I\ensuremath{\gamma}=100 \textit{6} in \href{https://www.nndc.bnl.gov/nsr/nsrlink.jsp?2005Ca02,B}{2005Ca02}.\vspace{0.1cm}}&\\
\multicolumn{1}{r@{}}{726}&\multicolumn{1}{@{.}l}{9 {\it 5}}&\multicolumn{1}{r@{}}{100}&\multicolumn{1}{@{ }l}{{\it 6}}&\multicolumn{1}{r@{}}{1101}&\multicolumn{1}{@{.}l}{4}&\multicolumn{1}{l}{(13/2\ensuremath{^{-}})}&\multicolumn{1}{r@{}}{374}&\multicolumn{1}{@{.}l}{2 }&\multicolumn{1}{@{}l}{(9/2\ensuremath{^{-}})}&\parbox[t][0.3cm]{9.80624cm}{\raggedright E\ensuremath{_{\gamma}},I\ensuremath{_{\gamma}}: E\ensuremath{\gamma}=727.2 keV \textit{2}, I\ensuremath{\gamma}=90 \textit{7} in \href{https://www.nndc.bnl.gov/nsr/nsrlink.jsp?2005Ca02,B}{2005Ca02}.\vspace{0.1cm}}&\\
\end{longtable}
\parbox[b][0.3cm]{17.7cm}{\makebox[1ex]{\ensuremath{^{\hypertarget{PT5GAMMA0}{\dagger}}}} From \href{https://www.nndc.bnl.gov/nsr/nsrlink.jsp?2011St21,B}{2011St21}. \ensuremath{\Delta}E\ensuremath{\gamma} were estimated by the evaluator.}\\
\vspace{0.5cm}
\clearpage
\begin{figure}[h]
\begin{center}
\includegraphics{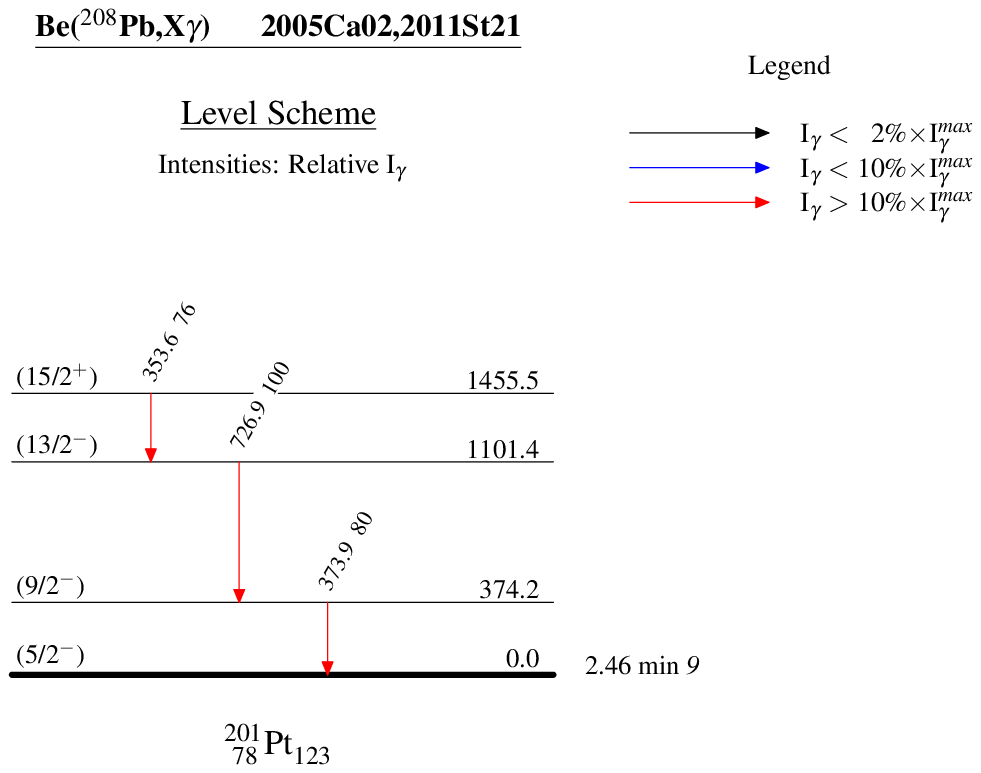}\\
\end{center}
\end{figure}
\clearpage
%ADOPTED LEVELS, GAMMAS
\section[\ensuremath{^{201}_{\ 79}}Au\ensuremath{_{122}^{~}}]{ }
\vspace{-30pt}
\setcounter{chappage}{1}
\subsection[\hspace{-0.2cm}Adopted Levels, Gammas]{ }
\vspace{-20pt}
\vspace{0.3cm}
\hypertarget{AU6}{{\bf \small \underline{Adopted \hyperlink{201AU_LEVEL}{Levels}, \hyperlink{201AU_GAMMA}{Gammas}}}}\\
\vspace{4pt}
\vspace{8pt}
\parbox[b][0.3cm]{17.7cm}{\addtolength{\parindent}{-0.2in}Q(\ensuremath{\beta^-})=1262 {\it 3}; S(n)=7232 {\it 27}; S(p)=7090 {\it 20}; Q(\ensuremath{\alpha})=$-$561 {\it 20}\hspace{0.2in}\href{https://www.nndc.bnl.gov/nsr/nsrlink.jsp?2021Wa16,B}{2021Wa16}}\\

\vspace{12pt}
\hypertarget{201AU_LEVEL}{\underline{$^{201}$Au Levels}}\\
\begin{longtable}[c]{ll}
\multicolumn{2}{c}{\underline{Cross Reference (XREF) Flags}}\\
 \\
\hyperlink{PT7}{\texttt{A }}& \ensuremath{^{\textnormal{201}}}Pt \ensuremath{\beta}\ensuremath{^{-}} decay\\
\hyperlink{AU8}{\texttt{B }}& \ensuremath{^{\textnormal{202}}}Hg(t,\ensuremath{\alpha})\\
\hyperlink{AU9}{\texttt{C }}& \ensuremath{^{\textnormal{9}}}Be(\ensuremath{^{\textnormal{208}}}Pb,X\ensuremath{\gamma})\\
\end{longtable}
\vspace{-0.5cm}
\begin{longtable}{ccccccc@{\extracolsep{\fill}}c}
\multicolumn{2}{c}{E(level)$^{{\hyperlink{AU6LEVEL0}{\dagger}}}$}&J$^{\pi}$$^{{\hyperlink{AU6LEVEL1}{\ddagger}}}$&\multicolumn{2}{c}{T$_{1/2}$$^{}$}&XREF&Comments&\\[-.2cm]
\multicolumn{2}{c}{\hrulefill}&\hrulefill&\multicolumn{2}{c}{\hrulefill}&\hrulefill&\hrulefill&
\endfirsthead
\multicolumn{1}{r@{}}{0}&\multicolumn{1}{@{}l}{\ensuremath{^{{\hyperlink{AU6LEVEL2}{\#}}}}}&\multicolumn{1}{l}{3/2\ensuremath{^{+}}}&\multicolumn{1}{r@{}}{26}&\multicolumn{1}{@{.}l}{0 min {\it 8}}&\multicolumn{1}{l}{\texttt{\hyperlink{PT7}{A}\hyperlink{AU8}{B}\hyperlink{AU9}{C}} }&\parbox[t][0.3cm]{10.673161cm}{\raggedright \%\ensuremath{\beta}\ensuremath{^{-}}=100\vspace{0.1cm}}&\\
&&&&&&\parbox[t][0.3cm]{10.673161cm}{\raggedright T\ensuremath{_{1/2}}: Weighted average of 26 min \textit{1} (\href{https://www.nndc.bnl.gov/nsr/nsrlink.jsp?1957Er24,B}{1957Er24}), 27 min \textit{2} (\href{https://www.nndc.bnl.gov/nsr/nsrlink.jsp?1962Fa06,B}{1962Fa06}), 22\vspace{0.1cm}}&\\
&&&&&&\parbox[t][0.3cm]{10.673161cm}{\raggedright {\ }{\ }{\ }min \textit{3} (\href{https://www.nndc.bnl.gov/nsr/nsrlink.jsp?1962Eu01,B}{1962Eu01}) and 26.4 min \textit{16} (\href{https://www.nndc.bnl.gov/nsr/nsrlink.jsp?1972Pa24,B}{1972Pa24}). Other: 26 min (\href{https://www.nndc.bnl.gov/nsr/nsrlink.jsp?1952Bu80,B}{1952Bu80}).\vspace{0.1cm}}&\\
\multicolumn{1}{r@{}}{101}&\multicolumn{1}{@{}l}{\ensuremath{^{{\hyperlink{AU6LEVEL3}{@}}}} {\it 5}}&\multicolumn{1}{l}{1/2\ensuremath{^{+}}}&&&\multicolumn{1}{l}{\texttt{\hyperlink{PT7}{A}\hyperlink{AU8}{B}\ } }&\parbox[t][0.3cm]{10.673161cm}{\raggedright XREF: A(70?).\vspace{0.1cm}}&\\
\multicolumn{1}{r@{}}{359}&\multicolumn{1}{@{ }l}{{\it 5}}&\multicolumn{1}{l}{3/2\ensuremath{^{+}}}&&&\multicolumn{1}{l}{\texttt{\hyperlink{PT7}{A}\hyperlink{AU8}{B}\ } }&\parbox[t][0.3cm]{10.673161cm}{\raggedright XREF: A(230?).\vspace{0.1cm}}&\\
\multicolumn{1}{r@{}}{553}&\multicolumn{1}{@{.}l}{0 {\it 10}}&\multicolumn{1}{l}{(7/2\ensuremath{^{+}})}&&&\multicolumn{1}{l}{\texttt{\ \hyperlink{AU8}{B}\hyperlink{AU9}{C}} }&\parbox[t][0.3cm]{10.673161cm}{\raggedright XREF: B(549?).\vspace{0.1cm}}&\\
\end{longtable}
\parbox[b][0.3cm]{17.7cm}{\makebox[1ex]{\ensuremath{^{\hypertarget{AU6LEVEL0}{\dagger}}}} From \ensuremath{^{\textnormal{202}}}Hg(t,\ensuremath{\alpha}) (\href{https://www.nndc.bnl.gov/nsr/nsrlink.jsp?1981Fl05,B}{1981Fl05}) and from a least-squares fir to E\ensuremath{\gamma} when \ensuremath{\gamma}-ray data are available.}\\
\parbox[b][0.3cm]{17.7cm}{\makebox[1ex]{\ensuremath{^{\hypertarget{AU6LEVEL1}{\ddagger}}}} From \ensuremath{\sigma}(\ensuremath{\theta},pol) and analyzing powers in \ensuremath{^{\textnormal{202}}}Hg(t,\ensuremath{\alpha}) (\href{https://www.nndc.bnl.gov/nsr/nsrlink.jsp?1981Fl05,B}{1981Fl05}), unless otherwise stated.}\\
\parbox[b][0.3cm]{17.7cm}{\makebox[1ex]{\ensuremath{^{\hypertarget{AU6LEVEL2}{\#}}}} Main configuration=\ensuremath{\pi} d\ensuremath{_{\textnormal{3/2}}^{\textnormal{$-$1}}}.}\\
\parbox[b][0.3cm]{17.7cm}{\makebox[1ex]{\ensuremath{^{\hypertarget{AU6LEVEL3}{@}}}} Main configuration=\ensuremath{\pi} s\ensuremath{_{\textnormal{1/2}}^{\textnormal{$-$1}}}.}\\
\parbox[b][0.3cm]{17.7cm}{\makebox[1ex]{\ensuremath{^{\hypertarget{AU6LEVEL4}{\&}}}} Main configuration=\ensuremath{\pi} h\ensuremath{_{\textnormal{11/2}}^{\textnormal{$-$1}}}.}\\
\parbox[b][0.3cm]{17.7cm}{\makebox[1ex]{\ensuremath{^{\hypertarget{AU6LEVEL5}{a}}}} Probable configuration=\ensuremath{\pi} (h\ensuremath{_{\textnormal{11/2}}^{\textnormal{$-$1}}})\ensuremath{\otimes}2\ensuremath{^{\textnormal{+}}}.}\\
\vspace{0.5cm}
\clearpage
\vspace{0.3cm}
\vspace*{-0.5cm}
{\bf \small \underline{Adopted \hyperlink{201AU_LEVEL}{Levels}, \hyperlink{201AU_GAMMA}{Gammas} (continued)}}\\
\vspace{0.3cm}
\hypertarget{201AU_GAMMA}{\underline{$\gamma$($^{201}$Au)}}\\
\begin{longtable}{ccccccccc@{}ccccc@{\extracolsep{\fill}}c}
\multicolumn{2}{c}{E\ensuremath{_{i}}(level)}&J\ensuremath{^{\pi}_{i}}&\multicolumn{2}{c}{E\ensuremath{_{\gamma}}\ensuremath{^{\hyperlink{AU6GAMMA0}{\dagger}}}}&\multicolumn{2}{c}{I\ensuremath{_{\gamma}}\ensuremath{^{\hyperlink{AU6GAMMA0}{\dagger}}}}&\multicolumn{2}{c}{E\ensuremath{_{f}}}&J\ensuremath{^{\pi}_{f}}&Mult.&\multicolumn{2}{c}{\ensuremath{\alpha}\ensuremath{^{\hyperlink{AU6GAMMA3}{@}}}}&Comments&\\[-.2cm]
\multicolumn{2}{c}{\hrulefill}&\hrulefill&\multicolumn{2}{c}{\hrulefill}&\multicolumn{2}{c}{\hrulefill}&\multicolumn{2}{c}{\hrulefill}&\hrulefill&\hrulefill&\multicolumn{2}{c}{\hrulefill}&\hrulefill&
\endfirsthead
\multicolumn{1}{r@{}}{553}&\multicolumn{1}{@{.}l}{0}&\multicolumn{1}{l}{(7/2\ensuremath{^{+}})}&\multicolumn{1}{r@{}}{553}&\multicolumn{1}{@{ }l}{{\it 1}}&\multicolumn{1}{r@{}}{100}&\multicolumn{1}{@{}l}{}&\multicolumn{1}{r@{}}{0}&\multicolumn{1}{@{}l}{}&\multicolumn{1}{@{}l}{3/2\ensuremath{^{+}}}&&\multicolumn{1}{r@{}}{}&\multicolumn{1}{@{}l}{}&&\\
\multicolumn{1}{r@{}}{594}&\multicolumn{1}{@{}l}{}&\multicolumn{1}{l}{11/2\ensuremath{^{-}}}&\multicolumn{1}{r@{}}{(41}&\multicolumn{1}{@{ }l}{{\it 5})}&\multicolumn{1}{r@{}}{100}&\multicolumn{1}{@{}l}{}&\multicolumn{1}{r@{}}{553}&\multicolumn{1}{@{.}l}{0 }&\multicolumn{1}{@{}l}{(7/2\ensuremath{^{+}})}&\multicolumn{1}{l}{[M2]}&\multicolumn{1}{r@{}}{1.1\ensuremath{\times10^{3}}}&\multicolumn{1}{@{ }l}{{\it 9}}&\parbox[t][0.3cm]{6.192141cm}{\raggedright B(M2)(W.u.)=0.021 \textit{+79{\textminus}17}\vspace{0.1cm}}&\\
&&&&&&&&&&&&&\parbox[t][0.3cm]{6.192141cm}{\raggedright E\ensuremath{_{\gamma}}: From level energy differences; not\vspace{0.1cm}}&\\
&&&&&&&&&&&&&\parbox[t][0.3cm]{6.192141cm}{\raggedright {\ }{\ }{\ }observed in \href{https://www.nndc.bnl.gov/nsr/nsrlink.jsp?2011St21,B}{2011St21}.\vspace{0.1cm}}&\\
\multicolumn{1}{r@{}}{1232}&\multicolumn{1}{@{}l}{}&\multicolumn{1}{l}{(15/2\ensuremath{^{-}})}&\multicolumn{1}{r@{}}{638}&\multicolumn{1}{@{.}l}{0\ensuremath{^{\hyperlink{AU6GAMMA1}{\ddagger}}} {\it 5}}&\multicolumn{1}{r@{}}{100}&\multicolumn{1}{@{}l}{}&\multicolumn{1}{r@{}}{594}&\multicolumn{1}{@{ }l}{}&\multicolumn{1}{@{}l}{11/2\ensuremath{^{-}}}&&&&&\\
\end{longtable}
\parbox[b][0.3cm]{17.7cm}{\makebox[1ex]{\ensuremath{^{\hypertarget{AU6GAMMA0}{\dagger}}}} From \ensuremath{^{\textnormal{9}}}Be(\ensuremath{^{\textnormal{208}}}Pb,X\ensuremath{\gamma}) (\href{https://www.nndc.bnl.gov/nsr/nsrlink.jsp?2011St21,B}{2011St21}), unless otherwise stated.}\\
\parbox[b][0.3cm]{17.7cm}{\makebox[1ex]{\ensuremath{^{\hypertarget{AU6GAMMA1}{\ddagger}}}} This \ensuremath{\gamma} ray shows the 5.0{\textminus}\ensuremath{\mu}s isomer half-life, but the ordering and placement in the decay scheme are not experimentally known.}\\
\parbox[b][0.3cm]{17.7cm}{{\ }{\ }The interpretation is made by the evaluator.}\\
\parbox[b][0.3cm]{17.7cm}{\makebox[1ex]{\ensuremath{^{\hypertarget{AU6GAMMA2}{\#}}}} From \ensuremath{^{\textnormal{201}}}Pt \ensuremath{\beta}\ensuremath{^{-}} decay (\href{https://www.nndc.bnl.gov/nsr/nsrlink.jsp?1963Go06,B}{1963Go06}).}\\
\parbox[b][0.3cm]{17.7cm}{\makebox[1ex]{\ensuremath{^{\hypertarget{AU6GAMMA3}{@}}}} Total theoretical internal conversion coefficients, calculated using the BrIcc code (\href{https://www.nndc.bnl.gov/nsr/nsrlink.jsp?2008Ki07,B}{2008Ki07}) with Frozen orbital approximation}\\
\parbox[b][0.3cm]{17.7cm}{{\ }{\ }based on \ensuremath{\gamma}-ray energies, assigned multipolarities, and mixing ratios, unless otherwise specified.}\\
\vspace{0.5cm}
\begin{figure}[h]
\begin{center}
\includegraphics{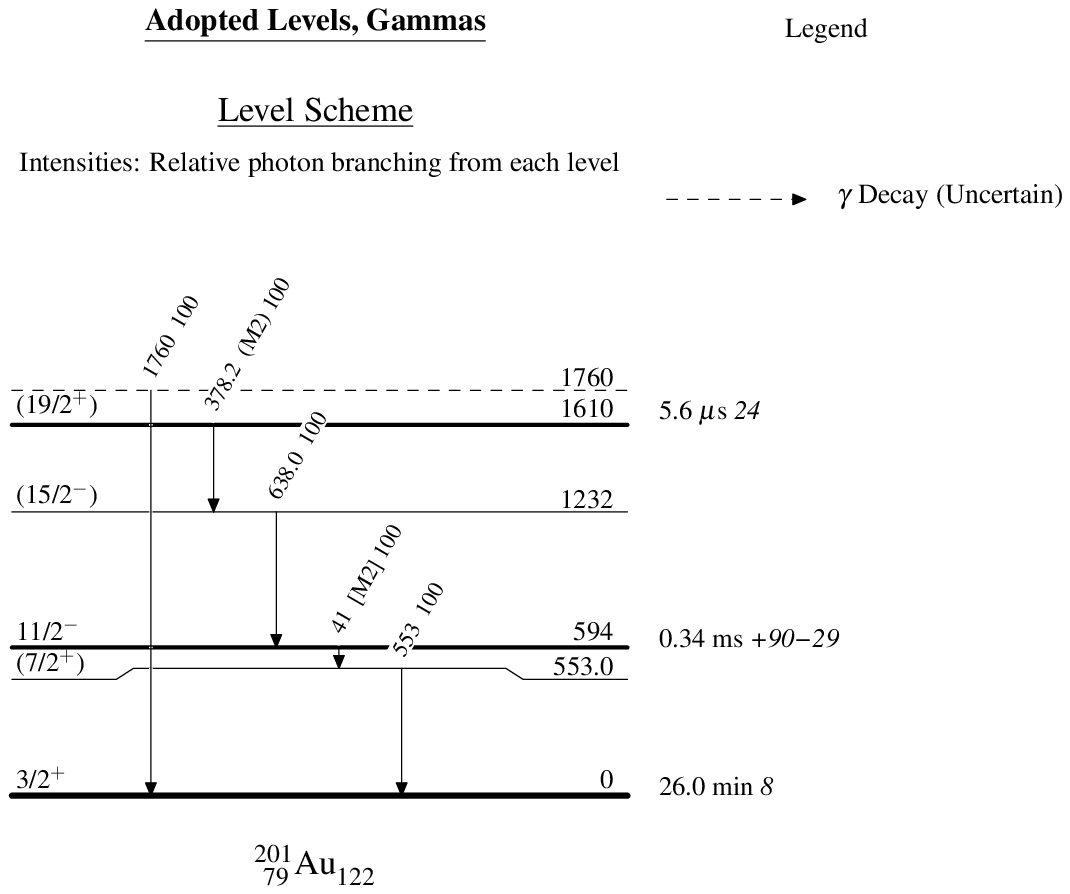}\\
\end{center}
\end{figure}
\clearpage
%201PT B- DECAY
\subsection[\hspace{-0.2cm}\ensuremath{^{\textnormal{201}}}Pt \ensuremath{\beta}\ensuremath{^{-}} decay]{ }
\vspace{-27pt}
\vspace{0.3cm}
\hypertarget{PT7}{{\bf \small \underline{\ensuremath{^{\textnormal{201}}}Pt \ensuremath{\beta}\ensuremath{^{-}} decay\hspace{0.2in}\href{https://www.nndc.bnl.gov/nsr/nsrlink.jsp?1963Go06,B}{1963Go06}}}}\\
\vspace{4pt}
\vspace{8pt}
\parbox[b][0.3cm]{17.7cm}{\addtolength{\parindent}{-0.2in}Parent: $^{201}$Pt: E=0; J$^{\pi}$=(5/2\ensuremath{^{-}}); T$_{1/2}$=2.46 min {\it 9}; Q(\ensuremath{\beta}\ensuremath{^{-}})=2660 {\it 50}; \%\ensuremath{\beta}\ensuremath{^{-}} decay=100.0

}\\
\parbox[b][0.3cm]{17.7cm}{\addtolength{\parindent}{-0.2in}\ensuremath{^{201}}Pt-Source produced using the \ensuremath{^{\textnormal{204}}}Hg(n,\ensuremath{\alpha}) reaction (\ensuremath{\sigma}=2.5 mb relative to that for \ensuremath{^{\textnormal{58}}}Ni(n,p) in \href{https://www.nndc.bnl.gov/nsr/nsrlink.jsp?1963Go06,B}{1963Go06}), following radiochemical}\\
\parbox[b][0.3cm]{17.7cm}{separation. Detectors: NaI(Tl) for gammas and scintillation spectrometer for \ensuremath{\beta}. Measured: \ensuremath{\beta}, \ensuremath{\gamma}, \ensuremath{\beta}\ensuremath{\gamma} coin.}\\
\vspace{12pt}
\underline{$^{201}$Au Levels}\\
\begin{longtable}{ccccc@{\extracolsep{\fill}}c}
\multicolumn{2}{c}{E(level)$^{{\hyperlink{AU7LEVEL0}{\dagger}}}$}&J$^{\pi}$$^{{\hyperlink{AU7LEVEL1}{\ddagger}}}$&\multicolumn{2}{c}{T$_{1/2}$$^{{\hyperlink{AU7LEVEL1}{\ddagger}}}$}&\\[-.2cm]
\multicolumn{2}{c}{\hrulefill}&\hrulefill&\multicolumn{2}{c}{\hrulefill}&
\endfirsthead
\multicolumn{1}{r@{}}{0}&\multicolumn{1}{@{}l}{}&\multicolumn{1}{l}{3/2\ensuremath{^{+}}}&\multicolumn{1}{r@{}}{26}&\multicolumn{1}{@{.}l}{0 min {\it 8}}&\\
\multicolumn{1}{r@{}}{70?}&\multicolumn{1}{@{ }l}{{\it 5}}&&&&\\
\multicolumn{1}{r@{}}{230?}&\multicolumn{1}{@{ }l}{{\it 20}}&&&&\\
\multicolumn{1}{r@{}}{1760?}&\multicolumn{1}{@{ }l}{{\it 20}}&&&&\\
\end{longtable}
\parbox[b][0.3cm]{17.7cm}{\makebox[1ex]{\ensuremath{^{\hypertarget{AU7LEVEL0}{\dagger}}}} From the observed E\ensuremath{\gamma} in \ensuremath{\beta}\ensuremath{\gamma}-coin data.}\\
\parbox[b][0.3cm]{17.7cm}{\makebox[1ex]{\ensuremath{^{\hypertarget{AU7LEVEL1}{\ddagger}}}} From Adopted Levels.}\\
\vspace{0.5cm}

\underline{\ensuremath{\beta^-} radiations}\\
\begin{longtable}{ccccc@{\extracolsep{\fill}}c}
\multicolumn{2}{c}{E(decay)$$}&\multicolumn{2}{c}{E(level)}&Comments&\\[-.2cm]
\multicolumn{2}{c}{\hrulefill}&\multicolumn{2}{c}{\hrulefill}&\hrulefill&
\endfirsthead
\multicolumn{1}{r@{}}{(9.0\ensuremath{\times10^{2}}}&\multicolumn{1}{@{ }l}{{\it 5})}&\multicolumn{1}{r@{}}{1760?}&\multicolumn{1}{@{}l}{}&&\\
\multicolumn{1}{r@{}}{(2.43\ensuremath{\times10^{3}}}&\multicolumn{1}{@{ }l}{{\it 5})}&\multicolumn{1}{r@{}}{230?}&\multicolumn{1}{@{}l}{}&&\\
\multicolumn{1}{r@{}}{2660}&\multicolumn{1}{@{ }l}{{\it 50}}&\multicolumn{1}{r@{}}{0}&\multicolumn{1}{@{}l}{}&\parbox[t][0.3cm]{14.21664cm}{\raggedright E(decay): From \href{https://www.nndc.bnl.gov/nsr/nsrlink.jsp?1963Go06,B}{1963Go06}.\vspace{0.1cm}}&\\
\end{longtable}
\underline{$\gamma$($^{201}$Au)}\\
\begin{longtable}{cccccc@{}cc@{\extracolsep{\fill}}c}
\multicolumn{2}{c}{E\ensuremath{_{\gamma}}\ensuremath{^{\hyperlink{PT7GAMMA0}{\dagger}}}}&\multicolumn{2}{c}{E\ensuremath{_{i}}(level)}&\multicolumn{2}{c}{E\ensuremath{_{f}}}&J\ensuremath{^{\pi}_{f}}&Comments&\\[-.2cm]
\multicolumn{2}{c}{\hrulefill}&\multicolumn{2}{c}{\hrulefill}&\multicolumn{2}{c}{\hrulefill}&\hrulefill&\hrulefill&
\endfirsthead
\multicolumn{1}{r@{}}{70}&\multicolumn{1}{@{}l}{\ensuremath{^{\hyperlink{PT7GAMMA1}{\ddagger}\hyperlink{PT7GAMMA2}{\#}}} {\it 5}}&\multicolumn{1}{r@{}}{70?}&\multicolumn{1}{@{}l}{}&\multicolumn{1}{r@{}}{0}&\multicolumn{1}{@{}l}{}&\multicolumn{1}{@{}l}{3/2\ensuremath{^{+}}}&\parbox[t][0.3cm]{12.88748cm}{\raggedright E\ensuremath{_{\gamma}}: Energy close to Au K\ensuremath{\alpha}{} x ray.\vspace{0.1cm}}&\\
\multicolumn{1}{r@{}}{150}&\multicolumn{1}{@{}l}{\ensuremath{^{\hyperlink{PT7GAMMA1}{\ddagger}}} {\it 10}}&\multicolumn{1}{r@{}}{230?}&\multicolumn{1}{@{}l}{}&\multicolumn{1}{r@{}}{70?}&\multicolumn{1}{@{ }l}{}&&&\\
\multicolumn{1}{r@{}}{230}&\multicolumn{1}{@{}l}{\ensuremath{^{\hyperlink{PT7GAMMA1}{\ddagger}}} {\it 20}}&\multicolumn{1}{r@{}}{230?}&\multicolumn{1}{@{}l}{}&\multicolumn{1}{r@{}}{0}&\multicolumn{1}{@{}l}{}&\multicolumn{1}{@{}l}{3/2\ensuremath{^{+}}}&&\\
\multicolumn{1}{r@{}}{1760}&\multicolumn{1}{@{ }l}{{\it 20}}&\multicolumn{1}{r@{}}{1760?}&\multicolumn{1}{@{}l}{}&\multicolumn{1}{r@{}}{0}&\multicolumn{1}{@{}l}{}&\multicolumn{1}{@{}l}{3/2\ensuremath{^{+}}}&\parbox[t][0.3cm]{12.88748cm}{\raggedright E\ensuremath{_{\gamma}}: No coin with other gammas were observed. It shows the parent (\ensuremath{^{\textnormal{201}}}Pt) half-life of 2.5 m\vspace{0.1cm}}&\\
&&&&&&&\parbox[t][0.3cm]{12.88748cm}{\raggedright {\ }{\ }{\ }\textit{1}.\vspace{0.1cm}}&\\
\end{longtable}
\parbox[b][0.3cm]{17.7cm}{\makebox[1ex]{\ensuremath{^{\hypertarget{PT7GAMMA0}{\dagger}}}} From \href{https://www.nndc.bnl.gov/nsr/nsrlink.jsp?1963Go06,B}{1963Go06}.}\\
\parbox[b][0.3cm]{17.7cm}{\makebox[1ex]{\ensuremath{^{\hypertarget{PT7GAMMA1}{\ddagger}}}} Observed only in the \ensuremath{\beta}\ensuremath{\gamma}-coin data when E(\ensuremath{\beta})\ensuremath{>}1600 keV.}\\
\parbox[b][0.3cm]{17.7cm}{\makebox[1ex]{\ensuremath{^{\hypertarget{PT7GAMMA2}{\#}}}} Placement of transition in the level scheme is uncertain.}\\
\vspace{0.5cm}
\clearpage
\begin{figure}[h]
\begin{center}
\includegraphics{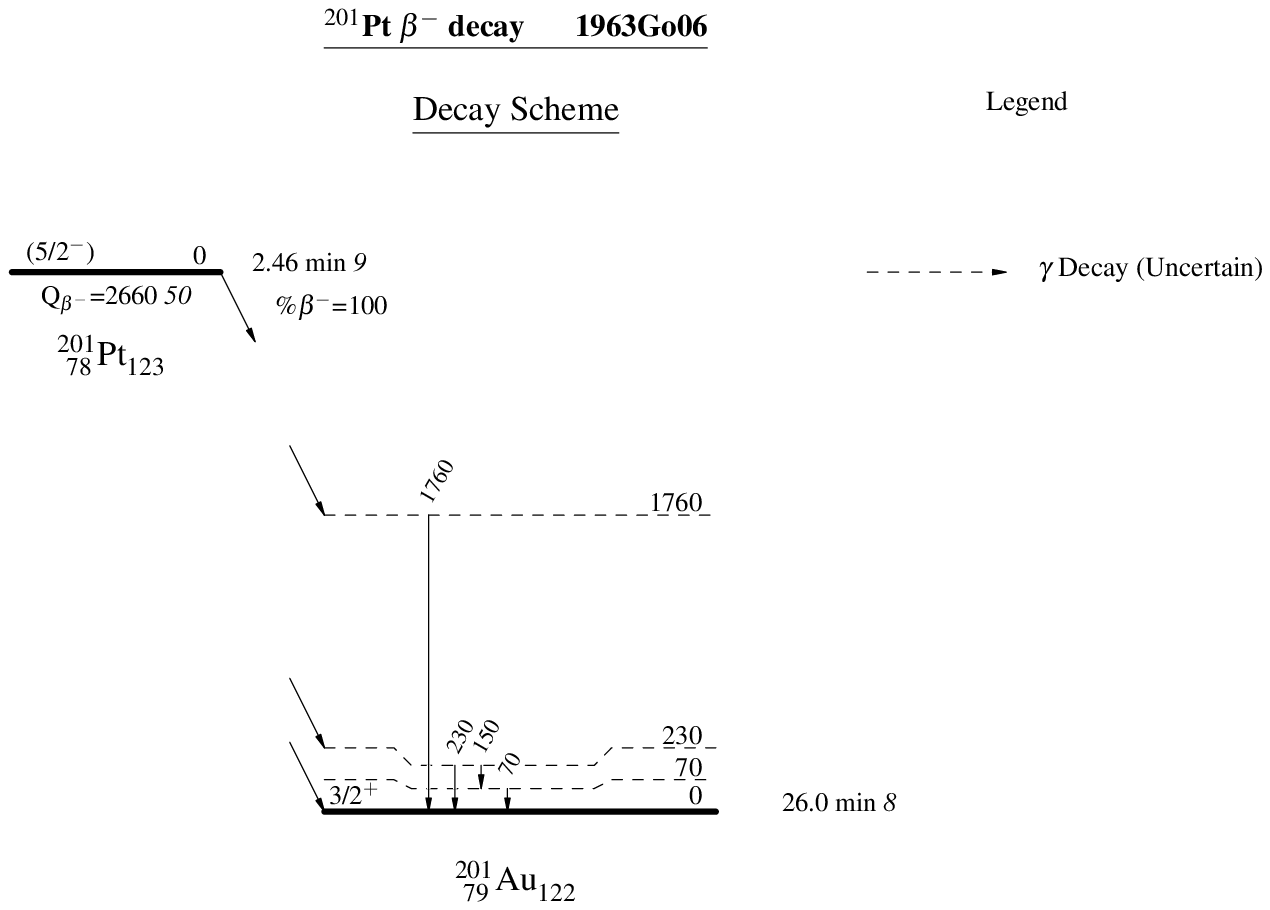}\\
\end{center}
\end{figure}
\clearpage
%202HG(T,A)
\subsection[\hspace{-0.2cm}\ensuremath{^{\textnormal{202}}}Hg(t,\ensuremath{\alpha})]{ }
\vspace{-27pt}
\vspace{0.3cm}
\hypertarget{AU8}{{\bf \small \underline{\ensuremath{^{\textnormal{202}}}Hg(t,\ensuremath{\alpha})\hspace{0.2in}\href{https://www.nndc.bnl.gov/nsr/nsrlink.jsp?1981Fl05,B}{1981Fl05}}}}\\
\vspace{4pt}
\vspace{8pt}
\parbox[b][0.3cm]{17.7cm}{\addtolength{\parindent}{-0.2in}\href{https://www.nndc.bnl.gov/nsr/nsrlink.jsp?1981Fl05,B}{1981Fl05}: 17-MeV polarized triton$'$s beam; Detectors: magnetic spectrograph with FWHM=15-18 keV; Measured: \ensuremath{\sigma}(\ensuremath{\theta},pol); DWBA}\\
\parbox[b][0.3cm]{17.7cm}{analysis relative to \ensuremath{^{\textnormal{208}}}Pb(t,\ensuremath{\alpha}).}\\
\vspace{12pt}
\underline{$^{201}$Au Levels}\\
\begin{longtable}{ccccc|ccccc|ccccc|cc@{\extracolsep{\fill}}c}
\multicolumn{2}{c}{E(level)$^{{\hyperlink{AU8LEVEL0}{\dagger}}}$}&J$^{\pi}$$^{{\hyperlink{AU8LEVEL1}{\ddagger}}}$&\multicolumn{2}{c|}{S$^{{\hyperlink{AU8LEVEL2}{\#}}}$}&\multicolumn{2}{c}{E(level)$^{{\hyperlink{AU8LEVEL0}{\dagger}}}$}&J$^{\pi}$$^{{\hyperlink{AU8LEVEL1}{\ddagger}}}$&\multicolumn{2}{c|}{S$^{{\hyperlink{AU8LEVEL2}{\#}}}$}&\multicolumn{2}{c}{E(level)$^{{\hyperlink{AU8LEVEL0}{\dagger}}}$}&J$^{\pi}$$^{{\hyperlink{AU8LEVEL1}{\ddagger}}}$&\multicolumn{2}{c|}{S$^{{\hyperlink{AU8LEVEL2}{\#}}}$}&\multicolumn{2}{c}{E(level)$^{{\hyperlink{AU8LEVEL0}{\dagger}}}$}&\\[-.2cm]
\multicolumn{2}{c}{\hrulefill}&\hrulefill&\multicolumn{2}{c|}{\hrulefill}&\multicolumn{2}{c}{\hrulefill}&\hrulefill&\multicolumn{2}{c|}{\hrulefill}&\multicolumn{2}{c}{\hrulefill}&\hrulefill&\multicolumn{2}{c|}{\hrulefill}&\multicolumn{2}{c}{\hrulefill}&
\endfirsthead
\multicolumn{1}{r@{}}{0}&\multicolumn{1}{@{}l}{\ensuremath{^{{\hyperlink{AU8LEVEL3}{@}}}}}&\multicolumn{1}{l}{3/2\ensuremath{^{+}}}&\multicolumn{1}{r@{}}{0}&\multicolumn{1}{@{.}l|}{44}&\multicolumn{1}{r@{}}{653}&\multicolumn{1}{@{ }l}{{\it 5}}&\multicolumn{1}{l}{5/2\ensuremath{^{+}}}&\multicolumn{1}{r@{}}{0}&\multicolumn{1}{@{.}l|}{27}&\multicolumn{1}{r@{}}{1242}&\multicolumn{1}{@{ }l}{{\it 5}}&\multicolumn{1}{l}{(5/2\ensuremath{^{+}})}&\multicolumn{1}{r@{}}{0}&\multicolumn{1}{@{.}l|}{07}&\multicolumn{1}{r@{}}{1698}&\multicolumn{1}{@{ }l}{{\it 5}}&\\
\multicolumn{1}{r@{}}{101}&\multicolumn{1}{@{}l}{\ensuremath{^{{\hyperlink{AU8LEVEL4}{\&}}}} {\it 5}}&\multicolumn{1}{l}{1/2\ensuremath{^{+}}}&\multicolumn{1}{r@{}}{0}&\multicolumn{1}{@{.}l|}{23}&\multicolumn{1}{r@{}}{810}&\multicolumn{1}{@{ }l}{{\it 5}}&\multicolumn{1}{l}{5/2\ensuremath{^{+}}}&\multicolumn{1}{r@{}}{0}&\multicolumn{1}{@{.}l|}{03}&\multicolumn{1}{r@{}}{1465}&\multicolumn{1}{@{ }l}{{\it 5}}&\multicolumn{1}{l}{5/2\ensuremath{^{+}}}&\multicolumn{1}{r@{}}{0}&\multicolumn{1}{@{.}l|}{20}&\multicolumn{1}{r@{}}{1900}&\multicolumn{1}{@{ }l}{{\it 5}}&\\
\multicolumn{1}{r@{}}{359}&\multicolumn{1}{@{ }l}{{\it 5}}&\multicolumn{1}{l}{3/2\ensuremath{^{+}}}&\multicolumn{1}{r@{}}{0}&\multicolumn{1}{@{.}l|}{05}&\multicolumn{1}{r@{}}{897}&\multicolumn{1}{@{ }l}{{\it 5}}&\multicolumn{1}{l}{1/2\ensuremath{^{+}}}&\multicolumn{1}{r@{}}{0}&\multicolumn{1}{@{.}l|}{09}&\multicolumn{1}{r@{}}{1506}&\multicolumn{1}{@{ }l}{{\it 5}}&\multicolumn{1}{l}{11/2\ensuremath{^{-}}}&\multicolumn{1}{r@{}}{0}&\multicolumn{1}{@{.}l|}{29}&\multicolumn{1}{r@{}}{1941}&\multicolumn{1}{@{ }l}{{\it 5}}&\\
\multicolumn{1}{r@{}}{549?}&\multicolumn{1}{@{ }l}{{\it 5}}&&&&\multicolumn{1}{r@{}}{1055}&\multicolumn{1}{@{ }l}{{\it 5}}&\multicolumn{1}{l}{3/2\ensuremath{^{+}}}&\multicolumn{1}{r@{}}{0}&\multicolumn{1}{@{.}l|}{07}&\multicolumn{1}{r@{}}{1548}&\multicolumn{1}{@{ }l}{{\it 5}}&&&&\multicolumn{1}{r@{}}{1981}&\multicolumn{1}{@{ }l}{{\it 5}}&\\
\end{longtable}
\parbox[b][0.3cm]{17.7cm}{\makebox[1ex]{\ensuremath{^{\hypertarget{AU8LEVEL0}{\dagger}}}} From \href{https://www.nndc.bnl.gov/nsr/nsrlink.jsp?1981Fl05,B}{1981Fl05}.}\\
\parbox[b][0.3cm]{17.7cm}{\makebox[1ex]{\ensuremath{^{\hypertarget{AU8LEVEL1}{\ddagger}}}} Based on angular distributions and analyzing powers in \href{https://www.nndc.bnl.gov/nsr/nsrlink.jsp?1981Fl05,B}{1981Fl05}.}\\
\parbox[b][0.3cm]{17.7cm}{\makebox[1ex]{\ensuremath{^{\hypertarget{AU8LEVEL2}{\#}}}} Spectroscopic factors relative to \ensuremath{^{\textnormal{208}}}Pb(t,\ensuremath{\alpha})\ensuremath{^{\textnormal{207}}}Tl. Values indicate larger fragmentation of the proton-hole strength compared to}\\
\parbox[b][0.3cm]{17.7cm}{{\ }{\ }\ensuremath{^{\textnormal{203}}}Au, which is interpreted as a result of the larger collectivity (deformation) of the \ensuremath{^{\textnormal{202}}}Hg core relative to \ensuremath{^{\textnormal{204}}}Hg one}\\
\parbox[b][0.3cm]{17.7cm}{{\ }{\ }(\href{https://www.nndc.bnl.gov/nsr/nsrlink.jsp?1981Fl05,B}{1981Fl05}).}\\
\parbox[b][0.3cm]{17.7cm}{\makebox[1ex]{\ensuremath{^{\hypertarget{AU8LEVEL3}{@}}}} Main configuration=\ensuremath{\pi} d\ensuremath{_{\textnormal{3/2}}^{\textnormal{$-$1}}}.}\\
\parbox[b][0.3cm]{17.7cm}{\makebox[1ex]{\ensuremath{^{\hypertarget{AU8LEVEL4}{\&}}}} Main configuration=\ensuremath{\pi} s\ensuremath{_{\textnormal{1/2}}^{\textnormal{$-$1}}}.}\\
\parbox[b][0.3cm]{17.7cm}{\makebox[1ex]{\ensuremath{^{\hypertarget{AU8LEVEL5}{a}}}} Main configuration=\ensuremath{\pi} h\ensuremath{_{\textnormal{11/2}}^{\textnormal{$-$1}}}.}\\
\vspace{0.5cm}
\clearpage
%9BE(208PB,XG)
\subsection[\hspace{-0.2cm}\ensuremath{^{\textnormal{9}}}Be(\ensuremath{^{\textnormal{208}}}Pb,X\ensuremath{\gamma})]{ }
\vspace{-27pt}
\vspace{0.3cm}
\hypertarget{AU9}{{\bf \small \underline{\ensuremath{^{\textnormal{9}}}Be(\ensuremath{^{\textnormal{208}}}Pb,X\ensuremath{\gamma})\hspace{0.2in}\href{https://www.nndc.bnl.gov/nsr/nsrlink.jsp?2011St21,B}{2011St21}}}}\\
\vspace{4pt}
\vspace{8pt}
\parbox[b][0.3cm]{17.7cm}{\addtolength{\parindent}{-0.2in}\href{https://www.nndc.bnl.gov/nsr/nsrlink.jsp?2011St21,B}{2011St21}: in-flight fragmentation of \ensuremath{^{\textnormal{208}}}Pb beam at 1 GeV/A on a 2.526 g/cm\ensuremath{^{\textnormal{2}}} Be target, backed by 0.223 g/cm\ensuremath{^{\textnormal{2}}}-thick \ensuremath{^{\textnormal{93}}}Nb foil.}\\
\parbox[b][0.3cm]{17.7cm}{Fragment Recoil Separator at GSI. Measured E\ensuremath{\gamma}, I\ensuremath{\gamma}, \ensuremath{\gamma}\ensuremath{\gamma}, \ensuremath{\gamma}\ensuremath{\gamma}(t) using the RISING \ensuremath{\gamma}-ray spectrometer.}\\
\vspace{12pt}
\underline{$^{201}$Au Levels}\\
\begin{longtable}{cccccc@{\extracolsep{\fill}}c}
\multicolumn{2}{c}{E(level)$^{{\hyperlink{AU9LEVEL0}{\dagger}}}$}&J$^{\pi}$$^{{\hyperlink{AU9LEVEL1}{\ddagger}}}$&\multicolumn{2}{c}{T$_{1/2}$$^{{\hyperlink{AU9LEVEL1}{\ddagger}}}$}&Comments&\\[-.2cm]
\multicolumn{2}{c}{\hrulefill}&\hrulefill&\multicolumn{2}{c}{\hrulefill}&\hrulefill&
\endfirsthead
\multicolumn{1}{r@{}}{0}&\multicolumn{1}{@{}l}{}&\multicolumn{1}{l}{3/2\ensuremath{^{+}}}&\multicolumn{1}{r@{}}{26}&\multicolumn{1}{@{.}l}{0 min {\it 8}}&&\\
\multicolumn{1}{r@{}}{553}&\multicolumn{1}{@{.}l}{0 {\it 10}}&\multicolumn{1}{l}{(7/2\ensuremath{^{+}})}&&&&\\
\multicolumn{1}{r@{}}{594}&\multicolumn{1}{@{ }l}{{\it 5}}&\multicolumn{1}{l}{11/2\ensuremath{^{-}}}&\multicolumn{1}{r@{}}{0}&\multicolumn{1}{@{.}l}{34 ms {\it +90\textminus29}}&\parbox[t][0.3cm]{11.89278cm}{\raggedright E(level): From Adopted Levels.\vspace{0.1cm}}&\\
&&&&&\parbox[t][0.3cm]{11.89278cm}{\raggedright T\ensuremath{_{1/2}}: From 553\ensuremath{\gamma}(t) in \href{https://www.nndc.bnl.gov/nsr/nsrlink.jsp?2011St21,B}{2011St21}.\vspace{0.1cm}}&\\
&&&&&\parbox[t][0.3cm]{11.89278cm}{\raggedright Experimental isomeric state population ratio=13\% \textit{+36{\textminus}11}.\vspace{0.1cm}}&\\
\multicolumn{1}{r@{}}{1232}&\multicolumn{1}{@{ }l}{{\it 6}}&\multicolumn{1}{l}{(15/2\ensuremath{^{-}})}&&&\parbox[t][0.3cm]{11.89278cm}{\raggedright E(level): Introduced by the evaluator by assuming that 638\ensuremath{\gamma} directly feeds the\vspace{0.1cm}}&\\
&&&&&\parbox[t][0.3cm]{11.89278cm}{\raggedright {\ }{\ }{\ }\ensuremath{J^{\pi}}=11/2\ensuremath{^{-}} isomer. The existence is tentative.\vspace{0.1cm}}&\\
\multicolumn{1}{r@{}}{1610}&\multicolumn{1}{@{ }l}{{\it 6}}&\multicolumn{1}{l}{(19/2\ensuremath{^{+}})}&\multicolumn{1}{r@{}}{5}&\multicolumn{1}{@{.}l}{6 \ensuremath{\mu}s {\it 24}}&\parbox[t][0.3cm]{11.89278cm}{\raggedright E(level): Introduced by the evaluator by assuming that 378.2\ensuremath{\gamma} precedes the 638.0\ensuremath{\gamma}.\vspace{0.1cm}}&\\
\end{longtable}
\parbox[b][0.3cm]{17.7cm}{\makebox[1ex]{\ensuremath{^{\hypertarget{AU9LEVEL0}{\dagger}}}} From a least-squares fit to E\ensuremath{\gamma}, unless otherwise stated.}\\
\parbox[b][0.3cm]{17.7cm}{\makebox[1ex]{\ensuremath{^{\hypertarget{AU9LEVEL1}{\ddagger}}}} From Adopted Levels, unless otherwise stated.}\\
\vspace{0.5cm}
\underline{$\gamma$($^{201}$Au)}\\
\begin{longtable}{ccccccccc@{}ccccc@{\extracolsep{\fill}}c}
\multicolumn{2}{c}{E\ensuremath{_{\gamma}}\ensuremath{^{\hyperlink{AU9GAMMA0}{\dagger}}}}&\multicolumn{2}{c}{I\ensuremath{_{\gamma}}\ensuremath{^{\hyperlink{AU9GAMMA0}{\dagger}}}}&\multicolumn{2}{c}{E\ensuremath{_{i}}(level)}&J\ensuremath{^{\pi}_{i}}&\multicolumn{2}{c}{E\ensuremath{_{f}}}&J\ensuremath{^{\pi}_{f}}&Mult.&\multicolumn{2}{c}{\ensuremath{\alpha}\ensuremath{^{\hyperlink{AU9GAMMA2}{\#}}}}&Comments&\\[-.2cm]
\multicolumn{2}{c}{\hrulefill}&\multicolumn{2}{c}{\hrulefill}&\multicolumn{2}{c}{\hrulefill}&\hrulefill&\multicolumn{2}{c}{\hrulefill}&\hrulefill&\hrulefill&\multicolumn{2}{c}{\hrulefill}&\hrulefill&
\endfirsthead
\multicolumn{1}{r@{}}{(41}&\multicolumn{1}{@{ }l}{{\it 5})}&\multicolumn{1}{r@{}}{}&\multicolumn{1}{@{}l}{}&\multicolumn{1}{r@{}}{594}&\multicolumn{1}{@{}l}{}&\multicolumn{1}{l}{11/2\ensuremath{^{-}}}&\multicolumn{1}{r@{}}{553}&\multicolumn{1}{@{.}l}{0 }&\multicolumn{1}{@{}l}{(7/2\ensuremath{^{+}})}&&\multicolumn{1}{r@{}}{}&\multicolumn{1}{@{}l}{}&\parbox[t][0.3cm]{6.8833604cm}{\raggedright E\ensuremath{_{\gamma}}: From level energy differences; not observed\vspace{0.1cm}}&\\
&&&&&&&&&&&&&\parbox[t][0.3cm]{6.8833604cm}{\raggedright {\ }{\ }{\ }in \href{https://www.nndc.bnl.gov/nsr/nsrlink.jsp?2011St21,B}{2011St21}.\vspace{0.1cm}}&\\
\multicolumn{1}{r@{}}{378}&\multicolumn{1}{@{.}l}{2\ensuremath{^{\hyperlink{AU9GAMMA1}{\ddagger}}} {\it 5}}&\multicolumn{1}{r@{}}{58}&\multicolumn{1}{@{ }l}{{\it 17}}&\multicolumn{1}{r@{}}{1610}&\multicolumn{1}{@{}l}{}&\multicolumn{1}{l}{(19/2\ensuremath{^{+}})}&\multicolumn{1}{r@{}}{1232}&\multicolumn{1}{@{ }l}{}&\multicolumn{1}{@{}l}{(15/2\ensuremath{^{-}})}&\multicolumn{1}{l}{(M2)}&\multicolumn{1}{r@{}}{0}&\multicolumn{1}{@{.}l}{601 {\it 9}}&\parbox[t][0.3cm]{6.8833604cm}{\raggedright Mult.: From \ensuremath{\alpha}(tot,378\ensuremath{\gamma}) \ensuremath{\approx} 0.7 from total\vspace{0.1cm}}&\\
&&&&&&&&&&&&&\parbox[t][0.3cm]{6.8833604cm}{\raggedright {\ }{\ }{\ }intensity balance between 638-keV and 378-keV\vspace{0.1cm}}&\\
&&&&&&&&&&&&&\parbox[t][0.3cm]{6.8833604cm}{\raggedright {\ }{\ }{\ }\ensuremath{\gamma} rays.\vspace{0.1cm}}&\\
\multicolumn{1}{r@{}}{553}&\multicolumn{1}{@{ }l}{{\it 1}}&\multicolumn{1}{r@{}}{100}&\multicolumn{1}{@{ }l}{{\it 51}}&\multicolumn{1}{r@{}}{553}&\multicolumn{1}{@{.}l}{0}&\multicolumn{1}{l}{(7/2\ensuremath{^{+}})}&\multicolumn{1}{r@{}}{0}&\multicolumn{1}{@{}l}{}&\multicolumn{1}{@{}l}{3/2\ensuremath{^{+}}}&&\multicolumn{1}{r@{}}{}&\multicolumn{1}{@{}l}{}&&\\
\end{longtable}
\parbox[b][0.3cm]{17.7cm}{\makebox[1ex]{\ensuremath{^{\hypertarget{AU9GAMMA0}{\dagger}}}} From \href{https://www.nndc.bnl.gov/nsr/nsrlink.jsp?2011St21,B}{2011St21}. \ensuremath{\Delta}E\ensuremath{\gamma} were estimated by the evaluator.}\\
\parbox[b][0.3cm]{17.7cm}{\makebox[1ex]{\ensuremath{^{\hypertarget{AU9GAMMA1}{\ddagger}}}} This \ensuremath{\gamma} ray shows the 5.0{\textminus}\ensuremath{\mu}s isomer half-life, but the ordering and placement in the decay scheme are not experimentally known.}\\
\parbox[b][0.3cm]{17.7cm}{{\ }{\ }The interpretation is made by the evaluator.}\\
\parbox[b][0.3cm]{17.7cm}{\makebox[1ex]{\ensuremath{^{\hypertarget{AU9GAMMA2}{\#}}}} Total theoretical internal conversion coefficients, calculated using the BrIcc code (\href{https://www.nndc.bnl.gov/nsr/nsrlink.jsp?2008Ki07,B}{2008Ki07}) with Frozen orbital approximation}\\
\parbox[b][0.3cm]{17.7cm}{{\ }{\ }based on \ensuremath{\gamma}-ray energies, assigned multipolarities, and mixing ratios, unless otherwise specified.}\\
\vspace{0.5cm}
\clearpage
\begin{figure}[h]
\begin{center}
\includegraphics{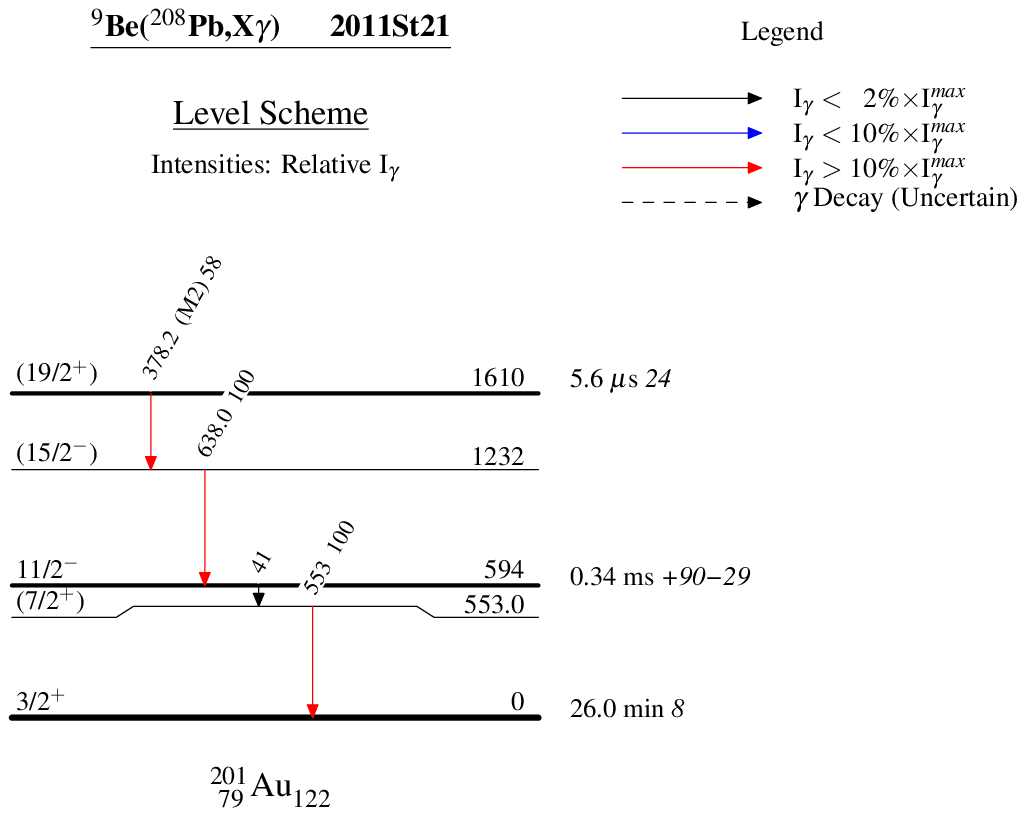}\\
\end{center}
\end{figure}
\clearpage
%ADOPTED LEVELS, GAMMAS
\section[\ensuremath{^{201}_{\ 80}}Hg\ensuremath{_{121}^{~}}]{ }
\vspace{-30pt}
\setcounter{chappage}{1}
\subsection[\hspace{-0.2cm}Adopted Levels, Gammas]{ }
\vspace{-20pt}
\vspace{0.3cm}
\hypertarget{HG10}{{\bf \small \underline{Adopted \hyperlink{201HG_LEVEL}{Levels}, \hyperlink{201HG_GAMMA}{Gammas}}}}\\
\vspace{4pt}
\vspace{8pt}
\parbox[b][0.3cm]{17.7cm}{\addtolength{\parindent}{-0.2in}Q(\ensuremath{\beta^-})=$-$482 {\it 14}; S(n)=6230.6 {\it 6}; S(p)=7711 {\it 27}; Q(\ensuremath{\alpha})=332.3 {\it 8}\hspace{0.2in}\href{https://www.nndc.bnl.gov/nsr/nsrlink.jsp?2021Wa16,B}{2021Wa16}}\\

\vspace{12pt}
\hypertarget{201HG_LEVEL}{\underline{$^{201}$Hg Levels}}\\
\begin{longtable}[c]{llll}
\multicolumn{4}{c}{\underline{Cross Reference (XREF) Flags}}\\
 \\
\hyperlink{AU11}{\texttt{A }}& \ensuremath{^{\textnormal{201}}}Au \ensuremath{\beta}\ensuremath{^{-}} decay & \hyperlink{HG16}{\texttt{F }}& \ensuremath{^{\textnormal{201}}}Hg(d,d\ensuremath{'}),\ensuremath{^{\textnormal{201}}}Hg(p,p\ensuremath{'})\\
\hyperlink{HG12}{\texttt{B }}& \ensuremath{^{\textnormal{201}}}Hg IT decay & \hyperlink{HG17}{\texttt{G }}& Coulomb excitation\\
\hyperlink{TL13}{\texttt{C }}& \ensuremath{^{\textnormal{201}}}Tl \ensuremath{\varepsilon} decay & \hyperlink{HG18}{\texttt{H }}& \ensuremath{^{\textnormal{202}}}Hg(d,t)\\
\hyperlink{HG14}{\texttt{D }}& \ensuremath{^{\textnormal{200}}}Hg(d,p) & \hyperlink{HG19}{\texttt{I }}& \ensuremath{^{\textnormal{203}}}Tl(\ensuremath{\mu},X\ensuremath{\gamma})\\
\hyperlink{HG15}{\texttt{E }}& \ensuremath{^{\textnormal{201}}}Hg(\ensuremath{\gamma},\ensuremath{\gamma}\ensuremath{'}) & \\
\end{longtable}
\vspace{-0.5cm}
\begin{longtable}{ccccccc@{\extracolsep{\fill}}c}
\multicolumn{2}{c}{E(level)$^{{\hyperlink{HG10LEVEL0}{\dagger}}}$}&J$^{\pi}$$^{}$&\multicolumn{2}{c}{T$_{1/2}$$^{}$}&XREF&Comments&\\[-.2cm]
\multicolumn{2}{c}{\hrulefill}&\hrulefill&\multicolumn{2}{c}{\hrulefill}&\hrulefill&\hrulefill&
\endfirsthead
\multicolumn{1}{r@{}}{0}&\multicolumn{1}{@{}l}{\ensuremath{^{{\hyperlink{HG10LEVEL4}{\&}}}}}&\multicolumn{1}{l}{3/2\ensuremath{^{-}}}&\multicolumn{1}{r@{}}{sta}&\multicolumn{1}{@{}l}{ble}&\multicolumn{1}{l}{\texttt{\hyperlink{AU11}{A}\hyperlink{HG12}{B}\hyperlink{TL13}{C}\hyperlink{HG14}{D}\hyperlink{HG15}{E}\hyperlink{HG16}{F}\hyperlink{HG17}{G}\hyperlink{HG18}{H}\hyperlink{HG19}{I}} }&\parbox[t][0.3cm]{10.67033cm}{\raggedright \ensuremath{\mu}={\textminus}0.5602257 \textit{14} (\href{https://www.nndc.bnl.gov/nsr/nsrlink.jsp?1973Re04,B}{1973Re04},\href{https://www.nndc.bnl.gov/nsr/nsrlink.jsp?2014StZZ,B}{2014StZZ})\vspace{0.1cm}}&\\
&&&&&&\parbox[t][0.3cm]{10.67033cm}{\raggedright Q=+0.387 \textit{6} (\href{https://www.nndc.bnl.gov/nsr/nsrlink.jsp?2005Bi03,B}{2005Bi03},\href{https://www.nndc.bnl.gov/nsr/nsrlink.jsp?2016St14,B}{2016St14})\vspace{0.1cm}}&\\
&&&&&&\parbox[t][0.3cm]{10.67033cm}{\raggedright J\ensuremath{^{\pi}}: Optical spectroscopy (\href{https://www.nndc.bnl.gov/nsr/nsrlink.jsp?1931Sc03,B}{1931Sc03},\href{https://www.nndc.bnl.gov/nsr/nsrlink.jsp?1935Sc04,B}{1935Sc04}); \ensuremath{\pi} from \ensuremath{\mu} and L(d,t)=1.\vspace{0.1cm}}&\\
&&&&&&\parbox[t][0.3cm]{10.67033cm}{\raggedright \ensuremath{\mu}: Measured using the nuclear magnetic resonance using optically pumped\vspace{0.1cm}}&\\
&&&&&&\parbox[t][0.3cm]{10.67033cm}{\raggedright {\ }{\ }{\ }ions technique; Other: 0.560226 \textit{3} (\href{https://www.nndc.bnl.gov/nsr/nsrlink.jsp?1961Ca21,B}{1961Ca21}).\vspace{0.1cm}}&\\
&&&&&&\parbox[t][0.3cm]{10.67033cm}{\raggedright Q: Measured using the muonic x-ray hyperfine structure technique; Others:\vspace{0.1cm}}&\\
\end{longtable}
\begin{textblock}{29}(0,27.3)
Continued on next page (footnotes at end of table)
\end{textblock}
\clearpage
\begin{longtable}{ccccccc@{\extracolsep{\fill}}c}
\\[-.4cm]
\multicolumn{8}{c}{{\bf \small \underline{Adopted \hyperlink{201HG_LEVEL}{Levels}, \hyperlink{201HG_GAMMA}{Gammas} (continued)}}}\\
\multicolumn{8}{c}{~}\\
\multicolumn{8}{c}{\underline{\ensuremath{^{201}}Hg Levels (continued)}}\\
\multicolumn{8}{c}{~}\\
\multicolumn{2}{c}{E(level)$^{{\hyperlink{HG10LEVEL0}{\dagger}}}$}&J$^{\pi}$$^{}$&\multicolumn{2}{c}{T$_{1/2}$$^{}$}&XREF&Comments&\\[-.2cm]
\multicolumn{2}{c}{\hrulefill}&\hrulefill&\multicolumn{2}{c}{\hrulefill}&\hrulefill&\hrulefill&
\endhead
&&&&&&\parbox[t][0.3cm]{9.58535cm}{\raggedright J\ensuremath{^{\pi}}: 352.42\ensuremath{\gamma} M1(+E2) to 3/2\ensuremath{^{-}}; 358.36\ensuremath{\gamma} M1+E2 to 5/2\ensuremath{^{-}}; population in\vspace{0.1cm}}&\\
&&&&&&\parbox[t][0.3cm]{9.58535cm}{\raggedright {\ }{\ }{\ }Coulomb excitation.\vspace{0.1cm}}&\\
&&&&&&\parbox[t][0.3cm]{9.58535cm}{\raggedright B(E2)\ensuremath{\uparrow}=0.085 \textit{5} (\href{https://www.nndc.bnl.gov/nsr/nsrlink.jsp?1983Sc38,B}{1983Sc38}) in Coulomb excitation.\vspace{0.1cm}}&\\
\multicolumn{1}{r@{}}{414}&\multicolumn{1}{@{.}l}{543 {\it 18}}&\multicolumn{1}{l}{7/2\ensuremath{^{-}}}&\multicolumn{1}{r@{}}{21}&\multicolumn{1}{@{.}l}{0 ps {\it 9}}&\multicolumn{1}{l}{\texttt{\ \ \ \hyperlink{HG14}{D}\ \hyperlink{HG16}{F}\hyperlink{HG17}{G}\ \hyperlink{HG19}{I}} }&\parbox[t][0.3cm]{9.58535cm}{\raggedright XREF: D(417)F(412).\vspace{0.1cm}}&\\
&&&&&&\parbox[t][0.3cm]{9.58535cm}{\raggedright J\ensuremath{^{\pi}}: 414.49\ensuremath{\gamma} E2 to 3/2\ensuremath{^{-}}, 388.26\ensuremath{\gamma} M1+E2 to 5/2\ensuremath{^{-}}.\vspace{0.1cm}}&\\
&&&&&&\parbox[t][0.3cm]{9.58535cm}{\raggedright T\ensuremath{_{1/2}}: From B(E2)\ensuremath{\uparrow}=0.152 \textit{6} (\href{https://www.nndc.bnl.gov/nsr/nsrlink.jsp?1980Bo05,B}{1980Bo05}) in Coulomb excitation.\vspace{0.1cm}}&\\
&&&&&&\parbox[t][0.3cm]{9.58535cm}{\raggedright B(E2)\ensuremath{\uparrow}=0.152 \textit{6} (\href{https://www.nndc.bnl.gov/nsr/nsrlink.jsp?1980Bo05,B}{1980Bo05}) in Coulomb excitation.\vspace{0.1cm}}&\\
\multicolumn{1}{r@{}}{464}&\multicolumn{1}{@{.}l}{41 {\it 4}}&\multicolumn{1}{l}{5/2\ensuremath{^{-}}}&&&\multicolumn{1}{l}{\texttt{\hyperlink{AU11}{A}\ \ \hyperlink{HG14}{D}\ \hyperlink{HG16}{F}\hyperlink{HG17}{G}\hyperlink{HG18}{H}\ } }&\parbox[t][0.3cm]{9.58535cm}{\raggedright XREF: D(466)F(465).\vspace{0.1cm}}&\\
\end{longtable}
\begin{textblock}{29}(0,27.3)
Continued on next page (footnotes at end of table)
\end{textblock}
\clearpage
\begin{longtable}{ccccc@{\extracolsep{\fill}}c}
\\[-.4cm]
\multicolumn{6}{c}{{\bf \small \underline{Adopted \hyperlink{201HG_LEVEL}{Levels}, \hyperlink{201HG_GAMMA}{Gammas} (continued)}}}\\
\multicolumn{6}{c}{~}\\
\multicolumn{6}{c}{\underline{\ensuremath{^{201}}Hg Levels (continued)}}\\
\multicolumn{6}{c}{~}\\
\multicolumn{2}{c}{E(level)$^{{\hyperlink{HG10LEVEL0}{\dagger}}}$}&J$^{\pi}$$^{}$&XREF&Comments&\\[-.2cm]
\multicolumn{2}{c}{\hrulefill}&\hrulefill&\hrulefill&\hrulefill&
\endhead
\multicolumn{1}{r@{}}{2663?}&\multicolumn{1}{@{}l}{\ensuremath{^{{\hyperlink{HG10LEVEL1}{\ddagger}}}}}&&\multicolumn{1}{l}{\texttt{\ \ \ \hyperlink{HG14}{D}\ \ \ \hyperlink{HG18}{H}\ } }&\parbox[t][0.3cm]{13.45306cm}{\raggedright XREF: D(2660).\vspace{0.1cm}}&\\
\multicolumn{1}{r@{}}{2681}&\multicolumn{1}{@{}l}{\ensuremath{^{{\hyperlink{HG10LEVEL3}{@}}}} {\it 11}}&&\multicolumn{1}{l}{\texttt{\ \ \ \ \ \hyperlink{HG16}{F}\ \ \ } }&&\\
\multicolumn{1}{r@{}}{2795}&\multicolumn{1}{@{}l}{\ensuremath{^{{\hyperlink{HG10LEVEL2}{\#}}}} {\it 11}}&\multicolumn{1}{l}{(\ensuremath{^{+}})}&\multicolumn{1}{l}{\texttt{\ \ \ \hyperlink{HG14}{D}\ \ \ \ \ } }&\parbox[t][0.3cm]{13.45306cm}{\raggedright J\ensuremath{^{\pi}}: L(d,p)=(4,2). \ensuremath{J^{\pi}}=7/2\ensuremath{^{+}}, 9/2\ensuremath{^{+}} or 3/2\ensuremath{^{+}}, 5/2\ensuremath{^{+}}.\vspace{0.1cm}}&\\
\multicolumn{1}{r@{}}{2863}&\multicolumn{1}{@{}l}{\ensuremath{^{{\hyperlink{HG10LEVEL2}{\#}}}} {\it 12}}&&\multicolumn{1}{l}{\texttt{\ \ \ \hyperlink{HG14}{D}\ \ \ \ \ } }&&\\
\multicolumn{1}{r@{}}{2890}&\multicolumn{1}{@{}l}{\ensuremath{^{{\hyperlink{HG10LEVEL2}{\#}}}} {\it 12}}&&\multicolumn{1}{l}{\texttt{\ \ \ \hyperlink{HG14}{D}\ \hyperlink{HG16}{F}\ \ \ } }&\parbox[t][0.3cm]{13.45306cm}{\raggedright XREF: F(2891).\vspace{0.1cm}}&\\
\multicolumn{1}{r@{}}{2911}&\multicolumn{1}{@{}l}{\ensuremath{^{{\hyperlink{HG10LEVEL2}{\#}}}} {\it 12}}&&\multicolumn{1}{l}{\texttt{\ \ \ \hyperlink{HG14}{D}\ \ \ \ \ } }&&\\
\multicolumn{1}{r@{}}{2938}&\multicolumn{1}{@{}l}{\ensuremath{^{{\hyperlink{HG10LEVEL2}{\#}}}} {\it 12}}&&\multicolumn{1}{l}{\texttt{\ \ \ \hyperlink{HG14}{D}\ \ \ \ \ } }&&\\
\end{longtable}
\begin{textblock}{29}(0,27.3)
Continued on next page (footnotes at end of table)
\end{textblock}
\clearpage
\vspace*{-0.5cm}
{\bf \small \underline{Adopted \hyperlink{201HG_LEVEL}{Levels}, \hyperlink{201HG_GAMMA}{Gammas} (continued)}}\\
\vspace{0.3cm}
\underline{$^{201}$Hg Levels (continued)}\\
\vspace{0.3cm}
\parbox[b][0.3cm]{17.7cm}{\makebox[1ex]{\ensuremath{^{\hypertarget{HG10LEVEL0}{\dagger}}}} From a least-squares fit to E\ensuremath{\gamma}, unless otherwise stated.}\\
\parbox[b][0.3cm]{17.7cm}{\makebox[1ex]{\ensuremath{^{\hypertarget{HG10LEVEL1}{\ddagger}}}} From \ensuremath{^{\textnormal{202}}}Hg(d,t).}\\
\parbox[b][0.3cm]{17.7cm}{\makebox[1ex]{\ensuremath{^{\hypertarget{HG10LEVEL2}{\#}}}} From \ensuremath{^{\textnormal{200}}}Hg(d,p).}\\
\parbox[b][0.3cm]{17.7cm}{\makebox[1ex]{\ensuremath{^{\hypertarget{HG10LEVEL3}{@}}}} From \ensuremath{^{\textnormal{201}}}Hg(d,d\ensuremath{'}), \ensuremath{^{\textnormal{201}}}Hg(p,p\ensuremath{'}).}\\
\parbox[b][0.3cm]{17.7cm}{\makebox[1ex]{\ensuremath{^{\hypertarget{HG10LEVEL4}{\&}}}} Configuration=\ensuremath{\nu} p\ensuremath{_{\textnormal{3/2}}^{\textnormal{$-$1}}}.}\\
\parbox[b][0.3cm]{17.7cm}{\makebox[1ex]{\ensuremath{^{\hypertarget{HG10LEVEL5}{a}}}} Configuration=\ensuremath{\nu} p\ensuremath{_{\textnormal{1/2}}^{\textnormal{$-$1}}}.}\\
\parbox[b][0.3cm]{17.7cm}{\makebox[1ex]{\ensuremath{^{\hypertarget{HG10LEVEL6}{b}}}} Dominant configuration=\ensuremath{\nu} f\ensuremath{_{\textnormal{5/2}}^{\textnormal{$-$1}}}.}\\
\parbox[b][0.3cm]{17.7cm}{\makebox[1ex]{\ensuremath{^{\hypertarget{HG10LEVEL7}{c}}}} Configuration=\ensuremath{\nu} i\ensuremath{_{\textnormal{13/2}}^{\textnormal{$-$1}}}.}\\
\vspace{0.5cm}
\clearpage
\vspace{0.3cm}
\begin{landscape}
\vspace*{-0.5cm}
{\bf \small \underline{Adopted \hyperlink{201HG_LEVEL}{Levels}, \hyperlink{201HG_GAMMA}{Gammas} (continued)}}\\
\vspace{0.3cm}
\hypertarget{201HG_GAMMA}{\underline{$\gamma$($^{201}$Hg)}}\\
\begin{longtable}{ccccccccc@{}ccccccc@{\extracolsep{\fill}}c}
\multicolumn{2}{c}{E\ensuremath{_{i}}(level)}&J\ensuremath{^{\pi}_{i}}&\multicolumn{2}{c}{E\ensuremath{_{\gamma}}\ensuremath{^{\hyperlink{HG10GAMMA0}{\dagger}}}}&\multicolumn{2}{c}{I\ensuremath{_{\gamma}}\ensuremath{^{\hyperlink{HG10GAMMA0}{\dagger}}}}&\multicolumn{2}{c}{E\ensuremath{_{f}}}&J\ensuremath{^{\pi}_{f}}&Mult.\ensuremath{^{\hyperlink{HG10GAMMA0}{\dagger}}}&\multicolumn{2}{c}{\ensuremath{\delta}\ensuremath{^{\hyperlink{HG10GAMMA0}{\dagger}}}}&\multicolumn{2}{c}{\ensuremath{\alpha}\ensuremath{^{\hyperlink{HG10GAMMA4}{\&}}}}&Comments&\\[-.2cm]
\multicolumn{2}{c}{\hrulefill}&\hrulefill&\multicolumn{2}{c}{\hrulefill}&\multicolumn{2}{c}{\hrulefill}&\multicolumn{2}{c}{\hrulefill}&\hrulefill&\hrulefill&\multicolumn{2}{c}{\hrulefill}&\multicolumn{2}{c}{\hrulefill}&\hrulefill&
\endfirsthead
\multicolumn{1}{r@{}}{1}&\multicolumn{1}{@{.}l}{5648}&\multicolumn{1}{l}{1/2\ensuremath{^{-}}}&\multicolumn{1}{r@{}}{1}&\multicolumn{1}{@{.}l}{5648 {\it 10}}&\multicolumn{1}{r@{}}{100}&\multicolumn{1}{@{}l}{}&\multicolumn{1}{r@{}}{0}&\multicolumn{1}{@{}l}{}&\multicolumn{1}{@{}l}{3/2\ensuremath{^{-}}}&\multicolumn{1}{l}{M1+E2}&\multicolumn{1}{r@{}}{0}&\multicolumn{1}{@{.}l}{0105 {\it 14}}&\multicolumn{1}{r@{}}{4.7\ensuremath{\times10^{4}}}&\multicolumn{1}{@{ }l}{{\it 7}}&\parbox[t][0.3cm]{8.618859cm}{\raggedright B(M1)(W.u.)=0.00151 \textit{+29{\textminus}21}; B(E2)(W.u.)=25 \textit{+9{\textminus}7}\vspace{0.1cm}}&\\
&&&&&&&&&&&&&&&\parbox[t][0.3cm]{8.618859cm}{\raggedright E\ensuremath{_{\gamma}}: From \href{https://www.nndc.bnl.gov/nsr/nsrlink.jsp?1997Ge09,B}{1997Ge09}. Other: 1.565 keV \textit{6} (\href{https://www.nndc.bnl.gov/nsr/nsrlink.jsp?1987Dr06,B}{1987Dr06}).\vspace{0.1cm}}&\\
&&&&&&&&&&&&&&&\parbox[t][0.3cm]{8.618859cm}{\raggedright Mult.: From N1/N2=0.94 \textit{31}, N1/N3=0.60 \textit{20}, N2/N3=0.64 \textit{5},\vspace{0.1cm}}&\\
&&&&&&&&&&&&&&&\parbox[t][0.3cm]{8.618859cm}{\raggedright {\ }{\ }{\ }N4/N3=0.042 \textit{18}, N5/N3=0.043 \textit{23}, N4/N5=0.98 \textit{32},\vspace{0.1cm}}&\\
&&&&&&&&&&&&&&&\parbox[t][0.3cm]{8.618859cm}{\raggedright {\ }{\ }{\ }O1/O2=0.81 \textit{16}, O2/N3=0.158 \textit{30}, O3/N3=0.20 \textit{4},\vspace{0.1cm}}&\\
&&&&&&&&&&&&&&&\parbox[t][0.3cm]{8.618859cm}{\raggedright {\ }{\ }{\ }O1/O3=0.64 \textit{12}, O1/N3=0.128 \textit{30} and O2/O3=0.79 \textit{19}\vspace{0.1cm}}&\\
&&&&&&&&&&&&&&&\parbox[t][0.3cm]{8.618859cm}{\raggedright {\ }{\ }{\ }subshell ratios in \href{https://www.nndc.bnl.gov/nsr/nsrlink.jsp?1997Ge09,B}{1997Ge09} and N1/N2=1.2 \textit{2}, N1/N3=1.1 \textit{2},\vspace{0.1cm}}&\\
&&&&&&&&&&&&&&&\parbox[t][0.3cm]{8.618859cm}{\raggedright {\ }{\ }{\ }N2/N3=0.92 \textit{15}, N4/N3=0.03 \textit{2} and N5/N3=0.04 \textit{2} subshell\vspace{0.1cm}}&\\
\end{longtable}
\clearpage
\begin{longtable}{ccccccccc@{}ccccccc@{\extracolsep{\fill}}c}
\\[-.4cm]
\multicolumn{17}{c}{{\bf \small \underline{Adopted \hyperlink{201HG_LEVEL}{Levels}, \hyperlink{201HG_GAMMA}{Gammas} (continued)}}}\\
\multicolumn{17}{c}{~}\\
\multicolumn{17}{c}{\underline{$\gamma$($^{201}$Hg) (continued)}}\\
\multicolumn{17}{c}{~~~}\\
\multicolumn{2}{c}{E\ensuremath{_{i}}(level)}&J\ensuremath{^{\pi}_{i}}&\multicolumn{2}{c}{E\ensuremath{_{\gamma}}\ensuremath{^{\hyperlink{HG10GAMMA0}{\dagger}}}}&\multicolumn{2}{c}{I\ensuremath{_{\gamma}}\ensuremath{^{\hyperlink{HG10GAMMA0}{\dagger}}}}&\multicolumn{2}{c}{E\ensuremath{_{f}}}&J\ensuremath{^{\pi}_{f}}&Mult.\ensuremath{^{\hyperlink{HG10GAMMA0}{\dagger}}}&\multicolumn{2}{c}{\ensuremath{\delta}\ensuremath{^{\hyperlink{HG10GAMMA0}{\dagger}}}}&\multicolumn{2}{c}{\ensuremath{\alpha}\ensuremath{^{\hyperlink{HG10GAMMA4}{\&}}}}&Comments&\\[-.2cm]
\multicolumn{2}{c}{\hrulefill}&\hrulefill&\multicolumn{2}{c}{\hrulefill}&\multicolumn{2}{c}{\hrulefill}&\multicolumn{2}{c}{\hrulefill}&\hrulefill&\hrulefill&\multicolumn{2}{c}{\hrulefill}&\multicolumn{2}{c}{\hrulefill}&\hrulefill&
\endhead
&&&&&&&&&&&&&&&\parbox[t][0.3cm]{7.1658cm}{\raggedright \ensuremath{\delta}: From K:L1:L2:L3 in \href{https://www.nndc.bnl.gov/nsr/nsrlink.jsp?1960He05,B}{1960He05} and the\vspace{0.1cm}}&\\
&&&&&&&&&&&&&&&\parbox[t][0.3cm]{7.1658cm}{\raggedright {\ }{\ }{\ }briccmixing program.\vspace{0.1cm}}&\\
\multicolumn{1}{r@{}}{384}&\multicolumn{1}{@{.}l}{604}&\multicolumn{1}{l}{(5/2)\ensuremath{^{-}}}&\multicolumn{1}{r@{}}{352}&\multicolumn{1}{@{.}l}{42\ensuremath{^{\hyperlink{HG10GAMMA1}{\ddagger}}} {\it 5}}&\multicolumn{1}{r@{}}{100}&\multicolumn{1}{@{.}l}{0\ensuremath{^{\hyperlink{HG10GAMMA1}{\ddagger}}} {\it 16}}&\multicolumn{1}{r@{}}{32}&\multicolumn{1}{@{.}l}{155 }&\multicolumn{1}{@{}l}{3/2\ensuremath{^{-}}}&\multicolumn{1}{l}{M1(+E2)}&\multicolumn{1}{r@{}}{+0}&\multicolumn{1}{@{.}l}{07 {\it 7}}&\multicolumn{1}{r@{}}{0}&\multicolumn{1}{@{.}l}{232 {\it 4}}&\parbox[t][0.3cm]{7.1658cm}{\raggedright Mult.,\ensuremath{\delta}: From \ensuremath{\gamma}(\ensuremath{\theta}) (A\ensuremath{_{\textnormal{2}}}={\textminus}0.047 \textit{24}) in Coulomb\vspace{0.1cm}}&\\
&&&&&&&&&&&&&&&\parbox[t][0.3cm]{7.1658cm}{\raggedright {\ }{\ }{\ }excitation (\href{https://www.nndc.bnl.gov/nsr/nsrlink.jsp?1980Bo05,B}{1980Bo05}). other (alternative): \ensuremath{\delta}={\textminus}4.5\vspace{0.1cm}}&\\
&&&&&&&&&&&&&&&\parbox[t][0.3cm]{7.1658cm}{\raggedright {\ }{\ }{\ }\textit{+1.2{\textminus}1.7} (\href{https://www.nndc.bnl.gov/nsr/nsrlink.jsp?1980Bo05,B}{1980Bo05}).\vspace{0.1cm}}&\\
\end{longtable}
\clearpage
\begin{longtable}{cccccccc@{}cc@{\extracolsep{\fill}}c}
\\[-.4cm]
\multicolumn{11}{c}{{\bf \small \underline{Adopted \hyperlink{201HG_LEVEL}{Levels}, \hyperlink{201HG_GAMMA}{Gammas} (continued)}}}\\
\multicolumn{11}{c}{~}\\
\multicolumn{11}{c}{\underline{$\gamma$($^{201}$Hg) (continued)}}\\
\multicolumn{11}{c}{~~~}\\
\multicolumn{2}{c}{E\ensuremath{_{i}}(level)}&\multicolumn{2}{c}{E\ensuremath{_{\gamma}}\ensuremath{^{\hyperlink{HG10GAMMA0}{\dagger}}}}&\multicolumn{2}{c}{I\ensuremath{_{\gamma}}\ensuremath{^{\hyperlink{HG10GAMMA0}{\dagger}}}}&\multicolumn{2}{c}{E\ensuremath{_{f}}}&J\ensuremath{^{\pi}_{f}}&Comments&\\[-.2cm]
\multicolumn{2}{c}{\hrulefill}&\multicolumn{2}{c}{\hrulefill}&\multicolumn{2}{c}{\hrulefill}&\multicolumn{2}{c}{\hrulefill}&\hrulefill&\hrulefill&
\endhead
&&&&&&&&&\parbox[t][0.3cm]{15.10876cm}{\raggedright \ensuremath{\delta}: From K/L=3.2 \textit{6}, L/M+=3.4 \textit{9} (\href{https://www.nndc.bnl.gov/nsr/nsrlink.jsp?1990Lo17,B}{1990Lo17}) and the briccmixing program.\vspace{0.1cm}}&\\
\multicolumn{1}{r@{}}{1187}&\multicolumn{1}{@{.}l}{8}&\multicolumn{1}{r@{}}{645}&\multicolumn{1}{@{.}l}{0\ensuremath{^{\hyperlink{HG10GAMMA2}{\#}}} {\it 4}}&\multicolumn{1}{r@{}}{100}&\multicolumn{1}{@{}l}{\ensuremath{^{\hyperlink{HG10GAMMA2}{\#}}}}&\multicolumn{1}{r@{}}{542}&\multicolumn{1}{@{.}l}{81 }&\multicolumn{1}{@{}l}{1/2\ensuremath{^{-}},3/2,5/2}&&\\
\end{longtable}
\parbox[b][0.3cm]{21.881866cm}{\makebox[1ex]{\ensuremath{^{\hypertarget{HG10GAMMA0}{\dagger}}}} From \ensuremath{^{\textnormal{201}}}Tl \ensuremath{\varepsilon} decay, unless otherwise stated.}\\
\parbox[b][0.3cm]{21.881866cm}{\makebox[1ex]{\ensuremath{^{\hypertarget{HG10GAMMA1}{\ddagger}}}} From Coulomb excitation.}\\
\parbox[b][0.3cm]{21.881866cm}{\makebox[1ex]{\ensuremath{^{\hypertarget{HG10GAMMA2}{\#}}}} From \ensuremath{^{\textnormal{201}}}Au \ensuremath{\beta}\ensuremath{^{-}} decay.}\\
\parbox[b][0.3cm]{21.881866cm}{\makebox[1ex]{\ensuremath{^{\hypertarget{HG10GAMMA3}{@}}}} From \ensuremath{^{\textnormal{201}}}Hg IT decay.}\\
\parbox[b][0.3cm]{21.881866cm}{\makebox[1ex]{\ensuremath{^{\hypertarget{HG10GAMMA4}{\&}}}} Total theoretical internal conversion coefficients, calculated using the BrIcc code (\href{https://www.nndc.bnl.gov/nsr/nsrlink.jsp?2008Ki07,B}{2008Ki07}) with Frozen orbital approximation based on \ensuremath{\gamma}-ray energies,}\\
\parbox[b][0.3cm]{21.881866cm}{{\ }{\ }assigned multipolarities, and mixing ratios, unless otherwise specified.}\\
\parbox[b][0.3cm]{21.881866cm}{\makebox[1ex]{\ensuremath{^{\hypertarget{HG10GAMMA5}{a}}}} Placement of transition in the level scheme is uncertain.}\\
\vspace{0.5cm}
\end{landscape}\clearpage
\clearpage
\begin{figure}[h]
\begin{center}
\includegraphics{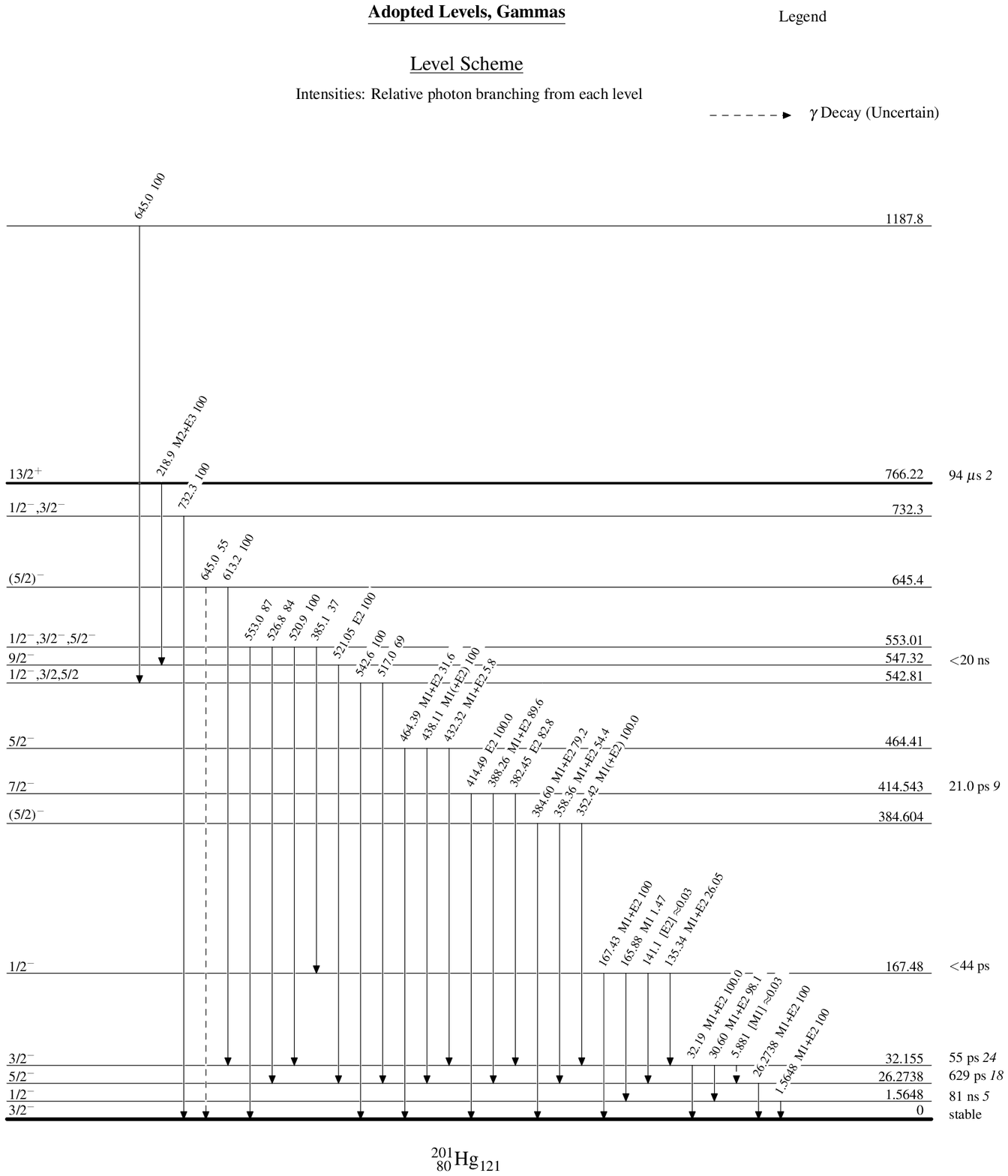}\\
\end{center}
\end{figure}
\clearpage
%201AU B- DECAY
\subsection[\hspace{-0.2cm}\ensuremath{^{\textnormal{201}}}Au \ensuremath{\beta}\ensuremath{^{-}} decay]{ }
\vspace{-27pt}
\vspace{0.3cm}
\hypertarget{AU11}{{\bf \small \underline{\ensuremath{^{\textnormal{201}}}Au \ensuremath{\beta}\ensuremath{^{-}} decay\hspace{0.2in}\href{https://www.nndc.bnl.gov/nsr/nsrlink.jsp?1972Pa24,B}{1972Pa24}}}}\\
\vspace{4pt}
\vspace{8pt}
\parbox[b][0.3cm]{17.7cm}{\addtolength{\parindent}{-0.2in}Parent: $^{201}$Au: E=0; J$^{\pi}$=3/2\ensuremath{^{+}}; T$_{1/2}$=26.0 min {\it 8}; Q(\ensuremath{\beta}\ensuremath{^{-}})=1262 {\it 3}; \%\ensuremath{\beta}\ensuremath{^{-}} decay=100.0

}\\
\parbox[b][0.3cm]{17.7cm}{\addtolength{\parindent}{-0.2in}\href{https://www.nndc.bnl.gov/nsr/nsrlink.jsp?1972Pa24,B}{1972Pa24}: Source produced by irradiating natural mercury targets of 10-100 g with 14.5 MeV neutrons; Detectors: Ge(Li), Si(Li)}\\
\parbox[b][0.3cm]{17.7cm}{and NaI(Tl); Measured: \ensuremath{\gamma} singles, \ensuremath{\gamma}\ensuremath{\gamma} coin, \ensuremath{\beta}\ensuremath{^{-}}, E\ensuremath{\gamma}, I\ensuremath{\gamma}.}\\
\vspace{12pt}
\underline{$^{201}$Hg Levels}\\
\begin{longtable}{ccc|ccc|ccc@{\extracolsep{\fill}}c}
\multicolumn{2}{c}{E(level)$^{{\hyperlink{HG11LEVEL0}{\dagger}}}$}&J$^{\pi}$$^{{\hyperlink{HG11LEVEL1}{\ddagger}}}$&\multicolumn{2}{c}{E(level)$^{{\hyperlink{HG11LEVEL0}{\dagger}}}$}&J$^{\pi}$$^{{\hyperlink{HG11LEVEL1}{\ddagger}}}$&\multicolumn{2}{c}{E(level)$^{{\hyperlink{HG11LEVEL0}{\dagger}}}$}&J$^{\pi}$$^{{\hyperlink{HG11LEVEL1}{\ddagger}}}$&\\[-.2cm]
\multicolumn{2}{c}{\hrulefill}&\hrulefill&\multicolumn{2}{c}{\hrulefill}&\hrulefill&\multicolumn{2}{c}{\hrulefill}&\hrulefill&
\endfirsthead
\multicolumn{1}{r@{}}{0}&\multicolumn{1}{@{.}l}{0}&\multicolumn{1}{l|}{3/2\ensuremath{^{-}}}&\multicolumn{1}{r@{}}{167}&\multicolumn{1}{@{.}l}{48 {\it 4}}&\multicolumn{1}{l|}{1/2\ensuremath{^{-}}}&\multicolumn{1}{r@{}}{552}&\multicolumn{1}{@{.}l}{89 {\it 13}}&\multicolumn{1}{l}{1/2\ensuremath{^{-}},3/2\ensuremath{^{-}},5/2\ensuremath{^{-}}}&\\
\multicolumn{1}{r@{}}{1}&\multicolumn{1}{@{.}l}{5648 {\it 10}}&\multicolumn{1}{l|}{1/2\ensuremath{^{-}}}&\multicolumn{1}{r@{}}{384}&\multicolumn{1}{@{.}l}{605 {\it 17}}&\multicolumn{1}{l|}{(5/2)\ensuremath{^{-}}}&\multicolumn{1}{r@{}}{645}&\multicolumn{1}{@{.}l}{4 {\it 3}}&\multicolumn{1}{l}{(5/2)\ensuremath{^{-}}}&\\
\multicolumn{1}{r@{}}{26}&\multicolumn{1}{@{.}l}{2738 {\it 3}}&\multicolumn{1}{l|}{5/2\ensuremath{^{-}}}&\multicolumn{1}{r@{}}{464}&\multicolumn{1}{@{.}l}{41 {\it 4}}&\multicolumn{1}{l|}{5/2\ensuremath{^{-}}}&\multicolumn{1}{r@{}}{732}&\multicolumn{1}{@{.}l}{3 {\it 4}}&\multicolumn{1}{l}{1/2\ensuremath{^{-}},3/2\ensuremath{^{-}}}&\\
\multicolumn{1}{r@{}}{32}&\multicolumn{1}{@{.}l}{168 {\it 14}}&\multicolumn{1}{l|}{3/2\ensuremath{^{-}}}&\multicolumn{1}{r@{}}{542}&\multicolumn{1}{@{.}l}{81 {\it 17}}&\multicolumn{1}{l|}{1/2\ensuremath{^{-}},3/2,5/2}&\multicolumn{1}{r@{}}{1187}&\multicolumn{1}{@{.}l}{8 {\it 5}}&&\\
\end{longtable}
\parbox[b][0.3cm]{17.7cm}{\makebox[1ex]{\ensuremath{^{\hypertarget{HG11LEVEL0}{\dagger}}}} From a least-squares fit to E\ensuremath{\gamma}.}\\
\parbox[b][0.3cm]{17.7cm}{\makebox[1ex]{\ensuremath{^{\hypertarget{HG11LEVEL1}{\ddagger}}}} From Adopted Levels.}\\
\vspace{0.5cm}

\underline{\ensuremath{\beta^-} radiations}\\
%
\parbox[b][0.3cm]{17.7cm}{\makebox[1ex]{\ensuremath{^{\hypertarget{HG11DECAY0}{\dagger}}}} From intensity balances, as explained in the text.}\\
\parbox[b][0.3cm]{17.7cm}{\makebox[1ex]{\ensuremath{^{\hypertarget{HG11DECAY1}{\ddagger}}}} Absolute intensity per 100 decays.}\\
\vspace{0.5cm}
\clearpage
\vspace{0.3cm}
\begin{landscape}
\vspace*{-0.5cm}
{\bf \small \underline{\ensuremath{^{\textnormal{201}}}Au \ensuremath{\beta}\ensuremath{^{-}} decay\hspace{0.2in}\href{https://www.nndc.bnl.gov/nsr/nsrlink.jsp?1972Pa24,B}{1972Pa24} (continued)}}\\
\vspace{0.3cm}
\underline{$\gamma$($^{201}$Hg)}\\
\vspace{0.34cm}
\parbox[b][0.3cm]{21.881866cm}{\addtolength{\parindent}{-0.254cm}I\ensuremath{\gamma} normalization: From I\ensuremath{\beta}(167.47 keV level)=3.5\% (\href{https://www.nndc.bnl.gov/nsr/nsrlink.jsp?1972Pa24,B}{1972Pa24}) and I(\ensuremath{\gamma}+ce) deduced from the decay scheme. Evaluator assigns 10\% uncertainty to the}\\
\parbox[b][0.3cm]{21.881866cm}{I\ensuremath{\beta}(167.47-keV level) value.}\\
\vspace{0.34cm}

\parbox[b][0.3cm]{21.881866cm}{\makebox[1ex]{\ensuremath{^{\hypertarget{AU11GAMMA0}{\dagger}}}} From \href{https://www.nndc.bnl.gov/nsr/nsrlink.jsp?1972Pa24,B}{1972Pa24}, unless otherwise stated. K\ensuremath{\alpha}{} x ray \ensuremath{\approx}90, L\ensuremath{_{\ensuremath{\alpha}}} x ray \ensuremath{\approx}85 and L\ensuremath{_{\ensuremath{\beta}}} x ray \ensuremath{\approx}55 (\href{https://www.nndc.bnl.gov/nsr/nsrlink.jsp?1972Pa24,B}{1972Pa24}).}\\
\parbox[b][0.3cm]{21.881866cm}{\makebox[1ex]{\ensuremath{^{\hypertarget{AU11GAMMA1}{\ddagger}}}} From adopted gammas.}\\
\parbox[b][0.3cm]{21.881866cm}{\makebox[1ex]{\ensuremath{^{\hypertarget{AU11GAMMA2}{\#}}}} For absolute intensity per 100 decays, multiply by 0.0182 \textit{22}.}\\
\parbox[b][0.3cm]{21.881866cm}{\makebox[1ex]{\ensuremath{^{\hypertarget{AU11GAMMA3}{@}}}} Total theoretical internal conversion coefficients, calculated using the BrIcc code (\href{https://www.nndc.bnl.gov/nsr/nsrlink.jsp?2008Ki07,B}{2008Ki07}) with Frozen orbital approximation based on \ensuremath{\gamma}-ray energies,}\\
\parbox[b][0.3cm]{21.881866cm}{{\ }{\ }assigned multipolarities, and mixing ratios, unless otherwise specified.}\\
\parbox[b][0.3cm]{21.881866cm}{\makebox[1ex]{\ensuremath{^{\hypertarget{AU11GAMMA4}{\&}}}} Placement of transition in the level scheme is uncertain.}\\
\vspace{0.5cm}
\end{landscape}\clearpage
\clearpage
\begin{figure}[h]
\begin{center}
\includegraphics{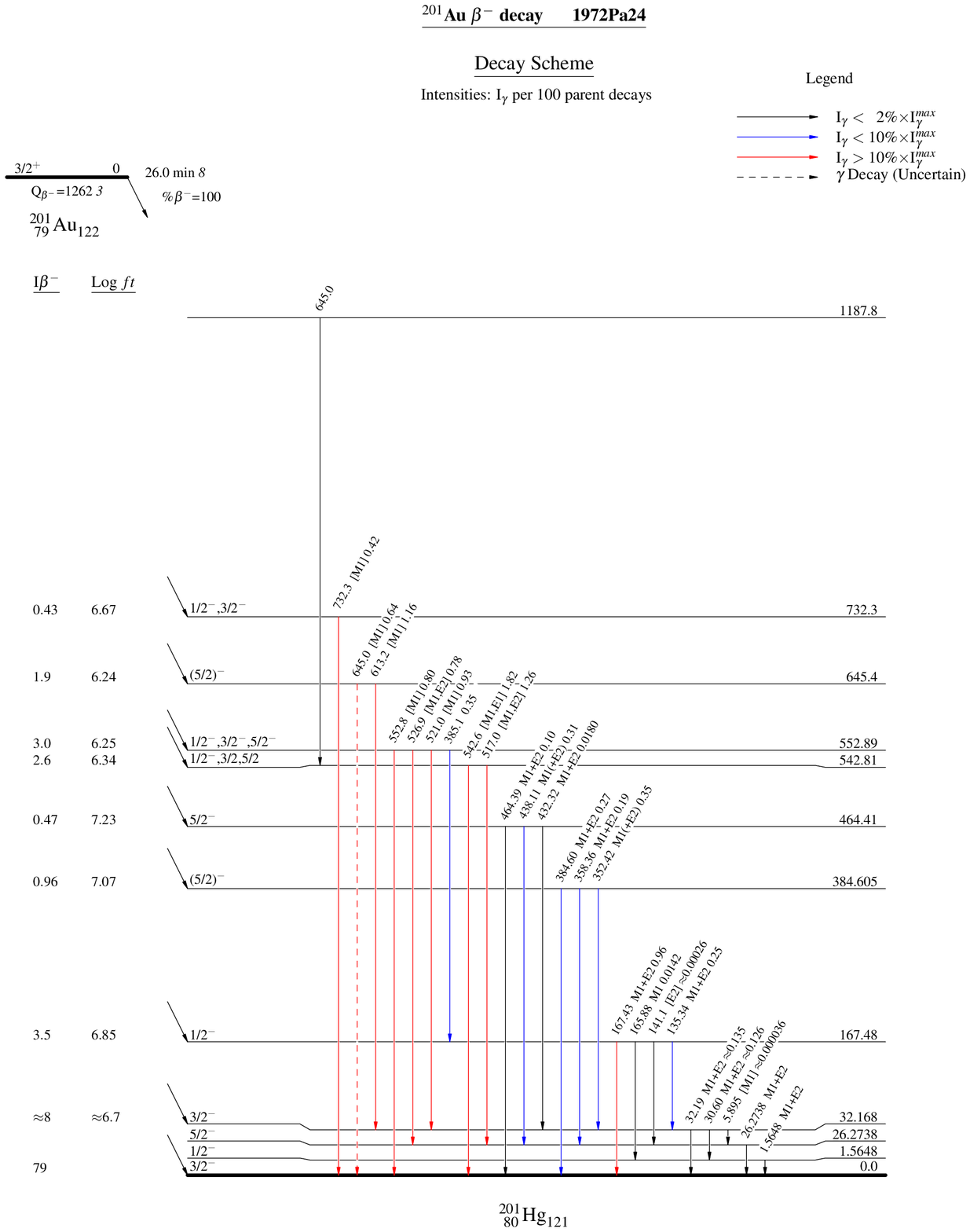}\\
\end{center}
\end{figure}
\clearpage
%201HG IT DECAY
\subsection[\hspace{-0.2cm}\ensuremath{^{\textnormal{201}}}Hg IT decay]{ }
\vspace{-27pt}
\vspace{0.3cm}
\hypertarget{HG12}{{\bf \small \underline{\ensuremath{^{\textnormal{201}}}Hg IT decay\hspace{0.2in}\href{https://www.nndc.bnl.gov/nsr/nsrlink.jsp?1990Lo17,B}{1990Lo17}}}}\\
\vspace{4pt}
\vspace{8pt}
\parbox[b][0.3cm]{17.7cm}{\addtolength{\parindent}{-0.2in}Parent: $^{201}$Hg: E=766.22 {\it 15}; J$^{\pi}$=13/2\ensuremath{^{+}}; T$_{1/2}$=94 \ensuremath{\mu}s {\it 2}; \%IT decay=100.0

}\\
\parbox[b][0.3cm]{17.7cm}{\addtolength{\parindent}{-0.2in}\href{https://www.nndc.bnl.gov/nsr/nsrlink.jsp?1990Lo17,B}{1990Lo17}: \ensuremath{^{\textnormal{198}}}Pt(\ensuremath{\alpha},n\ensuremath{\gamma}); E(\ensuremath{\alpha})=18.1 MeV; Target: 1.55 mg/cm\ensuremath{^{\textnormal{2}}} thick enriched to 95.8\% in \ensuremath{^{\textnormal{198}}}Pt; Detector: HPGE, electron}\\
\parbox[b][0.3cm]{17.7cm}{spectrometer; Measured: \ensuremath{\gamma} singles, \ensuremath{\gamma}\ensuremath{\gamma} coin, \ensuremath{\gamma}(t), E\ensuremath{\gamma}, I\ensuremath{\gamma}, ce; Deduced: \ensuremath{\alpha}(K)exp, level scheme.}\\
\parbox[b][0.3cm]{17.7cm}{\addtolength{\parindent}{-0.2in}Others: \href{https://www.nndc.bnl.gov/nsr/nsrlink.jsp?1976Uy01,B}{1976Uy01}, \href{https://www.nndc.bnl.gov/nsr/nsrlink.jsp?1964Br27,B}{1964Br27}, \href{https://www.nndc.bnl.gov/nsr/nsrlink.jsp?1962Eu01,B}{1962Eu01}, \href{https://www.nndc.bnl.gov/nsr/nsrlink.jsp?1961Kr01,B}{1961Kr01}.}\\
\vspace{12pt}
\underline{$^{201}$Hg Levels}\\

\parbox[b][0.3cm]{17.7cm}{\makebox[1ex]{\ensuremath{^{\hypertarget{HG12LEVEL0}{\dagger}}}} From a least-squares fit to E\ensuremath{\gamma}.}\\
\parbox[b][0.3cm]{17.7cm}{\makebox[1ex]{\ensuremath{^{\hypertarget{HG12LEVEL1}{\ddagger}}}} From deduced transition multipolarities, unless otherwise stated.}\\
\parbox[b][0.3cm]{17.7cm}{\makebox[1ex]{\ensuremath{^{\hypertarget{HG12LEVEL2}{\#}}}} From Adopted Levels.}\\
\vspace{0.5cm}
\underline{$\gamma$($^{201}$Hg)}\\
%
\parbox[b][0.3cm]{17.7cm}{\makebox[1ex]{\ensuremath{^{\hypertarget{HG12GAMMA0}{\dagger}}}} From \href{https://www.nndc.bnl.gov/nsr/nsrlink.jsp?1976Uy01,B}{1976Uy01}, unless otherwise stated.}\\
\parbox[b][0.3cm]{17.7cm}{\makebox[1ex]{\ensuremath{^{\hypertarget{HG12GAMMA1}{\ddagger}}}} From \ensuremath{\alpha} and by assuming I(\ensuremath{\gamma}+ce)=100 for each \ensuremath{\gamma}. I(K{} x ray):I(219\ensuremath{\gamma}):I(521\ensuremath{\gamma})=100:26:134 (\href{https://www.nndc.bnl.gov/nsr/nsrlink.jsp?1964Br27,B}{1964Br27}).}\\
\parbox[b][0.3cm]{17.7cm}{\makebox[1ex]{\ensuremath{^{\hypertarget{HG12GAMMA2}{\#}}}} Absolute intensity per 100 decays.}\\
\parbox[b][0.3cm]{17.7cm}{\makebox[1ex]{\ensuremath{^{\hypertarget{HG12GAMMA3}{@}}}} Total theoretical internal conversion coefficients, calculated using the BrIcc code (\href{https://www.nndc.bnl.gov/nsr/nsrlink.jsp?2008Ki07,B}{2008Ki07}) with Frozen orbital approximation}\\
\parbox[b][0.3cm]{17.7cm}{{\ }{\ }based on \ensuremath{\gamma}-ray energies, assigned multipolarities, and mixing ratios, unless otherwise specified.}\\
\vspace{0.5cm}
\clearpage
\begin{figure}[h]
\begin{center}
\includegraphics{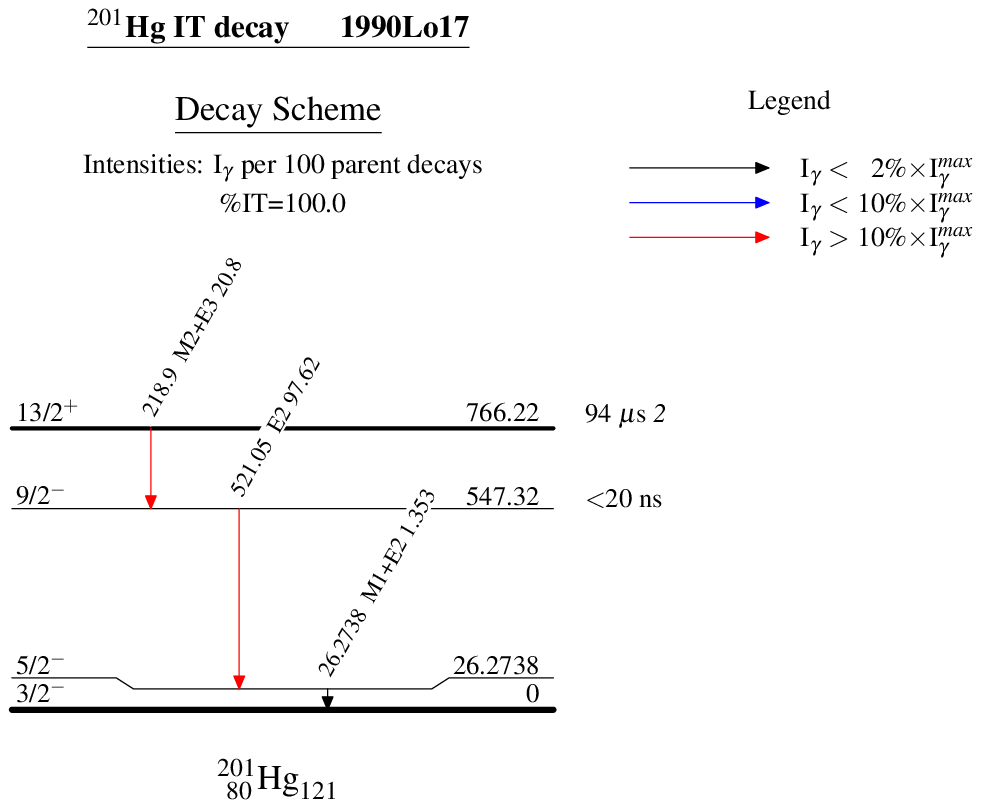}\\
\end{center}
\end{figure}
\clearpage
%201TL EC DECAY
\subsection[\hspace{-0.2cm}\ensuremath{^{\textnormal{201}}}Tl \ensuremath{\varepsilon} decay]{ }
\vspace{-27pt}
\vspace{0.3cm}
\hypertarget{TL13}{{\bf \small \underline{\ensuremath{^{\textnormal{201}}}Tl \ensuremath{\varepsilon} decay\hspace{0.2in}\href{https://www.nndc.bnl.gov/nsr/nsrlink.jsp?1990Co07,B}{1990Co07},\href{https://www.nndc.bnl.gov/nsr/nsrlink.jsp?1990Ka08,B}{1990Ka08},\href{https://www.nndc.bnl.gov/nsr/nsrlink.jsp?1979De42,B}{1979De42}}}}\\
\vspace{4pt}
\vspace{8pt}
\parbox[b][0.3cm]{17.7cm}{\addtolength{\parindent}{-0.2in}Parent: $^{201}$Tl: E=0.0; J$^{\pi}$=1/2\ensuremath{^{+}}; T$_{1/2}$=3.0421 d {\it 8}; Q(\ensuremath{\varepsilon})=482 {\it 14}; \%\ensuremath{\varepsilon} decay=100.0

}\\
\parbox[b][0.3cm]{17.7cm}{\addtolength{\parindent}{-0.2in}\href{https://www.nndc.bnl.gov/nsr/nsrlink.jsp?1990Co07,B}{1990Co07}: inter-comparison data performed at NIST, NPL and PTB metrology labs using samples produced by the same solution}\\
\parbox[b][0.3cm]{17.7cm}{of \ensuremath{^{\textnormal{201}}}Tl and the 4\ensuremath{\pi}-\ensuremath{\gamma} coincidence systems. In each case, corrections were applied for the presence of \ensuremath{^{\textnormal{200}}}Tl and \ensuremath{^{\textnormal{202}}}Tl}\\
\parbox[b][0.3cm]{17.7cm}{contaminants.}\\
\parbox[b][0.3cm]{17.7cm}{\addtolength{\parindent}{-0.2in}Others: \href{https://www.nndc.bnl.gov/nsr/nsrlink.jsp?2007Me12,B}{2007Me12}, \href{https://www.nndc.bnl.gov/nsr/nsrlink.jsp?2004De02,B}{2004De02}, \href{https://www.nndc.bnl.gov/nsr/nsrlink.jsp?1989Pl04,B}{1989Pl04}, \href{https://www.nndc.bnl.gov/nsr/nsrlink.jsp?1991Dr09,B}{1991Dr09}, \href{https://www.nndc.bnl.gov/nsr/nsrlink.jsp?1987Dr06,B}{1987Dr06}, \href{https://www.nndc.bnl.gov/nsr/nsrlink.jsp?1987Fu08,B}{1987Fu08}, \href{https://www.nndc.bnl.gov/nsr/nsrlink.jsp?1983Fu22,B}{1983Fu22}, \href{https://www.nndc.bnl.gov/nsr/nsrlink.jsp?1983SC38,B}{1983SC38}, \href{https://www.nndc.bnl.gov/nsr/nsrlink.jsp?1978No06,B}{1978No06}, \href{https://www.nndc.bnl.gov/nsr/nsrlink.jsp?1977Na31,B}{1977Na31}, \href{https://www.nndc.bnl.gov/nsr/nsrlink.jsp?1975Ho08,B}{1975Ho08},}\\
\parbox[b][0.3cm]{17.7cm}{\href{https://www.nndc.bnl.gov/nsr/nsrlink.jsp?1976HiZN,B}{1976HiZN}, \href{https://www.nndc.bnl.gov/nsr/nsrlink.jsp?1960Gu05,B}{1960Gu05}, \href{https://www.nndc.bnl.gov/nsr/nsrlink.jsp?1960He05,B}{1960He05}.}\\
\vspace{12pt}
\underline{$^{201}$Hg Levels}\\

\parbox[b][0.3cm]{17.7cm}{\makebox[1ex]{\ensuremath{^{\hypertarget{HG13LEVEL0}{\dagger}}}} From a least-squares fit to E\ensuremath{\gamma}.}\\
\parbox[b][0.3cm]{17.7cm}{\makebox[1ex]{\ensuremath{^{\hypertarget{HG13LEVEL1}{\ddagger}}}} From Adopted Levels.}\\
\vspace{0.5cm}

\underline{\ensuremath{\varepsilon} radiations}\\
%
\parbox[b][0.3cm]{17.7cm}{\makebox[1ex]{\ensuremath{^{\hypertarget{HG13DECAY0}{\dagger}}}} Estimated by the evaluator from intensity balances and the adopted decay scheme, unless otherwise stated.}\\
\parbox[b][0.3cm]{17.7cm}{\makebox[1ex]{\ensuremath{^{\hypertarget{HG13DECAY1}{\ddagger}}}} Absolute intensity per 100 decays.}\\
\vspace{0.5cm}
\clearpage
\vspace{0.3cm}
\begin{landscape}
\vspace*{-0.5cm}
{\bf \small \underline{\ensuremath{^{\textnormal{201}}}Tl \ensuremath{\varepsilon} decay\hspace{0.2in}\href{https://www.nndc.bnl.gov/nsr/nsrlink.jsp?1990Co07,B}{1990Co07},\href{https://www.nndc.bnl.gov/nsr/nsrlink.jsp?1990Ka08,B}{1990Ka08},\href{https://www.nndc.bnl.gov/nsr/nsrlink.jsp?1979De42,B}{1979De42} (continued)}}\\
\vspace{0.3cm}
\underline{$\gamma$($^{201}$Hg)}\\
\vspace{0.34cm}
\parbox[b][0.3cm]{21.881866cm}{\addtolength{\parindent}{-0.254cm}I\ensuremath{\gamma} normalization: From I\ensuremath{\gamma}(167\ensuremath{\gamma})=10.00\% \textit{6} (\href{https://www.nndc.bnl.gov/nsr/nsrlink.jsp?1990Co07,B}{1990Co07}), weighted average of 9.88\% \textit{8} (NIST), 10.05\% \textit{17} (NPL) and 10.18\% \textit{10} (PTB). Others: I\ensuremath{\gamma}(167\ensuremath{\gamma}):}\\
\parbox[b][0.3cm]{21.881866cm}{9.81\% \textit{12} (\href{https://www.nndc.bnl.gov/nsr/nsrlink.jsp?1990Ka08,B}{1990Ka08}), 10.60\% \textit{15} (\href{https://www.nndc.bnl.gov/nsr/nsrlink.jsp?1989Pl04,B}{1989Pl04}), 10.25\% \textit{10} (\href{https://www.nndc.bnl.gov/nsr/nsrlink.jsp?1983Fu22,B}{1983Fu22}), 10.60\% \textit{12} (\href{https://www.nndc.bnl.gov/nsr/nsrlink.jsp?1979De42,B}{1979De42}), 10.00\% \textit{17} (\href{https://www.nndc.bnl.gov/nsr/nsrlink.jsp?1976HiZN,B}{1976HiZN}), 10.00\% (\href{https://www.nndc.bnl.gov/nsr/nsrlink.jsp?1975Ho08,B}{1975Ho08}) and 8.4\% \textit{4}}\\
\parbox[b][0.3cm]{21.881866cm}{(\href{https://www.nndc.bnl.gov/nsr/nsrlink.jsp?1960He05,B}{1960He05}). The total energy realized in \ensuremath{^{\textnormal{201}}}Tl \ensuremath{\varepsilon} decay is calculated using RADLST as 471 keV \textit{14}. It is in a good agreement with Q(g.s.)=482 keV \textit{14}.}\\
\vspace{0.34cm}

\raggedright\texttt{}\\
\raggedright\texttt{\ \ \ \ \ \ \ \ \ \ \ \ \ \ \ x-ray\ \ \ \ \ \ \ \ \ \ \ \ \ \ \ \ \ E\ensuremath{\gamma}\ \ \ \ \ \ \ \ \ \ \ \ \ \ \ \ \ \ \ \ \ I\ensuremath{\gamma}}\\
\raggedright\texttt{\ \ \ \ \ \ \ \ \ \ \ \ \ \ \ \ \ \ \ \ \ \ \ \ \ \ \ \ \ \ \ \ \ \ \ \ \ keV\ \ \ \ \ \ \ \ \ \ \ \ \ per\ 100\ \hspace{-0.04cm}\ensuremath{\varepsilon}\ \ decays}\\
\raggedright\texttt{\ \ \ \ \ \ \ \ \ \ \ --------------\ \ \ \ \ \ \ \ \ \ -------\ \ \ \ \ \ \ \ \ \ \ -----------------}\\
\raggedright\texttt{\ \ \ \ \ \ \ \ \ \ \ K\ensuremath{\alpha}\ensuremath{_{\textnormal{1}}}{}\ \ \ \ \ \ x\ ray\ \ \ \ \ \ \ \ \ \ \ 70.8\ \ \ \ \ \ \ \ \ \ \ \ \ \ \ \ 44.6\ \ \ \textit{5}}\\
\raggedright\texttt{\ \ \ \ \ \ \ \ \ \ \ K\ensuremath{\alpha}\ensuremath{_{\textnormal{2}}}{}\ \ \ \ \ \ x\ ray\ \ \ \ \ \ \ \ \ \ \ 68.9\ \ \ \ \ \ \ \ \ \ \ \ \ \ \ \ 26.3\ \ \ \textit{3}}\\
\raggedright\texttt{\ \ \ \ \ \ \ \ \ \ \ K\ensuremath{\alpha}{}\ \ \ \ \ \ \ x\ ray\ \ \ \ \ \ \ \ \ \ \ \ \ \ \ \ \ \ \ \ \ \ \ \ \ \ \ \ \ \ \ 71.1\ \ \ \textit{5}}\\
\raggedright\texttt{\ \ \ \ \ \ \ \ \ \ \ K\ensuremath{\beta}\ensuremath{_{\textnormal{1}}}$'${}\ \ \ \ \ \ x\ ray\ \ \ \ \ \ \ \ \ \ \ 80.2\ \ \ \ \ \ \ \ \ \ \ \ \ \ \ \ 15.3\ \ \ \textit{4}}\\
\raggedright\texttt{\ \ \ \ \ \ \ \ \ \ \ K\ensuremath{\beta}\ensuremath{_{\textnormal{2}}}$'${}\ \ \ \ \ \ x\ ray\ \ \ \ \ \ \ \ \ \ \ 80.5\ \ \ \ \ \ \ \ \ \ \ \ \ \ \ \ 4.59\ \ \textit{15}}\\
\raggedright\texttt{\ \ \ \ \ \ \ \ \ \ \ K\ensuremath{\beta}{}\ \ \ \ \ \ \ x\ ray\ \ \ \ \ \ \ \ \ \ \ \ \ \ \ \ \ \ \ \ \ \ \ \ \ \ \ \ \ \ \ 20.0\ \ \ \textit{3}}\\
\raggedright\texttt{}\\
\raggedright\texttt{\ I\ensuremath{\gamma}\ \ -\ Weighted\ average\ of\ values\ given\ in\ \href{https://www.nndc.bnl.gov/nsr/nsrlink.jsp?1976HiZN,B}{1976HiZN},\ \href{https://www.nndc.bnl.gov/nsr/nsrlink.jsp?1979De42,B}{1979De42},\ \href{https://www.nndc.bnl.gov/nsr/nsrlink.jsp?1983Fu22,B}{1983Fu22}}\\
\raggedright\texttt{\ \ \ \ \ \ and\ \href{https://www.nndc.bnl.gov/nsr/nsrlink.jsp?1990Ka08,B}{1990Ka08}.}\\
\raggedright\texttt{}\\
\centering \texttt{}\\
\vspace{-0.5cm}

\parbox[b][0.3cm]{21.881866cm}{\makebox[1ex]{\ensuremath{^{\hypertarget{TL13GAMMA0}{\dagger}}}} From \href{https://www.nndc.bnl.gov/nsr/nsrlink.jsp?1960He05,B}{1960He05}, unless otherwise stated.}\\
\parbox[b][0.3cm]{21.881866cm}{\makebox[1ex]{\ensuremath{^{\hypertarget{TL13GAMMA1}{\ddagger}}}} For absolute intensity per 100 decays, multiply by 0.1000 \textit{6}.}\\
\parbox[b][0.3cm]{21.881866cm}{\makebox[1ex]{\ensuremath{^{\hypertarget{TL13GAMMA2}{\#}}}} Total theoretical internal conversion coefficients, calculated using the BrIcc code (\href{https://www.nndc.bnl.gov/nsr/nsrlink.jsp?2008Ki07,B}{2008Ki07}) with Frozen orbital approximation based on \ensuremath{\gamma}-ray energies,}\\
\parbox[b][0.3cm]{21.881866cm}{{\ }{\ }assigned multipolarities, and mixing ratios, unless otherwise specified.}\\
\vspace{0.5cm}
\end{landscape}\clearpage
\clearpage
\begin{figure}[h]
\begin{center}
\includegraphics{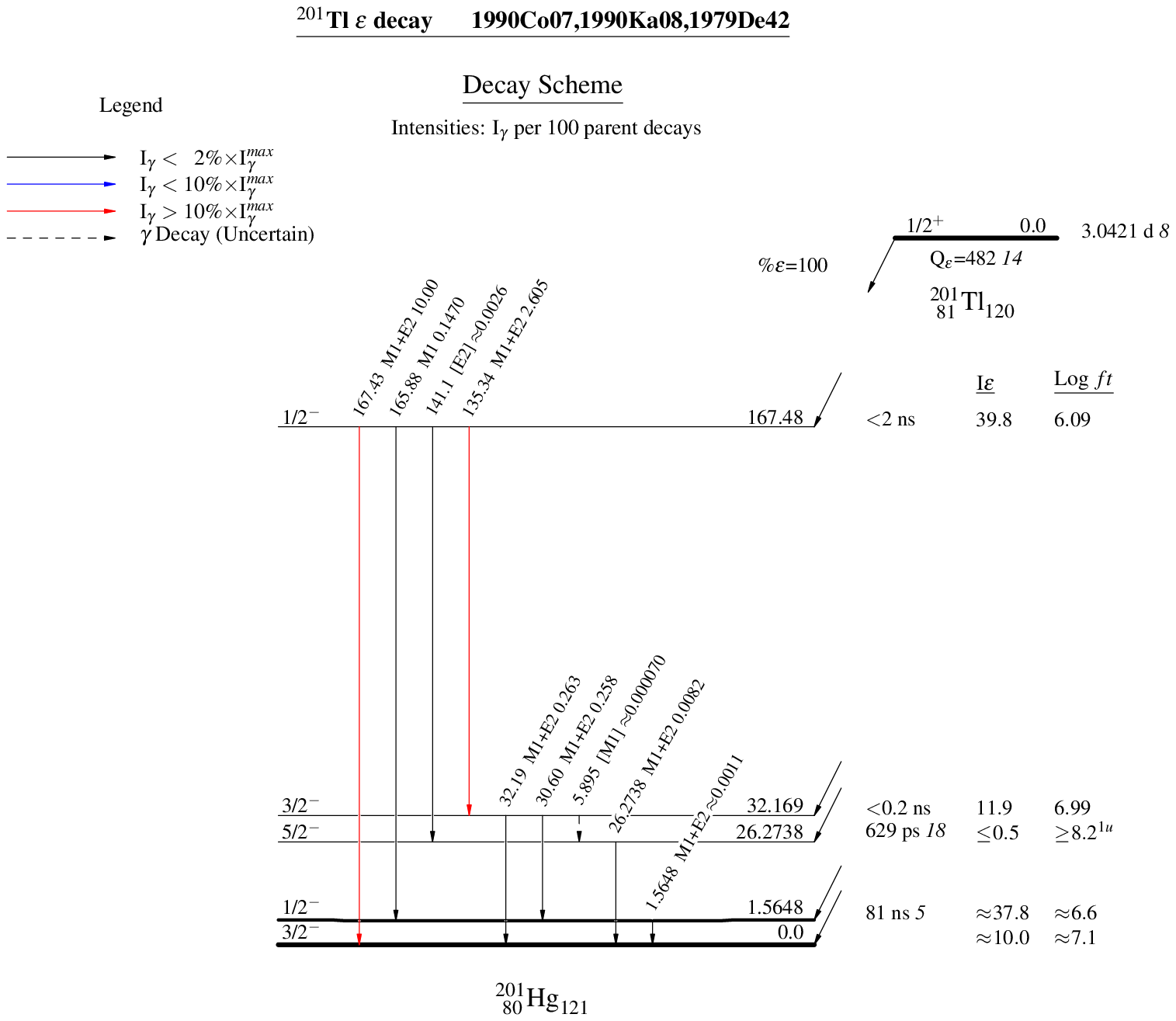}\\
\end{center}
\end{figure}
\clearpage
%200HG(D,P)
\subsection[\hspace{-0.2cm}\ensuremath{^{\textnormal{200}}}Hg(d,p)]{ }
\vspace{-27pt}
\vspace{0.3cm}
\hypertarget{HG14}{{\bf \small \underline{\ensuremath{^{\textnormal{200}}}Hg(d,p)\hspace{0.2in}\href{https://www.nndc.bnl.gov/nsr/nsrlink.jsp?1972Mo12,B}{1972Mo12}}}}\\
\vspace{4pt}
\vspace{8pt}
\parbox[b][0.3cm]{17.7cm}{\addtolength{\parindent}{-0.2in}Beam: E(d)=17 MeV; Target: \ensuremath{^{\textnormal{202}}}Hg, but isotopic purity is unknown; Detectors: photographic emulsions, split-pole spectrograph,}\\
\parbox[b][0.3cm]{17.7cm}{FWHM=10-14 keV.}\\
\vspace{12pt}
\underline{$^{201}$Hg Levels}\\

\begin{textblock}{29}(0,27.3)
Continued on next page (footnotes at end of table)
\end{textblock}
\clearpage
%
\parbox[b][0.3cm]{17.7cm}{\makebox[1ex]{\ensuremath{^{\hypertarget{HG14LEVEL0}{\dagger}}}} From \href{https://www.nndc.bnl.gov/nsr/nsrlink.jsp?1972Mo12,B}{1972Mo12}. \ensuremath{\Delta}E=0.4\% for well-resolved peaks.}\\
\parbox[b][0.3cm]{17.7cm}{\makebox[1ex]{\ensuremath{^{\hypertarget{HG14LEVEL1}{\ddagger}}}} From the deduced L values and spectroscopic factors (\href{https://www.nndc.bnl.gov/nsr/nsrlink.jsp?1972Mo12,B}{1972Mo12}), unless otherwise stated.}\\
\parbox[b][0.3cm]{17.7cm}{\makebox[1ex]{\ensuremath{^{\hypertarget{HG14LEVEL2}{\#}}}} \ensuremath{\Delta}S\ensuremath{\approx}\ensuremath{\pm}50\%. S=N*(d\ensuremath{\sigma}/d\ensuremath{\Omega})\ensuremath{_{\textnormal{expt}}}/(d\ensuremath{\sigma}/d\ensuremath{\Omega})\ensuremath{_{\textnormal{DWBA}}}. N=(1/1.5)/(2J+1).}\\
\parbox[b][0.3cm]{17.7cm}{\makebox[1ex]{\ensuremath{^{\hypertarget{HG14LEVEL3}{@}}}} Dominant configuration=\ensuremath{\nu} p\ensuremath{_{\textnormal{3/2}}^{\textnormal{$-$1}}}.}\\
\parbox[b][0.3cm]{17.7cm}{\makebox[1ex]{\ensuremath{^{\hypertarget{HG14LEVEL4}{\&}}}} Dominant configuration=\ensuremath{\nu} f\ensuremath{_{\textnormal{5/2}}^{\textnormal{$-$1}}}.}\\
\vspace{0.5cm}
\clearpage
%201HG(G,G')
\subsection[\hspace{-0.2cm}\ensuremath{^{\textnormal{201}}}Hg(\ensuremath{\gamma},\ensuremath{\gamma}\ensuremath{'})]{ }
\vspace{-27pt}
\vspace{0.3cm}
\hypertarget{HG15}{{\bf \small \underline{\ensuremath{^{\textnormal{201}}}Hg(\ensuremath{\gamma},\ensuremath{\gamma}\ensuremath{'})\hspace{0.2in}\href{https://www.nndc.bnl.gov/nsr/nsrlink.jsp?1971Wa17,B}{1971Wa17},\href{https://www.nndc.bnl.gov/nsr/nsrlink.jsp?2005Is19,B}{2005Is19},\href{https://www.nndc.bnl.gov/nsr/nsrlink.jsp?2018Yo02,B}{2018Yo02}}}}\\
\vspace{4pt}
\vspace{8pt}
\parbox[b][0.3cm]{17.7cm}{\addtolength{\parindent}{-0.2in}\href{https://www.nndc.bnl.gov/nsr/nsrlink.jsp?1971Wa17,B}{1971Wa17}: Mossbauer transmission measurement with 32.2-keV \ensuremath{\gamma} from \ensuremath{^{\textnormal{201}}}Tl \ensuremath{\varepsilon} decay source.}\\
\parbox[b][0.3cm]{17.7cm}{\addtolength{\parindent}{-0.2in}\href{https://www.nndc.bnl.gov/nsr/nsrlink.jsp?2005Is19,B}{2005Is19}: synchrotron-based nuclear resonant scattering experiment at the Spring-8 facility.}\\
\parbox[b][0.3cm]{17.7cm}{\addtolength{\parindent}{-0.2in}\href{https://www.nndc.bnl.gov/nsr/nsrlink.jsp?2018Yo02,B}{2018Yo02}: synchrotron-based nuclear resonant scattering experiment at the Spring-8 facility.}\\
\vspace{12pt}
\underline{$^{201}$Hg Levels}\\

\parbox[b][0.3cm]{17.7cm}{\makebox[1ex]{\ensuremath{^{\hypertarget{HG15LEVEL0}{\dagger}}}} From Adopted Levels, unless otherwise stated.}\\
\vspace{0.5cm}
\underline{$\gamma$($^{201}$Hg)}\\
%
\parbox[b][0.3cm]{17.7cm}{\makebox[1ex]{\ensuremath{^{\hypertarget{HG15GAMMA0}{\dagger}}}} From adopted gammas.}\\
\vspace{0.5cm}
\begin{figure}[h]
\begin{center}
\includegraphics{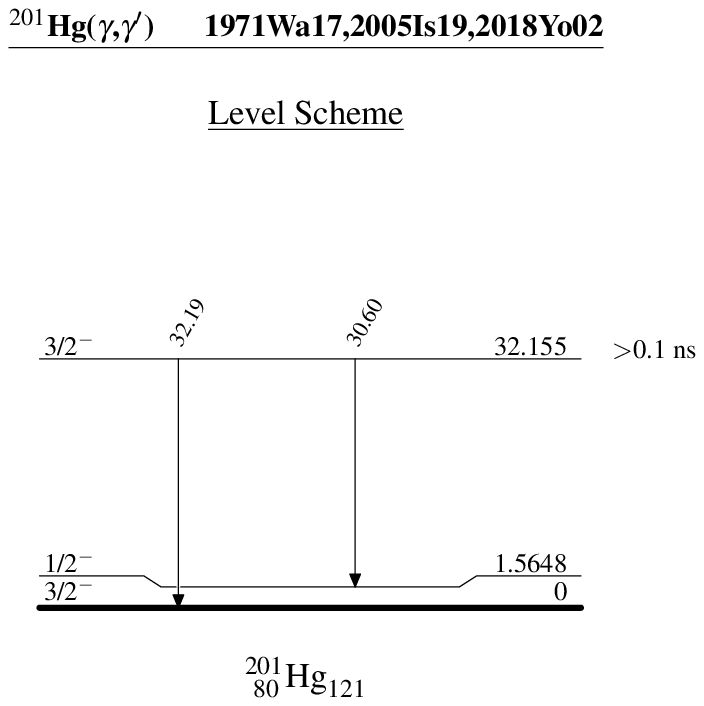}\\
\end{center}
\end{figure}
\clearpage
%201HG(D,D'),201HG(P,P')
\subsection[\hspace{-0.2cm}\ensuremath{^{\textnormal{201}}}Hg(d,d\ensuremath{'}),\ensuremath{^{\textnormal{201}}}Hg(p,p\ensuremath{'})]{ }
\vspace{-27pt}
\vspace{0.3cm}
\hypertarget{HG16}{{\bf \small \underline{\ensuremath{^{\textnormal{201}}}Hg(d,d\ensuremath{'}),\ensuremath{^{\textnormal{201}}}Hg(p,p\ensuremath{'})\hspace{0.2in}\href{https://www.nndc.bnl.gov/nsr/nsrlink.jsp?1972Mo12,B}{1972Mo12}}}}\\
\vspace{4pt}
\vspace{8pt}
\parbox[b][0.3cm]{17.7cm}{\addtolength{\parindent}{-0.2in}Beam: E(d)=17 MeV, \ensuremath{\theta}=90\ensuremath{^\circ}; \ensuremath{^{\textnormal{201}}}Hg(p,p\ensuremath{'}), \ensuremath{\theta}=75\ensuremath{^\circ},90\ensuremath{^\circ} Target: enriched \ensuremath{^{\textnormal{202}}}Hg, but isotopic purity is unknown; Detectors:}\\
\parbox[b][0.3cm]{17.7cm}{photographic emulsions, split-pole spectrograph, FWHM=8-9 keV.}\\
\vspace{12pt}
\underline{$^{201}$Hg Levels}\\
\begin{longtable}{ccc|ccc|ccc|cc@{\extracolsep{\fill}}c}
\multicolumn{2}{c}{E(level)$^{{\hyperlink{HG16LEVEL0}{\dagger}}}$}&J$^{\pi}$$^{{\hyperlink{HG16LEVEL1}{\ddagger}}}$&\multicolumn{2}{c}{E(level)$^{{\hyperlink{HG16LEVEL0}{\dagger}}}$}&J$^{\pi}$$^{{\hyperlink{HG16LEVEL1}{\ddagger}}}$&\multicolumn{2}{c}{E(level)$^{{\hyperlink{HG16LEVEL0}{\dagger}}}$}&J$^{\pi}$$^{{\hyperlink{HG16LEVEL1}{\ddagger}}}$&\multicolumn{2}{c}{E(level)$^{{\hyperlink{HG16LEVEL0}{\dagger}}}$}&\\[-.2cm]
\multicolumn{2}{c}{\hrulefill}&\hrulefill&\multicolumn{2}{c}{\hrulefill}&\hrulefill&\multicolumn{2}{c}{\hrulefill}&\hrulefill&\multicolumn{2}{c}{\hrulefill}&
\endfirsthead
\multicolumn{1}{r@{}}{0}&\multicolumn{1}{@{}l}{}&\multicolumn{1}{l|}{3/2\ensuremath{^{-}}}&\multicolumn{1}{r@{}}{412}&\multicolumn{1}{@{}l}{}&\multicolumn{1}{l|}{7/2\ensuremath{^{-}}}&\multicolumn{1}{r@{}}{1707}&\multicolumn{1}{@{}l}{}&\multicolumn{1}{l|}{(5/2\ensuremath{^{-}},7/2\ensuremath{^{-}})}&\multicolumn{1}{r@{}}{2891?}&\multicolumn{1}{@{}l}{}&\\
\multicolumn{1}{r@{}}{$\approx$32}&\multicolumn{1}{@{}l}{}&\multicolumn{1}{l|}{3/2\ensuremath{^{-}}}&\multicolumn{1}{r@{}}{465}&\multicolumn{1}{@{}l}{}&\multicolumn{1}{l|}{5/2\ensuremath{^{-}}}&\multicolumn{1}{r@{}}{2526}&\multicolumn{1}{@{}l}{}&&\multicolumn{1}{r@{}}{3735}&\multicolumn{1}{@{}l}{}&\\
\multicolumn{1}{r@{}}{163}&\multicolumn{1}{@{}l}{}&\multicolumn{1}{l|}{1/2\ensuremath{^{-}}}&\multicolumn{1}{r@{}}{1325}&\multicolumn{1}{@{}l}{}&&\multicolumn{1}{r@{}}{2629}&\multicolumn{1}{@{}l}{}&\multicolumn{1}{l|}{7/2\ensuremath{^{+}},9/2\ensuremath{^{+}}}&\multicolumn{1}{r@{}}{3965}&\multicolumn{1}{@{}l}{}&\\
\multicolumn{1}{r@{}}{382}&\multicolumn{1}{@{}l}{}&\multicolumn{1}{l|}{(5/2)\ensuremath{^{-}}}&\multicolumn{1}{r@{}}{1505}&\multicolumn{1}{@{}l}{}&&\multicolumn{1}{r@{}}{2681}&\multicolumn{1}{@{}l}{}&&&&\\
\end{longtable}
\parbox[b][0.3cm]{17.7cm}{\makebox[1ex]{\ensuremath{^{\hypertarget{HG16LEVEL0}{\dagger}}}} From \href{https://www.nndc.bnl.gov/nsr/nsrlink.jsp?1972Mo12,B}{1972Mo12}. \ensuremath{\Delta}E=0.4\% for well-resolved peaks.}\\
\parbox[b][0.3cm]{17.7cm}{\makebox[1ex]{\ensuremath{^{\hypertarget{HG16LEVEL1}{\ddagger}}}} From Adopted Levels.}\\
\vspace{0.5cm}
\clearpage
%COULOMB EXCITATION
\subsection[\hspace{-0.2cm}Coulomb excitation]{ }
\vspace{-27pt}
\vspace{0.3cm}
\hypertarget{HG17}{{\bf \small \underline{Coulomb excitation\hspace{0.2in}\href{https://www.nndc.bnl.gov/nsr/nsrlink.jsp?1980Bo05,B}{1980Bo05},\href{https://www.nndc.bnl.gov/nsr/nsrlink.jsp?1983Sc38,B}{1983Sc38}}}}\\
\vspace{4pt}
\vspace{8pt}
\parbox[b][0.3cm]{17.7cm}{\addtolength{\parindent}{-0.2in}\href{https://www.nndc.bnl.gov/nsr/nsrlink.jsp?1980Bo05,B}{1980Bo05}: (\ensuremath{^{\textnormal{16}}}O,\ensuremath{^{\textnormal{16}}}O\ensuremath{'}), E(\ensuremath{^{\textnormal{16}}}O)=35 to 64 MeV; (\ensuremath{\alpha},\ensuremath{\alpha}\ensuremath{'}), E(\ensuremath{^{\textnormal{4}}}He)=15 MeV. 81\% \ensuremath{^{\textnormal{201}}}Hg target; Detectors: Ge(Li); Measured: \ensuremath{\gamma}-ray}\\
\parbox[b][0.3cm]{17.7cm}{yield, \ensuremath{\gamma}(\ensuremath{\theta}); Deduced: B(E2), \ensuremath{\delta}.}\\
\parbox[b][0.3cm]{17.7cm}{\addtolength{\parindent}{-0.2in}\href{https://www.nndc.bnl.gov/nsr/nsrlink.jsp?1983Sc38,B}{1983Sc38}: (\ensuremath{\alpha},\ensuremath{\alpha}\ensuremath{'}), E(\ensuremath{^{\textnormal{4}}}He)=16 MeV magnetic spectrograph, FWHM=14 keV. B(E2) values for levels up to 167 keV measured;}\\
\parbox[b][0.3cm]{17.7cm}{normalization based on B(E2) values from \href{https://www.nndc.bnl.gov/nsr/nsrlink.jsp?1980Bo05,B}{1980Bo05} for the 414 and 464 levels.}\\
\vspace{12pt}
\underline{$^{201}$Hg Levels}\\

\parbox[b][0.3cm]{17.7cm}{\makebox[1ex]{\ensuremath{^{\hypertarget{HG17LEVEL0}{\dagger}}}} From a least-squares fit to E\ensuremath{\gamma}.}\\
\parbox[b][0.3cm]{17.7cm}{\makebox[1ex]{\ensuremath{^{\hypertarget{HG17LEVEL1}{\ddagger}}}} From \ensuremath{\gamma}(\ensuremath{\theta}) and direct excitation in Coulomb excitation (\href{https://www.nndc.bnl.gov/nsr/nsrlink.jsp?1980Bo05,B}{1980Bo05}), unless otherwise stated.}\\
\parbox[b][0.3cm]{17.7cm}{\makebox[1ex]{\ensuremath{^{\hypertarget{HG17LEVEL2}{\#}}}} From Adopted Levels.}\\
\vspace{0.5cm}
\underline{$\gamma$($^{201}$Hg)}\\

\begin{textblock}{29}(0,27.3)
Continued on next page (footnotes at end of table)
\end{textblock}
\clearpage
\vspace*{-0.5cm}
{\bf \small \underline{Coulomb excitation\hspace{0.2in}\href{https://www.nndc.bnl.gov/nsr/nsrlink.jsp?1980Bo05,B}{1980Bo05},\href{https://www.nndc.bnl.gov/nsr/nsrlink.jsp?1983Sc38,B}{1983Sc38} (continued)}}\\
\vspace{0.3cm}
\underline{$\gamma$($^{201}$Hg) (continued)}\\
\vspace{0.3cm}
\parbox[b][0.3cm]{17.7cm}{\makebox[1ex]{\ensuremath{^{\hypertarget{HG17GAMMA0}{\dagger}}}} From \href{https://www.nndc.bnl.gov/nsr/nsrlink.jsp?1980Bo05,B}{1980Bo05}, unless otherwise stated.}\\
\parbox[b][0.3cm]{17.7cm}{\makebox[1ex]{\ensuremath{^{\hypertarget{HG17GAMMA1}{\ddagger}}}} From adopted gammas.}\\
\parbox[b][0.3cm]{17.7cm}{\makebox[1ex]{\ensuremath{^{\hypertarget{HG17GAMMA2}{\#}}}} Branching intensity from each level in \% from \href{https://www.nndc.bnl.gov/nsr/nsrlink.jsp?1980Bo05,B}{1980Bo05}.}\\
\parbox[b][0.3cm]{17.7cm}{\makebox[1ex]{\ensuremath{^{\hypertarget{HG17GAMMA3}{@}}}} Total theoretical internal conversion coefficients, calculated using the BrIcc code (\href{https://www.nndc.bnl.gov/nsr/nsrlink.jsp?2008Ki07,B}{2008Ki07}) with Frozen orbital approximation}\\
\parbox[b][0.3cm]{17.7cm}{{\ }{\ }based on \ensuremath{\gamma}-ray energies, assigned multipolarities, and mixing ratios, unless otherwise specified.}\\
\vspace{0.5cm}
\clearpage
\begin{figure}[h]
\begin{center}
\includegraphics{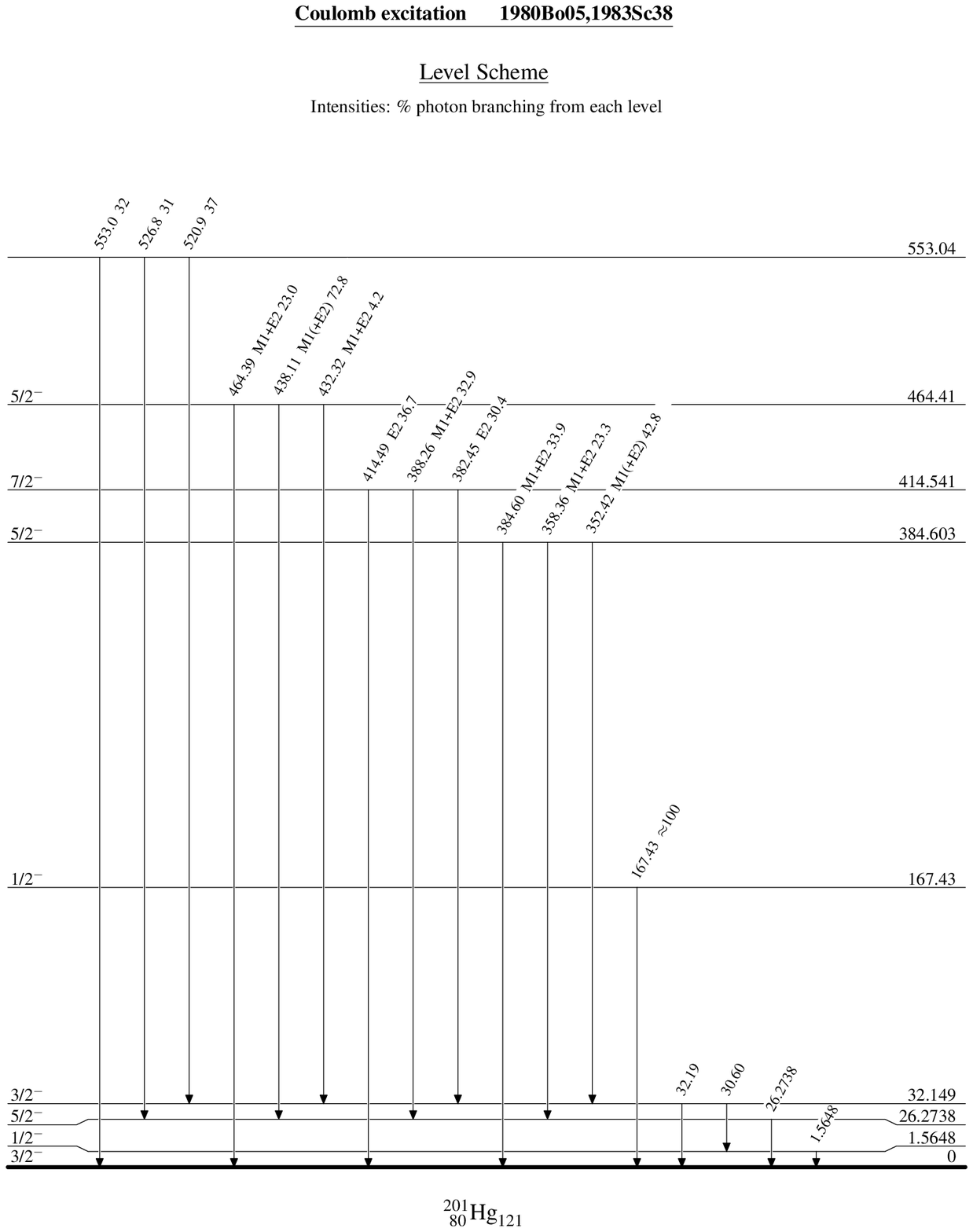}\\
\end{center}
\end{figure}
\clearpage
%202HG(D,T)
\subsection[\hspace{-0.2cm}\ensuremath{^{\textnormal{202}}}Hg(d,t)]{ }
\vspace{-27pt}
\vspace{0.3cm}
\hypertarget{HG18}{{\bf \small \underline{\ensuremath{^{\textnormal{202}}}Hg(d,t)\hspace{0.2in}\href{https://www.nndc.bnl.gov/nsr/nsrlink.jsp?1972Mo12,B}{1972Mo12}}}}\\
\vspace{4pt}
\vspace{8pt}
\parbox[b][0.3cm]{17.7cm}{\addtolength{\parindent}{-0.2in}Beam: E(d)=17 MeV; Target: enriched \ensuremath{^{\textnormal{202}}}Hg, but isotopic purity is unknown; Detectors: photographic emulsions, split-pole}\\
\parbox[b][0.3cm]{17.7cm}{spectrograph, FWHM=8-12 keV.}\\
\vspace{12pt}
\underline{$^{201}$Hg Levels}\\

\parbox[b][0.3cm]{17.7cm}{\makebox[1ex]{\ensuremath{^{\hypertarget{HG18LEVEL0}{\dagger}}}} From \href{https://www.nndc.bnl.gov/nsr/nsrlink.jsp?1972Mo12,B}{1972Mo12}. \ensuremath{\Delta}E=0.4\% for well-resolved peaks.}\\
\parbox[b][0.3cm]{17.7cm}{\makebox[1ex]{\ensuremath{^{\hypertarget{HG18LEVEL1}{\ddagger}}}} From the deduced L values and spectroscopic factors (\href{https://www.nndc.bnl.gov/nsr/nsrlink.jsp?1972Mo12,B}{1972Mo12}), unless otherwise specified.}\\
\parbox[b][0.3cm]{17.7cm}{\makebox[1ex]{\ensuremath{^{\hypertarget{HG18LEVEL2}{\#}}}} \ensuremath{\Delta}S\ensuremath{\approx}\ensuremath{\pm}50\%. S=N*(d\ensuremath{\sigma}/d\ensuremath{\Omega})\ensuremath{_{\textnormal{expt}}}/(d\ensuremath{\sigma}/d\ensuremath{\Omega})\ensuremath{_{\textnormal{DWBA}}}. N=1/3.33.}\\
\parbox[b][0.3cm]{17.7cm}{\makebox[1ex]{\ensuremath{^{\hypertarget{HG18LEVEL3}{@}}}} Dominant configuration=\ensuremath{\nu} p\ensuremath{_{\textnormal{3/2}}^{\textnormal{$-$1}}}.}\\
\parbox[b][0.3cm]{17.7cm}{\makebox[1ex]{\ensuremath{^{\hypertarget{HG18LEVEL4}{\&}}}} Dominant configuration=\ensuremath{\nu} f\ensuremath{_{\textnormal{5/2}}^{\textnormal{$-$1}}}.}\\
\parbox[b][0.3cm]{17.7cm}{\makebox[1ex]{\ensuremath{^{\hypertarget{HG18LEVEL5}{a}}}} Configuration=\ensuremath{\nu}i\ensuremath{_{\textnormal{13/2}}^{\textnormal{$-$1}}}.}\\
\vspace{0.5cm}
\clearpage
%203TL(MU,XG)
\subsection[\hspace{-0.2cm}\ensuremath{^{\textnormal{203}}}Tl(\ensuremath{\mu},X\ensuremath{\gamma})]{ }
\vspace{-27pt}
\vspace{0.3cm}
\hypertarget{HG19}{{\bf \small \underline{\ensuremath{^{\textnormal{203}}}Tl(\ensuremath{\mu},X\ensuremath{\gamma})\hspace{0.2in}\href{https://www.nndc.bnl.gov/nsr/nsrlink.jsp?1972Ba53,B}{1972Ba53}}}}\\
\vspace{4pt}
\vspace{8pt}
\parbox[b][0.3cm]{17.7cm}{\addtolength{\parindent}{-0.2in}Target: natural Tl; Detectors: Ge(Li); Measured: delayed \ensuremath{\gamma}$'$s in muonic Tl; E\ensuremath{\gamma}, I\ensuremath{\gamma}.}\\
\vspace{12pt}
\underline{$^{201}$Hg Levels}\\

\parbox[b][0.3cm]{17.7cm}{\makebox[1ex]{\ensuremath{^{\hypertarget{HG19LEVEL0}{\dagger}}}} From a least-squares fit to E\ensuremath{\gamma}, unless otherwise stated.}\\
\parbox[b][0.3cm]{17.7cm}{\makebox[1ex]{\ensuremath{^{\hypertarget{HG19LEVEL1}{\ddagger}}}} From Adopted Levels.}\\
\vspace{0.5cm}
\underline{$\gamma$($^{201}$Hg)}\\
%
\parbox[b][0.3cm]{17.7cm}{\makebox[1ex]{\ensuremath{^{\hypertarget{HG19GAMMA0}{\dagger}}}} From \href{https://www.nndc.bnl.gov/nsr/nsrlink.jsp?1972Ba53,B}{1972Ba53}. I\ensuremath{\gamma} is per 100 \ensuremath{\mu}\ensuremath{^{-}} stopped in natural Tl. Assignment to \ensuremath{^{\textnormal{201}}}Hg was made by the evaluator based on Adopted Levels}\\
\parbox[b][0.3cm]{17.7cm}{{\ }{\ }levels and gammas properties and on syst of structures populated via (\ensuremath{\mu},xn\ensuremath{\gamma}) reactions in neighboring nuclei.}\\
\vspace{0.5cm}
\clearpage
\begin{figure}[h]
\begin{center}
\includegraphics{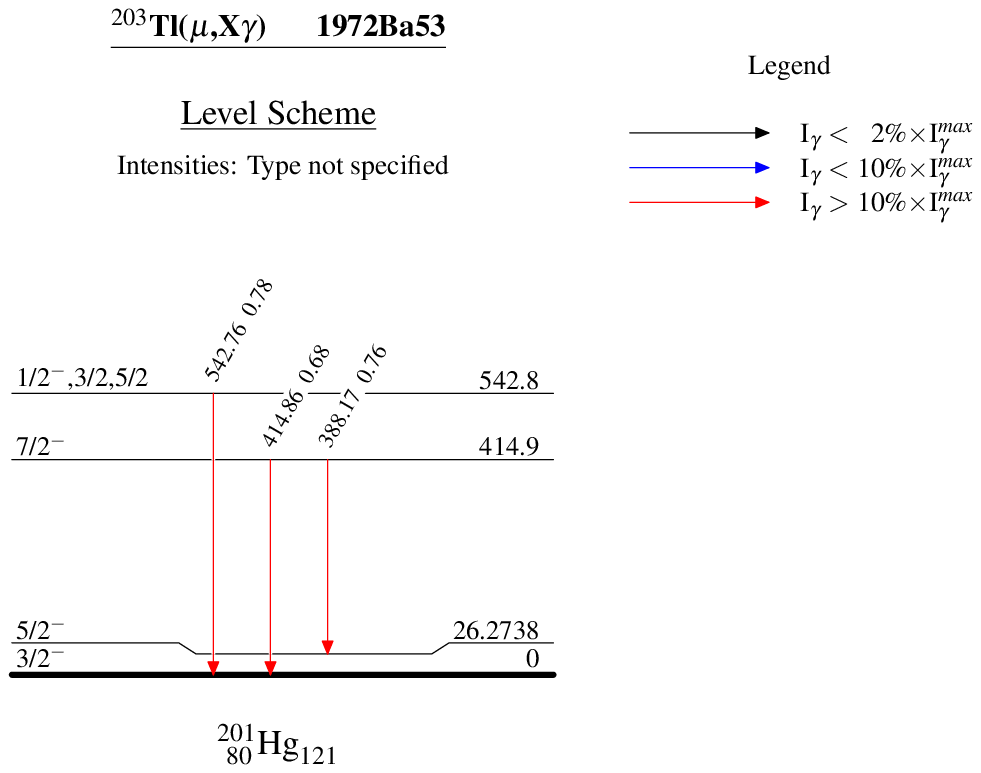}\\
\end{center}
\end{figure}
\clearpage
%ADOPTED LEVELS, GAMMAS
\section[\ensuremath{^{201}_{\ 81}}Tl\ensuremath{_{120}^{~}}]{ }
\vspace{-30pt}
\setcounter{chappage}{1}
\subsection[\hspace{-0.2cm}Adopted Levels, Gammas]{ }
\vspace{-20pt}
\vspace{0.3cm}
\hypertarget{TL20}{{\bf \small \underline{Adopted \hyperlink{201TL_LEVEL}{Levels}, \hyperlink{201TL_GAMMA}{Gammas}}}}\\
\vspace{4pt}
\vspace{8pt}
\parbox[b][0.3cm]{17.7cm}{\addtolength{\parindent}{-0.2in}Q(\ensuremath{\beta^-})=$-$1910 {\it 19}; S(n)=8205 {\it 15}; S(p)=4966 {\it 14}; Q(\ensuremath{\alpha})=1534 {\it 14}\hspace{0.2in}\href{https://www.nndc.bnl.gov/nsr/nsrlink.jsp?2021Wa16,B}{2021Wa16}}\\

\vspace{12pt}
\hypertarget{201TL_LEVEL}{\underline{$^{201}$Tl Levels}}\\

\begin{textblock}{29}(0,27.3)
Continued on next page (footnotes at end of table)
\end{textblock}
\clearpage
%
\begin{textblock}{29}(0,27.3)
Continued on next page (footnotes at end of table)
\end{textblock}
\clearpage
%
\parbox[b][0.3cm]{17.7cm}{\makebox[1ex]{\ensuremath{^{\hypertarget{TL20LEVEL0}{\dagger}}}} From a least-squares fit to E\ensuremath{\gamma}, unless otherwise stated.}\\
\begin{textblock}{29}(0,27.3)
Continued on next page (footnotes at end of table)
\end{textblock}
\clearpage
\vspace*{-0.5cm}
{\bf \small \underline{Adopted \hyperlink{201TL_LEVEL}{Levels}, \hyperlink{201TL_GAMMA}{Gammas} (continued)}}\\
\vspace{0.3cm}
\underline{$^{201}$Tl Levels (continued)}\\
\vspace{0.3cm}
\parbox[b][0.3cm]{17.7cm}{\makebox[1ex]{\ensuremath{^{\hypertarget{TL20LEVEL1}{\ddagger}}}} From \ensuremath{^{\textnormal{204}}}Pb(p,\ensuremath{\alpha}), but values were lowered by 15 keV to account for differences between excitation energies in \ensuremath{^{\textnormal{204}}}Pb(p,\ensuremath{\alpha}) and}\\
\parbox[b][0.3cm]{17.7cm}{{\ }{\ }these in the Adopted Levels for E(levels) below 1600 keV.}\\
\parbox[b][0.3cm]{17.7cm}{\makebox[1ex]{\ensuremath{^{\hypertarget{TL20LEVEL2}{\#}}}} From \ensuremath{^{\textnormal{203}}}Tl(p,t).}\\
\parbox[b][0.3cm]{17.7cm}{\makebox[1ex]{\ensuremath{^{\hypertarget{TL20LEVEL3}{@}}}} Configuration=Dominant \ensuremath{\pi} s\ensuremath{_{\textnormal{1/2}}^{\textnormal{$-$1}}}.}\\
\parbox[b][0.3cm]{17.7cm}{\makebox[1ex]{\ensuremath{^{\hypertarget{TL20LEVEL4}{\&}}}} Configuration=Dominant \ensuremath{\pi} d\ensuremath{_{\textnormal{3/2}}^{\textnormal{$-$1}}}.}\\
\parbox[b][0.3cm]{17.7cm}{\makebox[1ex]{\ensuremath{^{\hypertarget{TL20LEVEL5}{a}}}} Configuration=Dominant \ensuremath{\pi} (s\ensuremath{_{\textnormal{1/2}}^{\textnormal{$-$1}}})\ensuremath{\otimes}2\ensuremath{^{\textnormal{+}}}.}\\
\parbox[b][0.3cm]{17.7cm}{\makebox[1ex]{\ensuremath{^{\hypertarget{TL20LEVEL6}{b}}}} Configuration=\ensuremath{\pi} h\ensuremath{_{\textnormal{11/2}}^{\textnormal{$-$1}}}.}\\
\vspace{0.5cm}
\clearpage
\vspace{0.3cm}
\begin{landscape}
\vspace*{-0.5cm}
{\bf \small \underline{Adopted \hyperlink{201TL_LEVEL}{Levels}, \hyperlink{201TL_GAMMA}{Gammas} (continued)}}\\
\vspace{0.3cm}
\hypertarget{201TL_GAMMA}{\underline{$\gamma$($^{201}$Tl)}}\\

\clearpage
\vspace*{-0.5cm}
{\bf \small \underline{Adopted \hyperlink{201TL_LEVEL}{Levels}, \hyperlink{201TL_GAMMA}{Gammas} (continued)}}\\
\vspace{0.3cm}
\underline{$\gamma$($^{201}$Tl) (continued)}\\
\vspace{0.3cm}
\parbox[b][0.3cm]{21.881866cm}{\makebox[1ex]{\ensuremath{^{\hypertarget{TL20GAMMA0}{\dagger}}}} From \ensuremath{^{\textnormal{201}}}Pb \ensuremath{\varepsilon} decay, unless otherwise stated.}\\
\parbox[b][0.3cm]{21.881866cm}{\makebox[1ex]{\ensuremath{^{\hypertarget{TL20GAMMA1}{\ddagger}}}} From \ensuremath{^{\textnormal{202}}}Hg(d,3n\ensuremath{\gamma}).}\\
\parbox[b][0.3cm]{21.881866cm}{\makebox[1ex]{\ensuremath{^{\hypertarget{TL20GAMMA2}{\#}}}} From \ensuremath{^{\textnormal{198}}}Pt(\ensuremath{^{\textnormal{7}}}Li,4n\ensuremath{\gamma}).}\\
\parbox[b][0.3cm]{21.881866cm}{\makebox[1ex]{\ensuremath{^{\hypertarget{TL20GAMMA3}{@}}}} From \ensuremath{\alpha}(K)exp, \ensuremath{\alpha}(L)exp, K/L and \ensuremath{\gamma}\ensuremath{\gamma}(\ensuremath{\theta}) in \ensuremath{^{\textnormal{201}}}Pb \ensuremath{\varepsilon} decay, \ensuremath{\gamma}(\ensuremath{\theta}) in \ensuremath{^{\textnormal{202}}}Hg(d,3n\ensuremath{\gamma}), DCO and POL in \ensuremath{^{\textnormal{198}}}Pt(\ensuremath{^{\textnormal{7}}}Li,4n\ensuremath{\gamma}), unless otherwise stated.}\\
\parbox[b][0.3cm]{21.881866cm}{\makebox[1ex]{\ensuremath{^{\hypertarget{TL20GAMMA4}{\&}}}} Total theoretical internal conversion coefficients, calculated using the BrIcc code (\href{https://www.nndc.bnl.gov/nsr/nsrlink.jsp?2008Ki07,B}{2008Ki07}) with Frozen orbital approximation based on \ensuremath{\gamma}-ray energies,}\\
\parbox[b][0.3cm]{21.881866cm}{{\ }{\ }assigned multipolarities, and mixing ratios, unless otherwise specified.}\\
\parbox[b][0.3cm]{21.881866cm}{\makebox[1ex]{\ensuremath{^{\hypertarget{TL20GAMMA5}{a}}}} Placement of transition in the level scheme is uncertain.}\\
\vspace{0.5cm}
\end{landscape}\clearpage
\clearpage
\begin{figure}[h]
\begin{center}
\includegraphics{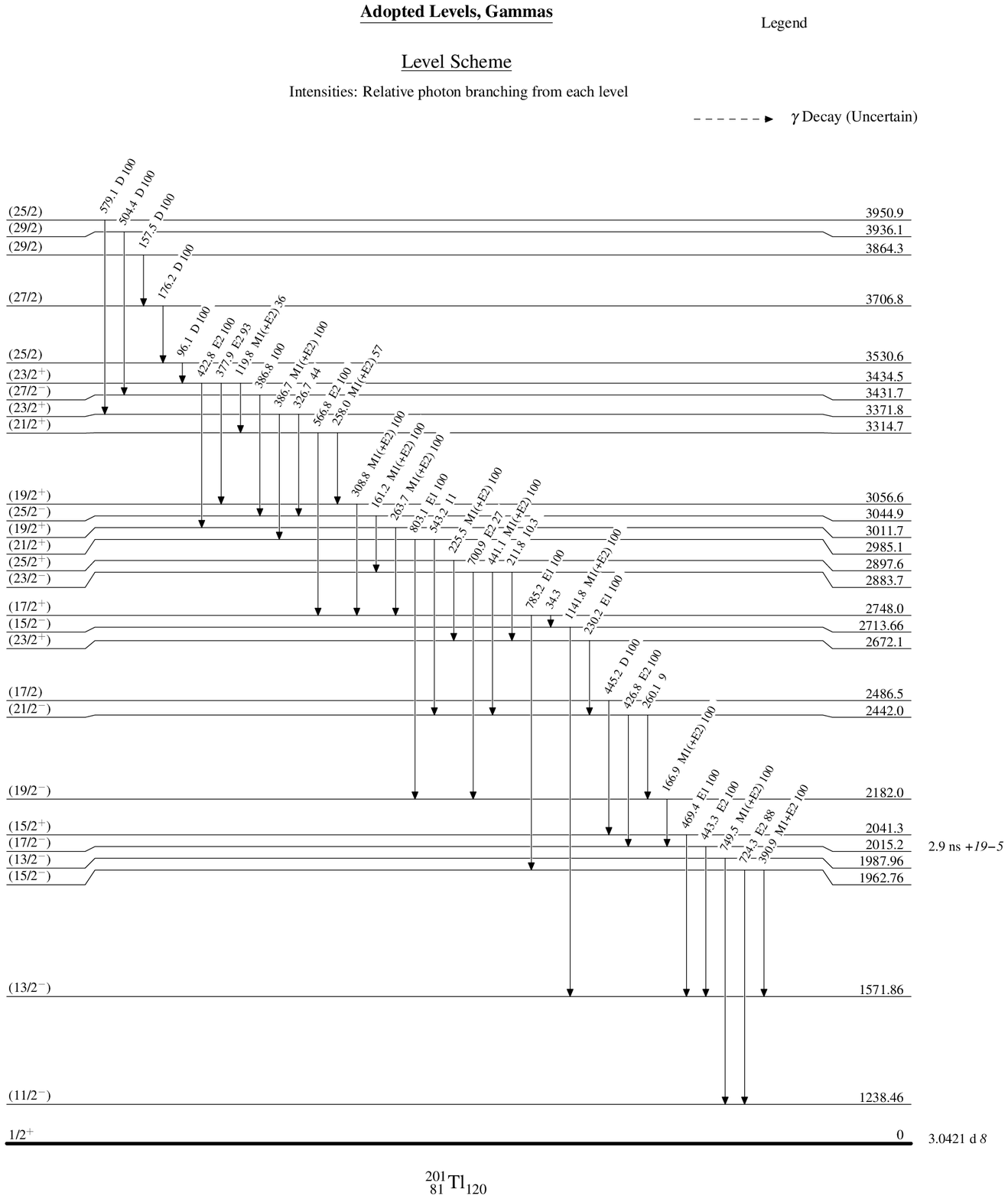}\\
\end{center}
\end{figure}
\clearpage
\begin{figure}[h]
\begin{center}
\includegraphics{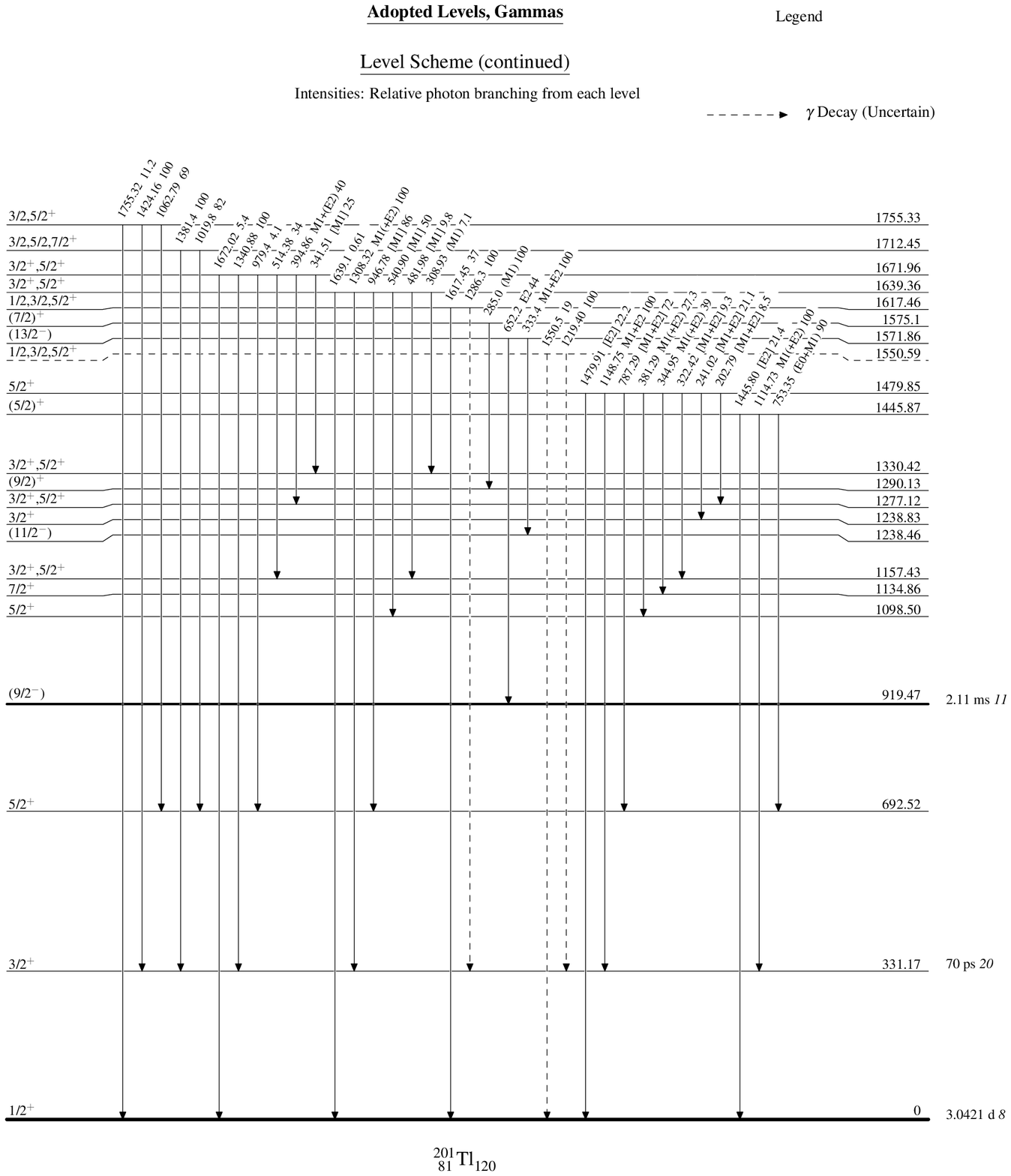}\\
\end{center}
\end{figure}
\clearpage
\begin{figure}[h]
\begin{center}
\includegraphics{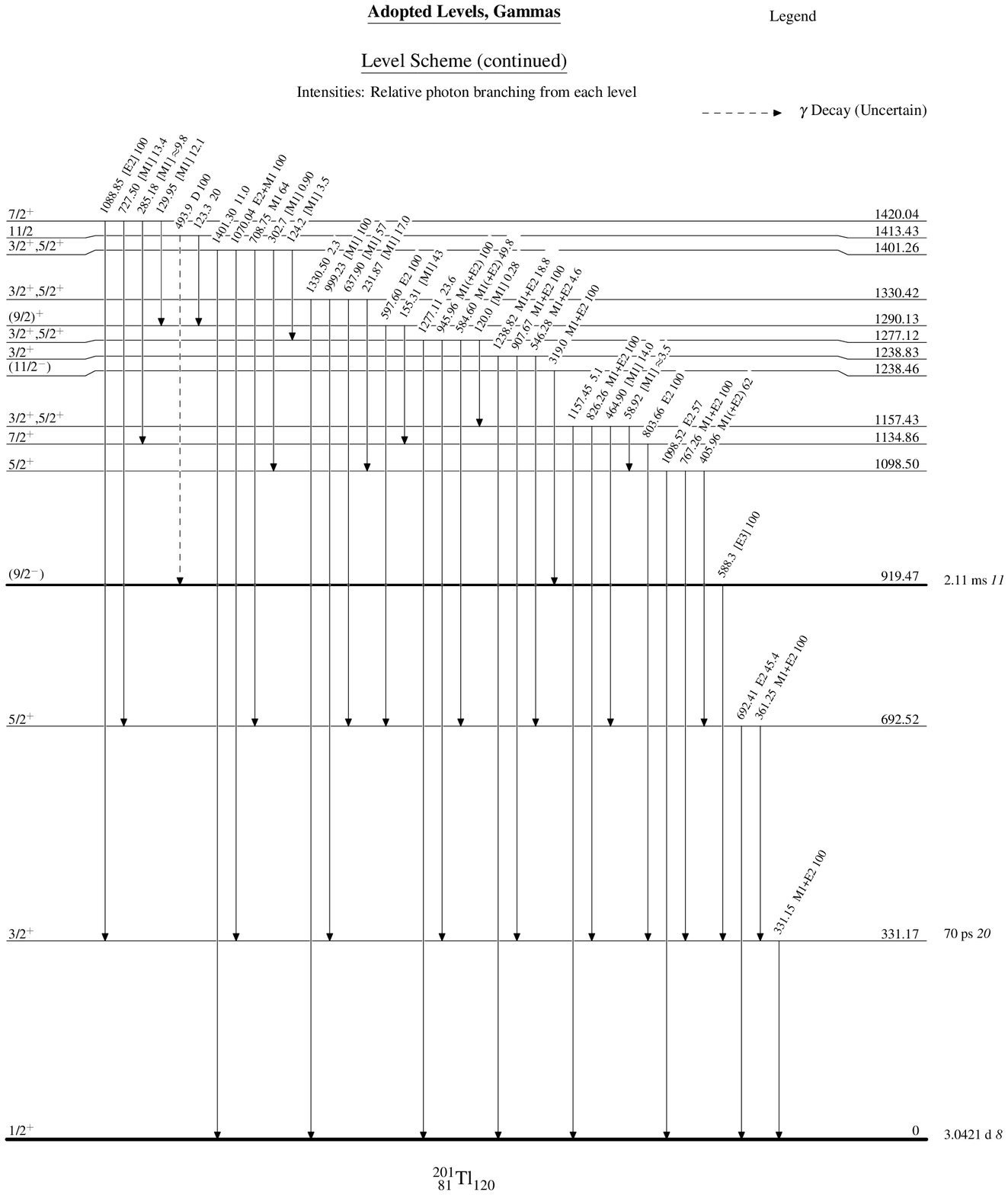}\\
\end{center}
\end{figure}
\clearpage
%201PB EC DECAY
\subsection[\hspace{-0.2cm}\ensuremath{^{\textnormal{201}}}Pb \ensuremath{\varepsilon} decay]{ }
\vspace{-27pt}
\vspace{0.3cm}
\hypertarget{PB21}{{\bf \small \underline{\ensuremath{^{\textnormal{201}}}Pb \ensuremath{\varepsilon} decay\hspace{0.2in}\href{https://www.nndc.bnl.gov/nsr/nsrlink.jsp?1979Do09,B}{1979Do09}}}}\\
\vspace{4pt}
\vspace{8pt}
\parbox[b][0.3cm]{17.7cm}{\addtolength{\parindent}{-0.2in}Parent: $^{201}$Pb: E=0.0; J$^{\pi}$=5/2\ensuremath{^{-}}; T$_{1/2}$=9.33 h {\it 5}; Q(\ensuremath{\varepsilon})=1910 {\it 19}; \%\ensuremath{\varepsilon}+\%\ensuremath{\beta^{+}} decay=100.0

}\\
\parbox[b][0.3cm]{17.7cm}{\addtolength{\parindent}{-0.2in}\href{https://www.nndc.bnl.gov/nsr/nsrlink.jsp?1979Do09,B}{1979Do09}: \ensuremath{^{\textnormal{201}}}Pb source produced using \ensuremath{^{\textnormal{203}}}Tl(p,3n) reaction; E(p)=27 MeV; Target: natural thallium; Detectors: Ge(Li) and NaI;}\\
\parbox[b][0.3cm]{17.7cm}{Compton suppressed; Measured: E\ensuremath{\gamma}, I\ensuremath{\gamma}, \ensuremath{\gamma} singles, \ensuremath{\gamma}\ensuremath{\gamma} coin; Deduced: \ensuremath{\alpha}(K)exp, \ensuremath{\alpha}(L)exp, subshell ratios, \ensuremath{J^{\pi}}, T\ensuremath{_{\textnormal{1/2}}}, level scheme.}\\
\parbox[b][0.3cm]{17.7cm}{\addtolength{\parindent}{-0.2in}Others: \href{https://www.nndc.bnl.gov/nsr/nsrlink.jsp?1974Ha18,B}{1974Ha18}, \href{https://www.nndc.bnl.gov/nsr/nsrlink.jsp?1971Hn04,B}{1971Hn04}, \href{https://www.nndc.bnl.gov/nsr/nsrlink.jsp?1970DoZT,B}{1970DoZT}, \href{https://www.nndc.bnl.gov/nsr/nsrlink.jsp?1964Aa01,B}{1964Aa01}, \href{https://www.nndc.bnl.gov/nsr/nsrlink.jsp?1961Pe05,B}{1961Pe05}, \href{https://www.nndc.bnl.gov/nsr/nsrlink.jsp?1960Li08,B}{1960Li08}.}\\
\vspace{12pt}
\underline{$^{201}$Tl Levels}\\

\parbox[b][0.3cm]{17.7cm}{\makebox[1ex]{\ensuremath{^{\hypertarget{TL21LEVEL0}{\dagger}}}} From a least-squares fit to E\ensuremath{\gamma}.}\\
\parbox[b][0.3cm]{17.7cm}{\makebox[1ex]{\ensuremath{^{\hypertarget{TL21LEVEL1}{\ddagger}}}} From Adopted Levels, unless otherwise stated.}\\
\vspace{0.5cm}

\underline{\ensuremath{\varepsilon,\beta^+} radiations}\\
%
\begin{textblock}{29}(0,27.3)
Continued on next page (footnotes at end of table)
\end{textblock}
\clearpage
%
\parbox[b][0.3cm]{17.7cm}{\makebox[1ex]{\ensuremath{^{\hypertarget{TL21DECAY0}{\dagger}}}} From intensity balances and the decay scheme, unless otherwise stated.}\\
\parbox[b][0.3cm]{17.7cm}{\makebox[1ex]{\ensuremath{^{\hypertarget{TL21DECAY1}{\ddagger}}}} Absolute intensity per 100 decays.}\\
\parbox[b][0.3cm]{17.7cm}{\makebox[1ex]{\ensuremath{^{\hypertarget{TL21DECAY2}{\#}}}} Existence of this branch is questionable.}\\
\vspace{0.5cm}
\clearpage
\vspace{0.3cm}
\begin{landscape}
\vspace*{-0.5cm}
{\bf \small \underline{\ensuremath{^{\textnormal{201}}}Pb \ensuremath{\varepsilon} decay\hspace{0.2in}\href{https://www.nndc.bnl.gov/nsr/nsrlink.jsp?1979Do09,B}{1979Do09} (continued)}}\\
\vspace{0.3cm}
\underline{$\gamma$($^{201}$Tl)}\\
\vspace{0.34cm}
\parbox[b][0.3cm]{21.881866cm}{\addtolength{\parindent}{-0.254cm}I\ensuremath{\gamma} normalization: Deduced using \ensuremath{\Sigma}(I(\ensuremath{\gamma}+ce)[g.s. \ensuremath{^{\textnormal{201}}}Tl])=100 {\textminus} I\ensuremath{\beta}\ensuremath{_{\textnormal{0}}}, with I\ensuremath{\beta}\ensuremath{_{\textnormal{0}}}=0.7\% \textit{4}.}\\
\vspace{0.34cm}

\parbox[b][0.3cm]{21.881866cm}{\makebox[1ex]{\ensuremath{^{\hypertarget{PB21GAMMA0}{\dagger}}}} From \href{https://www.nndc.bnl.gov/nsr/nsrlink.jsp?1979Do09,B}{1979Do09}, unless otherwise stated.}\\
\parbox[b][0.3cm]{21.881866cm}{\makebox[1ex]{\ensuremath{^{\hypertarget{PB21GAMMA1}{\ddagger}}}} From singles measurements in \href{https://www.nndc.bnl.gov/nsr/nsrlink.jsp?1979Do09,B}{1979Do09}, unless otherwise stated. I\ensuremath{\gamma}(x-ray)=4980 \textit{250} and I\ensuremath{\gamma}(\ensuremath{\gamma}\ensuremath{^{\ensuremath{\pm}}})=6 \textit{1} in \href{https://www.nndc.bnl.gov/nsr/nsrlink.jsp?1979Do09,B}{1979Do09}.}\\
\parbox[b][0.3cm]{21.881866cm}{\makebox[1ex]{\ensuremath{^{\hypertarget{PB21GAMMA2}{\#}}}} From \ensuremath{\alpha}(K)exp, \ensuremath{\alpha}(L)exp, K/L, \ensuremath{\gamma}\ensuremath{\gamma}(\ensuremath{\theta}) and multiple decay branches in \href{https://www.nndc.bnl.gov/nsr/nsrlink.jsp?1979Do09,B}{1979Do09}, \href{https://www.nndc.bnl.gov/nsr/nsrlink.jsp?1960Li08,B}{1960Li08}, \href{https://www.nndc.bnl.gov/nsr/nsrlink.jsp?1961Pe05,B}{1961Pe05}, \href{https://www.nndc.bnl.gov/nsr/nsrlink.jsp?1964Aa01,B}{1964Aa01} and \href{https://www.nndc.bnl.gov/nsr/nsrlink.jsp?1974Ha18,B}{1974Ha18}, unless otherwise stated.}\\
\parbox[b][0.3cm]{21.881866cm}{\makebox[1ex]{\ensuremath{^{\hypertarget{PB21GAMMA3}{@}}}} From \ensuremath{\alpha}(K)exp and sub-shell ratios in \href{https://www.nndc.bnl.gov/nsr/nsrlink.jsp?1979Do09,B}{1979Do09}, \href{https://www.nndc.bnl.gov/nsr/nsrlink.jsp?1974Ha18,B}{1974Ha18}, \href{https://www.nndc.bnl.gov/nsr/nsrlink.jsp?1971Hn04,B}{1971Hn04}, \href{https://www.nndc.bnl.gov/nsr/nsrlink.jsp?1964Aa01,B}{1964Aa01}, and\hphantom{a}\href{https://www.nndc.bnl.gov/nsr/nsrlink.jsp?1961Pe05,B}{1961Pe05} and the briccmixing program, unless otherwise stated.}\\
\parbox[b][0.3cm]{21.881866cm}{\makebox[1ex]{\ensuremath{^{\hypertarget{PB21GAMMA4}{\&}}}} Assignment to \ensuremath{^{\textnormal{201}}}Pb \ensuremath{\varepsilon} decay is uncertain (\href{https://www.nndc.bnl.gov/nsr/nsrlink.jsp?1979Do09,B}{1979Do09}).}\\
\parbox[b][0.3cm]{21.881866cm}{\makebox[1ex]{\ensuremath{^{\hypertarget{PB21GAMMA5}{a}}}} The authors in \href{https://www.nndc.bnl.gov/nsr/nsrlink.jsp?1979Do09,B}{1979Do09} report a transition with E\ensuremath{\gamma}=285.04 keV \textit{7} and I\ensuremath{\gamma}=10.3 \textit{10} doubly placed from the 1420 and 1575 keV levels with roughly equal}\\
\parbox[b][0.3cm]{21.881866cm}{{\ }{\ }intensities. \ensuremath{\alpha}(K)exp=0.37 \textit{8}, assuming Mult=M1 for the doublet. The transition is not included in the least-squares fit. For placement from the 1420 keV level,}\\
\parbox[b][0.3cm]{21.881866cm}{{\ }{\ }the evaluator chooses E\ensuremath{\gamma}=285.18 keV 13, as given from the levels energy difference in the least-squares fit. For placement from the 1575 keV level, where the}\\
\parbox[b][0.3cm]{21.881866cm}{{\ }{\ }285\ensuremath{\gamma} is the only deexciting transition, the evaluator adopts E\ensuremath{\gamma}=285.0 keV \textit{10}. The evaluator adopts I\ensuremath{\gamma}\ensuremath{\approx}5.2 for each placement. Both transitions involve \ensuremath{\Delta}J=0}\\
\parbox[b][0.3cm]{21.881866cm}{{\ }{\ }or 1 and \ensuremath{\Delta}\ensuremath{\pi}=no, and since I\ensuremath{\gamma}$'$s are roughly equal, Mult.=M1 can be assigned to both placements.}\\
\parbox[b][0.3cm]{21.881866cm}{\makebox[1ex]{\ensuremath{^{\hypertarget{PB21GAMMA6}{b}}}} For absolute intensity per 100 decays, multiply by 0.0170 \textit{8}.}\\
\parbox[b][0.3cm]{21.881866cm}{\makebox[1ex]{\ensuremath{^{\hypertarget{PB21GAMMA7}{c}}}} Total theoretical internal conversion coefficients, calculated using the BrIcc code (\href{https://www.nndc.bnl.gov/nsr/nsrlink.jsp?2008Ki07,B}{2008Ki07}) with Frozen orbital approximation based on \ensuremath{\gamma}-ray energies,}\\
\parbox[b][0.3cm]{21.881866cm}{{\ }{\ }assigned multipolarities, and mixing ratios, unless otherwise specified.}\\
\parbox[b][0.3cm]{21.881866cm}{\makebox[1ex]{\ensuremath{^{\hypertarget{PB21GAMMA8}{d}}}} Placement of transition in the level scheme is uncertain.}\\
\parbox[b][0.3cm]{21.881866cm}{\makebox[1ex]{\ensuremath{^{\hypertarget{PB21GAMMA9}{x}}}} \ensuremath{\gamma} ray not placed in level scheme.}\\
\vspace{0.5cm}
\end{landscape}\clearpage
\clearpage
\begin{figure}[h]
\begin{center}
\includegraphics{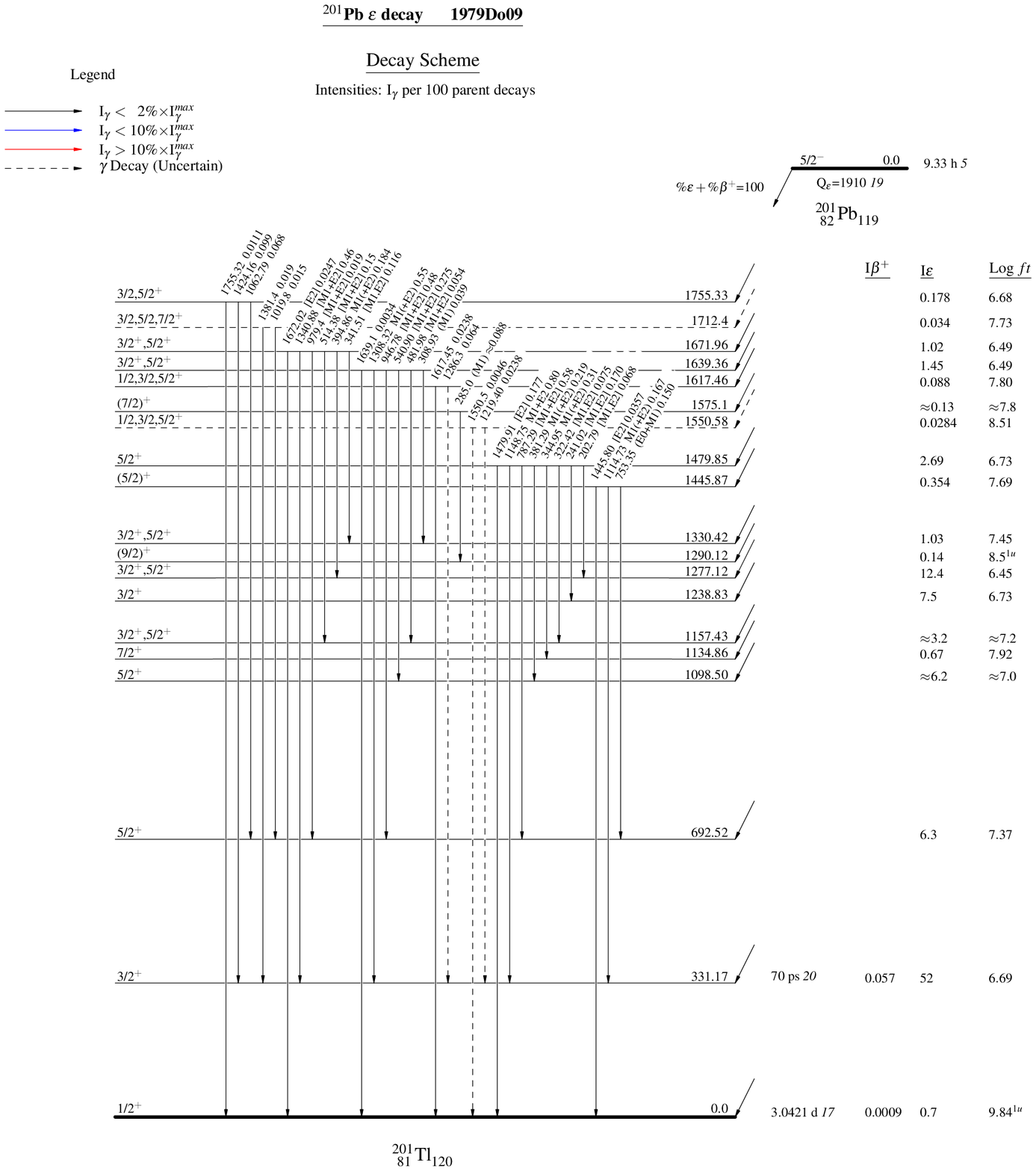}\\
\end{center}
\end{figure}
\clearpage
\begin{figure}[h]
\begin{center}
\includegraphics{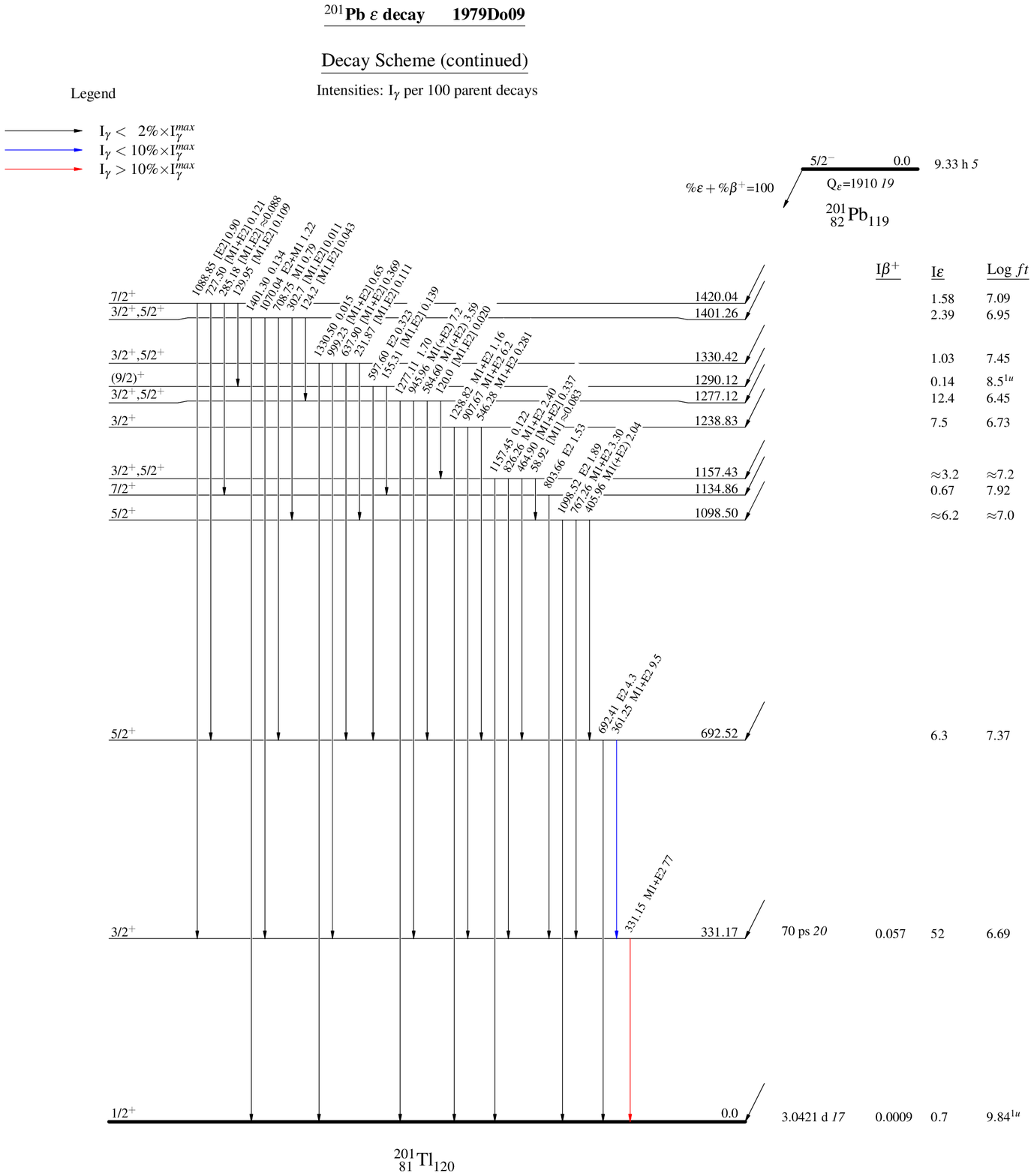}\\
\end{center}
\end{figure}
\clearpage
%201TL IT DECAY
\subsection[\hspace{-0.2cm}\ensuremath{^{\textnormal{201}}}Tl IT decay]{ }
\vspace{-27pt}
\vspace{0.3cm}
\hypertarget{TL22}{{\bf \small \underline{\ensuremath{^{\textnormal{201}}}Tl IT decay\hspace{0.2in}\href{https://www.nndc.bnl.gov/nsr/nsrlink.jsp?1975Uy01,B}{1975Uy01}}}}\\
\vspace{4pt}
\vspace{8pt}
\parbox[b][0.3cm]{17.7cm}{\addtolength{\parindent}{-0.2in}Parent: $^{201}$Tl: E=919.47 {\it 11}; J$^{\pi}$=(9/2\ensuremath{^{-}}); T$_{1/2}$=2.11 ms {\it 11}; \%IT decay=100.0

}\\
\parbox[b][0.3cm]{17.7cm}{\addtolength{\parindent}{-0.2in}\href{https://www.nndc.bnl.gov/nsr/nsrlink.jsp?1975Uy01,B}{1975Uy01}: Populated following (\ensuremath{\gamma},2n) reaction using a pulsed bremsstrahlung beam with E(\ensuremath{\gamma})=25-32 MeV; Target: natural}\\
\parbox[b][0.3cm]{17.7cm}{thallium; Detectors: Ge(Li); Measured: E\ensuremath{\gamma}, \ensuremath{\gamma}(t).}\\
\parbox[b][0.3cm]{17.7cm}{\addtolength{\parindent}{-0.2in}Others: \href{https://www.nndc.bnl.gov/nsr/nsrlink.jsp?1962Mo19,B}{1962Mo19}, \href{https://www.nndc.bnl.gov/nsr/nsrlink.jsp?1963De38,B}{1963De38}, \href{https://www.nndc.bnl.gov/nsr/nsrlink.jsp?1964Br27,B}{1964Br27}, \href{https://www.nndc.bnl.gov/nsr/nsrlink.jsp?1965Gr04,B}{1965Gr04}, \href{https://www.nndc.bnl.gov/nsr/nsrlink.jsp?1967Co20,B}{1967Co20}, \href{https://www.nndc.bnl.gov/nsr/nsrlink.jsp?1977KoZH,B}{1977KoZH}, \href{https://www.nndc.bnl.gov/nsr/nsrlink.jsp?1977Sl01,B}{1977Sl01}.}\\
\vspace{12pt}
\underline{$^{201}$Tl Levels}\\
\begin{longtable}{cccccc@{\extracolsep{\fill}}c}
\multicolumn{2}{c}{E(level)$^{{\hyperlink{TL22LEVEL0}{\dagger}}}$}&J$^{\pi}$$^{{\hyperlink{TL22LEVEL1}{\ddagger}}}$&\multicolumn{2}{c}{T$_{1/2}$$^{{\hyperlink{TL22LEVEL1}{\ddagger}}}$}&Comments&\\[-.2cm]
\multicolumn{2}{c}{\hrulefill}&\hrulefill&\multicolumn{2}{c}{\hrulefill}&\hrulefill&
\endfirsthead
\multicolumn{1}{r@{}}{0}&\multicolumn{1}{@{.}l}{0}&\multicolumn{1}{l}{1/2\ensuremath{^{+}}}&\multicolumn{1}{r@{}}{3}&\multicolumn{1}{@{.}l}{0421 d {\it 8}}&&\\
\multicolumn{1}{r@{}}{331}&\multicolumn{1}{@{.}l}{10 {\it 20}}&\multicolumn{1}{l}{3/2\ensuremath{^{+}}}&&&&\\
\multicolumn{1}{r@{}}{919}&\multicolumn{1}{@{.}l}{40 {\it 23}}&\multicolumn{1}{l}{(9/2\ensuremath{^{-}})}&\multicolumn{1}{r@{}}{2}&\multicolumn{1}{@{.}l}{11 ms {\it 11}}&\parbox[t][0.3cm]{12.89002cm}{\raggedright T\ensuremath{_{1/2}}: Unweighted average of 1.8 ms \textit{1} (\href{https://www.nndc.bnl.gov/nsr/nsrlink.jsp?1962Mo19,B}{1962Mo19}), 2.3 ms \textit{2} (\href{https://www.nndc.bnl.gov/nsr/nsrlink.jsp?1963De38,B}{1963De38}), 2.1 ms \textit{2}\vspace{0.1cm}}&\\
&&&&&\parbox[t][0.3cm]{12.89002cm}{\raggedright {\ }{\ }{\ }(\href{https://www.nndc.bnl.gov/nsr/nsrlink.jsp?1964Br27,B}{1964Br27}), 1.8 ms \textit{1} (\href{https://www.nndc.bnl.gov/nsr/nsrlink.jsp?1965Gr04,B}{1965Gr04}), 2.65 ms \textit{20} (\href{https://www.nndc.bnl.gov/nsr/nsrlink.jsp?1967Co20,B}{1967Co20}), 2.1 ms \textit{1} (\href{https://www.nndc.bnl.gov/nsr/nsrlink.jsp?1975Uy01,B}{1975Uy01}) and 2.035\vspace{0.1cm}}&\\
&&&&&\parbox[t][0.3cm]{12.89002cm}{\raggedright {\ }{\ }{\ }ms \textit{7} (\href{https://www.nndc.bnl.gov/nsr/nsrlink.jsp?1977KoZH,B}{1977KoZH}). Others: \ensuremath{>}60 ns \href{https://www.nndc.bnl.gov/nsr/nsrlink.jsp?1977Sl01,B}{1977Sl01}.\vspace{0.1cm}}&\\
&&&&&\parbox[t][0.3cm]{12.89002cm}{\raggedright configuration: \ensuremath{\pi} 9/2[505] (h\ensuremath{_{\textnormal{9/2}}}) orbital at oblate deformation.\vspace{0.1cm}}&\\
\end{longtable}
\parbox[b][0.3cm]{17.7cm}{\makebox[1ex]{\ensuremath{^{\hypertarget{TL22LEVEL0}{\dagger}}}} From a least-squares fit to E\ensuremath{\gamma}.}\\
\parbox[b][0.3cm]{17.7cm}{\makebox[1ex]{\ensuremath{^{\hypertarget{TL22LEVEL1}{\ddagger}}}} From Adopted Levels, unless otherwise stated.}\\
\vspace{0.5cm}
\underline{$\gamma$($^{201}$Tl)}\\
\vspace{0.34cm}
\parbox[b][0.3cm]{17.7cm}{\addtolength{\parindent}{-0.254cm}Note that \href{https://www.nndc.bnl.gov/nsr/nsrlink.jsp?1965Gr04,B}{1965Gr04} (scin detector system) reported 225 keV \textit{10} transition depopulating the isomer to the 5/2\ensuremath{^{+}} level at 692.5 keV.}\\
\parbox[b][0.3cm]{17.7cm}{Since, no 361.3\ensuremath{\gamma} was reported by \href{https://www.nndc.bnl.gov/nsr/nsrlink.jsp?1965Gr04,B}{1965Gr04} (this E\ensuremath{\gamma} should follow 225\ensuremath{\gamma} in the cascade), coupled with the fact that subsequent}\\
\parbox[b][0.3cm]{17.7cm}{measurements high-resolution Ge(Li) detectors (\href{https://www.nndc.bnl.gov/nsr/nsrlink.jsp?1975Uy01,B}{1975Uy01} and \href{https://www.nndc.bnl.gov/nsr/nsrlink.jsp?1977Sl01,B}{1977Sl01}) have not confirmed the \href{https://www.nndc.bnl.gov/nsr/nsrlink.jsp?1965Gr04,B}{1965Gr04} observation, the level}\\
\parbox[b][0.3cm]{17.7cm}{scheme proposed in \href{https://www.nndc.bnl.gov/nsr/nsrlink.jsp?1965Gr04,B}{1965Gr04} was rejected by the evaluator.}\\
\vspace{0.34cm}

\parbox[b][0.3cm]{17.7cm}{\makebox[1ex]{\ensuremath{^{\hypertarget{TL22GAMMA0}{\dagger}}}} From \href{https://www.nndc.bnl.gov/nsr/nsrlink.jsp?1975Uy01,B}{1975Uy01}, unless otherwise stated.}\\
\parbox[b][0.3cm]{17.7cm}{\makebox[1ex]{\ensuremath{^{\hypertarget{TL22GAMMA1}{\ddagger}}}} From Adopted Levels.}\\
\parbox[b][0.3cm]{17.7cm}{\makebox[1ex]{\ensuremath{^{\hypertarget{TL22GAMMA2}{\#}}}} From I(\ensuremath{\gamma}+ce)=100 and \ensuremath{\alpha}.}\\
\parbox[b][0.3cm]{17.7cm}{\makebox[1ex]{\ensuremath{^{\hypertarget{TL22GAMMA3}{@}}}} Absolute intensity per 100 decays.}\\
\parbox[b][0.3cm]{17.7cm}{\makebox[1ex]{\ensuremath{^{\hypertarget{TL22GAMMA4}{\&}}}} Total theoretical internal conversion coefficients, calculated using the BrIcc code (\href{https://www.nndc.bnl.gov/nsr/nsrlink.jsp?2008Ki07,B}{2008Ki07}) with Frozen orbital approximation}\\
\parbox[b][0.3cm]{17.7cm}{{\ }{\ }based on \ensuremath{\gamma}-ray energies, assigned multipolarities, and mixing ratios, unless otherwise specified.}\\
\vspace{0.5cm}
\clearpage
\begin{figure}[h]
\begin{center}
\includegraphics{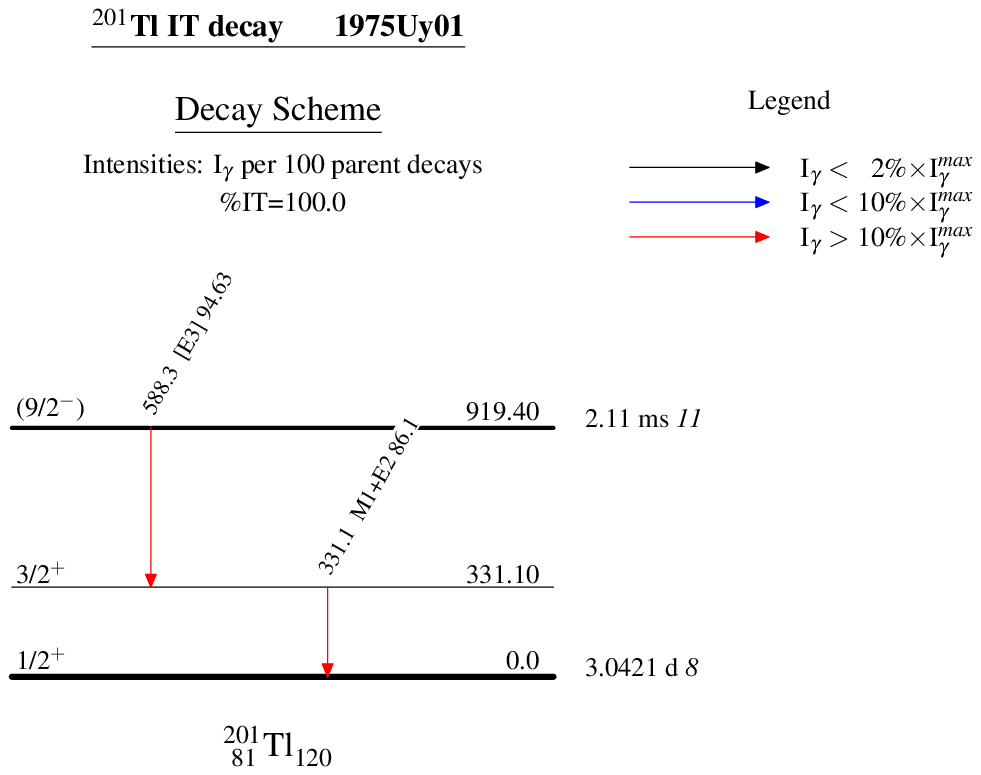}\\
\end{center}
\end{figure}
\clearpage
%202HG(D,3NG)
\subsection[\hspace{-0.2cm}\ensuremath{^{\textnormal{202}}}Hg(d,3n\ensuremath{\gamma})]{ }
\vspace{-27pt}
\vspace{0.3cm}
\hypertarget{TL23}{{\bf \small \underline{\ensuremath{^{\textnormal{202}}}Hg(d,3n\ensuremath{\gamma})\hspace{0.2in}\href{https://www.nndc.bnl.gov/nsr/nsrlink.jsp?1977Sl01,B}{1977Sl01}}}}\\
\vspace{4pt}
\vspace{8pt}
\parbox[b][0.3cm]{17.7cm}{\addtolength{\parindent}{-0.2in}E(d)=18-25 MeV; Target: metal oxide powder, 76.8\% enriched in \ensuremath{^{\textnormal{202}}}Hg; Detectors: Ge(Li), liquid scintillator; Measured:}\\
\parbox[b][0.3cm]{17.7cm}{excitation functions, E\ensuremath{\gamma}, I\ensuremath{\gamma}, \ensuremath{\gamma}\ensuremath{\gamma} coin, n\ensuremath{\gamma} coin, \ensuremath{\gamma}(t), \ensuremath{\gamma}(\ensuremath{\theta}); Deduced: \ensuremath{J^{\pi}}, T\ensuremath{_{\textnormal{1/2}}}, \ensuremath{\delta}.}\\
\vspace{12pt}
\underline{$^{201}$Tl Levels}\\

\parbox[b][0.3cm]{17.7cm}{\makebox[1ex]{\ensuremath{^{\hypertarget{TL23LEVEL0}{\dagger}}}} From a least-squares fit to E\ensuremath{\gamma}.}\\
\parbox[b][0.3cm]{17.7cm}{\makebox[1ex]{\ensuremath{^{\hypertarget{TL23LEVEL1}{\ddagger}}}} From the deduced \ensuremath{\gamma}-ray transition multipolarities using \ensuremath{\gamma}(\ensuremath{\theta}) in \href{https://www.nndc.bnl.gov/nsr/nsrlink.jsp?1977Sl01,B}{1977Sl01} and the apparent band structures, unless otherwise}\\
\parbox[b][0.3cm]{17.7cm}{{\ }{\ }stated.}\\
\vspace{0.5cm}
\underline{$\gamma$($^{201}$Tl)}\\

\begin{textblock}{29}(0,27.3)
Continued on next page (footnotes at end of table)
\end{textblock}
\clearpage
%
\parbox[b][0.3cm]{17.7cm}{\makebox[1ex]{\ensuremath{^{\hypertarget{TL23GAMMA0}{\dagger}}}} From \href{https://www.nndc.bnl.gov/nsr/nsrlink.jsp?1977Sl01,B}{1977Sl01}. I\ensuremath{\gamma} are from the E(d)=24 MeV data and were corrected for angular distribution effect.}\\
\parbox[b][0.3cm]{17.7cm}{\makebox[1ex]{\ensuremath{^{\hypertarget{TL23GAMMA1}{\ddagger}}}} Assignment to \ensuremath{^{\textnormal{201}}}Tl is tentative.}\\
\parbox[b][0.3cm]{17.7cm}{\makebox[1ex]{\ensuremath{^{\hypertarget{TL23GAMMA2}{\#}}}} From \ensuremath{\gamma}(\ensuremath{\theta}) in \href{https://www.nndc.bnl.gov/nsr/nsrlink.jsp?1977Sl01,B}{1977Sl01} and the apparent band structures.}\\
\parbox[b][0.3cm]{17.7cm}{\makebox[1ex]{\ensuremath{^{\hypertarget{TL23GAMMA3}{@}}}} Placement of transition in the level scheme is uncertain.}\\
\vspace{0.5cm}
\clearpage
\begin{figure}[h]
\begin{center}
\includegraphics{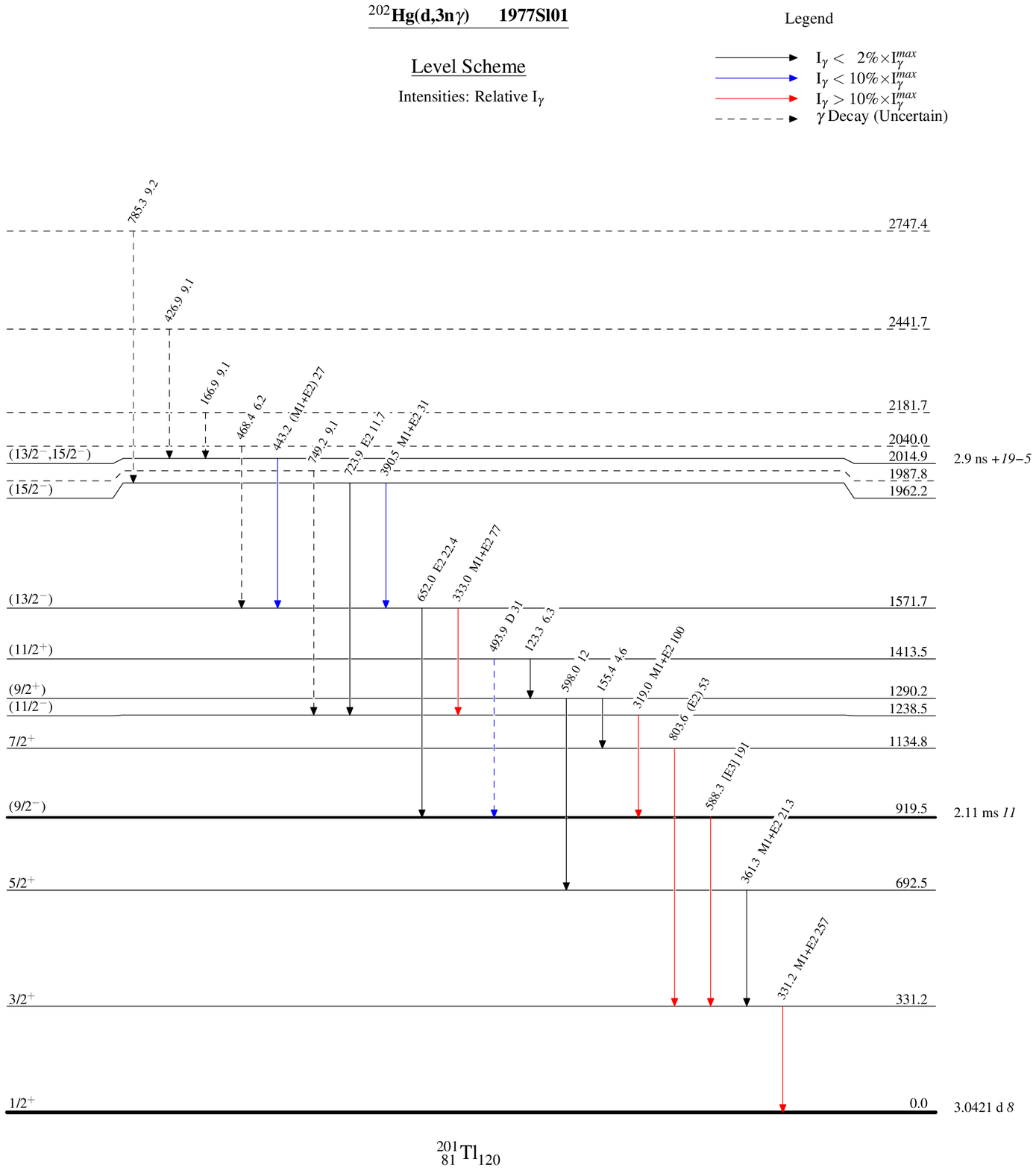}\\
\end{center}
\end{figure}
\clearpage
%198PT(7LI,4NG)
\subsection[\hspace{-0.2cm}\ensuremath{^{\textnormal{198}}}Pt(\ensuremath{^{\textnormal{7}}}Li,4n\ensuremath{\gamma})]{ }
\vspace{-27pt}
\vspace{0.3cm}
\hypertarget{TL24}{{\bf \small \underline{\ensuremath{^{\textnormal{198}}}Pt(\ensuremath{^{\textnormal{7}}}Li,4n\ensuremath{\gamma})\hspace{0.2in}\href{https://www.nndc.bnl.gov/nsr/nsrlink.jsp?2013Da15,B}{2013Da15}}}}\\
\vspace{4pt}
\vspace{8pt}
\parbox[b][0.3cm]{17.7cm}{\addtolength{\parindent}{-0.2in}\href{https://www.nndc.bnl.gov/nsr/nsrlink.jsp?2013Da15,B}{2013Da15}: reaction \ensuremath{^{\textnormal{198}}}Pt(\ensuremath{^{\textnormal{7}}}Li,4n\ensuremath{\gamma}); E(\ensuremath{^{\textnormal{7}}}Li)=45 MeV; Target: 1.3 mg/cm\ensuremath{^{\textnormal{2}}}-thick self-supporting foil, 95.7\% enriched in \ensuremath{^{\textnormal{198}}}Pt;}\\
\parbox[b][0.3cm]{17.7cm}{INGA array configured with 15 Compton suppressed clover high purity germanium (HPGe) detectors; Measured: E\ensuremath{\gamma}, I\ensuremath{\gamma}, \ensuremath{\gamma}\ensuremath{\gamma}\ensuremath{\gamma}}\\
\parbox[b][0.3cm]{17.7cm}{coin, \ensuremath{\gamma}\ensuremath{\gamma}(\ensuremath{\theta}), DCO ratios and \ensuremath{\gamma}-ray polarization; Deduced: level scheme, \ensuremath{J^{\pi}}.}\\
\vspace{12pt}
\underline{$^{201}$Tl Levels}\\

\parbox[b][0.3cm]{17.7cm}{\makebox[1ex]{\ensuremath{^{\hypertarget{TL24LEVEL0}{\dagger}}}} From least-squares fit to E\ensuremath{\gamma}.}\\
\parbox[b][0.3cm]{17.7cm}{\makebox[1ex]{\ensuremath{^{\hypertarget{TL24LEVEL1}{\ddagger}}}} From \href{https://www.nndc.bnl.gov/nsr/nsrlink.jsp?2013Da15,B}{2013Da15}, unless otherwise stated.}\\
\parbox[b][0.3cm]{17.7cm}{\makebox[1ex]{\ensuremath{^{\hypertarget{TL24LEVEL2}{\#}}}} From Adopted Levels.}\\
\vspace{0.5cm}
\underline{$\gamma$($^{201}$Tl)}\\
%
\begin{textblock}{29}(0,27.3)
Continued on next page (footnotes at end of table)
\end{textblock}
\clearpage
%
\begin{textblock}{29}(0,27.3)
Continued on next page (footnotes at end of table)
\end{textblock}
\clearpage
%
\parbox[b][0.3cm]{17.7cm}{\makebox[1ex]{\ensuremath{^{\hypertarget{TL24GAMMA0}{\dagger}}}} From \href{https://www.nndc.bnl.gov/nsr/nsrlink.jsp?2013Da15,B}{2013Da15}, unless otherwise stated.}\\
\parbox[b][0.3cm]{17.7cm}{\makebox[1ex]{\ensuremath{^{\hypertarget{TL24GAMMA1}{\ddagger}}}} From adopted gammas.}\\
\parbox[b][0.3cm]{17.7cm}{\makebox[1ex]{\ensuremath{^{\hypertarget{TL24GAMMA2}{\#}}}} DCO using 785.2 keV (E1) gate.}\\
\parbox[b][0.3cm]{17.7cm}{\makebox[1ex]{\ensuremath{^{\hypertarget{TL24GAMMA3}{@}}}} DCO using 443.3 keV (E2) gate.}\\
\parbox[b][0.3cm]{17.7cm}{\makebox[1ex]{\ensuremath{^{\hypertarget{TL24GAMMA4}{\&}}}} DCO using 803.1 keV (E1) gate.}\\
\parbox[b][0.3cm]{17.7cm}{\makebox[1ex]{\ensuremath{^{\hypertarget{TL24GAMMA5}{a}}}} DCO using 652.2 keV (E2) gate.}\\
\parbox[b][0.3cm]{17.7cm}{\makebox[1ex]{\ensuremath{^{\hypertarget{TL24GAMMA6}{b}}}} DCO using 319.0 keV (M1+E2) gate.}\\
\parbox[b][0.3cm]{17.7cm}{\makebox[1ex]{\ensuremath{^{\hypertarget{TL24GAMMA7}{c}}}} DCO using 724.3 keV (E2) gate.}\\
\vspace{0.5cm}
\clearpage
\begin{figure}[h]
\begin{center}
\includegraphics{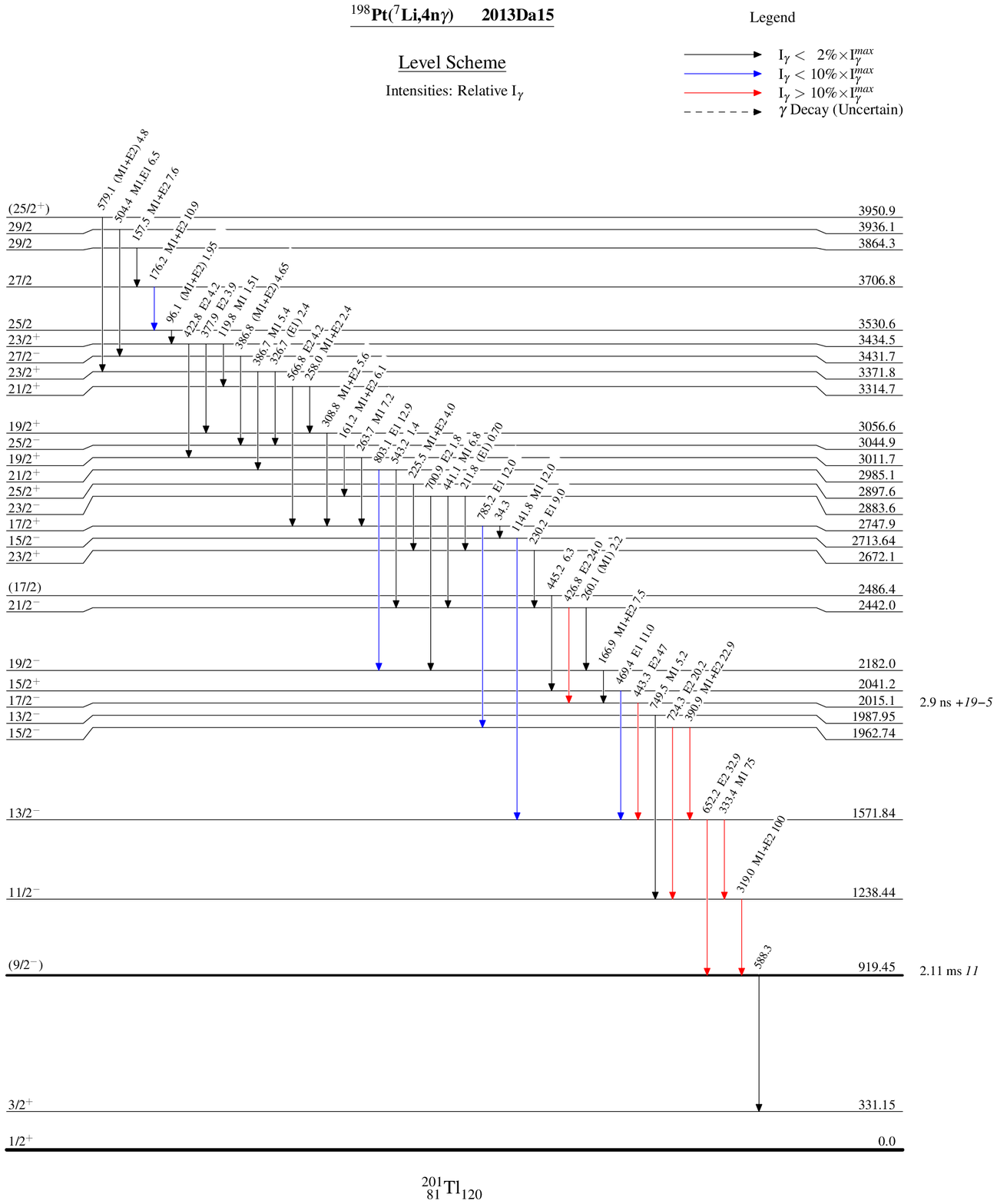}\\
\end{center}
\end{figure}
\clearpage
%9BE(238U,XG)
\subsection[\hspace{-0.2cm}\ensuremath{^{\textnormal{9}}}Be(\ensuremath{^{\textnormal{238}}}U,X\ensuremath{\gamma})]{ }
\vspace{-27pt}
\vspace{0.3cm}
\hypertarget{TL25}{{\bf \small \underline{\ensuremath{^{\textnormal{9}}}Be(\ensuremath{^{\textnormal{238}}}U,X\ensuremath{\gamma})\hspace{0.2in}\href{https://www.nndc.bnl.gov/nsr/nsrlink.jsp?2012BoZQ,B}{2012BoZQ},\href{https://www.nndc.bnl.gov/nsr/nsrlink.jsp?2013Bo18,B}{2013Bo18},\href{https://www.nndc.bnl.gov/nsr/nsrlink.jsp?2013BoZT,B}{2013BoZT}}}}\\
\vspace{4pt}
\vspace{8pt}
\parbox[b][0.3cm]{17.7cm}{\addtolength{\parindent}{-0.2in}\href{https://www.nndc.bnl.gov/nsr/nsrlink.jsp?2013Bo18,B}{2013Bo18},\href{https://www.nndc.bnl.gov/nsr/nsrlink.jsp?2012BoZQ,B}{2012BoZQ},\href{https://www.nndc.bnl.gov/nsr/nsrlink.jsp?2013BoZT,B}{2013BoZT}: E(\ensuremath{^{\textnormal{238}}}U)=1000 MeV/A from the SIS-18 synchrotron (GSI), pulsed beam 3-4 s beam-on with 2 s}\\
\parbox[b][0.3cm]{17.7cm}{beam-off periods; Target: \ensuremath{^{\textnormal{9}}}Be, 1.63 g/cm\ensuremath{^{\textnormal{2}}}-thick; Fragments were identified in flight by the Fragment Separator (FRS), based on}\\
\parbox[b][0.3cm]{17.7cm}{time of flight, B\ensuremath{\rho} and energy loss. Transmitted ions were slowed by a degrader and stopped in a passive ion 10-mm-thick Perspex}\\
\parbox[b][0.3cm]{17.7cm}{catcher, that was surrounded by the RISING \ensuremath{\gamma}-ray spectrometer. Measured E\ensuremath{\gamma}, I\ensuremath{\gamma}, delayed \ensuremath{\gamma} rays, isomer lifetime.}\\
\vspace{12pt}
\underline{$^{201}$Tl Levels}\\

\parbox[b][0.3cm]{17.7cm}{\makebox[1ex]{\ensuremath{^{\hypertarget{TL25LEVEL0}{\dagger}}}} From least-squares fit to E\ensuremath{\gamma}, unless otherwise stated.}\\
\parbox[b][0.3cm]{17.7cm}{\makebox[1ex]{\ensuremath{^{\hypertarget{TL25LEVEL1}{\ddagger}}}} From Adopted Levels, unless otherwise stated.}\\
\vspace{0.5cm}
\underline{$\gamma$($^{201}$Tl)}\\
%
\parbox[b][0.3cm]{17.7cm}{\makebox[1ex]{\ensuremath{^{\hypertarget{TL25GAMMA0}{\dagger}}}} From adopted gammas, unless otherwise stated.}\\
\parbox[b][0.3cm]{17.7cm}{\makebox[1ex]{\ensuremath{^{\hypertarget{TL25GAMMA1}{\ddagger}}}} Delayed \ensuremath{\gamma} rays observed in \href{https://www.nndc.bnl.gov/nsr/nsrlink.jsp?2012BoZQ,B}{2012BoZQ}.}\\
\parbox[b][0.3cm]{17.7cm}{\makebox[1ex]{\ensuremath{^{\hypertarget{TL25GAMMA2}{\#}}}} Delayed intensities from \href{https://www.nndc.bnl.gov/nsr/nsrlink.jsp?2013BoZT,B}{2013BoZT}.}\\
\parbox[b][0.3cm]{17.7cm}{\makebox[1ex]{\ensuremath{^{\hypertarget{TL25GAMMA3}{x}}}} \ensuremath{\gamma} ray not placed in level scheme.}\\
\vspace{0.5cm}
\clearpage
\begin{figure}[h]
\begin{center}
\includegraphics{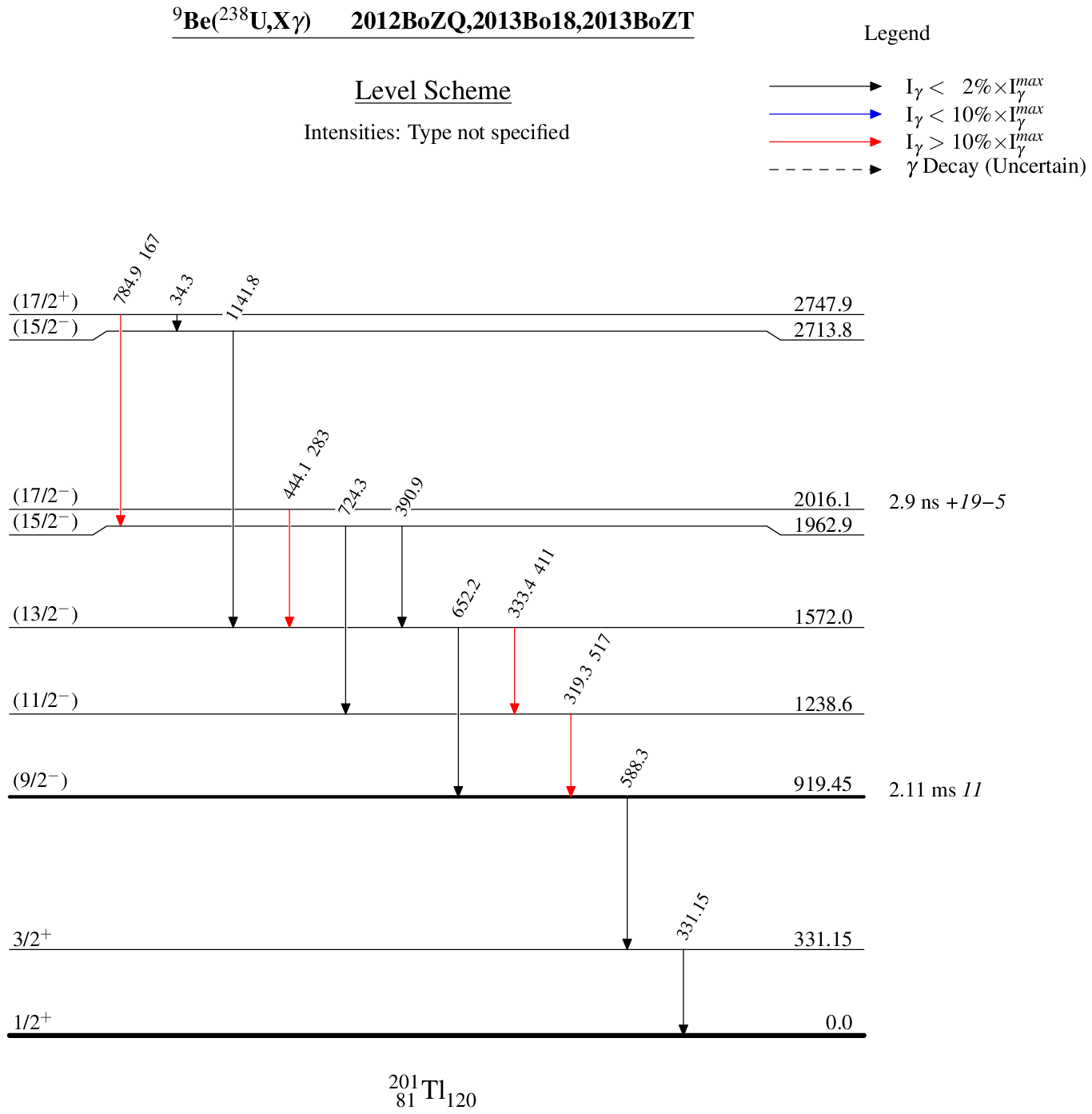}\\
\end{center}
\end{figure}
\clearpage
%203TL(P,T)
\subsection[\hspace{-0.2cm}\ensuremath{^{\textnormal{203}}}Tl(p,t)]{ }
\vspace{-27pt}
\vspace{0.3cm}
\hypertarget{TL26}{{\bf \small \underline{\ensuremath{^{\textnormal{203}}}Tl(p,t)\hspace{0.2in}\href{https://www.nndc.bnl.gov/nsr/nsrlink.jsp?1977Se15,B}{1977Se15}}}}\\
\vspace{4pt}
\vspace{8pt}
\parbox[b][0.3cm]{17.7cm}{\addtolength{\parindent}{-0.2in}\href{https://www.nndc.bnl.gov/nsr/nsrlink.jsp?1977Se15,B}{1977Se15}: E(p)=19 MeV; magnetic spectrograph; \ensuremath{\sigma}(\ensuremath{\theta}); \ensuremath{\Delta}E=5-7 keV; \ensuremath{J^{\pi}}(\ensuremath{^{\textnormal{203}}}Tl)=1/2\ensuremath{^{+}}.}\\
\vspace{12pt}
\underline{$^{201}$Tl Levels}\\

\parbox[b][0.3cm]{17.7cm}{\makebox[1ex]{\ensuremath{^{\hypertarget{TL26LEVEL0}{\dagger}}}} From \href{https://www.nndc.bnl.gov/nsr/nsrlink.jsp?1977Se15,B}{1977Se15}.}\\
\parbox[b][0.3cm]{17.7cm}{\makebox[1ex]{\ensuremath{^{\hypertarget{TL26LEVEL1}{\ddagger}}}} From L transfer value in \href{https://www.nndc.bnl.gov/nsr/nsrlink.jsp?1977Se15,B}{1977Se15}.}\\
\parbox[b][0.3cm]{17.7cm}{\makebox[1ex]{\ensuremath{^{\hypertarget{TL26LEVEL2}{\#}}}} From Adopted Levels.}\\
\parbox[b][0.3cm]{17.7cm}{\makebox[1ex]{\ensuremath{^{\hypertarget{TL26LEVEL3}{@}}}} From \ensuremath{\sigma}(\ensuremath{\theta}) in \href{https://www.nndc.bnl.gov/nsr/nsrlink.jsp?1977Se15,B}{1977Se15}.}\\
\vspace{0.5cm}
\clearpage
%204PB(P,A)
\subsection[\hspace{-0.2cm}\ensuremath{^{\textnormal{204}}}Pb(p,\ensuremath{\alpha})]{ }
\vspace{-27pt}
\vspace{0.3cm}
\hypertarget{TL27}{{\bf \small \underline{\ensuremath{^{\textnormal{204}}}Pb(p,\ensuremath{\alpha})\hspace{0.2in}\href{https://www.nndc.bnl.gov/nsr/nsrlink.jsp?1985Fi05,B}{1985Fi05}}}}\\
\vspace{4pt}
\vspace{8pt}
\parbox[b][0.3cm]{17.7cm}{\addtolength{\parindent}{-0.2in}E(p)=35 MeV; Target: enriched, 0.3 mg/cm\ensuremath{^{\textnormal{2}}} thick; Measured: magnetic spectrograph, \ensuremath{\sigma}(E(\ensuremath{\alpha}),\ensuremath{\theta}), FWHM(\ensuremath{\alpha})\ensuremath{\approx}35 keV; Deduced: E,}\\
\parbox[b][0.3cm]{17.7cm}{\ensuremath{J^{\pi}}. DWBA analysis.}\\
\vspace{12pt}
\underline{$^{201}$Tl Levels}\\

\parbox[b][0.3cm]{17.7cm}{\makebox[1ex]{\ensuremath{^{\hypertarget{TL27LEVEL0}{\dagger}}}} From \href{https://www.nndc.bnl.gov/nsr/nsrlink.jsp?1985Fi05,B}{1985Fi05}, but values were lowered by 15 keV for levels above 700 keV, since from comparison of the excitation energies}\\
\parbox[b][0.3cm]{17.7cm}{{\ }{\ }in \href{https://www.nndc.bnl.gov/nsr/nsrlink.jsp?1985Fi05,B}{1985Fi05} (up to 1600 keV) and these from the Adopted Levels, the former values appear to be {}~ 15 keV higher.}\\
\parbox[b][0.3cm]{17.7cm}{\makebox[1ex]{\ensuremath{^{\hypertarget{TL27LEVEL1}{\ddagger}}}} From comparison of measured \ensuremath{\sigma}(\ensuremath{\theta}) with cluster model DWBA calculations (\href{https://www.nndc.bnl.gov/nsr/nsrlink.jsp?1985Fi05,B}{1985Fi05}).}\\
\parbox[b][0.3cm]{17.7cm}{\makebox[1ex]{\ensuremath{^{\hypertarget{TL27LEVEL2}{\#}}}} From Adopted Levels.}\\
\vspace{0.5cm}
\clearpage
%207PB(MU,XG)
\subsection[\hspace{-0.2cm}\ensuremath{^{\textnormal{207}}}Pb(\ensuremath{\mu},X\ensuremath{\gamma})]{ }
\vspace{-27pt}
\vspace{0.3cm}
\hypertarget{TL28}{{\bf \small \underline{\ensuremath{^{\textnormal{207}}}Pb(\ensuremath{\mu},X\ensuremath{\gamma})\hspace{0.2in}\href{https://www.nndc.bnl.gov/nsr/nsrlink.jsp?1982Bu04,B}{1982Bu04},\href{https://www.nndc.bnl.gov/nsr/nsrlink.jsp?1983Bu02,B}{1983Bu02}}}}\\
\vspace{4pt}
\vspace{8pt}
\parbox[b][0.3cm]{17.7cm}{\addtolength{\parindent}{-0.2in}Target: \ensuremath{^{\textnormal{207}}}Pb, enriched to 92.77\%; Detectors: Ge(Li); Measured: E\ensuremath{\gamma},I\ensuremath{\gamma}.}\\
\vspace{12pt}
\underline{$^{201}$Tl Levels}\\
\begin{longtable}{ccc@{\extracolsep{\fill}}c}
\multicolumn{2}{c}{E(level)$^{{\hyperlink{TL28LEVEL0}{\dagger}}}$}&J$^{\pi}$$^{{\hyperlink{TL28LEVEL1}{\ddagger}}}$&\\[-.2cm]
\multicolumn{2}{c}{\hrulefill}&\hrulefill&
\endfirsthead
\multicolumn{1}{r@{}}{0}&\multicolumn{1}{@{}l}{}&\multicolumn{1}{l}{1/2\ensuremath{^{+}}}&\\
\multicolumn{1}{r@{}}{331}&\multicolumn{1}{@{.}l}{13 {\it 9}}&\multicolumn{1}{l}{3/2\ensuremath{^{+}}}&\\
\multicolumn{1}{r@{}}{1098}&\multicolumn{1}{@{.}l}{2 {\it 4}}&\multicolumn{1}{l}{5/2\ensuremath{^{+}}}&\\
\end{longtable}
\parbox[b][0.3cm]{17.7cm}{\makebox[1ex]{\ensuremath{^{\hypertarget{TL28LEVEL0}{\dagger}}}} From a least-squares fit to E\ensuremath{\gamma}.}\\
\parbox[b][0.3cm]{17.7cm}{\makebox[1ex]{\ensuremath{^{\hypertarget{TL28LEVEL1}{\ddagger}}}} From Adopted Levels.}\\
\vspace{0.5cm}
\underline{$\gamma$($^{201}$Tl)}\\
%
\parbox[b][0.3cm]{17.7cm}{\makebox[1ex]{\ensuremath{^{\hypertarget{TL28GAMMA0}{\dagger}}}} From \href{https://www.nndc.bnl.gov/nsr/nsrlink.jsp?1983Bu02,B}{1983Bu02}.}\\
\vspace{0.5cm}
\begin{figure}[h]
\begin{center}
\includegraphics{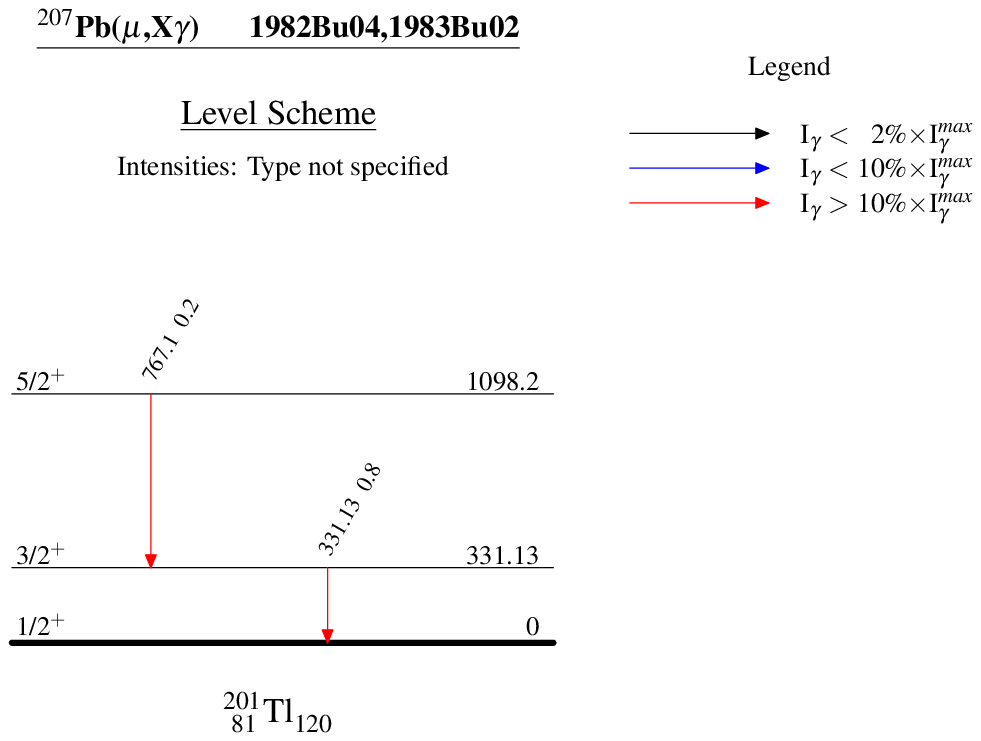}\\
\end{center}
\end{figure}
\clearpage
%ADOPTED LEVELS, GAMMAS
\section[\ensuremath{^{201}_{\ 82}}Pb\ensuremath{_{119}^{~}}]{ }
\vspace{-30pt}
\setcounter{chappage}{1}
\subsection[\hspace{-0.2cm}Adopted Levels, Gammas]{ }
\vspace{-20pt}
\vspace{0.3cm}
\hypertarget{PB29}{{\bf \small \underline{Adopted \hyperlink{201PB_LEVEL}{Levels}, \hyperlink{201PB_GAMMA}{Gammas}}}}\\
\vspace{4pt}
\vspace{8pt}
\parbox[b][0.3cm]{17.7cm}{\addtolength{\parindent}{-0.2in}Q(\ensuremath{\beta^-})=$-$3842 {\it 18}; S(n)=7091 {\it 17}; S(p)=5513 {\it 15}; Q(\ensuremath{\alpha})=2844 {\it 14}\hspace{0.2in}\href{https://www.nndc.bnl.gov/nsr/nsrlink.jsp?2021Wa16,B}{2021Wa16}}\\

\vspace{12pt}
\hypertarget{201PB_LEVEL}{\underline{$^{201}$Pb Levels}}\\

\begin{textblock}{29}(0,27.3)
Continued on next page (footnotes at end of table)
\end{textblock}
\clearpage
%
\begin{textblock}{29}(0,27.3)
Continued on next page (footnotes at end of table)
\end{textblock}
\clearpage
%
\begin{textblock}{29}(0,27.3)
Continued on next page (footnotes at end of table)
\end{textblock}
\clearpage
%
\parbox[b][0.3cm]{17.7cm}{\makebox[1ex]{\ensuremath{^{\hypertarget{PB29LEVEL0}{\dagger}}}} From a least squares fit to E\ensuremath{\gamma}.}\\
\parbox[b][0.3cm]{17.7cm}{\makebox[1ex]{\ensuremath{^{\hypertarget{PB29LEVEL1}{\ddagger}}}} Configuration=\ensuremath{\nu} f\ensuremath{_{\textnormal{5/2}}^{\textnormal{$-$1}}}.}\\
\parbox[b][0.3cm]{17.7cm}{\makebox[1ex]{\ensuremath{^{\hypertarget{PB29LEVEL2}{\#}}}} Configuration=\ensuremath{\nu} p\ensuremath{_{\textnormal{3/2}}^{\textnormal{$-$1}}}.}\\
\parbox[b][0.3cm]{17.7cm}{\makebox[1ex]{\ensuremath{^{\hypertarget{PB29LEVEL3}{@}}}} Configuration=\ensuremath{\nu} p\ensuremath{_{\textnormal{1/2}}^{\textnormal{$-$1}}}. The assignment is tentative.}\\
\parbox[b][0.3cm]{17.7cm}{\makebox[1ex]{\ensuremath{^{\hypertarget{PB29LEVEL4}{\&}}}} Configuration=\ensuremath{\nu} i\ensuremath{_{\textnormal{13/2}}^{\textnormal{$-$1}}}.}\\
\parbox[b][0.3cm]{17.7cm}{\makebox[1ex]{\ensuremath{^{\hypertarget{PB29LEVEL5}{a}}}} Configuration=\ensuremath{\nu} f\ensuremath{_{\textnormal{7/2}}^{\textnormal{$-$1}}}. The assignment is tentative.}\\
\parbox[b][0.3cm]{17.7cm}{\makebox[1ex]{\ensuremath{^{\hypertarget{PB29LEVEL6}{b}}}} Configuration=\ensuremath{\nu} (f\ensuremath{_{\textnormal{5/2}}^{\textnormal{$-$1}}})\ensuremath{\otimes}2\ensuremath{^{\textnormal{+}}}.}\\
\parbox[b][0.3cm]{17.7cm}{\makebox[1ex]{\ensuremath{^{\hypertarget{PB29LEVEL7}{c}}}} Configuration=\ensuremath{\nu} (p\ensuremath{_{\textnormal{3/2}}^{\textnormal{$-$1}}})\ensuremath{\otimes}2\ensuremath{^{\textnormal{+}}}.}\\
\parbox[b][0.3cm]{17.7cm}{\makebox[1ex]{\ensuremath{^{\hypertarget{PB29LEVEL8}{d}}}} Configuration=\ensuremath{\nu} (i\ensuremath{_{\textnormal{13/2}}^{\textnormal{$-$1}}})\ensuremath{\otimes}2\ensuremath{^{\textnormal{+}}}.}\\
\parbox[b][0.3cm]{17.7cm}{\makebox[1ex]{\ensuremath{^{\hypertarget{PB29LEVEL9}{e}}}} Probably an admixture of configuration=\ensuremath{\nu} (f\ensuremath{_{\textnormal{5/2}}^{\textnormal{$-$1}}},p\ensuremath{_{\textnormal{1/2}}^{\textnormal{$-$1}}},i\ensuremath{_{\textnormal{13/2}}^{\textnormal{$-$1}}})\ensuremath{\otimes}2\ensuremath{^{\textnormal{+}}} and configuration=\ensuremath{\nu} (i\ensuremath{_{\textnormal{13/2}}^{\textnormal{$-$1}}})\ensuremath{\otimes}2\ensuremath{^{\textnormal{+}}}.}\\
\parbox[b][0.3cm]{17.7cm}{\makebox[1ex]{\ensuremath{^{\hypertarget{PB29LEVEL10}{f}}}} Probably an admixture of configuration=\ensuremath{\nu} (f\ensuremath{_{\textnormal{5/2}}^{\textnormal{$-$1}}},p\ensuremath{_{\textnormal{1/2}}^{\textnormal{$-$1}}},i\ensuremath{_{\textnormal{13/2}}^{\textnormal{$-$1}}})\ensuremath{\otimes}4\ensuremath{^{\textnormal{+}}} and configuration=\ensuremath{\nu} (i\ensuremath{_{\textnormal{13/2}}^{\textnormal{$-$1}}})\ensuremath{\otimes}4\ensuremath{^{\textnormal{+}}}.}\\
\parbox[b][0.3cm]{17.7cm}{\makebox[1ex]{\ensuremath{^{\hypertarget{PB29LEVEL11}{g}}}} Configuration=\ensuremath{\nu} (f\ensuremath{_{\textnormal{5/2}}^{\textnormal{$-$2}}},i\ensuremath{_{\textnormal{13/2}}^{\textnormal{$-$1}}}).}\\
\parbox[b][0.3cm]{17.7cm}{\makebox[1ex]{\ensuremath{^{\hypertarget{PB29LEVEL12}{h}}}} Configuration=\ensuremath{\nu} [p\ensuremath{_{\textnormal{3/2}}^{\textnormal{$-$1}}},(i\ensuremath{_{\textnormal{13/2}}^{\textnormal{$-$2}}})\ensuremath{_{\textnormal{12+}}}].}\\
\parbox[b][0.3cm]{17.7cm}{\makebox[1ex]{\ensuremath{^{\hypertarget{PB29LEVEL13}{i}}}} Probably an admixture of configuration=\ensuremath{\nu} [f\ensuremath{_{\textnormal{5/2}}^{\textnormal{$-$1}}},(i\ensuremath{_{\textnormal{13/2}}^{\textnormal{$-$2}}})\ensuremath{_{\textnormal{10+}}}], configuration=\ensuremath{\nu} [p\ensuremath{_{\textnormal{3/2}}^{\textnormal{$-$1}}},(i\ensuremath{_{\textnormal{13/2}}^{\textnormal{$-$2}}})\ensuremath{_{\textnormal{12+}}}] and configuration=\ensuremath{\nu}}\\
\parbox[b][0.3cm]{17.7cm}{{\ }{\ }[p\ensuremath{_{\textnormal{1/2}}^{\textnormal{$-$1}}},(i\ensuremath{_{\textnormal{13/2}}^{\textnormal{$-$2}}})\ensuremath{_{\textnormal{12+}}}].}\\
\parbox[b][0.3cm]{17.7cm}{\makebox[1ex]{\ensuremath{^{\hypertarget{PB29LEVEL14}{j}}}} Configuration=\ensuremath{\nu} [f\ensuremath{_{\textnormal{5/2}}^{\textnormal{$-$1}}},(i\ensuremath{_{\textnormal{13/2}}^{\textnormal{$-$2}}})\ensuremath{_{\textnormal{12+}}}].}\\
\parbox[b][0.3cm]{17.7cm}{\makebox[1ex]{\ensuremath{^{\hypertarget{PB29LEVEL15}{k}}}} Configuration=\ensuremath{\nu} (i\ensuremath{_{\textnormal{13/2}}^{\textnormal{$-$3}}}).}\\
\parbox[b][0.3cm]{17.7cm}{\makebox[1ex]{\ensuremath{^{\hypertarget{PB29LEVEL16}{l}}}} Configuration=\ensuremath{\nu} (p\ensuremath{_{\textnormal{3/2}}^{\textnormal{$-$1}}},f\ensuremath{_{\textnormal{5/2}}^{\textnormal{$-$1}}},i\ensuremath{_{\textnormal{13/2}}^{\textnormal{$-$3}}}).}\\
\parbox[b][0.3cm]{17.7cm}{\makebox[1ex]{\ensuremath{^{\hypertarget{PB29LEVEL17}{m}}}} Band(A): configuration=\ensuremath{\nu} [p\ensuremath{_{\textnormal{3/2}}^{\textnormal{$-$1}}},(i\ensuremath{_{\textnormal{13/2}}})\ensuremath{^{\textnormal{$-$2}}})\ensuremath{_{\textnormal{12+}}}]\ensuremath{\otimes} \ensuremath{\pi} (h\ensuremath{_{\textnormal{9/2}}^{\textnormal{+1}}},i\ensuremath{_{\textnormal{13/2}}^{\textnormal{+1}}})\ensuremath{_{\textnormal{11$-$}}}.\hphantom{a}Band 2 in \href{https://www.nndc.bnl.gov/nsr/nsrlink.jsp?1995Ba70,B}{1995Ba70}.}\\
\parbox[b][0.3cm]{17.7cm}{\makebox[1ex]{\ensuremath{^{\hypertarget{PB29LEVEL18}{n}}}} Band(B): configuration=\ensuremath{\nu} (i\ensuremath{_{\textnormal{13/2}}^{\textnormal{$-$1}}}) \ensuremath{\otimes}\ensuremath{\pi} (h\ensuremath{_{\textnormal{9/2}}^{\textnormal{+1}}},i\ensuremath{_{\textnormal{13/2}}^{\textnormal{+1}}})\ensuremath{_{\textnormal{11$-$}}}. Band 1 in \href{https://www.nndc.bnl.gov/nsr/nsrlink.jsp?1995Ba70,B}{1995Ba70}.}\\
\parbox[b][0.3cm]{17.7cm}{\makebox[1ex]{\ensuremath{^{\hypertarget{PB29LEVEL19}{o}}}} Band(C): Band 3 in \href{https://www.nndc.bnl.gov/nsr/nsrlink.jsp?1995Ba70,B}{1995Ba70}.}\\
\parbox[b][0.3cm]{17.7cm}{\makebox[1ex]{\ensuremath{^{\hypertarget{PB29LEVEL20}{p}}}} Band(D): Band 4 in \href{https://www.nndc.bnl.gov/nsr/nsrlink.jsp?1995Ba70,B}{1995Ba70}.}\\
\parbox[b][0.3cm]{17.7cm}{\makebox[1ex]{\ensuremath{^{\hypertarget{PB29LEVEL21}{q}}}} Band(E): Band 5 in \href{https://www.nndc.bnl.gov/nsr/nsrlink.jsp?1995Ba70,B}{1995Ba70}.}\\
\vspace{0.5cm}
\clearpage
\vspace{0.3cm}
\begin{landscape}
\vspace*{-0.5cm}
{\bf \small \underline{Adopted \hyperlink{201PB_LEVEL}{Levels}, \hyperlink{201PB_GAMMA}{Gammas} (continued)}}\\
\vspace{0.3cm}
\hypertarget{201PB_GAMMA}{\underline{$\gamma$($^{201}$Pb)}}\\

\parbox[b][0.3cm]{21.881866cm}{\makebox[1ex]{\ensuremath{^{\hypertarget{PB29GAMMA0}{\dagger}}}} From \href{https://www.nndc.bnl.gov/nsr/nsrlink.jsp?1978Ri04,B}{1978Ri04} in \ensuremath{^{\textnormal{201}}}Bi \ensuremath{\varepsilon} decay, unless otherwise specified.}\\
\parbox[b][0.3cm]{21.881866cm}{\makebox[1ex]{\ensuremath{^{\hypertarget{PB29GAMMA1}{\ddagger}}}} From \ensuremath{^{\textnormal{192}}}Os(\ensuremath{^{\textnormal{14}}}C,5n\ensuremath{\gamma}).}\\
\parbox[b][0.3cm]{21.881866cm}{\makebox[1ex]{\ensuremath{^{\hypertarget{PB29GAMMA2}{\#}}}} From \ensuremath{^{\textnormal{200}}}Hg(\ensuremath{\alpha},3n\ensuremath{\gamma}).}\\
\parbox[b][0.3cm]{21.881866cm}{\makebox[1ex]{\ensuremath{^{\hypertarget{PB29GAMMA3}{@}}}} From \ensuremath{^{\textnormal{197}}}Au(\ensuremath{^{\textnormal{207}}}Pb,X\ensuremath{\gamma}).}\\
\parbox[b][0.3cm]{21.881866cm}{\makebox[1ex]{\ensuremath{^{\hypertarget{PB29GAMMA4}{\&}}}} From \ensuremath{\alpha}(K)exp, \ensuremath{\alpha}(L)exp and subshell ratios in \ensuremath{^{\textnormal{201}}}Bi \ensuremath{\varepsilon} decay (\href{https://www.nndc.bnl.gov/nsr/nsrlink.jsp?1978Ri04,B}{1978Ri04}), \ensuremath{\gamma}(\ensuremath{\theta}) in \ensuremath{^{\textnormal{200}}}Hg(\ensuremath{\alpha},3n\ensuremath{\gamma}) and DCO in \ensuremath{^{\textnormal{192}}}Os(\ensuremath{^{\textnormal{14}}}C,5n\ensuremath{\gamma}), coupled together with the}\\
\parbox[b][0.3cm]{21.881866cm}{{\ }{\ }observed multiple decay branches and band structures. For rotational band transitions whose multipolarity is determined from \ensuremath{\gamma}(\ensuremath{\theta}) or DCO, Mult.=(M1), instead}\\
\parbox[b][0.3cm]{21.881866cm}{{\ }{\ }of D, is assigned in this evaluation.}\\
\parbox[b][0.3cm]{21.881866cm}{\makebox[1ex]{\ensuremath{^{\hypertarget{PB29GAMMA5}{a}}}} From \ensuremath{\alpha}(K)exp, \ensuremath{\alpha}(L)exp and subshell ratios in \ensuremath{^{\textnormal{201}}}Bi \ensuremath{\varepsilon} decay (\href{https://www.nndc.bnl.gov/nsr/nsrlink.jsp?1978Ri04,B}{1978Ri04}) and the briccmixing program, unless otherwise stated.}\\
\parbox[b][0.3cm]{21.881866cm}{\makebox[1ex]{\ensuremath{^{\hypertarget{PB29GAMMA6}{b}}}} Total theoretical internal conversion coefficients, calculated using the BrIcc code (\href{https://www.nndc.bnl.gov/nsr/nsrlink.jsp?2008Ki07,B}{2008Ki07}) with Frozen orbital approximation based on \ensuremath{\gamma}-ray energies,}\\
\parbox[b][0.3cm]{21.881866cm}{{\ }{\ }assigned multipolarities, and mixing ratios, unless otherwise specified.}\\
\parbox[b][0.3cm]{21.881866cm}{\makebox[1ex]{\ensuremath{^{\hypertarget{PB29GAMMA7}{c}}}} Placement of transition in the level scheme is uncertain.}\\
\vspace{0.5cm}
\end{landscape}\clearpage
\clearpage
\begin{figure}[h]
\begin{center}
\includegraphics{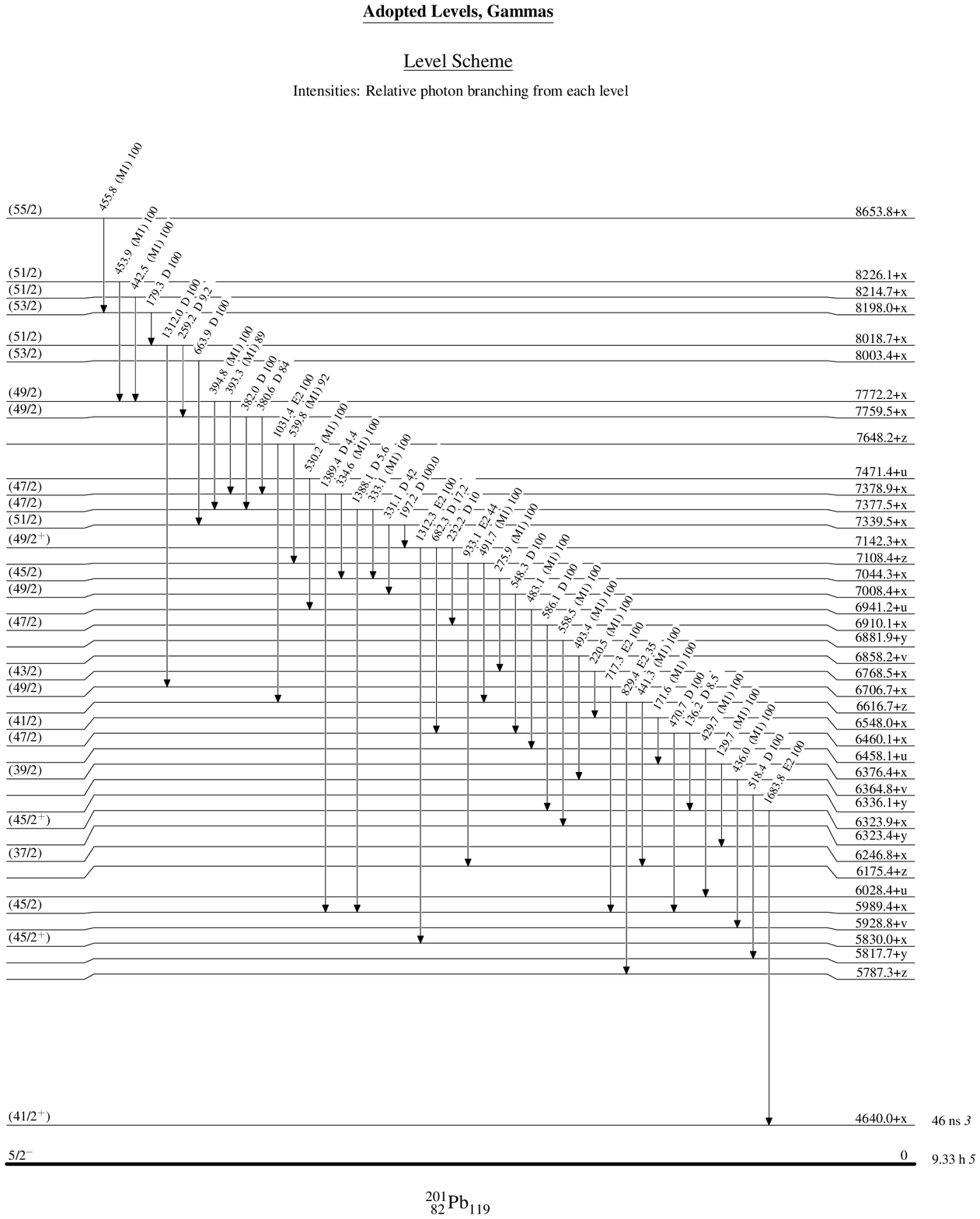}\\
\end{center}
\end{figure}
\clearpage
\begin{figure}[h]
\begin{center}
\includegraphics{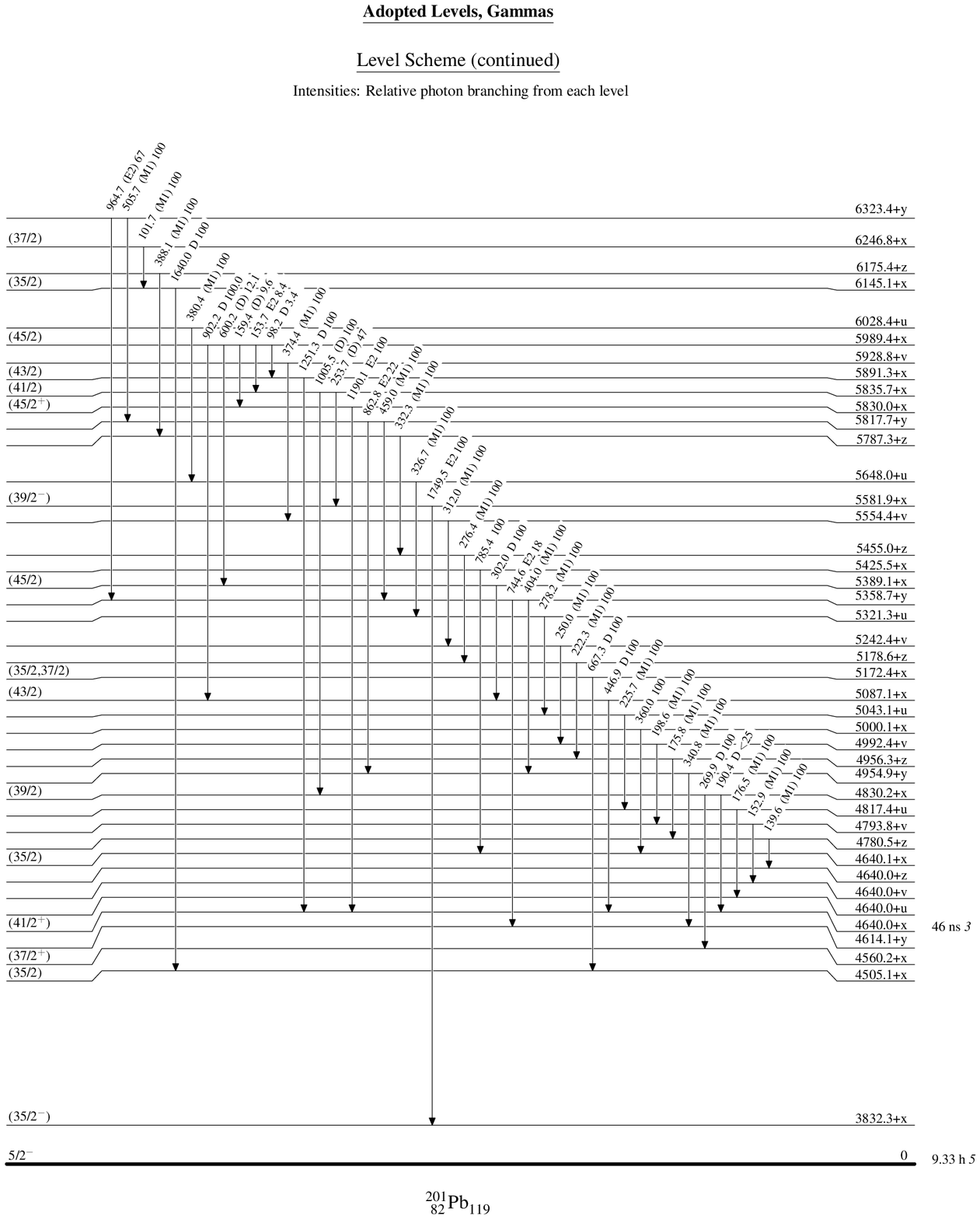}\\
\end{center}
\end{figure}
\clearpage
\begin{figure}[h]
\begin{center}
\includegraphics{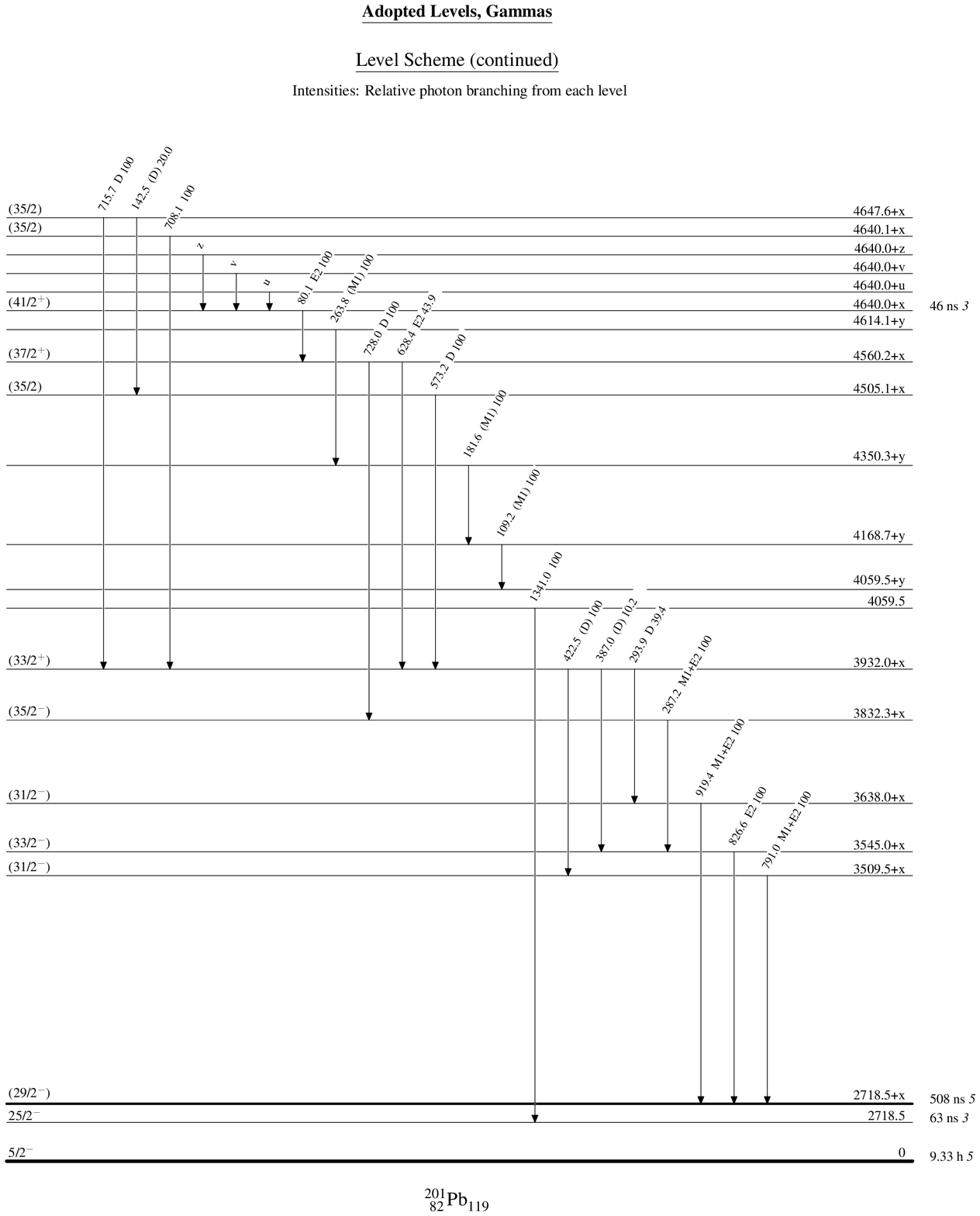}\\
\end{center}
\end{figure}
\clearpage
\begin{figure}[h]
\begin{center}
\includegraphics{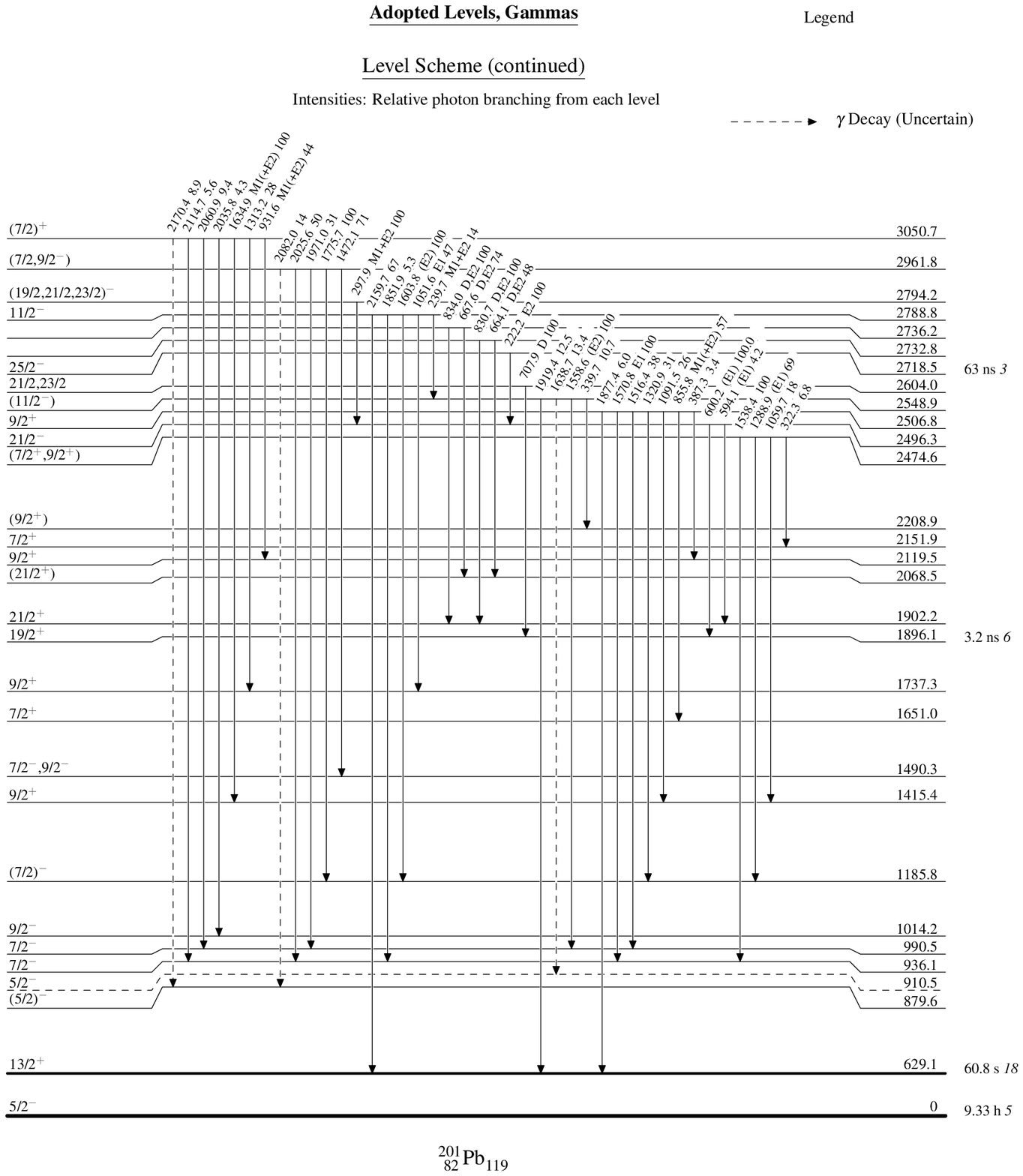}\\
\end{center}
\end{figure}
\clearpage
\begin{figure}[h]
\begin{center}
\includegraphics{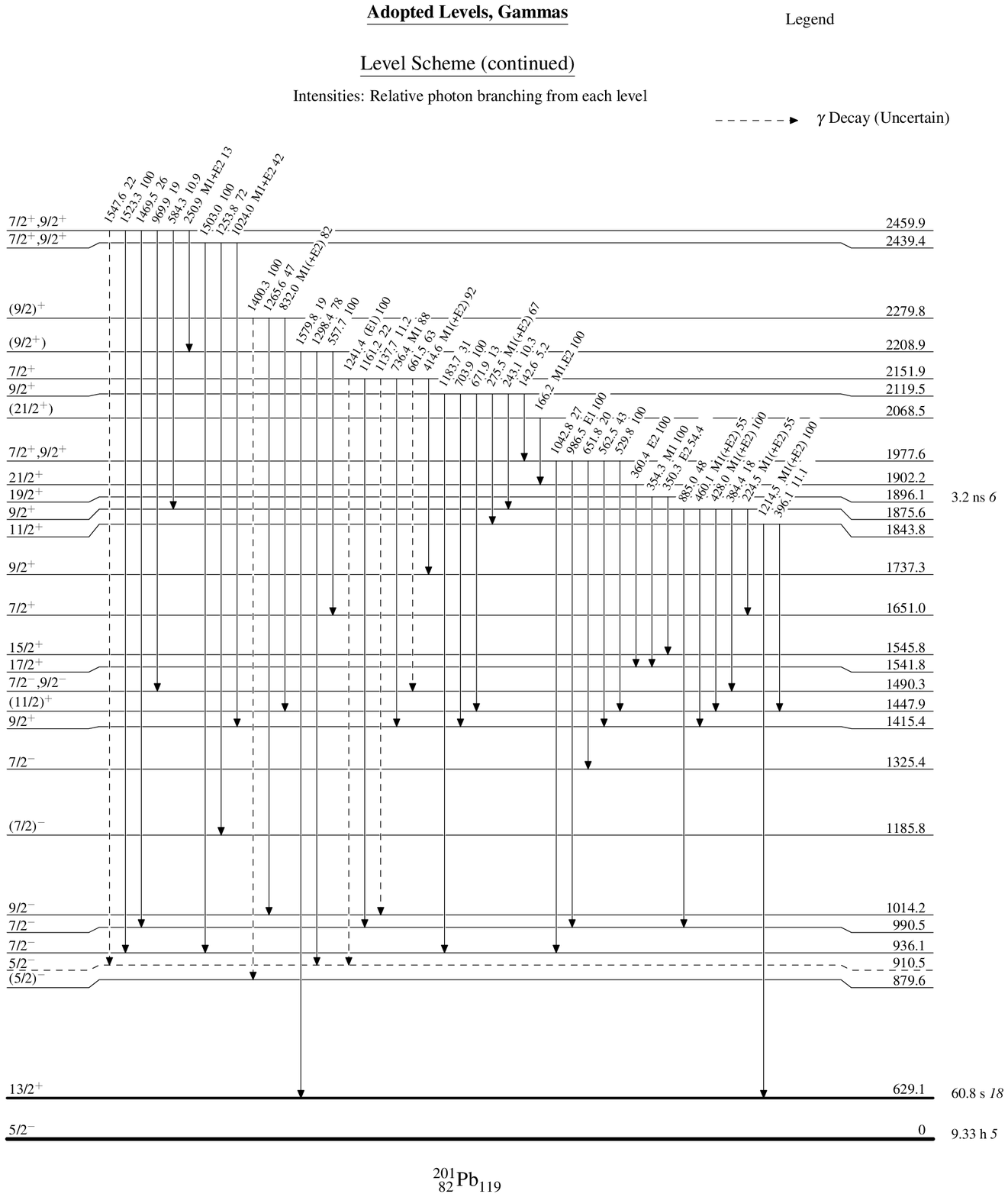}\\
\end{center}
\end{figure}
\clearpage
\begin{figure}[h]
\begin{center}
\includegraphics{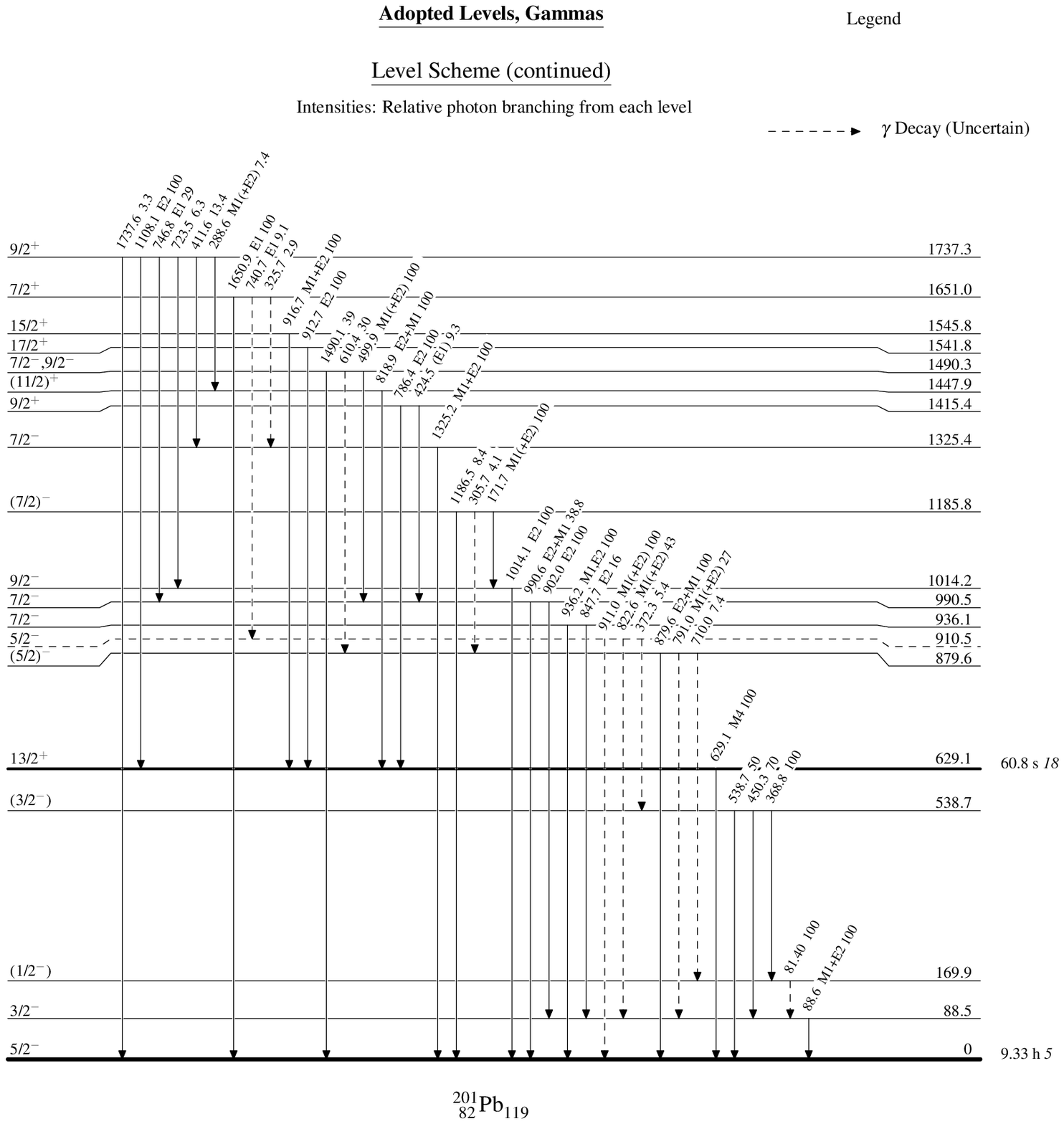}\\
\end{center}
\end{figure}
\clearpage
\clearpage
\begin{figure}[h]
\begin{center}
\includegraphics{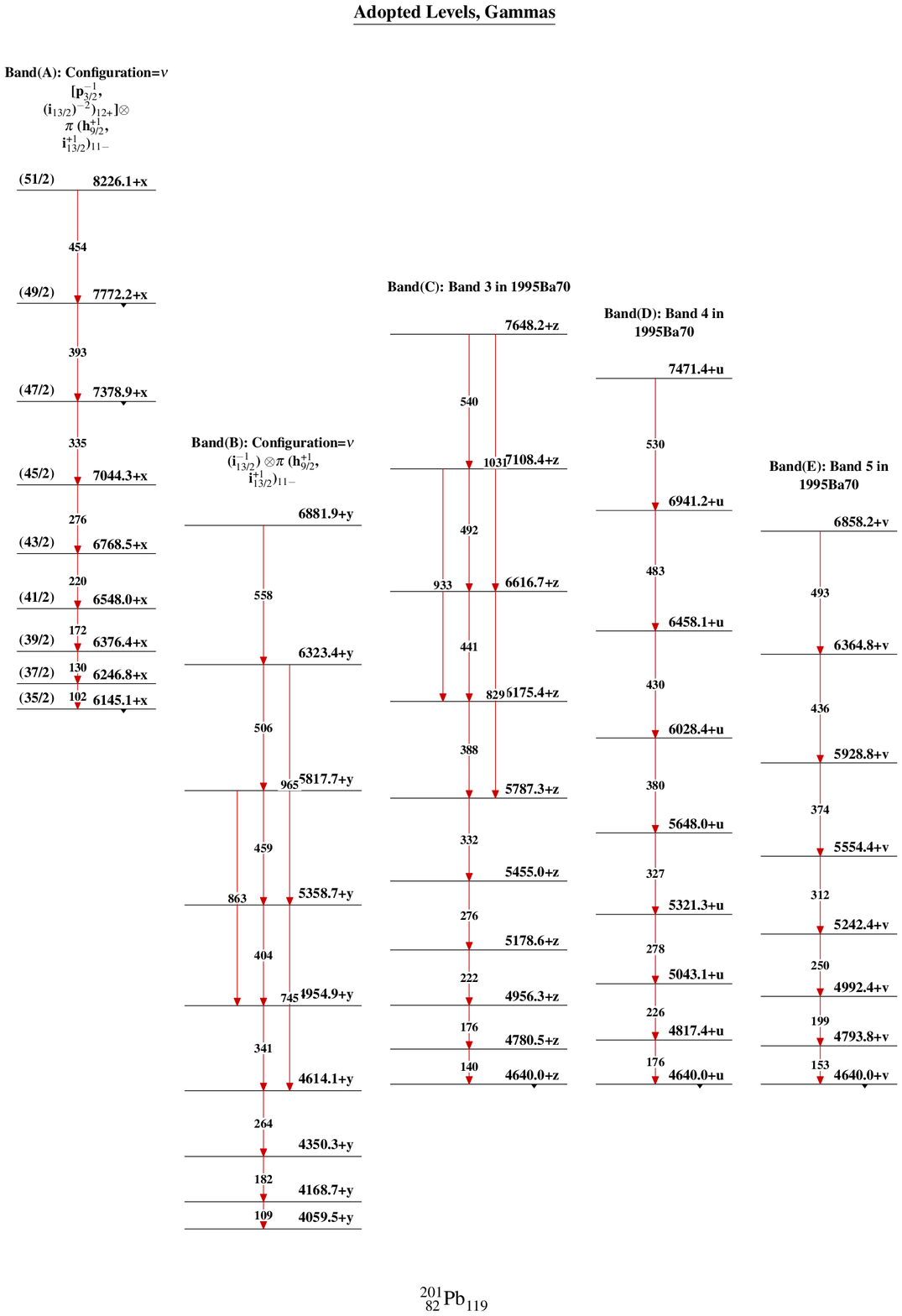}\\
\end{center}
\end{figure}
\clearpage
%201BI EC DECAY
\subsection[\hspace{-0.2cm}\ensuremath{^{\textnormal{201}}}Bi \ensuremath{\varepsilon} decay]{ }
\vspace{-27pt}
\vspace{0.3cm}
\hypertarget{BI30}{{\bf \small \underline{\ensuremath{^{\textnormal{201}}}Bi \ensuremath{\varepsilon} decay\hspace{0.2in}\href{https://www.nndc.bnl.gov/nsr/nsrlink.jsp?1978Ri04,B}{1978Ri04}}}}\\
\vspace{4pt}
\vspace{8pt}
\parbox[b][0.3cm]{17.7cm}{\addtolength{\parindent}{-0.2in}Parent: $^{201}$Bi: E=0; J$^{\pi}$=9/2\ensuremath{^{-}}; T$_{1/2}$=103.2 min {\it 17}; Q(\ensuremath{\varepsilon})=3842 {\it 18}; \%\ensuremath{\varepsilon}+\%\ensuremath{\beta^{+}} decay=100.0

}\\
\parbox[b][0.3cm]{17.7cm}{\addtolength{\parindent}{-0.2in}\href{https://www.nndc.bnl.gov/nsr/nsrlink.jsp?1978Ri04,B}{1978Ri04}: mass-separated source following (p,xn) reaction of 73-MeV protons on natural lead; Detectors: Ge(Li) and Si(Li);}\\
\parbox[b][0.3cm]{17.7cm}{Radiochemical separation of bismuth from its lead and thallium daughters; Measured: E\ensuremath{\gamma}, I\ensuremath{\gamma}, \ensuremath{\gamma}\ensuremath{\gamma} coin, \ensuremath{\alpha}(K)exp, \ensuremath{\alpha}(L)exp;}\\
\parbox[b][0.3cm]{17.7cm}{Deduced: \ensuremath{J^{\pi}}, level scheme.}\\
\parbox[b][0.3cm]{17.7cm}{\addtolength{\parindent}{-0.2in}Others: \href{https://www.nndc.bnl.gov/nsr/nsrlink.jsp?1956St05,B}{1956St05}, \href{https://www.nndc.bnl.gov/nsr/nsrlink.jsp?1970DaZM,B}{1970DaZM}, \href{https://www.nndc.bnl.gov/nsr/nsrlink.jsp?1970Jo26,B}{1970Jo26}.}\\
\vspace{12pt}
\underline{$^{201}$Pb Levels}\\

\parbox[b][0.3cm]{17.7cm}{\makebox[1ex]{\ensuremath{^{\hypertarget{PB30LEVEL0}{\dagger}}}} From a least squares fit to E\ensuremath{\gamma}.}\\
\parbox[b][0.3cm]{17.7cm}{\makebox[1ex]{\ensuremath{^{\hypertarget{PB30LEVEL1}{\ddagger}}}} Configuration=\ensuremath{\nu} f\ensuremath{_{\textnormal{5/2}}^{\textnormal{$-$1}}}.}\\
\parbox[b][0.3cm]{17.7cm}{\makebox[1ex]{\ensuremath{^{\hypertarget{PB30LEVEL2}{\#}}}} Configuration=\ensuremath{\nu} p\ensuremath{_{\textnormal{3/2}}^{\textnormal{$-$1}}}.}\\
\parbox[b][0.3cm]{17.7cm}{\makebox[1ex]{\ensuremath{^{\hypertarget{PB30LEVEL3}{@}}}} Configuration=\ensuremath{\nu} p\ensuremath{_{\textnormal{1/2}}^{\textnormal{$-$1}}}. The assignment is tentative.}\\
\parbox[b][0.3cm]{17.7cm}{\makebox[1ex]{\ensuremath{^{\hypertarget{PB30LEVEL4}{\&}}}} Configuration=\ensuremath{\nu} i\ensuremath{_{\textnormal{13/2}}^{\textnormal{$-$1}}}.}\\
\parbox[b][0.3cm]{17.7cm}{\makebox[1ex]{\ensuremath{^{\hypertarget{PB30LEVEL5}{a}}}} Configuration=\ensuremath{\nu} f\ensuremath{_{\textnormal{7/2}}^{\textnormal{$-$1}}}. The assignment is tentative.}\\
\parbox[b][0.3cm]{17.7cm}{\makebox[1ex]{\ensuremath{^{\hypertarget{PB30LEVEL6}{b}}}} Configuration=\ensuremath{\nu} (f\ensuremath{_{\textnormal{5/2}}^{\textnormal{$-$1}}})\ensuremath{\otimes}2\ensuremath{^{\textnormal{+}}}.}\\
\parbox[b][0.3cm]{17.7cm}{\makebox[1ex]{\ensuremath{^{\hypertarget{PB30LEVEL7}{c}}}} Configuration=\ensuremath{\nu} (p\ensuremath{_{\textnormal{3/2}}^{\textnormal{$-$1}}})\ensuremath{\otimes}2\ensuremath{^{\textnormal{+}}}.}\\
\parbox[b][0.3cm]{17.7cm}{\makebox[1ex]{\ensuremath{^{\hypertarget{PB30LEVEL8}{d}}}} Configuration=\ensuremath{\nu} (i\ensuremath{_{\textnormal{13/2}}^{\textnormal{$-$1}}})\ensuremath{\otimes}2\ensuremath{^{\textnormal{+}}}.}\\
\parbox[b][0.3cm]{17.7cm}{\makebox[1ex]{\ensuremath{^{\hypertarget{PB30LEVEL9}{e}}}} From Adopted Levels.}\\
\vspace{0.5cm}

\underline{\ensuremath{\varepsilon,\beta^+} radiations}\\

\begin{textblock}{29}(0,27.3)
Continued on next page (footnotes at end of table)
\end{textblock}
\clearpage
%
\parbox[b][0.3cm]{17.7cm}{\makebox[1ex]{\ensuremath{^{\hypertarget{PB30DECAY0}{\dagger}}}} Deduced from the decay scheme using intensity balances considerations and by assuming no direct feeding to the g.s.}\\
\parbox[b][0.3cm]{17.7cm}{\makebox[1ex]{\ensuremath{^{\hypertarget{PB30DECAY1}{\ddagger}}}} Absolute intensity per 100 decays.}\\
\vspace{0.5cm}
\clearpage
\vspace{0.3cm}
\begin{landscape}
\vspace*{-0.5cm}
{\bf \small \underline{\ensuremath{^{\textnormal{201}}}Bi \ensuremath{\varepsilon} decay\hspace{0.2in}\href{https://www.nndc.bnl.gov/nsr/nsrlink.jsp?1978Ri04,B}{1978Ri04} (continued)}}\\
\vspace{0.3cm}
\underline{$\gamma$($^{201}$Pb)}\\
\vspace{0.34cm}
\parbox[b][0.3cm]{21.881866cm}{\addtolength{\parindent}{-0.254cm}I\ensuremath{\gamma} normalization: Deduced using \ensuremath{\Sigma}(I(\ensuremath{\gamma}+ce)[g.s. \ensuremath{^{\textnormal{201}}}Pb])=100\% and by assuming that there is no direct feeding to the \ensuremath{^{\textnormal{201}}}Pb g.s. (\ensuremath{J^{\pi}}=5/2\ensuremath{^{-}}).}\\
\vspace{0.34cm}

\clearpage
\vspace*{-0.5cm}
{\bf \small \underline{\ensuremath{^{\textnormal{201}}}Bi \ensuremath{\varepsilon} decay\hspace{0.2in}\href{https://www.nndc.bnl.gov/nsr/nsrlink.jsp?1978Ri04,B}{1978Ri04} (continued)}}\\
\vspace{0.3cm}
\underline{$\gamma$($^{201}$Pb) (continued)}\\
\vspace{0.3cm}
\parbox[b][0.3cm]{21.881866cm}{\makebox[1ex]{\ensuremath{^{\hypertarget{BI30GAMMA0}{\dagger}}}} From \href{https://www.nndc.bnl.gov/nsr/nsrlink.jsp?1978Ri04,B}{1978Ri04} where \ensuremath{\Delta}E\ensuremath{\gamma}\ensuremath{\leq}0.5 keV for I\ensuremath{\gamma}\ensuremath{\geq}10 and \ensuremath{\Delta}E\ensuremath{\gamma}\ensuremath{\leq}1.0 keV for I\ensuremath{\gamma}\ensuremath{\leq}1 were reported. The evaluator assigns \ensuremath{\Delta}E\ensuremath{\gamma}=1 keV for I\ensuremath{\gamma}\ensuremath{<}10 and 0.5 keV for}\\
\parbox[b][0.3cm]{21.881866cm}{{\ }{\ }I\ensuremath{\gamma}\ensuremath{\geq}10.}\\
\parbox[b][0.3cm]{21.881866cm}{\makebox[1ex]{\ensuremath{^{\hypertarget{BI30GAMMA1}{\ddagger}}}} From \href{https://www.nndc.bnl.gov/nsr/nsrlink.jsp?1978Ri04,B}{1978Ri04} where \ensuremath{\Delta}I\ensuremath{\gamma}\ensuremath{\leq}5\% for I\ensuremath{\gamma}\ensuremath{\geq}10 and \ensuremath{\leq}20\% for I\ensuremath{\gamma}\ensuremath{\leq}1 were reported. The evaluator assigns \ensuremath{\Delta}I\ensuremath{\gamma}=5\% for I\ensuremath{\gamma}\ensuremath{\geq}10 and 20\% for I\ensuremath{\gamma}\ensuremath{<}10.}\\
\parbox[b][0.3cm]{21.881866cm}{\makebox[1ex]{\ensuremath{^{\hypertarget{BI30GAMMA2}{\#}}}} Based on \ensuremath{\alpha}(K)exp, \ensuremath{\alpha}(L)exp and subshell ratios in \href{https://www.nndc.bnl.gov/nsr/nsrlink.jsp?1978Ri04,B}{1978Ri04}, unless otherwise stated; \ensuremath{\alpha}(K)exp and \ensuremath{\alpha}(L)exp values were normalized using M4 mult. for 629.5\ensuremath{\gamma},}\\
\parbox[b][0.3cm]{21.881866cm}{{\ }{\ }as determined in \href{https://www.nndc.bnl.gov/nsr/nsrlink.jsp?1956St05,B}{1956St05}.}\\
\parbox[b][0.3cm]{21.881866cm}{\makebox[1ex]{\ensuremath{^{\hypertarget{BI30GAMMA3}{@}}}} From \ensuremath{\alpha}(K)exp, \ensuremath{\alpha}(L)exp and subshell ratios in \href{https://www.nndc.bnl.gov/nsr/nsrlink.jsp?1978Ri04,B}{1978Ri04} and the briccmixing program, unless otherwise stated.}\\
\parbox[b][0.3cm]{21.881866cm}{\makebox[1ex]{\ensuremath{^{\hypertarget{BI30GAMMA4}{\&}}}} For absolute intensity per 100 decays, multiply by 0.247 \textit{11}.}\\
\parbox[b][0.3cm]{21.881866cm}{\makebox[1ex]{\ensuremath{^{\hypertarget{BI30GAMMA5}{a}}}} Total theoretical internal conversion coefficients, calculated using the BrIcc code (\href{https://www.nndc.bnl.gov/nsr/nsrlink.jsp?2008Ki07,B}{2008Ki07}) with Frozen orbital approximation based on \ensuremath{\gamma}-ray energies,}\\
\parbox[b][0.3cm]{21.881866cm}{{\ }{\ }assigned multipolarities, and mixing ratios, unless otherwise specified.}\\
\parbox[b][0.3cm]{21.881866cm}{\makebox[1ex]{\ensuremath{^{\hypertarget{BI30GAMMA6}{b}}}} Placement of transition in the level scheme is uncertain.}\\
\parbox[b][0.3cm]{21.881866cm}{\makebox[1ex]{\ensuremath{^{\hypertarget{BI30GAMMA7}{x}}}} \ensuremath{\gamma} ray not placed in level scheme.}\\
\vspace{0.5cm}
\end{landscape}\clearpage
\clearpage
\begin{figure}[h]
\begin{center}
\includegraphics{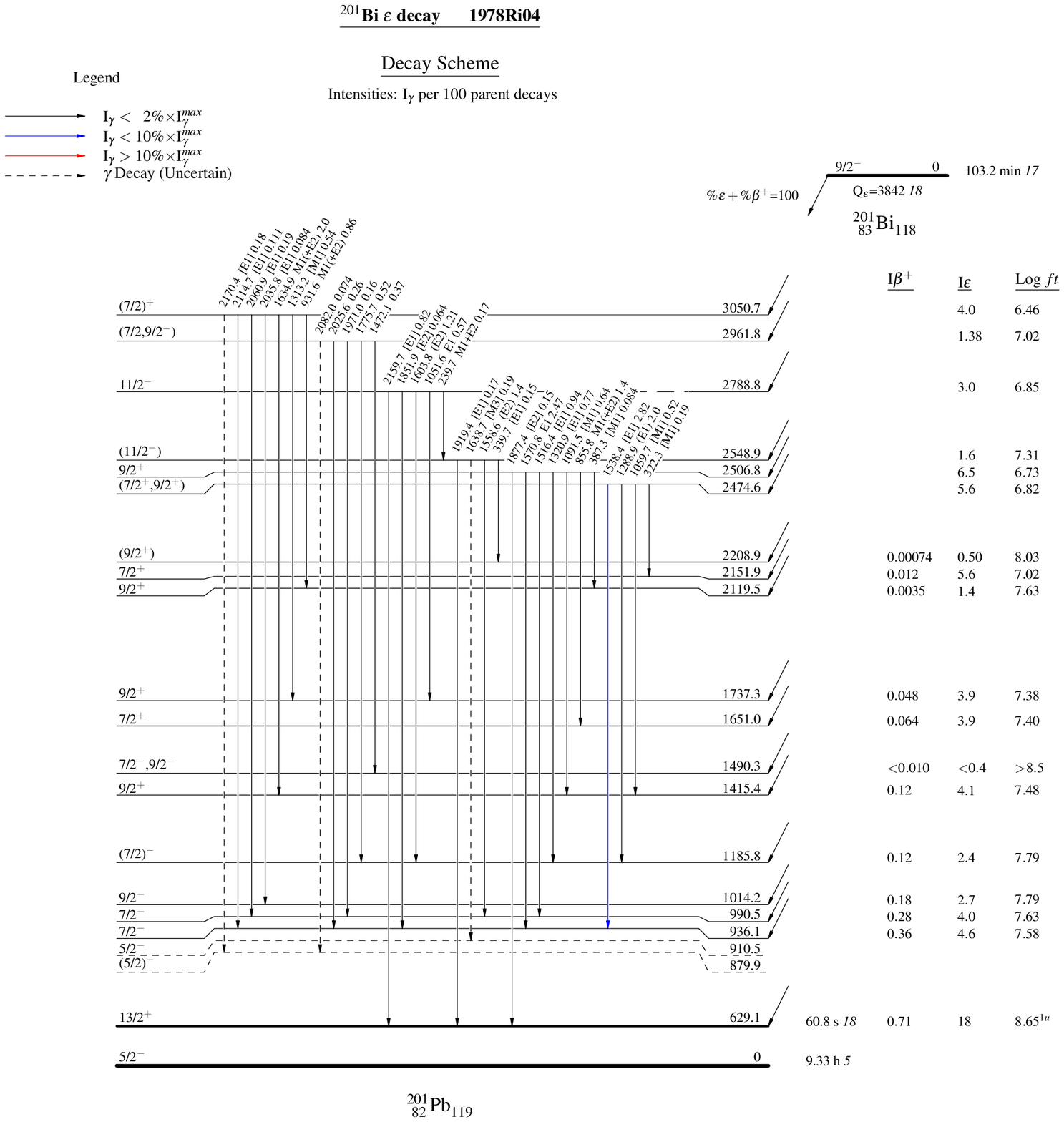}\\
\end{center}
\end{figure}
\clearpage
\begin{figure}[h]
\begin{center}
\includegraphics{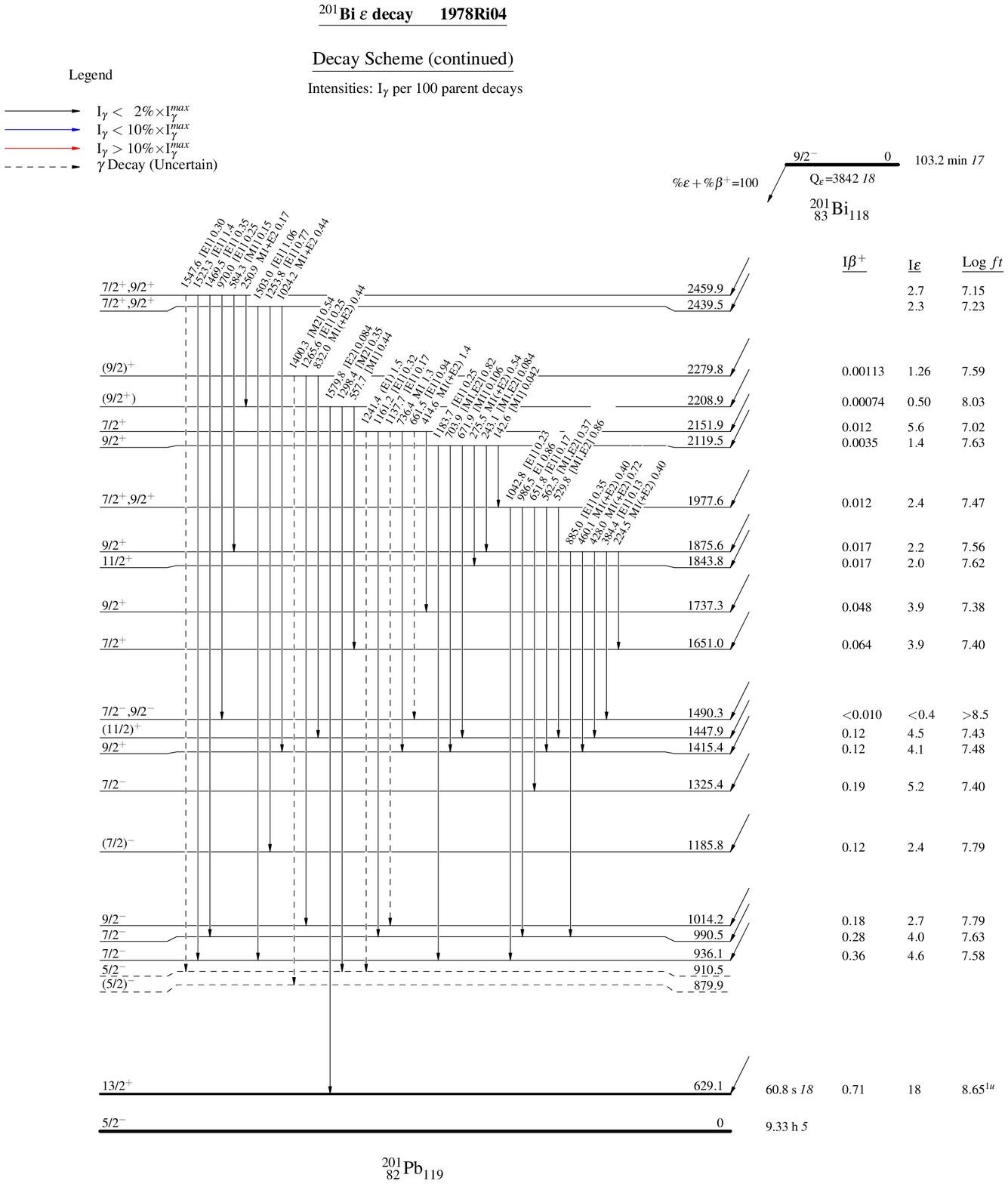}\\
\end{center}
\end{figure}
\clearpage
\begin{figure}[h]
\begin{center}
\includegraphics{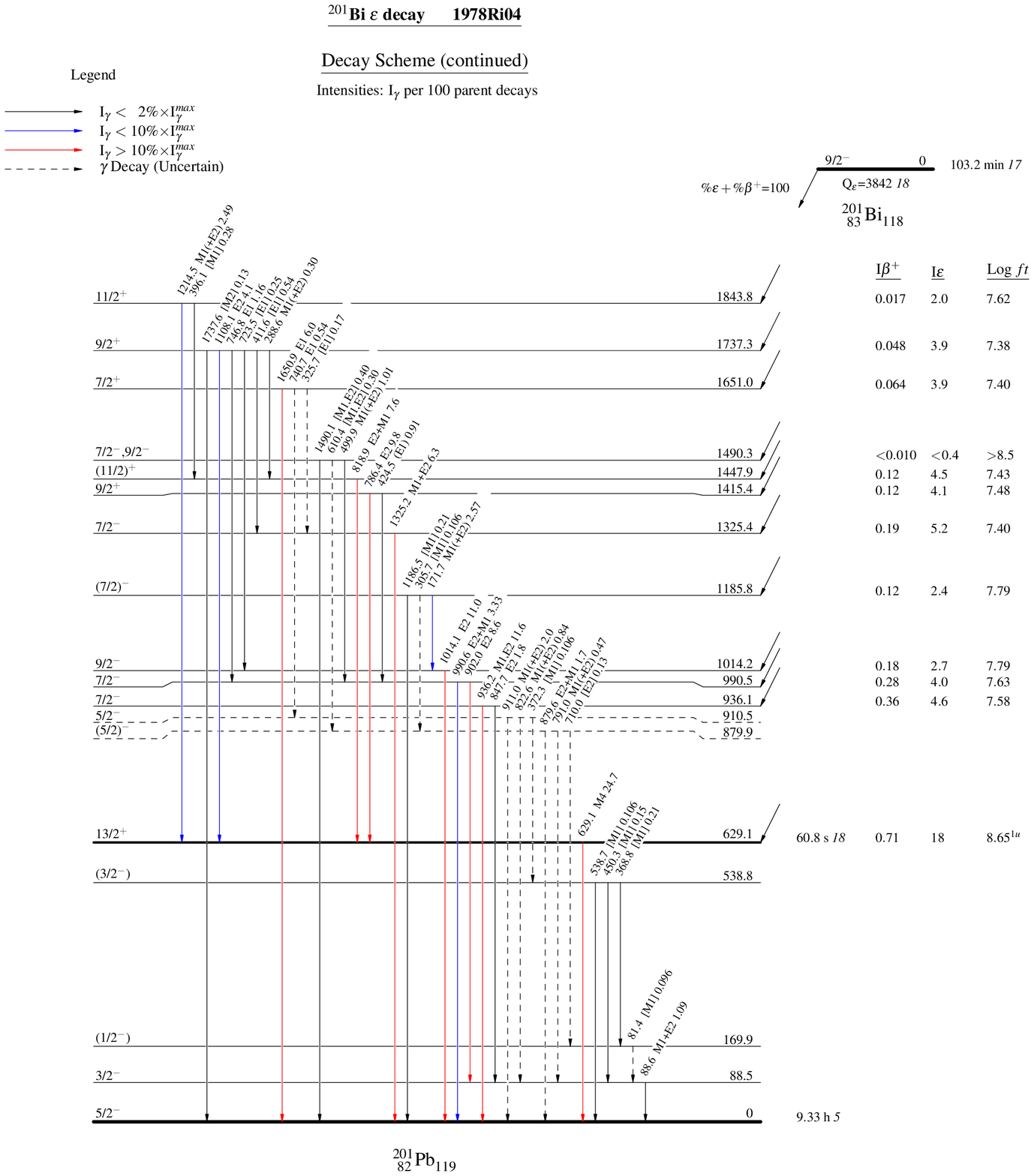}\\
\end{center}
\end{figure}
\clearpage
%205PO A DECAY
\subsection[\hspace{-0.2cm}\ensuremath{^{\textnormal{205}}}Po \ensuremath{\alpha} decay]{ }
\vspace{-27pt}
\vspace{0.3cm}
\hypertarget{PO31}{{\bf \small \underline{\ensuremath{^{\textnormal{205}}}Po \ensuremath{\alpha} decay\hspace{0.2in}\href{https://www.nndc.bnl.gov/nsr/nsrlink.jsp?1967Ti04,B}{1967Ti04},\href{https://www.nndc.bnl.gov/nsr/nsrlink.jsp?1970Jo26,B}{1970Jo26}}}}\\
\vspace{4pt}
\vspace{8pt}
\parbox[b][0.3cm]{17.7cm}{\addtolength{\parindent}{-0.2in}Parent: $^{205}$Po: E=0; J$^{\pi}$=5/2\ensuremath{^{-}}; T$_{1/2}$=1.74 h {\it 8}; Q(\ensuremath{\alpha})=5325 {\it 10}; \%\ensuremath{\alpha} decay=0.040 {\it 12}

}\\
\parbox[b][0.3cm]{17.7cm}{\addtolength{\parindent}{-0.2in}\ensuremath{^{205}}Po-J$^{\pi}$,T$_{1/2}$: From \href{https://www.nndc.bnl.gov/nsr/nsrlink.jsp?2020Ko17,B}{2020Ko17}.}\\
\parbox[b][0.3cm]{17.7cm}{\addtolength{\parindent}{-0.2in}\ensuremath{^{205}}Po-Q(\ensuremath{\alpha}): From \href{https://www.nndc.bnl.gov/nsr/nsrlink.jsp?2021Wa16,B}{2021Wa16}.}\\
\parbox[b][0.3cm]{17.7cm}{\addtolength{\parindent}{-0.2in}\ensuremath{^{205}}Po-\%\ensuremath{\alpha} decay: From \href{https://www.nndc.bnl.gov/nsr/nsrlink.jsp?2020Ko17,B}{2020Ko17}.}\\
\vspace{12pt}
\underline{$^{201}$Pb Levels}\\
\begin{longtable}{ccccc@{\extracolsep{\fill}}c}
\multicolumn{2}{c}{E(level)$^{}$}&J$^{\pi}$$^{{\hyperlink{PB31LEVEL0}{\dagger}}}$&\multicolumn{2}{c}{T$_{1/2}$$^{{\hyperlink{PB31LEVEL0}{\dagger}}}$}&\\[-.2cm]
\multicolumn{2}{c}{\hrulefill}&\hrulefill&\multicolumn{2}{c}{\hrulefill}&
\endfirsthead
\multicolumn{1}{r@{}}{0}&\multicolumn{1}{@{}l}{}&\multicolumn{1}{l}{5/2\ensuremath{^{-}}}&\multicolumn{1}{r@{}}{9}&\multicolumn{1}{@{.}l}{33 h {\it 5}}&\\
\end{longtable}
\parbox[b][0.3cm]{17.7cm}{\makebox[1ex]{\ensuremath{^{\hypertarget{PB31LEVEL0}{\dagger}}}} From Adopted Levels.}\\
\vspace{0.5cm}

\underline{\ensuremath{\alpha} radiations}\\
\begin{longtable}{ccccccccc@{\extracolsep{\fill}}c}
\multicolumn{2}{c}{E$\alpha^{{}}$}&\multicolumn{2}{c}{E(level)}&\multicolumn{2}{c}{I$\alpha^{{\hyperlink{PB31DECAY1}{\ddagger}}}$}&\multicolumn{2}{c}{HF$^{{\hyperlink{PB31DECAY0}{\dagger}}}$}&Comments&\\[-.2cm]
\multicolumn{2}{c}{\hrulefill}&\multicolumn{2}{c}{\hrulefill}&\multicolumn{2}{c}{\hrulefill}&\multicolumn{2}{c}{\hrulefill}&\hrulefill&
\endfirsthead
\multicolumn{1}{r@{}}{5220}&\multicolumn{1}{@{ }l}{{\it 10}}&\multicolumn{1}{r@{}}{0}&\multicolumn{1}{@{}l}{}&\multicolumn{1}{r@{}}{100}&\multicolumn{1}{@{}l}{}&\multicolumn{1}{r@{}}{1}&\multicolumn{1}{@{.}l}{1 {\it 4}}&\parbox[t][0.3cm]{12.8046cm}{\raggedright E$\alpha$: From \href{https://www.nndc.bnl.gov/nsr/nsrlink.jsp?1967Ti04,B}{1967Ti04}; Other: 5224 \textit{10} (\href{https://www.nndc.bnl.gov/nsr/nsrlink.jsp?1970Jo26,B}{1970Jo26}).\vspace{0.1cm}}&\\
\end{longtable}
\parbox[b][0.3cm]{17.7cm}{\makebox[1ex]{\ensuremath{^{\hypertarget{PB31DECAY0}{\dagger}}}} Using r\ensuremath{_{\textnormal{0}}}(\ensuremath{^{\textnormal{201}}}Pb)=1.4586 \textit{16} from \href{https://www.nndc.bnl.gov/nsr/nsrlink.jsp?2020Si16,B}{2020Si16}.}\\
\parbox[b][0.3cm]{17.7cm}{\makebox[1ex]{\ensuremath{^{\hypertarget{PB31DECAY1}{\ddagger}}}} For absolute intensity per 100 decays, multiply by 0.00040 \textit{12}.}\\
\vspace{0.5cm}
\clearpage
%200HG(A,3NG)
\subsection[\hspace{-0.2cm}\ensuremath{^{\textnormal{200}}}Hg(\ensuremath{\alpha},3n\ensuremath{\gamma})]{ }
\vspace{-27pt}
\vspace{0.3cm}
\hypertarget{PB32}{{\bf \small \underline{\ensuremath{^{\textnormal{200}}}Hg(\ensuremath{\alpha},3n\ensuremath{\gamma})\hspace{0.2in}\href{https://www.nndc.bnl.gov/nsr/nsrlink.jsp?1988Ro08,B}{1988Ro08},\href{https://www.nndc.bnl.gov/nsr/nsrlink.jsp?1981He07,B}{1981He07}}}}\\
\vspace{4pt}
\vspace{8pt}
\parbox[b][0.3cm]{17.7cm}{\addtolength{\parindent}{-0.2in}\href{https://www.nndc.bnl.gov/nsr/nsrlink.jsp?1988Ro08,B}{1988Ro08}: E(\ensuremath{\alpha})=53 MeV; Target: enriched liquid \ensuremath{^{\textnormal{200}}}Hg; Detectors: intrinsic germanium and Ge(Li); Measured: \ensuremath{\gamma}, \ensuremath{\gamma}(\ensuremath{\theta}),}\\
\parbox[b][0.3cm]{17.7cm}{\ensuremath{\gamma}(\ensuremath{\theta},\ensuremath{\beta},t) and \ensuremath{\gamma}\ensuremath{\gamma}(t).}\\
\parbox[b][0.3cm]{17.7cm}{\addtolength{\parindent}{-0.2in}\href{https://www.nndc.bnl.gov/nsr/nsrlink.jsp?1981He07,B}{1981He07}: E(\ensuremath{\alpha})=40 MeV; Target: enriched to 95.7\% \ensuremath{^{\textnormal{200}}}Hg oxide; Detectors: Ge(Li); Measured: \ensuremath{\gamma}, \ensuremath{\gamma}\ensuremath{\gamma}, \ensuremath{\gamma}(\ensuremath{\theta}), \ensuremath{\gamma}(t) and \ensuremath{\gamma}\ensuremath{\gamma}(t).}\\
\parbox[b][0.3cm]{17.7cm}{\addtolength{\parindent}{-0.2in}Other: \href{https://www.nndc.bnl.gov/nsr/nsrlink.jsp?1977He06,B}{1977He06}.}\\
\vspace{12pt}
\underline{$^{201}$Pb Levels}\\

\parbox[b][0.3cm]{17.7cm}{\makebox[1ex]{\ensuremath{^{\hypertarget{PB32LEVEL0}{\dagger}}}} From a least-squares fit to E\ensuremath{\gamma}.}\\
\parbox[b][0.3cm]{17.7cm}{\makebox[1ex]{\ensuremath{^{\hypertarget{PB32LEVEL1}{\ddagger}}}} From \href{https://www.nndc.bnl.gov/nsr/nsrlink.jsp?1988Ro08,B}{1988Ro08}, unless otherwise stated.}\\
\parbox[b][0.3cm]{17.7cm}{\makebox[1ex]{\ensuremath{^{\hypertarget{PB32LEVEL2}{\#}}}} From Adopted Levels, unless otherwise stated.}\\
\parbox[b][0.3cm]{17.7cm}{\makebox[1ex]{\ensuremath{^{\hypertarget{PB32LEVEL3}{@}}}} Configuration=\ensuremath{\nu} f\ensuremath{_{\textnormal{5/2}}^{\textnormal{$-$1}}}.}\\
\parbox[b][0.3cm]{17.7cm}{\makebox[1ex]{\ensuremath{^{\hypertarget{PB32LEVEL4}{\&}}}} Configuration=\ensuremath{\nu} p\ensuremath{_{\textnormal{3/2}}^{\textnormal{$-$1}}}.}\\
\begin{textblock}{29}(0,27.3)
Continued on next page (footnotes at end of table)
\end{textblock}
\clearpage
\vspace*{-0.5cm}
{\bf \small \underline{\ensuremath{^{\textnormal{200}}}Hg(\ensuremath{\alpha},3n\ensuremath{\gamma})\hspace{0.2in}\href{https://www.nndc.bnl.gov/nsr/nsrlink.jsp?1988Ro08,B}{1988Ro08},\href{https://www.nndc.bnl.gov/nsr/nsrlink.jsp?1981He07,B}{1981He07} (continued)}}\\
\vspace{0.3cm}
\underline{$^{201}$Pb Levels (continued)}\\
\vspace{0.3cm}
\parbox[b][0.3cm]{17.7cm}{\makebox[1ex]{\ensuremath{^{\hypertarget{PB32LEVEL5}{a}}}} Configuration=\ensuremath{\nu} p\ensuremath{_{\textnormal{1/2}}^{\textnormal{$-$1}}}. The assignment is tentative.}\\
\parbox[b][0.3cm]{17.7cm}{\makebox[1ex]{\ensuremath{^{\hypertarget{PB32LEVEL6}{b}}}} Configuration=\ensuremath{\nu} i\ensuremath{_{\textnormal{13/2}}^{\textnormal{$-$1}}}.}\\
\parbox[b][0.3cm]{17.7cm}{\makebox[1ex]{\ensuremath{^{\hypertarget{PB32LEVEL7}{c}}}} Probably an admixture of configuration=\ensuremath{\nu} (f\ensuremath{_{\textnormal{5/2}}^{\textnormal{$-$1}}},p\ensuremath{_{\textnormal{1/2}}^{\textnormal{$-$1}}},i\ensuremath{_{\textnormal{13/2}}^{\textnormal{$-$1}}})\ensuremath{\otimes}2\ensuremath{^{\textnormal{+}}} and configuration=\ensuremath{\nu} (i\ensuremath{_{\textnormal{13/2}}^{\textnormal{$-$1}}})\ensuremath{\otimes}2\ensuremath{^{\textnormal{+}}}.}\\
\parbox[b][0.3cm]{17.7cm}{\makebox[1ex]{\ensuremath{^{\hypertarget{PB32LEVEL8}{d}}}} Probably an admixture of configuration=\ensuremath{\nu} (f\ensuremath{_{\textnormal{5/2}}^{\textnormal{$-$1}}},p\ensuremath{_{\textnormal{1/2}}^{\textnormal{$-$1}}},i\ensuremath{_{\textnormal{13/2}}^{\textnormal{$-$1}}})\ensuremath{\otimes}4\ensuremath{^{\textnormal{+}}} and configuration=\ensuremath{\nu} (i\ensuremath{_{\textnormal{13/2}}^{\textnormal{$-$1}}})\ensuremath{\otimes}4\ensuremath{^{\textnormal{+}}}.}\\
\parbox[b][0.3cm]{17.7cm}{\makebox[1ex]{\ensuremath{^{\hypertarget{PB32LEVEL9}{e}}}} Configuration=\ensuremath{\nu} (f\ensuremath{_{\textnormal{5/2}}^{\textnormal{$-$2}}},i\ensuremath{_{\textnormal{13/2}}^{\textnormal{$-$1}}}).}\\
\parbox[b][0.3cm]{17.7cm}{\makebox[1ex]{\ensuremath{^{\hypertarget{PB32LEVEL10}{f}}}} Configuration=\ensuremath{\nu} [p\ensuremath{_{\textnormal{3/2}}^{\textnormal{$-$1}}},(i\ensuremath{_{\textnormal{13/2}}^{\textnormal{$-$2}}})\ensuremath{_{\textnormal{12+}}}].}\\
\parbox[b][0.3cm]{17.7cm}{\makebox[1ex]{\ensuremath{^{\hypertarget{PB32LEVEL11}{g}}}} Probably an admixture of configuration=\ensuremath{\nu} [f\ensuremath{_{\textnormal{5/2}}^{\textnormal{$-$1}}},(i\ensuremath{_{\textnormal{13/2}}^{\textnormal{$-$2}}})\ensuremath{_{\textnormal{10+}}}], configuration=\ensuremath{\nu} [p\ensuremath{_{\textnormal{3/2}}^{\textnormal{$-$1}}},(i\ensuremath{_{\textnormal{13/2}}^{\textnormal{$-$2}}})\ensuremath{_{\textnormal{12+}}}] and configuration=\ensuremath{\nu}}\\
\parbox[b][0.3cm]{17.7cm}{{\ }{\ }[p\ensuremath{_{\textnormal{1/2}}^{\textnormal{$-$1}}},(i\ensuremath{_{\textnormal{13/2}}^{\textnormal{$-$2}}})\ensuremath{_{\textnormal{12+}}}].}\\
\parbox[b][0.3cm]{17.7cm}{\makebox[1ex]{\ensuremath{^{\hypertarget{PB32LEVEL12}{h}}}} Configuration=\ensuremath{\nu} [f\ensuremath{_{\textnormal{5/2}}^{\textnormal{$-$1}}},(i\ensuremath{_{\textnormal{13/2}}^{\textnormal{$-$2}}})\ensuremath{_{\textnormal{12+}}}].}\\
\parbox[b][0.3cm]{17.7cm}{\makebox[1ex]{\ensuremath{^{\hypertarget{PB32LEVEL13}{i}}}} Configuration=\ensuremath{\nu} (i\ensuremath{_{\textnormal{13/2}}^{\textnormal{$-$3}}}).}\\
\parbox[b][0.3cm]{17.7cm}{\makebox[1ex]{\ensuremath{^{\hypertarget{PB32LEVEL14}{j}}}} Configuration=\ensuremath{\nu} (p\ensuremath{_{\textnormal{3/2}}^{\textnormal{$-$1}}},f\ensuremath{_{\textnormal{5/2}}^{\textnormal{$-$1}}},i\ensuremath{_{\textnormal{13/2}}^{\textnormal{$-$3}}}).}\\
\vspace{0.5cm}
\underline{$\gamma$($^{201}$Pb)}\\

\begin{textblock}{29}(0,27.3)
Continued on next page (footnotes at end of table)
\end{textblock}
\clearpage
%
\parbox[b][0.3cm]{17.7cm}{\makebox[1ex]{\ensuremath{^{\hypertarget{PB32GAMMA0}{\dagger}}}} From \href{https://www.nndc.bnl.gov/nsr/nsrlink.jsp?1988Ro08,B}{1988Ro08}, unless otherwise specified. Evaluator assigns a 0.5 keV uncertainty for E\ensuremath{\gamma} and a 10\% uncertainty for I\ensuremath{\gamma}.}\\
\parbox[b][0.3cm]{17.7cm}{\makebox[1ex]{\ensuremath{^{\hypertarget{PB32GAMMA1}{\ddagger}}}} From \href{https://www.nndc.bnl.gov/nsr/nsrlink.jsp?1981He07,B}{1981He07}.}\\
\parbox[b][0.3cm]{17.7cm}{\makebox[1ex]{\ensuremath{^{\hypertarget{PB32GAMMA2}{\#}}}} From adopted gammas.}\\
\begin{textblock}{29}(0,27.3)
Continued on next page (footnotes at end of table)
\end{textblock}
\clearpage
\vspace*{-0.5cm}
{\bf \small \underline{\ensuremath{^{\textnormal{200}}}Hg(\ensuremath{\alpha},3n\ensuremath{\gamma})\hspace{0.2in}\href{https://www.nndc.bnl.gov/nsr/nsrlink.jsp?1988Ro08,B}{1988Ro08},\href{https://www.nndc.bnl.gov/nsr/nsrlink.jsp?1981He07,B}{1981He07} (continued)}}\\
\vspace{0.3cm}
\underline{$\gamma$($^{201}$Pb) (continued)}\\
\vspace{0.3cm}
\parbox[b][0.3cm]{17.7cm}{\makebox[1ex]{\ensuremath{^{\hypertarget{PB32GAMMA3}{@}}}} From \ensuremath{\gamma}(\ensuremath{\theta}) in \href{https://www.nndc.bnl.gov/nsr/nsrlink.jsp?1981He07,B}{1981He07} and \href{https://www.nndc.bnl.gov/nsr/nsrlink.jsp?1988Ro08,B}{1988Ro08}, unless otherwise stated.}\\
\parbox[b][0.3cm]{17.7cm}{\makebox[1ex]{\ensuremath{^{\hypertarget{PB32GAMMA4}{\&}}}} Total theoretical internal conversion coefficients, calculated using the BrIcc code (\href{https://www.nndc.bnl.gov/nsr/nsrlink.jsp?2008Ki07,B}{2008Ki07}) with Frozen orbital approximation}\\
\parbox[b][0.3cm]{17.7cm}{{\ }{\ }based on \ensuremath{\gamma}-ray energies, assigned multipolarities, and mixing ratios, unless otherwise specified.}\\
\parbox[b][0.3cm]{17.7cm}{\makebox[1ex]{\ensuremath{^{\hypertarget{PB32GAMMA5}{a}}}} Placement of transition in the level scheme is uncertain.}\\
\vspace{0.5cm}
\clearpage
\begin{figure}[h]
\begin{center}
\includegraphics{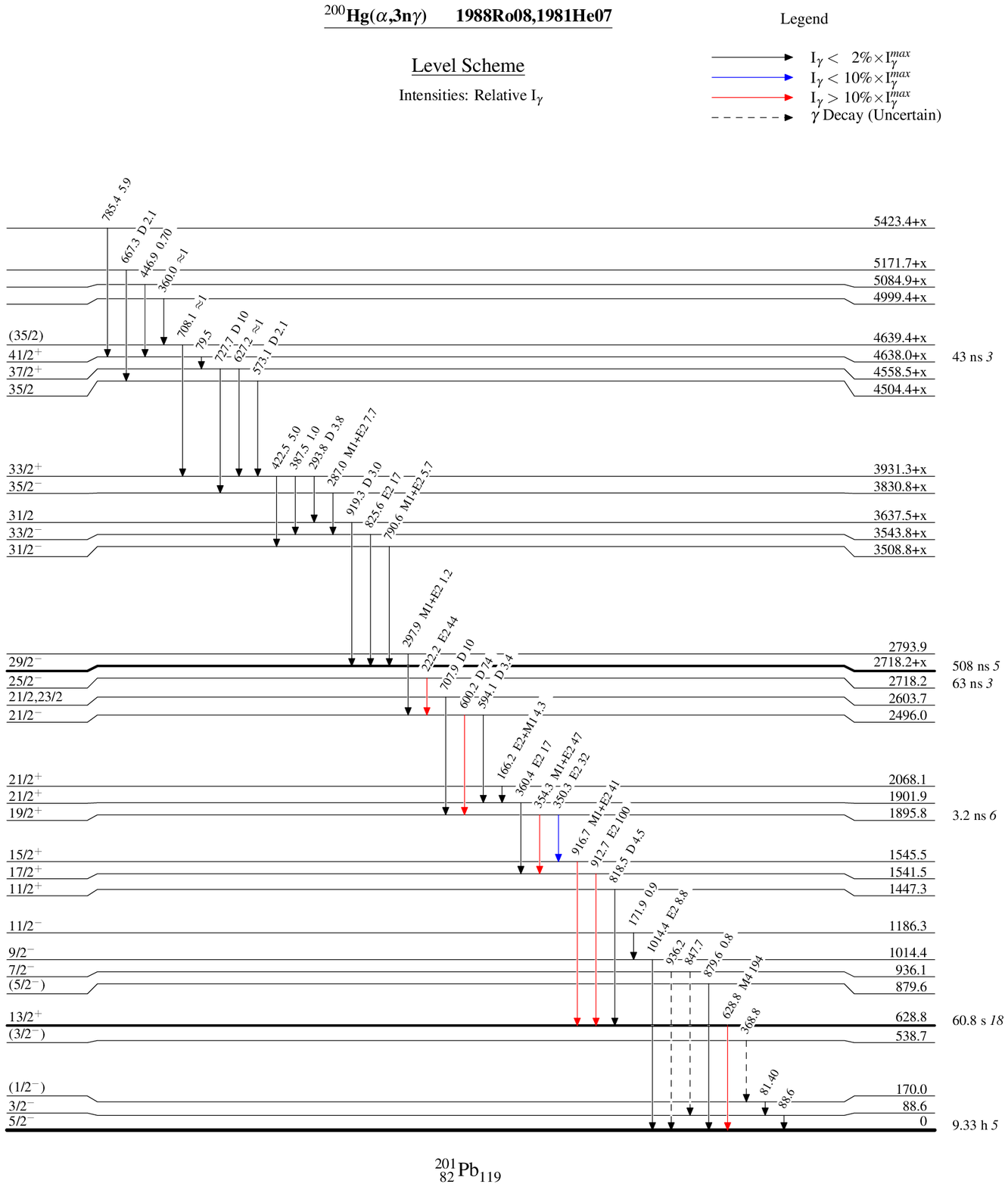}\\
\end{center}
\end{figure}
\clearpage
%192OS(14C,5NG)
\subsection[\hspace{-0.2cm}\ensuremath{^{\textnormal{192}}}Os(\ensuremath{^{\textnormal{14}}}C,5n\ensuremath{\gamma})]{ }
\vspace{-27pt}
\vspace{0.3cm}
\hypertarget{PB33}{{\bf \small \underline{\ensuremath{^{\textnormal{192}}}Os(\ensuremath{^{\textnormal{14}}}C,5n\ensuremath{\gamma})\hspace{0.2in}\href{https://www.nndc.bnl.gov/nsr/nsrlink.jsp?1995Ba70,B}{1995Ba70}}}}\\
\vspace{4pt}
\vspace{8pt}
\parbox[b][0.3cm]{17.7cm}{\addtolength{\parindent}{-0.2in}E(\ensuremath{^{\textnormal{14}}}C)=76 MeV; Target: \ensuremath{^{\textnormal{192}}}Os, enriched to 99\% with average thickness of 100 mg/cm\ensuremath{^{\textnormal{2}}}; Detectors: 12 Compton suppressed Ge}\\
\parbox[b][0.3cm]{17.7cm}{detectors and 48 BGO scintillation counters; Measured: E\ensuremath{\gamma}, I\ensuremath{\gamma}, \ensuremath{\gamma}\ensuremath{\gamma} coin, \ensuremath{\gamma}\ensuremath{\gamma}(\ensuremath{\theta})(DCO); Deduced: level scheme, \ensuremath{J^{\pi}}.}\\
\parbox[b][0.3cm]{17.7cm}{\addtolength{\parindent}{-0.2in}Other: \href{https://www.nndc.bnl.gov/nsr/nsrlink.jsp?1989Su12,B}{1989Su12}, \href{https://www.nndc.bnl.gov/nsr/nsrlink.jsp?1992Ba39,B}{1992Ba39} (superseded by \href{https://www.nndc.bnl.gov/nsr/nsrlink.jsp?1995Ba70,B}{1995Ba70}).}\\
\vspace{12pt}
\underline{$^{201}$Pb Levels}\\

\begin{textblock}{29}(0,27.3)
Continued on next page (footnotes at end of table)
\end{textblock}
\clearpage
%
\parbox[b][0.3cm]{17.7cm}{\makebox[1ex]{\ensuremath{^{\hypertarget{PB33LEVEL0}{\dagger}}}} From a least-squares fit to E\ensuremath{\gamma}. Energies are relative to E(13/2\ensuremath{^{+}})=629.11 \textit{18} keV. For levels labeled with +X, +Y, +Z, +U and}\\
\parbox[b][0.3cm]{17.7cm}{{\ }{\ }+V the excitation energies are relative to the 2719.6+Y, 4060.6+y, 0+Z, 0+U and 0+V states, respectively.}\\
\parbox[b][0.3cm]{17.7cm}{\makebox[1ex]{\ensuremath{^{\hypertarget{PB33LEVEL1}{\ddagger}}}} From deduced transition multipolarities and multiple decay branches in \href{https://www.nndc.bnl.gov/nsr/nsrlink.jsp?1995Ba70,B}{1995Ba70}, unless otherwise stated.}\\
\parbox[b][0.3cm]{17.7cm}{\makebox[1ex]{\ensuremath{^{\hypertarget{PB33LEVEL2}{\#}}}} Configuration=\ensuremath{\nu} i\ensuremath{_{\textnormal{13/2}}^{\textnormal{$-$1}}}.}\\
\parbox[b][0.3cm]{17.7cm}{\makebox[1ex]{\ensuremath{^{\hypertarget{PB33LEVEL3}{@}}}} Probably an admixture of configuration=\ensuremath{\nu} (f\ensuremath{_{\textnormal{5/2}}^{\textnormal{$-$1}}},p\ensuremath{_{\textnormal{1/2}}^{\textnormal{$-$1}}},i\ensuremath{_{\textnormal{13/2}}^{\textnormal{$-$1}}})\ensuremath{\otimes}2\ensuremath{^{\textnormal{+}}} and configuration=\ensuremath{\nu} (i\ensuremath{_{\textnormal{13/2}}^{\textnormal{$-$1}}})\ensuremath{\otimes}2\ensuremath{^{\textnormal{+}}}.}\\
\parbox[b][0.3cm]{17.7cm}{\makebox[1ex]{\ensuremath{^{\hypertarget{PB33LEVEL4}{\&}}}} Probably an admixture of configuration=\ensuremath{\nu} (f\ensuremath{_{\textnormal{5/2}}^{\textnormal{$-$1}}},p\ensuremath{_{\textnormal{1/2}}^{\textnormal{$-$1}}},i\ensuremath{_{\textnormal{13/2}}^{\textnormal{$-$1}}})\ensuremath{\otimes}4\ensuremath{^{\textnormal{+}}} and configuration=\ensuremath{\nu} (i\ensuremath{_{\textnormal{13/2}}^{\textnormal{$-$1}}})\ensuremath{\otimes}4\ensuremath{^{\textnormal{+}}}.}\\
\parbox[b][0.3cm]{17.7cm}{\makebox[1ex]{\ensuremath{^{\hypertarget{PB33LEVEL5}{a}}}} Configuration=\ensuremath{\nu} (f\ensuremath{_{\textnormal{5/2}}^{\textnormal{$-$2}}},i\ensuremath{_{\textnormal{13/2}}^{\textnormal{$-$1}}}).}\\
\parbox[b][0.3cm]{17.7cm}{\makebox[1ex]{\ensuremath{^{\hypertarget{PB33LEVEL6}{b}}}} Configuration=\ensuremath{\nu} [p\ensuremath{_{\textnormal{3/2}}^{\textnormal{$-$1}}},(i\ensuremath{_{\textnormal{13/2}}^{\textnormal{$-$2}}})\ensuremath{_{\textnormal{12+}}}].}\\
\parbox[b][0.3cm]{17.7cm}{\makebox[1ex]{\ensuremath{^{\hypertarget{PB33LEVEL7}{c}}}} Probably an admixture of configuration=\ensuremath{\nu} [f\ensuremath{_{\textnormal{5/2}}^{\textnormal{$-$1}}},(i\ensuremath{_{\textnormal{13/2}}^{\textnormal{$-$2}}})\ensuremath{_{\textnormal{10+}}}], configuration=\ensuremath{\nu} [p\ensuremath{_{\textnormal{3/2}}^{\textnormal{$-$1}}},(i\ensuremath{_{\textnormal{13/2}}^{\textnormal{$-$2}}})\ensuremath{_{\textnormal{12+}}}] and configuration=\ensuremath{\nu}}\\
\parbox[b][0.3cm]{17.7cm}{{\ }{\ }[p\ensuremath{_{\textnormal{1/2}}^{\textnormal{$-$1}}},(i\ensuremath{_{\textnormal{13/2}}^{\textnormal{$-$2}}})\ensuremath{_{\textnormal{12+}}}].}\\
\parbox[b][0.3cm]{17.7cm}{\makebox[1ex]{\ensuremath{^{\hypertarget{PB33LEVEL8}{d}}}} Configuration=\ensuremath{\nu} [f\ensuremath{_{\textnormal{5/2}}^{\textnormal{$-$1}}},(i\ensuremath{_{\textnormal{13/2}}^{\textnormal{$-$2}}})\ensuremath{_{\textnormal{12+}}}].}\\
\parbox[b][0.3cm]{17.7cm}{\makebox[1ex]{\ensuremath{^{\hypertarget{PB33LEVEL9}{e}}}} Configuration=\ensuremath{\nu} (i\ensuremath{_{\textnormal{13/2}}^{\textnormal{$-$3}}}).}\\
\parbox[b][0.3cm]{17.7cm}{\makebox[1ex]{\ensuremath{^{\hypertarget{PB33LEVEL10}{f}}}} Configuration=\ensuremath{\nu} (p\ensuremath{_{\textnormal{3/2}}^{\textnormal{$-$1}}},f\ensuremath{_{\textnormal{5/2}}^{\textnormal{$-$1}}},i\ensuremath{_{\textnormal{13/2}}^{\textnormal{$-$3}}}).}\\
\parbox[b][0.3cm]{17.7cm}{\makebox[1ex]{\ensuremath{^{\hypertarget{PB33LEVEL11}{g}}}} Band(A): configuration=\ensuremath{\nu} [p\ensuremath{_{\textnormal{3/2}}^{\textnormal{$-$1}}},(i\ensuremath{_{\textnormal{13/2}}})\ensuremath{^{\textnormal{$-$2}}})\ensuremath{_{\textnormal{12+}}}]\ensuremath{\otimes} \ensuremath{\pi} (h\ensuremath{_{\textnormal{9/2}}^{\textnormal{+1}}},i\ensuremath{_{\textnormal{13/2}}^{\textnormal{+1}}})\ensuremath{_{\textnormal{11$-$}}}.\hphantom{a}Band 2 in \href{https://www.nndc.bnl.gov/nsr/nsrlink.jsp?1995Ba70,B}{1995Ba70}.}\\
\parbox[b][0.3cm]{17.7cm}{\makebox[1ex]{\ensuremath{^{\hypertarget{PB33LEVEL12}{h}}}} Band(B): configuration=\ensuremath{\nu} (i\ensuremath{_{\textnormal{13/2}}^{\textnormal{$-$1}}}) \ensuremath{\otimes}\ensuremath{\pi} (h\ensuremath{_{\textnormal{9/2}}^{\textnormal{+1}}},i\ensuremath{_{\textnormal{13/2}}^{\textnormal{+1}}})\ensuremath{_{\textnormal{11$-$}}}. Band 1 in \href{https://www.nndc.bnl.gov/nsr/nsrlink.jsp?1995Ba70,B}{1995Ba70}.}\\
\parbox[b][0.3cm]{17.7cm}{\makebox[1ex]{\ensuremath{^{\hypertarget{PB33LEVEL13}{i}}}} Band(C): Band 3 in \href{https://www.nndc.bnl.gov/nsr/nsrlink.jsp?1995Ba70,B}{1995Ba70}.}\\
\parbox[b][0.3cm]{17.7cm}{\makebox[1ex]{\ensuremath{^{\hypertarget{PB33LEVEL14}{j}}}} Band(D): Band 4 in \href{https://www.nndc.bnl.gov/nsr/nsrlink.jsp?1995Ba70,B}{1995Ba70}.}\\
\parbox[b][0.3cm]{17.7cm}{\makebox[1ex]{\ensuremath{^{\hypertarget{PB33LEVEL15}{k}}}} Band(E): Band 5 in \href{https://www.nndc.bnl.gov/nsr/nsrlink.jsp?1995Ba70,B}{1995Ba70}.}\\
\vspace{0.5cm}
\clearpage
\vspace{0.3cm}
\begin{landscape}
\vspace*{-0.5cm}
{\bf \small \underline{\ensuremath{^{\textnormal{192}}}Os(\ensuremath{^{\textnormal{14}}}C,5n\ensuremath{\gamma})\hspace{0.2in}\href{https://www.nndc.bnl.gov/nsr/nsrlink.jsp?1995Ba70,B}{1995Ba70} (continued)}}\\
\vspace{0.3cm}
\underline{$\gamma$($^{201}$Pb)}\\

\parbox[b][0.3cm]{21.881866cm}{\makebox[1ex]{\ensuremath{^{\hypertarget{PB33GAMMA0}{\dagger}}}} From \href{https://www.nndc.bnl.gov/nsr/nsrlink.jsp?1995Ba70,B}{1995Ba70}, unless otherwise stated. \ensuremath{\Delta}E\ensuremath{\gamma}=0.5 keV estimated by the evaluator.}\\
\parbox[b][0.3cm]{21.881866cm}{\makebox[1ex]{\ensuremath{^{\hypertarget{PB33GAMMA1}{\ddagger}}}} From DCO ratios and multiple decay branches in \href{https://www.nndc.bnl.gov/nsr/nsrlink.jsp?1995Ba70,B}{1995Ba70}, unless otherwise stated.}\\
\parbox[b][0.3cm]{21.881866cm}{\makebox[1ex]{\ensuremath{^{\hypertarget{PB33GAMMA2}{\#}}}} Relative total intensity within each band from \href{https://www.nndc.bnl.gov/nsr/nsrlink.jsp?1995Ba70,B}{1995Ba70}. Values were corrected in \href{https://www.nndc.bnl.gov/nsr/nsrlink.jsp?1995Ba70,B}{1995Ba70} for internal electron conversion by assuming pure M1 and E2}\\
\parbox[b][0.3cm]{21.881866cm}{{\ }{\ }characters for the \ensuremath{\Delta}J=1 and \ensuremath{\Delta}J=2 transitions, respectively.}\\
\parbox[b][0.3cm]{21.881866cm}{\makebox[1ex]{\ensuremath{^{\hypertarget{PB33GAMMA3}{@}}}} Total theoretical internal conversion coefficients, calculated using the BrIcc code (\href{https://www.nndc.bnl.gov/nsr/nsrlink.jsp?2008Ki07,B}{2008Ki07}) with Frozen orbital approximation based on \ensuremath{\gamma}-ray energies,}\\
\parbox[b][0.3cm]{21.881866cm}{{\ }{\ }assigned multipolarities, and mixing ratios, unless otherwise specified.}\\
\parbox[b][0.3cm]{21.881866cm}{\makebox[1ex]{\ensuremath{^{\hypertarget{PB33GAMMA4}{x}}}} \ensuremath{\gamma} ray not placed in level scheme.}\\
\vspace{0.5cm}
\end{landscape}\clearpage
\clearpage
\begin{figure}[h]
\begin{center}
\includegraphics{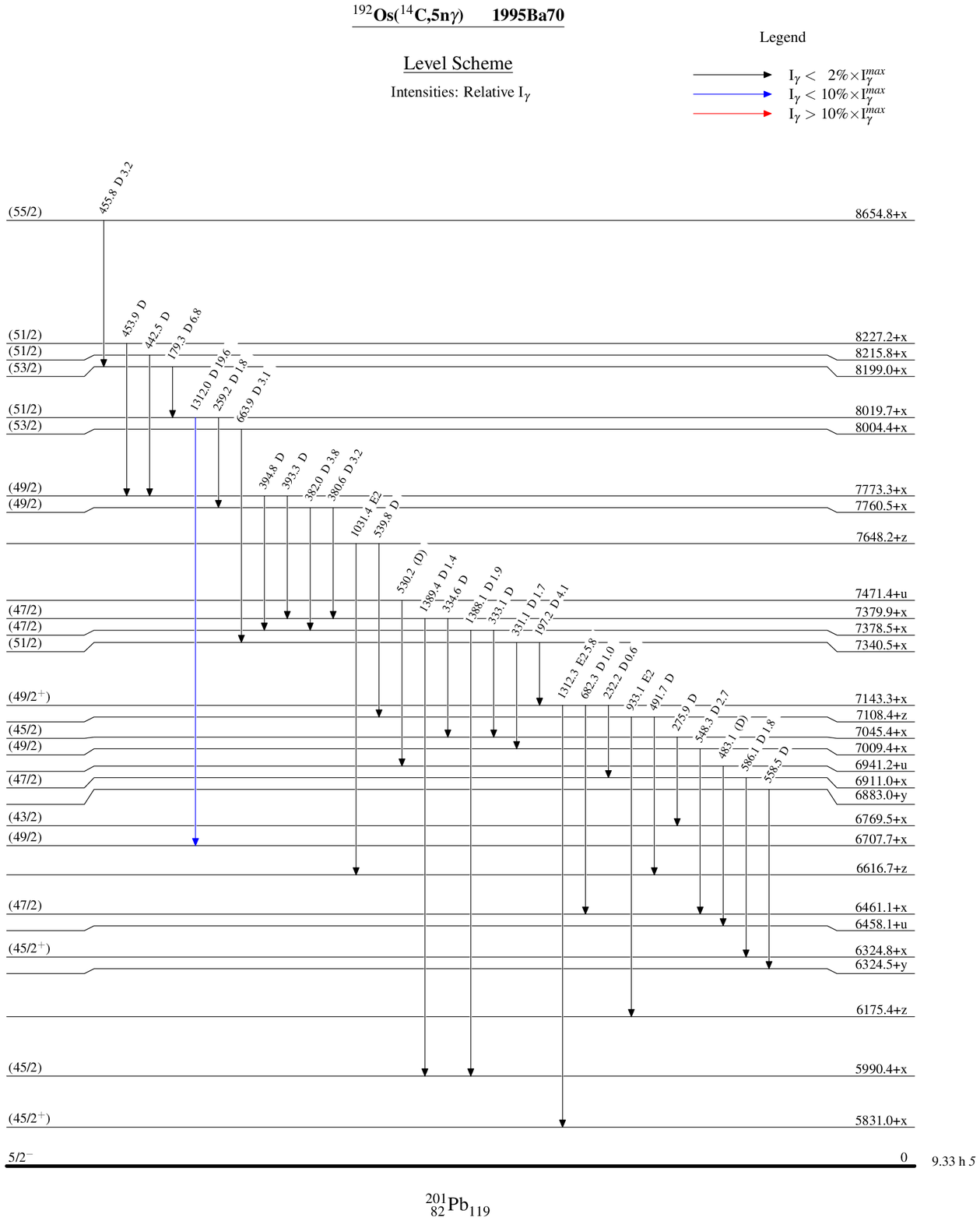}\\
\end{center}
\end{figure}
\clearpage
\begin{figure}[h]
\begin{center}
\includegraphics{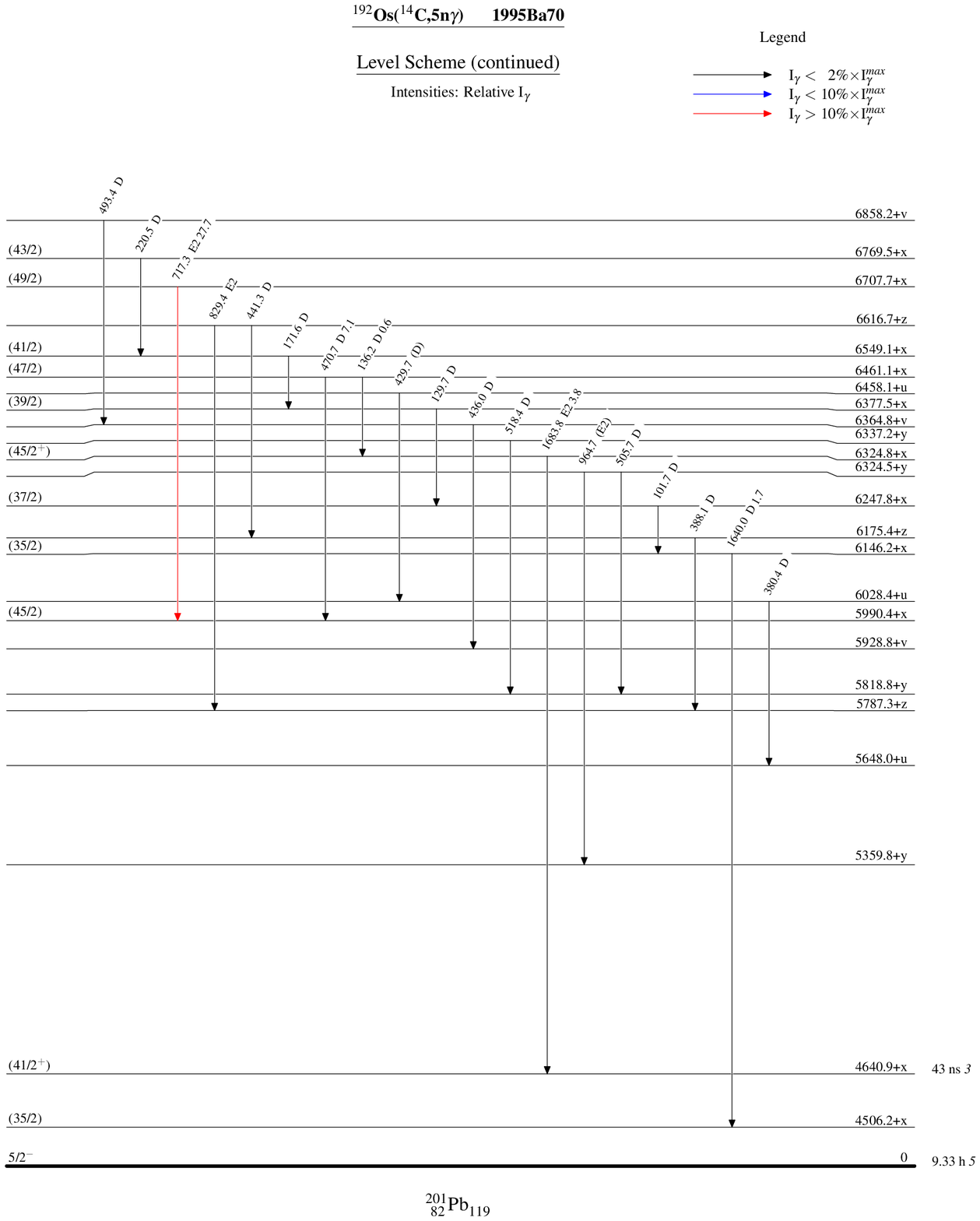}\\
\end{center}
\end{figure}
\clearpage
\begin{figure}[h]
\begin{center}
\includegraphics{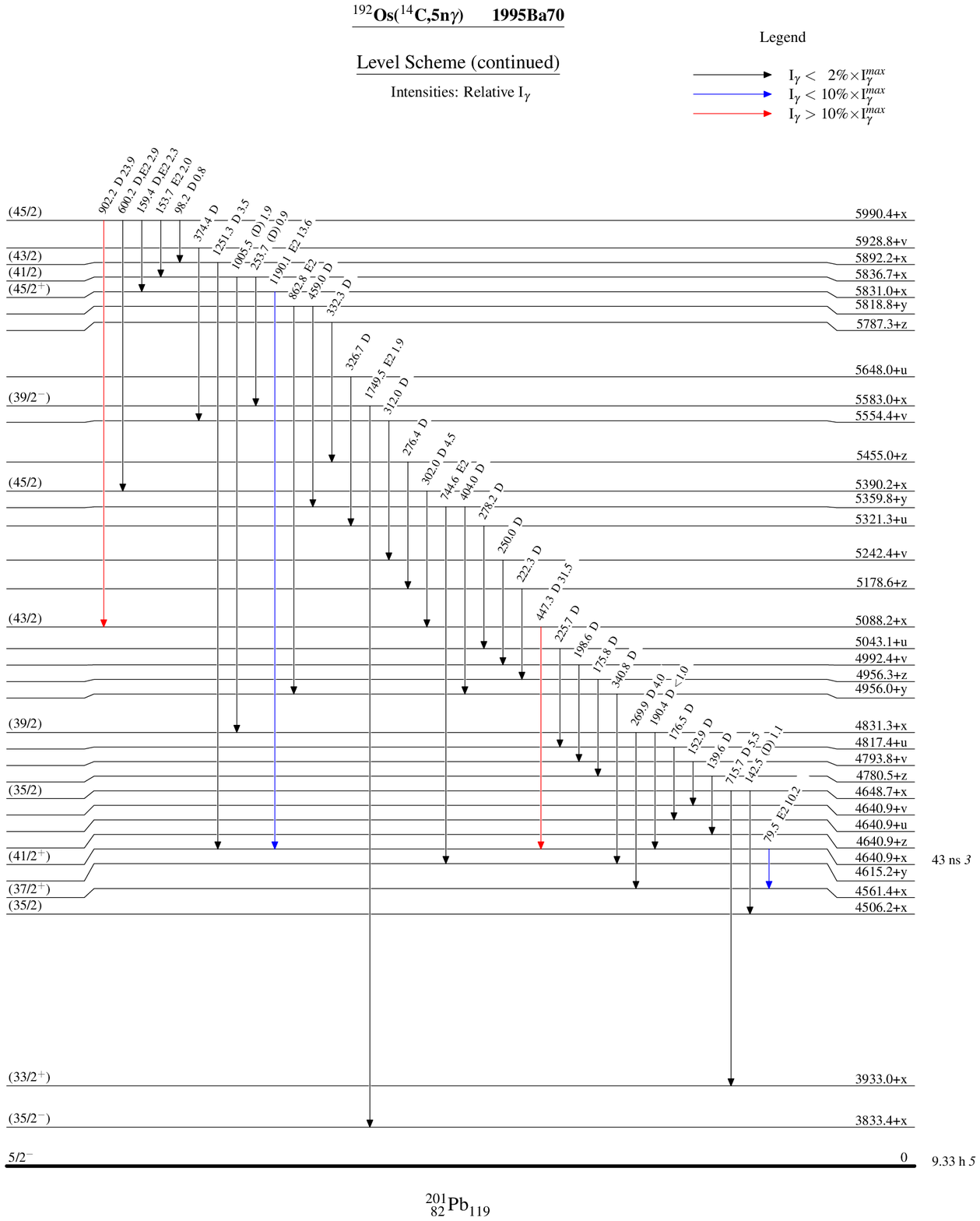}\\
\end{center}
\end{figure}
\clearpage
\begin{figure}[h]
\begin{center}
\includegraphics{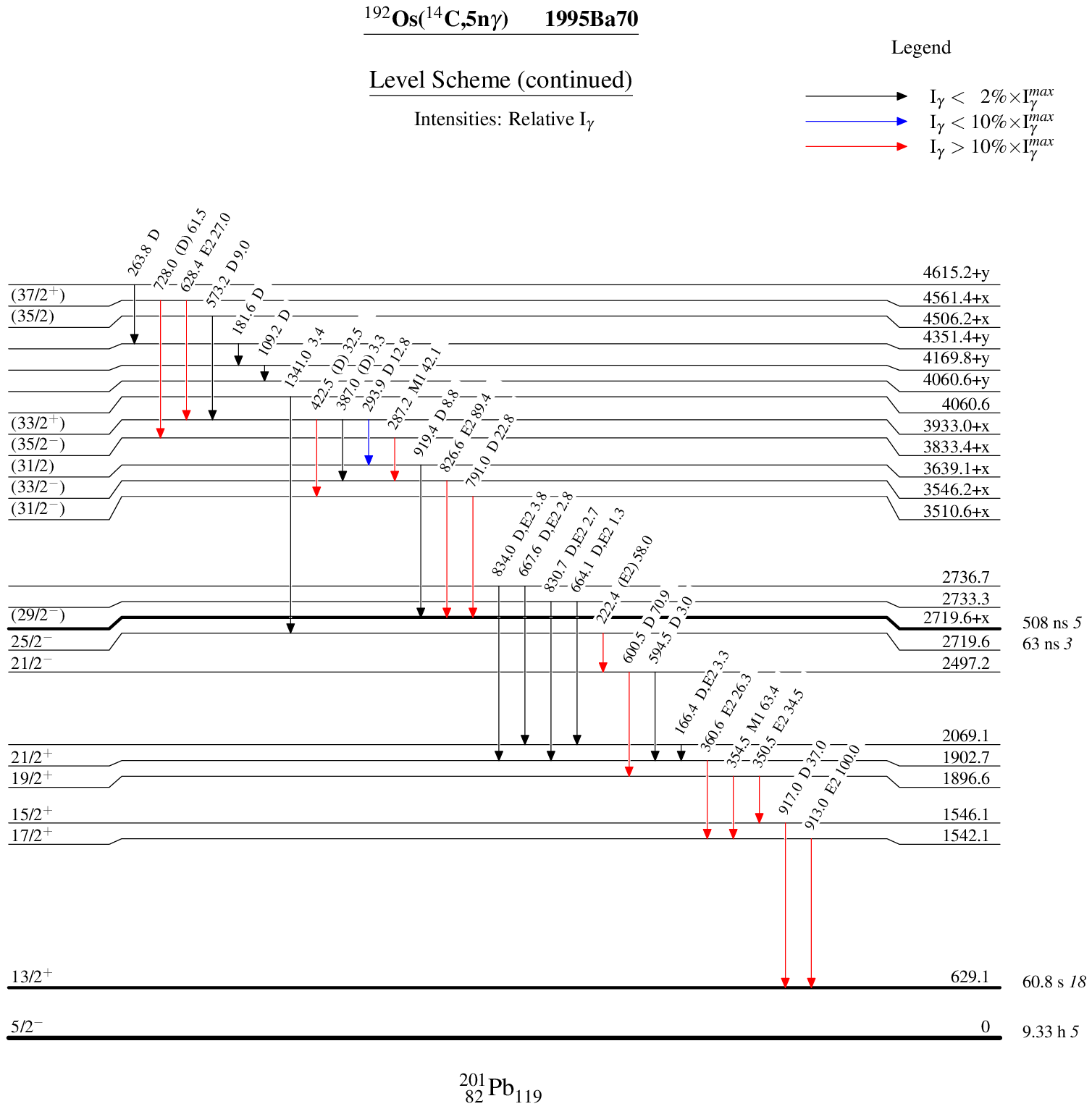}\\
\end{center}
\end{figure}
\clearpage
\clearpage
\begin{figure}[h]
\begin{center}
\includegraphics{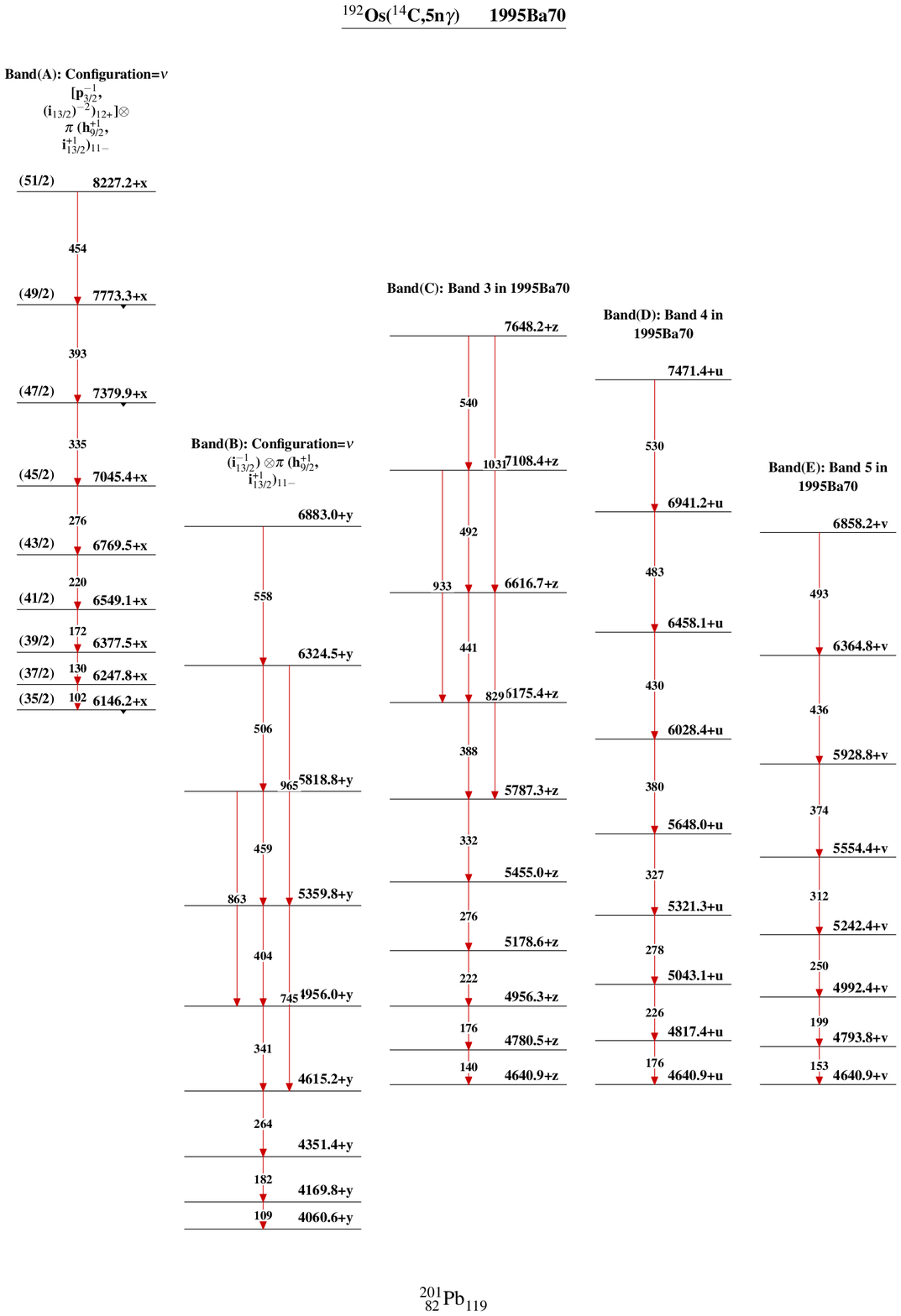}\\
\end{center}
\end{figure}
\clearpage
%197AU(207PB,XG)
\subsection[\hspace{-0.2cm}\ensuremath{^{\textnormal{197}}}Au(\ensuremath{^{\textnormal{207}}}Pb,X\ensuremath{\gamma})]{ }
\vspace{-27pt}
\vspace{0.3cm}
\hypertarget{PB34}{{\bf \small \underline{\ensuremath{^{\textnormal{197}}}Au(\ensuremath{^{\textnormal{207}}}Pb,X\ensuremath{\gamma})\hspace{0.2in}\href{https://www.nndc.bnl.gov/nsr/nsrlink.jsp?2019Ro12,B}{2019Ro12}}}}\\
\vspace{4pt}
\vspace{8pt}
\parbox[b][0.3cm]{17.7cm}{\addtolength{\parindent}{-0.2in}\href{https://www.nndc.bnl.gov/nsr/nsrlink.jsp?2019Ro12,B}{2019Ro12}: E=1430{\textminus}MeV \ensuremath{^{\textnormal{207}}}Pb beam was produced from the ATLAS accelerator at ANL. Target was \ensuremath{\approx}50 mg/cm\ensuremath{^{\textnormal{2}}} \ensuremath{^{\textnormal{197}}}Au. \ensuremath{\gamma} rays}\\
\parbox[b][0.3cm]{17.7cm}{were detected with the Gammasphere array comprising of 100 HPGe detectors. Measured: E\ensuremath{\gamma}, I\ensuremath{\gamma}, \ensuremath{\gamma}\ensuremath{\gamma}(t). Deduced: level scheme,}\\
\parbox[b][0.3cm]{17.7cm}{T\ensuremath{_{\textnormal{1/2}}}.}\\
\vspace{12pt}
\underline{$^{201}$Pb Levels}\\

\parbox[b][0.3cm]{17.7cm}{\makebox[1ex]{\ensuremath{^{\hypertarget{PB34LEVEL0}{\dagger}}}} From a least-squares fit to E\ensuremath{\gamma}, unless otherwise stated. \ensuremath{\Delta}E\ensuremath{\gamma}=1 keV were estimated by the evaluator.}\\
\parbox[b][0.3cm]{17.7cm}{\makebox[1ex]{\ensuremath{^{\hypertarget{PB34LEVEL1}{\ddagger}}}} From \href{https://www.nndc.bnl.gov/nsr/nsrlink.jsp?2019Ro12,B}{2019Ro12}, based on previous studies (\href{https://www.nndc.bnl.gov/nsr/nsrlink.jsp?1981He07,B}{1981He07},\href{https://www.nndc.bnl.gov/nsr/nsrlink.jsp?1988Ro08,B}{1988Ro08}, \href{https://www.nndc.bnl.gov/nsr/nsrlink.jsp?1995Ba70,B}{1995Ba70}), shell-model and systematics.}\\
\vspace{0.5cm}
\underline{$\gamma$($^{201}$Pb)}\\

\parbox[b][0.3cm]{17.7cm}{\makebox[1ex]{\ensuremath{^{\hypertarget{PB34GAMMA0}{\dagger}}}} From \href{https://www.nndc.bnl.gov/nsr/nsrlink.jsp?2019Ro12,B}{2019Ro12}.}\\
\vspace{0.5cm}
\clearpage
\begin{figure}[h]
\begin{center}
\includegraphics{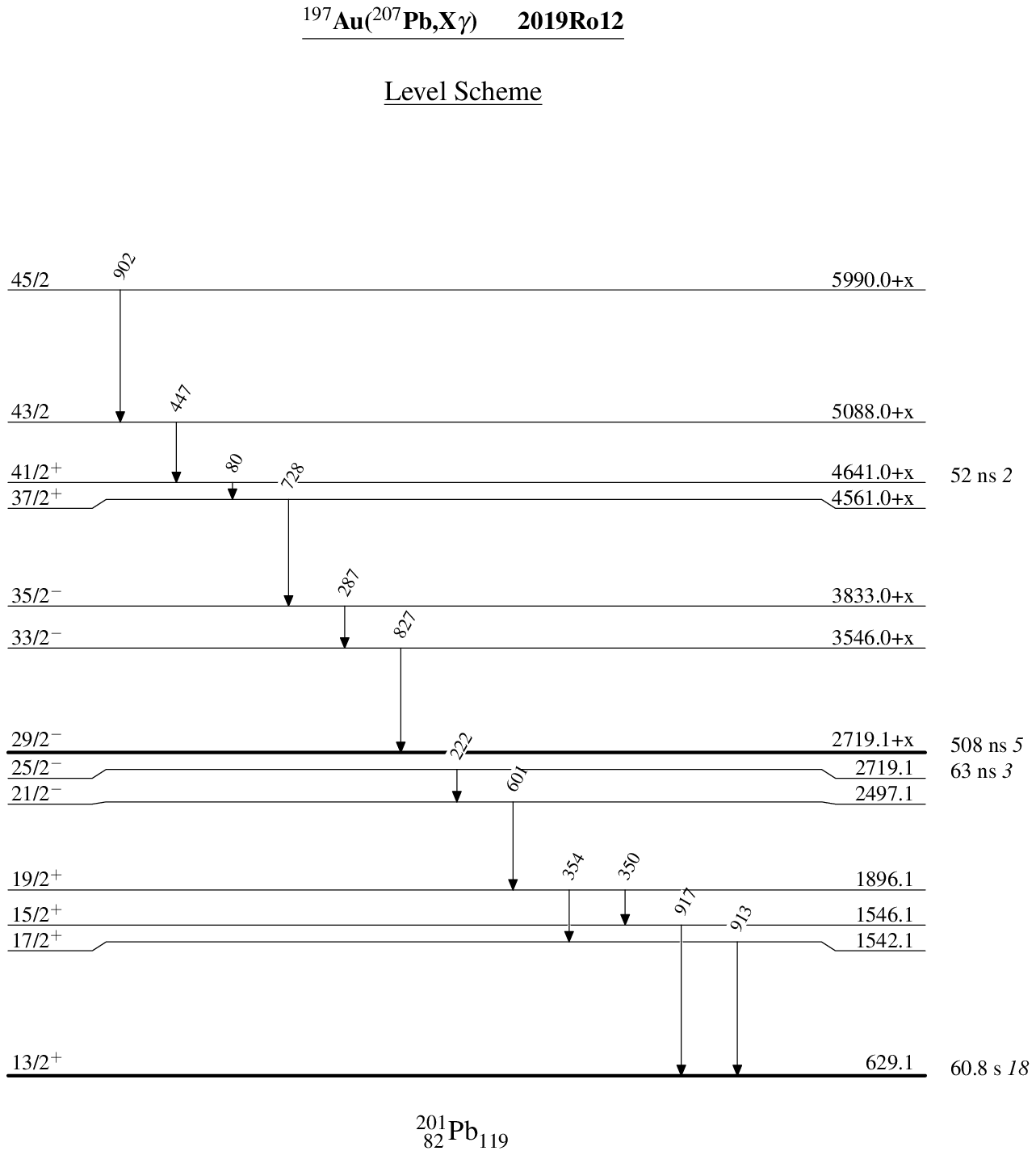}\\
\end{center}
\end{figure}
\clearpage
%ADOPTED LEVELS, GAMMAS
\section[\ensuremath{^{201}_{\ 83}}Bi\ensuremath{_{118}^{~}}]{ }
\vspace{-30pt}
\setcounter{chappage}{1}
\subsection[\hspace{-0.2cm}Adopted Levels, Gammas]{ }
\vspace{-20pt}
\vspace{0.3cm}
\hypertarget{BI35}{{\bf \small \underline{Adopted \hyperlink{201BI_LEVEL}{Levels}, \hyperlink{201BI_GAMMA}{Gammas}}}}\\
\vspace{4pt}
\vspace{8pt}
\parbox[b][0.3cm]{17.7cm}{\addtolength{\parindent}{-0.2in}Q(\ensuremath{\beta^-})=$-$4908 {\it 13}; S(n)=9130 {\it 26}; S(p)=2467 {\it 16}; Q(\ensuremath{\alpha})=4500 {\it 6}\hspace{0.2in}\href{https://www.nndc.bnl.gov/nsr/nsrlink.jsp?2021Wa16,B}{2021Wa16}}\\

\vspace{12pt}
\hypertarget{201BI_LEVEL}{\underline{$^{201}$Bi Levels}}\\

\begin{textblock}{29}(0,27.3)
Continued on next page (footnotes at end of table)
\end{textblock}
\clearpage
%
\begin{textblock}{29}(0,27.3)
Continued on next page (footnotes at end of table)
\end{textblock}
\clearpage
%
\parbox[b][0.3cm]{17.7cm}{\makebox[1ex]{\ensuremath{^{\hypertarget{BI35LEVEL0}{\dagger}}}} From a least squares fit to E\ensuremath{\gamma}.}\\
\parbox[b][0.3cm]{17.7cm}{\makebox[1ex]{\ensuremath{^{\hypertarget{BI35LEVEL1}{\ddagger}}}} From deduced \ensuremath{\gamma}-ray transition multipolarities using \ensuremath{\gamma}(\ensuremath{\theta}) and DCO in \ensuremath{^{\textnormal{196}}}Pt(\ensuremath{^{\textnormal{10}}}B,5n\ensuremath{\gamma}), \ensuremath{\gamma}(\ensuremath{\theta}) in\hphantom{a}\ensuremath{^{\textnormal{203}}}Tl(\ensuremath{\alpha},6n\ensuremath{\gamma}) and \ensuremath{\alpha}(K)exp,}\\
\parbox[b][0.3cm]{17.7cm}{{\ }{\ }\ensuremath{\alpha}(L)exp in \ensuremath{^{\textnormal{201}}}Po \ensuremath{\varepsilon} decay (15.6\hphantom{a}min) and \ensuremath{^{\textnormal{201}}}Po \ensuremath{\varepsilon} decay (8.96 min), unless otherwise stated.}\\
\parbox[b][0.3cm]{17.7cm}{\makebox[1ex]{\ensuremath{^{\hypertarget{BI35LEVEL2}{\#}}}} Configuration=\ensuremath{\pi} h\ensuremath{_{\textnormal{9/2}}^{\textnormal{+1}}}.}\\
\parbox[b][0.3cm]{17.7cm}{\makebox[1ex]{\ensuremath{^{\hypertarget{BI35LEVEL3}{@}}}} Configuration=\ensuremath{\pi} s\ensuremath{_{\textnormal{1/2}}^{\textnormal{$-$1}}}.}\\
\parbox[b][0.3cm]{17.7cm}{\makebox[1ex]{\ensuremath{^{\hypertarget{BI35LEVEL4}{\&}}}} Configuration=\ensuremath{\pi} (h\ensuremath{_{\textnormal{9/2}}^{\textnormal{+1}}})\ensuremath{\otimes}2\ensuremath{^{\textnormal{+}}}.}\\
\parbox[b][0.3cm]{17.7cm}{\makebox[1ex]{\ensuremath{^{\hypertarget{BI35LEVEL5}{a}}}} Configuration=\ensuremath{\pi}\hphantom{a}d\ensuremath{_{\textnormal{3/2}}^{\textnormal{$-$1}}}.}\\
\parbox[b][0.3cm]{17.7cm}{\makebox[1ex]{\ensuremath{^{\hypertarget{BI35LEVEL6}{b}}}} Configuration=\ensuremath{\pi} (h\ensuremath{_{\textnormal{9/2}}^{\textnormal{+1}}})\ensuremath{\otimes}4\ensuremath{^{\textnormal{+}}}.}\\
\parbox[b][0.3cm]{17.7cm}{\makebox[1ex]{\ensuremath{^{\hypertarget{BI35LEVEL7}{c}}}} Configuration=\ensuremath{\pi} f\ensuremath{_{\textnormal{7/2}}^{\textnormal{+1}}}.}\\
\parbox[b][0.3cm]{17.7cm}{\makebox[1ex]{\ensuremath{^{\hypertarget{BI35LEVEL8}{d}}}} Admixture of configuration= \ensuremath{\pi} (h\ensuremath{_{\textnormal{9/2}}^{\textnormal{+1}}}) \ensuremath{\nu} (f\ensuremath{_{\textnormal{5/2}}^{\textnormal{$-$1}}},i\ensuremath{_{\textnormal{13/2}}^{\textnormal{$-$1}}})\ensuremath{_{\textnormal{5$-$}}} and configuration=\ensuremath{\pi} (h\ensuremath{_{\textnormal{9/2}}^{\textnormal{+1}}}) \ensuremath{\nu} (p\ensuremath{_{\textnormal{3/2}}^{\textnormal{$-$1}}},i\ensuremath{_{\textnormal{13/2}}^{\textnormal{$-$1}}})\ensuremath{_{\textnormal{5$-$}}}.}\\
\parbox[b][0.3cm]{17.7cm}{\makebox[1ex]{\ensuremath{^{\hypertarget{BI35LEVEL9}{e}}}} Admixture of configuration= \ensuremath{\pi} (h\ensuremath{_{\textnormal{9/2}}^{\textnormal{+1}}}) \ensuremath{\nu} (f\ensuremath{_{\textnormal{5/2}}^{\textnormal{$-$1}}},i\ensuremath{_{\textnormal{13/2}}^{\textnormal{$-$1}}})\ensuremath{_{\textnormal{7$-$}}} and configuration=\ensuremath{\pi} (h\ensuremath{_{\textnormal{9/2}}^{\textnormal{+1}}}) \ensuremath{\nu} (p\ensuremath{_{\textnormal{3/2}}^{\textnormal{$-$1}}},i\ensuremath{_{\textnormal{13/2}}^{\textnormal{$-$1}}})\ensuremath{_{\textnormal{7$-$}}}.}\\
\parbox[b][0.3cm]{17.7cm}{\makebox[1ex]{\ensuremath{^{\hypertarget{BI35LEVEL10}{f}}}} Configuration=\ensuremath{\pi} (h\ensuremath{_{\textnormal{9/2}}^{\textnormal{+1}}}) \ensuremath{\nu} (f\ensuremath{_{\textnormal{5/2}}^{\textnormal{$-$1}}},i\ensuremath{_{\textnormal{13/2}}^{\textnormal{$-$1}}})\ensuremath{_{\textnormal{9$-$}}}.}\\
\parbox[b][0.3cm]{17.7cm}{\makebox[1ex]{\ensuremath{^{\hypertarget{BI35LEVEL11}{g}}}} Configuration=\ensuremath{\pi} (h\ensuremath{_{\textnormal{9/2}}^{\textnormal{+1}}}) \ensuremath{\nu} (i\ensuremath{_{\textnormal{13/2}}^{\textnormal{$-$2}}})\ensuremath{_{\textnormal{12+}}}.}\\
\vspace{0.5cm}
\clearpage
\vspace{0.3cm}
\begin{landscape}
\vspace*{-0.5cm}
{\bf \small \underline{Adopted \hyperlink{201BI_LEVEL}{Levels}, \hyperlink{201BI_GAMMA}{Gammas} (continued)}}\\
\vspace{0.3cm}
\hypertarget{201BI_GAMMA}{\underline{$\gamma$($^{201}$Bi)}}\\

\parbox[b][0.3cm]{21.881866cm}{\makebox[1ex]{\ensuremath{^{\hypertarget{BI35GAMMA0}{\dagger}}}} From \ensuremath{^{\textnormal{196}}}Pt(\ensuremath{^{\textnormal{10}}}B,5n\ensuremath{\gamma}).}\\
\parbox[b][0.3cm]{21.881866cm}{\makebox[1ex]{\ensuremath{^{\hypertarget{BI35GAMMA1}{\ddagger}}}} From \href{https://www.nndc.bnl.gov/nsr/nsrlink.jsp?1986Br28,B}{1986Br28} in \ensuremath{^{\textnormal{201}}}Po \ensuremath{\varepsilon} decay (15.6 min).}\\
\parbox[b][0.3cm]{21.881866cm}{\makebox[1ex]{\ensuremath{^{\hypertarget{BI35GAMMA2}{\#}}}} From \href{https://www.nndc.bnl.gov/nsr/nsrlink.jsp?1986Br28,B}{1986Br28} in \ensuremath{^{\textnormal{201}}}Po \ensuremath{\varepsilon} decay (8.96 min).}\\
\parbox[b][0.3cm]{21.881866cm}{\makebox[1ex]{\ensuremath{^{\hypertarget{BI35GAMMA3}{@}}}} Reported by \href{https://www.nndc.bnl.gov/nsr/nsrlink.jsp?1985Pi05,B}{1985Pi05} to have a delayed component with T\ensuremath{_{\textnormal{1/2}}}=14 ns \textit{3}. Similar delayed component (T\ensuremath{_{\textnormal{1/2}}}\ensuremath{>}10 ns) is reported in \href{https://www.nndc.bnl.gov/nsr/nsrlink.jsp?1982Br21,B}{1982Br21} above the 3812+X}\\
\parbox[b][0.3cm]{21.881866cm}{{\ }{\ }keV level.}\\
\parbox[b][0.3cm]{21.881866cm}{\makebox[1ex]{\ensuremath{^{\hypertarget{BI35GAMMA4}{\&}}}} Total theoretical internal conversion coefficients, calculated using the BrIcc code (\href{https://www.nndc.bnl.gov/nsr/nsrlink.jsp?2008Ki07,B}{2008Ki07}) with Frozen orbital approximation based on \ensuremath{\gamma}-ray energies,}\\
\parbox[b][0.3cm]{21.881866cm}{{\ }{\ }assigned multipolarities, and mixing ratios, unless otherwise specified.}\\
\parbox[b][0.3cm]{21.881866cm}{\makebox[1ex]{\ensuremath{^{\hypertarget{BI35GAMMA5}{a}}}} Placement of transition in the level scheme is uncertain.}\\
\vspace{0.5cm}
\end{landscape}\clearpage
\clearpage
\begin{figure}[h]
\begin{center}
\includegraphics{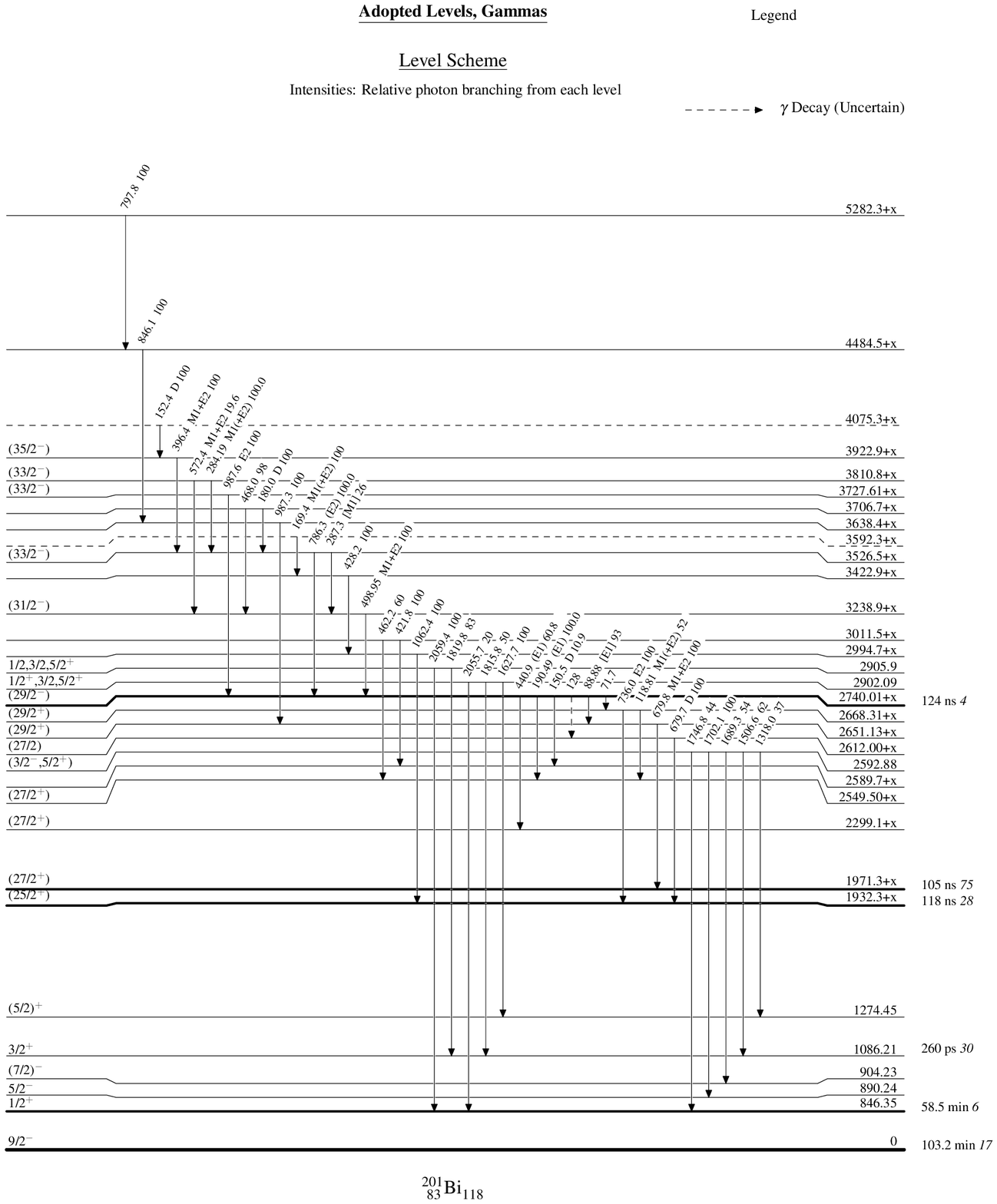}\\
\end{center}
\end{figure}
\clearpage
\begin{figure}[h]
\begin{center}
\includegraphics{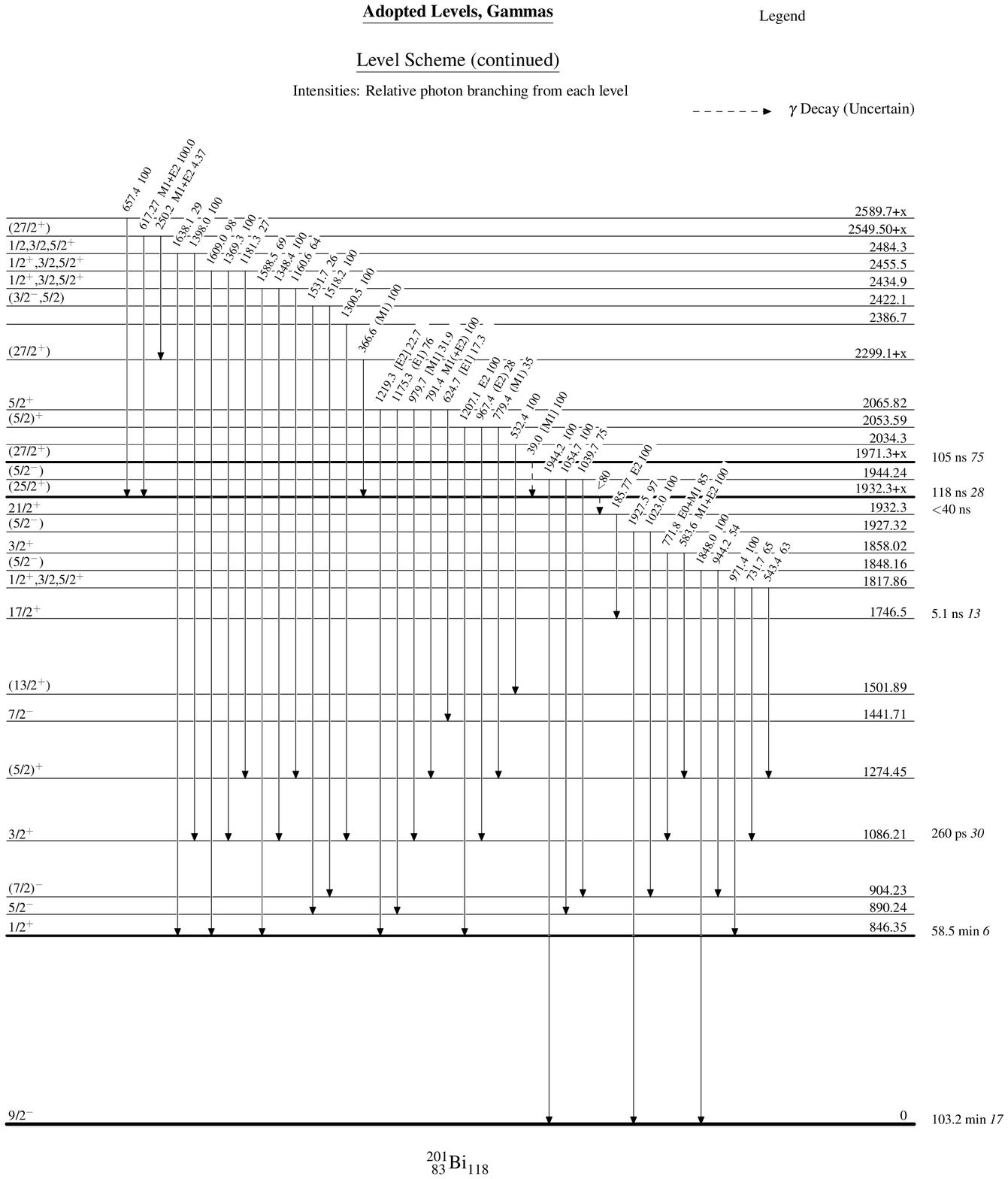}\\
\end{center}
\end{figure}
\clearpage
\begin{figure}[h]
\begin{center}
\includegraphics{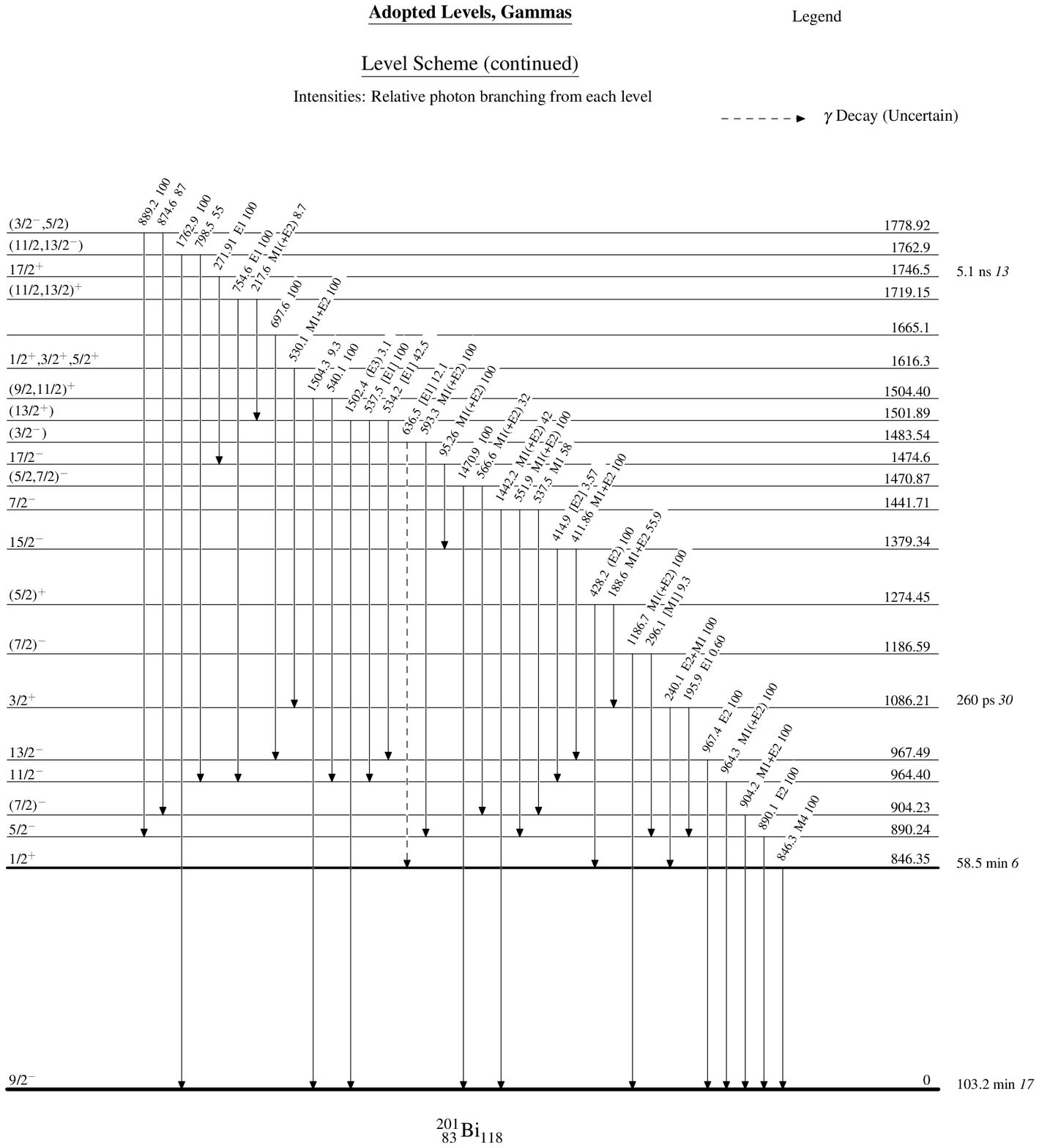}\\
\end{center}
\end{figure}
\clearpage
%201PO EC DECAY (15.50 M)
\subsection[\hspace{-0.2cm}\ensuremath{^{\textnormal{201}}}Po \ensuremath{\varepsilon} decay (15.50 min)]{ }
\vspace{-27pt}
\vspace{0.3cm}
\hypertarget{PO36}{{\bf \small \underline{\ensuremath{^{\textnormal{201}}}Po \ensuremath{\varepsilon} decay (15.50 min)\hspace{0.2in}\href{https://www.nndc.bnl.gov/nsr/nsrlink.jsp?1986Br28,B}{1986Br28}}}}\\
\vspace{4pt}
\vspace{8pt}
\parbox[b][0.3cm]{17.7cm}{\addtolength{\parindent}{-0.2in}Parent: $^{201}$Po: E=0; J$^{\pi}$=3/2\ensuremath{^{-}}; T$_{1/2}$=15.50 min {\it 14}; Q(\ensuremath{\varepsilon})=4908 {\it 13}; \%\ensuremath{\varepsilon}+\%\ensuremath{\beta^{+}} decay=100.0

}\\
\parbox[b][0.3cm]{17.7cm}{\addtolength{\parindent}{-0.2in}\href{https://www.nndc.bnl.gov/nsr/nsrlink.jsp?1986Br28,B}{1986Br28}:\ensuremath{^{\textnormal{193}}}Ir(\ensuremath{^{\textnormal{14}}}N,6n), E=116 MeV, mass separated source; Detectors: Ge(Li) and cooled Si(Li); Measured: \ensuremath{\gamma}, \ensuremath{\gamma}\ensuremath{\gamma} coin, \ensuremath{\gamma}\ensuremath{\gamma}(t),}\\
\parbox[b][0.3cm]{17.7cm}{\ensuremath{\gamma}(x-ray)(t), ce, and T\ensuremath{_{\textnormal{1/2}}}.}\\
\parbox[b][0.3cm]{17.7cm}{\addtolength{\parindent}{-0.2in}Others: \href{https://www.nndc.bnl.gov/nsr/nsrlink.jsp?1986Be07,B}{1986Be07}, \href{https://www.nndc.bnl.gov/nsr/nsrlink.jsp?1980Br23,B}{1980Br23}, \href{https://www.nndc.bnl.gov/nsr/nsrlink.jsp?1976Ko13,B}{1976Ko13}, \href{https://www.nndc.bnl.gov/nsr/nsrlink.jsp?1970DaZM,B}{1970DaZM}, \href{https://www.nndc.bnl.gov/nsr/nsrlink.jsp?1970Jo26,B}{1970Jo26}, \href{https://www.nndc.bnl.gov/nsr/nsrlink.jsp?1969Al10,B}{1969Al10}.}\\
\vspace{12pt}
\underline{$^{201}$Bi Levels}\\

\parbox[b][0.3cm]{17.7cm}{\makebox[1ex]{\ensuremath{^{\hypertarget{BI36LEVEL0}{\dagger}}}} From a least squares fit to E\ensuremath{\gamma}.}\\
\parbox[b][0.3cm]{17.7cm}{\makebox[1ex]{\ensuremath{^{\hypertarget{BI36LEVEL1}{\ddagger}}}} From Adopted Levels, unless otherwise stated.}\\
\parbox[b][0.3cm]{17.7cm}{\makebox[1ex]{\ensuremath{^{\hypertarget{BI36LEVEL2}{\#}}}} Configuration=\ensuremath{\pi} h\ensuremath{_{\textnormal{9/2}}^{\textnormal{+1}}}.}\\
\parbox[b][0.3cm]{17.7cm}{\makebox[1ex]{\ensuremath{^{\hypertarget{BI36LEVEL3}{@}}}} Configuration=\ensuremath{\pi} s\ensuremath{_{\textnormal{1/2}}^{\textnormal{$-$1}}}.}\\
\parbox[b][0.3cm]{17.7cm}{\makebox[1ex]{\ensuremath{^{\hypertarget{BI36LEVEL4}{\&}}}} Configuration=\ensuremath{\pi} (h\ensuremath{_{\textnormal{9/2}}^{\textnormal{+1}}})\ensuremath{\otimes}2\ensuremath{^{\textnormal{+}}}.}\\
\parbox[b][0.3cm]{17.7cm}{\makebox[1ex]{\ensuremath{^{\hypertarget{BI36LEVEL5}{a}}}} Configuration=\ensuremath{\pi} d\ensuremath{_{\textnormal{3/2}}^{\textnormal{$-$1}}}.}\\
\parbox[b][0.3cm]{17.7cm}{\makebox[1ex]{\ensuremath{^{\hypertarget{BI36LEVEL6}{b}}}} Configuration=\ensuremath{\pi} f\ensuremath{_{\textnormal{7/2}}^{\textnormal{+1}}}.}\\
\vspace{0.5cm}
\clearpage
\vspace{0.3cm}
\vspace*{-0.5cm}
{\bf \small \underline{\ensuremath{^{\textnormal{201}}}Po \ensuremath{\varepsilon} decay (15.50 min)\hspace{0.2in}\href{https://www.nndc.bnl.gov/nsr/nsrlink.jsp?1986Br28,B}{1986Br28} (continued)}}\\
\vspace{0.3cm}

\underline{\ensuremath{\varepsilon,\beta^+} radiations}\\
\vspace{0.34cm}
\parbox[b][0.3cm]{17.7cm}{\addtolength{\parindent}{-0.254cm}The I(\ensuremath{\beta}\ensuremath{^{\textnormal{+}}}+\ensuremath{\varepsilon}), I\ensuremath{\beta}, I\ensuremath{\varepsilon} and log{} \textit{ft} values are approximate, given the uncertain \%IT value for the 846 keV, \ensuremath{J^{\pi}}=1/2\ensuremath{^{+}} state.}\\
\vspace{0.34cm}

\parbox[b][0.3cm]{17.7cm}{\makebox[1ex]{\ensuremath{^{\hypertarget{BI36DECAY0}{\dagger}}}} Deduced from the decay scheme using intensity balances and by assuming no direct feeding to the g.s. (\ensuremath{J^{\pi}}=9/2\ensuremath{^{-}}).}\\
\parbox[b][0.3cm]{17.7cm}{\makebox[1ex]{\ensuremath{^{\hypertarget{BI36DECAY1}{\ddagger}}}} Absolute intensity per 100 decays.}\\
\parbox[b][0.3cm]{17.7cm}{\makebox[1ex]{\ensuremath{^{\hypertarget{BI36DECAY2}{\#}}}} Existence of this branch is questionable.}\\
\vspace{0.5cm}
\clearpage
\vspace{0.3cm}
\begin{landscape}
\vspace*{-0.5cm}
{\bf \small \underline{\ensuremath{^{\textnormal{201}}}Po \ensuremath{\varepsilon} decay (15.50 min)\hspace{0.2in}\href{https://www.nndc.bnl.gov/nsr/nsrlink.jsp?1986Br28,B}{1986Br28} (continued)}}\\
\vspace{0.3cm}
\underline{$\gamma$($^{201}$Bi)}\\
\vspace{0.34cm}
\parbox[b][0.3cm]{21.881866cm}{\addtolength{\parindent}{-0.254cm}I\ensuremath{\gamma} normalization: Using \ensuremath{\Sigma}I(\ensuremath{\gamma}+ce)(to g.s.)=100\% and by assuming that there is no direct feeding to the g.s. (\ensuremath{J^{\pi}}=9/2\ensuremath{^{-}}).}\\
\vspace{0.34cm}

\clearpage
\vspace*{-0.5cm}
{\bf \small \underline{\ensuremath{^{\textnormal{201}}}Po \ensuremath{\varepsilon} decay (15.50 min)\hspace{0.2in}\href{https://www.nndc.bnl.gov/nsr/nsrlink.jsp?1986Br28,B}{1986Br28} (continued)}}\\
\vspace{0.3cm}
\underline{$\gamma$($^{201}$Bi) (continued)}\\
\vspace{0.3cm}
\parbox[b][0.3cm]{21.881866cm}{\makebox[1ex]{\ensuremath{^{\hypertarget{PO36GAMMA0}{\dagger}}}} From \href{https://www.nndc.bnl.gov/nsr/nsrlink.jsp?1986Br28,B}{1986Br28}, unless otherwise stated.}\\
\parbox[b][0.3cm]{21.881866cm}{\makebox[1ex]{\ensuremath{^{\hypertarget{PO36GAMMA1}{\ddagger}}}} From the ce measurements in \href{https://www.nndc.bnl.gov/nsr/nsrlink.jsp?1986Br28,B}{1986Br28} and \href{https://www.nndc.bnl.gov/nsr/nsrlink.jsp?1969Al10,B}{1969Al10}, unless otherwise stated.}\\
\parbox[b][0.3cm]{21.881866cm}{\makebox[1ex]{\ensuremath{^{\hypertarget{PO36GAMMA2}{\#}}}} From \ensuremath{\alpha}(K)exp and subshell ratios in \href{https://www.nndc.bnl.gov/nsr/nsrlink.jsp?1986Br28,B}{1986Br28} and the briccmixing program, unless otherwise stated.}\\
\parbox[b][0.3cm]{21.881866cm}{\makebox[1ex]{\ensuremath{^{\hypertarget{PO36GAMMA3}{@}}}} Estimated from coincidence intensity.}\\
\parbox[b][0.3cm]{21.881866cm}{\makebox[1ex]{\ensuremath{^{\hypertarget{PO36GAMMA4}{\&}}}} For absolute intensity per 100 decays, multiply by 0.304 \textit{7}.}\\
\parbox[b][0.3cm]{21.881866cm}{\makebox[1ex]{\ensuremath{^{\hypertarget{PO36GAMMA5}{a}}}} Total theoretical internal conversion coefficients, calculated using the BrIcc code (\href{https://www.nndc.bnl.gov/nsr/nsrlink.jsp?2008Ki07,B}{2008Ki07}) with Frozen orbital approximation based on \ensuremath{\gamma}-ray energies,}\\
\parbox[b][0.3cm]{21.881866cm}{{\ }{\ }assigned multipolarities, and mixing ratios, unless otherwise specified.}\\
\parbox[b][0.3cm]{21.881866cm}{\makebox[1ex]{\ensuremath{^{\hypertarget{PO36GAMMA6}{b}}}} Placement of transition in the level scheme is uncertain.}\\
\parbox[b][0.3cm]{21.881866cm}{\makebox[1ex]{\ensuremath{^{\hypertarget{PO36GAMMA7}{x}}}} \ensuremath{\gamma} ray not placed in level scheme.}\\
\vspace{0.5cm}
\end{landscape}\clearpage
\clearpage
\begin{figure}[h]
\begin{center}
\includegraphics[angle=90]{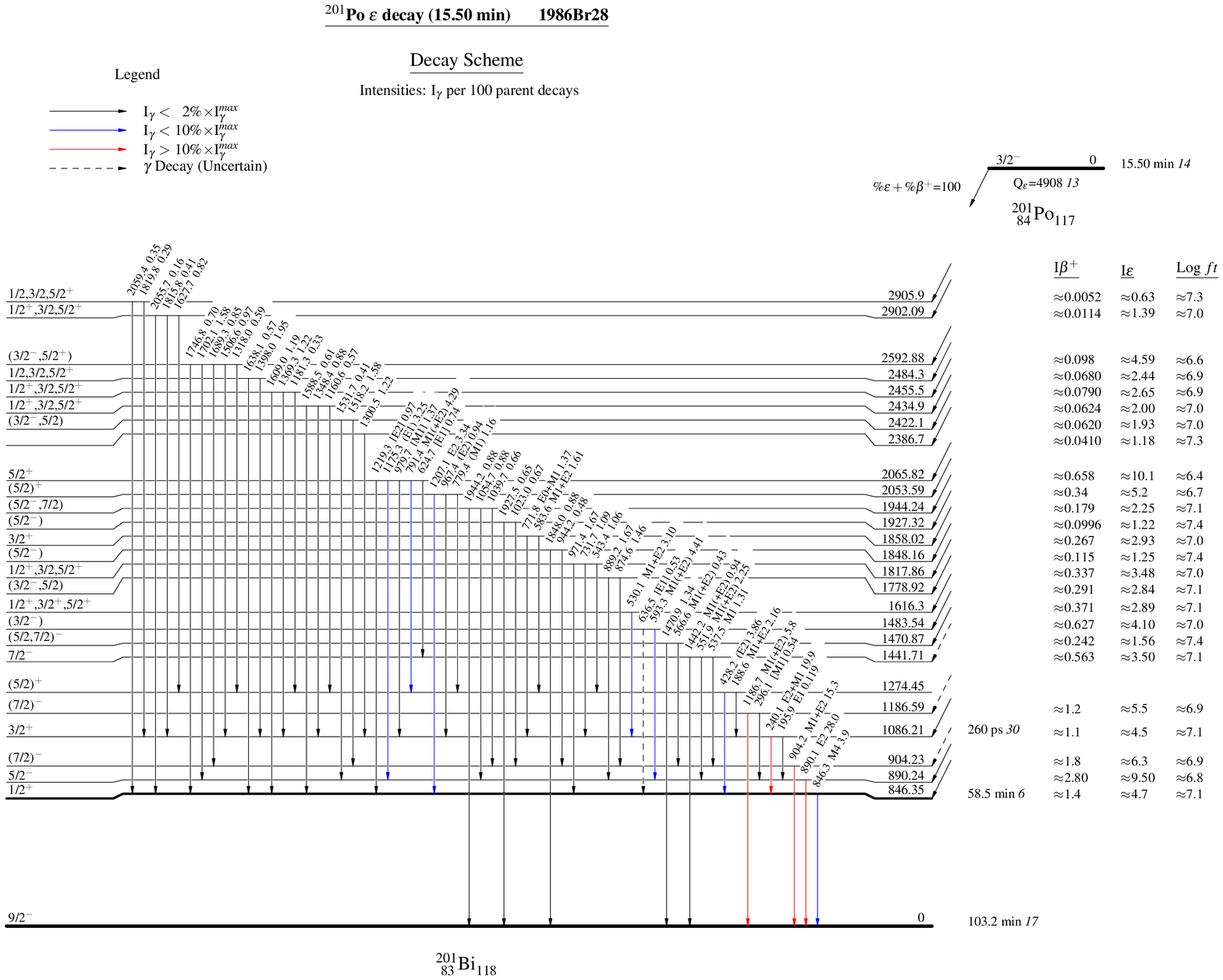}\\
\end{center}
\end{figure}
\clearpage
%201PO EC DECAY (8.96 M)
\subsection[\hspace{-0.2cm}\ensuremath{^{\textnormal{201}}}Po \ensuremath{\varepsilon} decay (8.96 min)]{ }
\vspace{-27pt}
\vspace{0.3cm}
\hypertarget{PO37}{{\bf \small \underline{\ensuremath{^{\textnormal{201}}}Po \ensuremath{\varepsilon} decay (8.96 min)\hspace{0.2in}\href{https://www.nndc.bnl.gov/nsr/nsrlink.jsp?1986Br28,B}{1986Br28},\href{https://www.nndc.bnl.gov/nsr/nsrlink.jsp?1976Ko13,B}{1976Ko13}}}}\\
\vspace{4pt}
\vspace{8pt}
\parbox[b][0.3cm]{17.7cm}{\addtolength{\parindent}{-0.2in}Parent: $^{201}$Po: E=423.41 {\it 22}; J$^{\pi}$=13/2\ensuremath{^{+}}; T$_{1/2}$=8.96 min {\it 12}; Q(\ensuremath{\varepsilon})=4908 {\it 13}; \%\ensuremath{\varepsilon}+\%\ensuremath{\beta^{+}} decay$\approx$55.0

}\\
\parbox[b][0.3cm]{17.7cm}{\addtolength{\parindent}{-0.2in}\href{https://www.nndc.bnl.gov/nsr/nsrlink.jsp?1986Br28,B}{1986Br28}:\ensuremath{^{\textnormal{193}}}Ir(\ensuremath{^{\textnormal{14}}}N,6n), E=116 MeV; Detectors: Ge(Li) and cooled Si(Li); Measured: \ensuremath{\gamma}, \ensuremath{\gamma}\ensuremath{\gamma} coin, \ensuremath{\gamma}\ensuremath{\gamma}(t), \ensuremath{\gamma}(x-ray)(t), ce, and}\\
\parbox[b][0.3cm]{17.7cm}{T\ensuremath{_{\textnormal{1/2}}}.}\\
\parbox[b][0.3cm]{17.7cm}{\addtolength{\parindent}{-0.2in}\href{https://www.nndc.bnl.gov/nsr/nsrlink.jsp?1976Ko13,B}{1976Ko13}: \ensuremath{^{\textnormal{197}}}Au(\ensuremath{^{\textnormal{10}}}B,6n), E(\ensuremath{^{\textnormal{10}}}B)\ensuremath{\approx}90 MeV; Ge(Li) and Si(Li); Measured \ensuremath{\gamma}, \ensuremath{\gamma}\ensuremath{\gamma}, I(ce), T\ensuremath{_{\textnormal{1/2}}}.}\\
\parbox[b][0.3cm]{17.7cm}{\addtolength{\parindent}{-0.2in}Other: \href{https://www.nndc.bnl.gov/nsr/nsrlink.jsp?1971Jo19,B}{1971Jo19}.}\\
\vspace{12pt}
\underline{$^{201}$Bi Levels}\\

\parbox[b][0.3cm]{17.7cm}{\makebox[1ex]{\ensuremath{^{\hypertarget{BI37LEVEL0}{\dagger}}}} From a least-squares fit to E\ensuremath{\gamma}.}\\
\parbox[b][0.3cm]{17.7cm}{\makebox[1ex]{\ensuremath{^{\hypertarget{BI37LEVEL1}{\ddagger}}}} From Adopted Levels.}\\
\vspace{0.5cm}

\underline{\ensuremath{\varepsilon,\beta^+} radiations}\\
%
\parbox[b][0.3cm]{17.7cm}{\makebox[1ex]{\ensuremath{^{\hypertarget{BI37DECAY0}{\dagger}}}} For absolute intensity per 100 decays, multiply by{ }\ensuremath{\approx}0.55.}\\
\parbox[b][0.3cm]{17.7cm}{\makebox[1ex]{\ensuremath{^{\hypertarget{BI37DECAY1}{\ddagger}}}} Existence of this branch is questionable.}\\
\vspace{0.5cm}
\clearpage
\vspace{0.3cm}
\begin{landscape}
\vspace*{-0.5cm}
{\bf \small \underline{\ensuremath{^{\textnormal{201}}}Po \ensuremath{\varepsilon} decay (8.96 min)\hspace{0.2in}\href{https://www.nndc.bnl.gov/nsr/nsrlink.jsp?1986Br28,B}{1986Br28},\href{https://www.nndc.bnl.gov/nsr/nsrlink.jsp?1976Ko13,B}{1976Ko13} (continued)}}\\
\vspace{0.3cm}
\underline{$\gamma$($^{201}$Bi)}\\
\vspace{0.34cm}
\parbox[b][0.3cm]{21.881866cm}{\addtolength{\parindent}{-0.254cm}I\ensuremath{\gamma} normalization: Using \ensuremath{\Sigma}I(\ensuremath{\gamma}+ce)(to g.s.)=100 {\textminus} I\ensuremath{\beta}(g.s.) where I\ensuremath{\beta}(g.s.) \ensuremath{\approx} 45\% was deduced by the evaluator from log \textit{ft}\ensuremath{^{\textnormal{I1u}}}=8.2 in \ensuremath{^{\textnormal{197}}}Pb \ensuremath{\varepsilon} decay (\ensuremath{J^{\pi}}=13/2\ensuremath{^{+}})}\\
\parbox[b][0.3cm]{21.881866cm}{to the \ensuremath{J^{\pi}}=9/2\ensuremath{^{-}} level in \ensuremath{^{\textnormal{193}}}Tl.}\\
\vspace{0.34cm}

\parbox[b][0.3cm]{21.881866cm}{\makebox[1ex]{\ensuremath{^{\hypertarget{PO37GAMMA0}{\dagger}}}} From adopted gammas, unless otherwise stated.}\\
\parbox[b][0.3cm]{21.881866cm}{\makebox[1ex]{\ensuremath{^{\hypertarget{PO37GAMMA1}{\ddagger}}}} From \href{https://www.nndc.bnl.gov/nsr/nsrlink.jsp?1986Br28,B}{1986Br28}.}\\
\parbox[b][0.3cm]{21.881866cm}{\makebox[1ex]{\ensuremath{^{\hypertarget{PO37GAMMA2}{\#}}}} Estimated from coincidence intensities in \href{https://www.nndc.bnl.gov/nsr/nsrlink.jsp?1986Br28,B}{1986Br28}.}\\
\parbox[b][0.3cm]{21.881866cm}{\makebox[1ex]{\ensuremath{^{\hypertarget{PO37GAMMA3}{@}}}} For absolute intensity per 100 decays, multiply by{ }\ensuremath{\approx}0.162.}\\
\parbox[b][0.3cm]{21.881866cm}{\makebox[1ex]{\ensuremath{^{\hypertarget{PO37GAMMA4}{\&}}}} Total theoretical internal conversion coefficients, calculated using the BrIcc code (\href{https://www.nndc.bnl.gov/nsr/nsrlink.jsp?2008Ki07,B}{2008Ki07}) with Frozen orbital approximation based on \ensuremath{\gamma}-ray energies,}\\
\parbox[b][0.3cm]{21.881866cm}{{\ }{\ }assigned multipolarities, and mixing ratios, unless otherwise specified.}\\
\vspace{0.5cm}
\end{landscape}\clearpage
\clearpage
\begin{figure}[h]
\begin{center}
\includegraphics{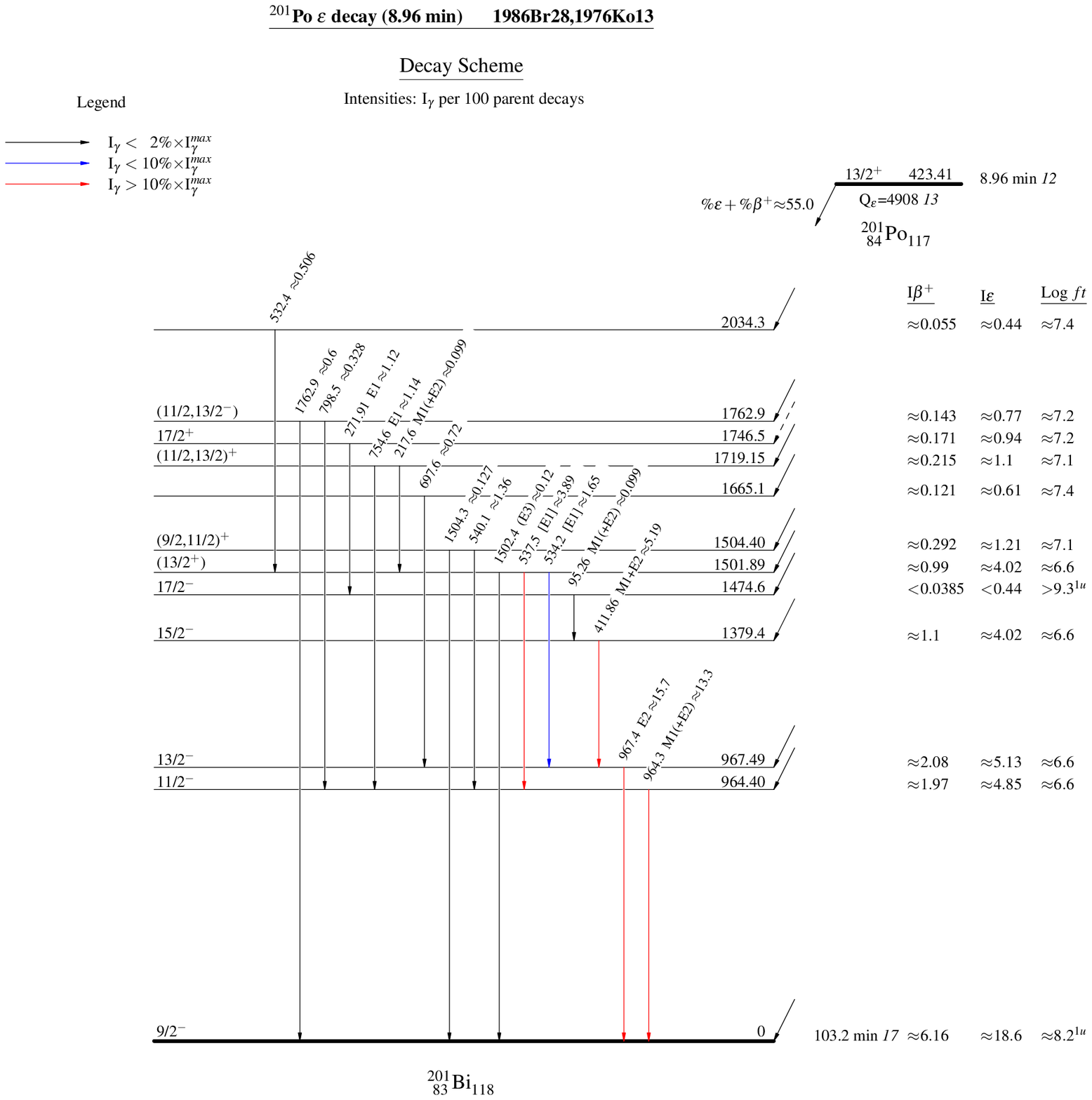}\\
\end{center}
\end{figure}
\clearpage
%205AT A DECAY
\subsection[\hspace{-0.2cm}\ensuremath{^{\textnormal{205}}}At \ensuremath{\alpha} decay]{ }
\vspace{-27pt}
\vspace{0.3cm}
\hypertarget{AT38}{{\bf \small \underline{\ensuremath{^{\textnormal{205}}}At \ensuremath{\alpha} decay}}}\\
\vspace{4pt}
\vspace{8pt}
\parbox[b][0.3cm]{17.7cm}{\addtolength{\parindent}{-0.2in}Parent: $^{205}$At: E=0.0; J$^{\pi}$=9/2\ensuremath{^{-}}; T$_{1/2}$=26.9 min {\it 8}; Q(\ensuremath{\alpha})=6019.6 {\it 17}; \%\ensuremath{\alpha} decay=10 {\it 2}

}\\
\parbox[b][0.3cm]{17.7cm}{\addtolength{\parindent}{-0.2in}\ensuremath{^{205}}At-\ensuremath{J^{\pi}}, T\ensuremath{_{\textnormal{1/2}}} and \%\ensuremath{\alpha} decay from \href{https://www.nndc.bnl.gov/nsr/nsrlink.jsp?2020Ko17,B}{2020Ko17}; Q(\ensuremath{\alpha}) from \href{https://www.nndc.bnl.gov/nsr/nsrlink.jsp?2021Wa16,B}{2021Wa16}.}\\
\vspace{12pt}
\underline{$^{201}$Bi Levels}\\

\parbox[b][0.3cm]{17.7cm}{\makebox[1ex]{\ensuremath{^{\hypertarget{BI38LEVEL0}{\dagger}}}} From Adopted Levels.}\\
\vspace{0.5cm}

\underline{\ensuremath{\alpha} radiations}\\
%
\parbox[b][0.3cm]{17.7cm}{\makebox[1ex]{\ensuremath{^{\hypertarget{BI38DECAY0}{\dagger}}}} Using r\ensuremath{_{\textnormal{0}}}(\ensuremath{^{\textnormal{201}}}Bi)=1.4771 \textit{24} from \href{https://www.nndc.bnl.gov/nsr/nsrlink.jsp?2020Si16,B}{2020Si16}.}\\
\parbox[b][0.3cm]{17.7cm}{\makebox[1ex]{\ensuremath{^{\hypertarget{BI38DECAY1}{\ddagger}}}} For absolute intensity per 100 decays, multiply by 0.10 \textit{2}.}\\
\vspace{0.5cm}
\clearpage
%203TL(A,6NG)
\subsection[\hspace{-0.2cm}\ensuremath{^{\textnormal{203}}}Tl(\ensuremath{\alpha},6n\ensuremath{\gamma})]{ }
\vspace{-27pt}
\vspace{0.3cm}
\hypertarget{BI39}{{\bf \small \underline{\ensuremath{^{\textnormal{203}}}Tl(\ensuremath{\alpha},6n\ensuremath{\gamma})\hspace{0.2in}\href{https://www.nndc.bnl.gov/nsr/nsrlink.jsp?1982Br21,B}{1982Br21}}}}\\
\vspace{4pt}
\vspace{8pt}
\parbox[b][0.3cm]{17.7cm}{\addtolength{\parindent}{-0.2in}\href{https://www.nndc.bnl.gov/nsr/nsrlink.jsp?1982Br21,B}{1982Br21}: E(\ensuremath{\alpha})=66, 70 and 77 MeV; Detectors: Ge(Li); Measured: \ensuremath{\gamma}\ensuremath{\gamma} coin, \ensuremath{\gamma}(t), \ensuremath{\gamma}(\ensuremath{\theta}); Deduced: level scheme, \ensuremath{J^{\pi}}, T\ensuremath{_{\textnormal{1/2}}}.}\\
\parbox[b][0.3cm]{17.7cm}{\addtolength{\parindent}{-0.2in}Other: \ensuremath{^{\textnormal{204}}}Pb(p,4n\ensuremath{\gamma}),\ensuremath{^{\textnormal{206}}}Pb(p,6n\ensuremath{\gamma}) in \href{https://www.nndc.bnl.gov/nsr/nsrlink.jsp?1975OHZZ,B}{1975OHZZ} (E(p)=33-52 MeV; Detectors: Ge(Li)), where no delayed \ensuremath{\gamma} rays with T\ensuremath{_{\textnormal{1/2}}}\ensuremath{\geq}1}\\
\parbox[b][0.3cm]{17.7cm}{\ensuremath{\mu}s were observed.}\\
\vspace{12pt}
\underline{$^{201}$Bi Levels}\\

\parbox[b][0.3cm]{17.7cm}{\makebox[1ex]{\ensuremath{^{\hypertarget{BI39LEVEL0}{\dagger}}}} From a least-squares fit to E\ensuremath{\gamma}.}\\
\parbox[b][0.3cm]{17.7cm}{\makebox[1ex]{\ensuremath{^{\hypertarget{BI39LEVEL1}{\ddagger}}}} Based on \ensuremath{\gamma}(\ensuremath{\theta}) and \ensuremath{\alpha}(exp) in \href{https://www.nndc.bnl.gov/nsr/nsrlink.jsp?1982Br21,B}{1982Br21}.}\\
\parbox[b][0.3cm]{17.7cm}{\makebox[1ex]{\ensuremath{^{\hypertarget{BI39LEVEL2}{\#}}}} T\ensuremath{_{\textnormal{1/2}}}\ensuremath{>}10 ns is reported in \href{https://www.nndc.bnl.gov/nsr/nsrlink.jsp?1982Br21,B}{1982Br21} for a level above 3810+X keV.}\\
\vspace{0.5cm}
\underline{$\gamma$($^{201}$Bi)}\\

\begin{textblock}{29}(0,27.3)
Continued on next page (footnotes at end of table)
\end{textblock}
\clearpage
%
\parbox[b][0.3cm]{17.7cm}{\makebox[1ex]{\ensuremath{^{\hypertarget{BI39GAMMA0}{\dagger}}}} From \href{https://www.nndc.bnl.gov/nsr/nsrlink.jsp?1982Br21,B}{1982Br21}.}\\
\parbox[b][0.3cm]{17.7cm}{\makebox[1ex]{\ensuremath{^{\hypertarget{BI39GAMMA1}{\ddagger}}}} From E(\ensuremath{\alpha})=77 MeV in \href{https://www.nndc.bnl.gov/nsr/nsrlink.jsp?1982Br21,B}{1982Br21}.}\\
\parbox[b][0.3cm]{17.7cm}{\makebox[1ex]{\ensuremath{^{\hypertarget{BI39GAMMA2}{\#}}}} Based on \ensuremath{\gamma}(\ensuremath{\theta}) in \href{https://www.nndc.bnl.gov/nsr/nsrlink.jsp?1982Br21,B}{1982Br21}, unless otherwise stated.}\\
\parbox[b][0.3cm]{17.7cm}{\makebox[1ex]{\ensuremath{^{\hypertarget{BI39GAMMA3}{@}}}} From \href{https://www.nndc.bnl.gov/nsr/nsrlink.jsp?1982Br21,B}{1982Br21}. Measured in the out-of-beam time region of 50 to 130 ns.}\\
\parbox[b][0.3cm]{17.7cm}{\makebox[1ex]{\ensuremath{^{\hypertarget{BI39GAMMA4}{\&}}}} Placement of transition in the level scheme is uncertain.}\\
\parbox[b][0.3cm]{17.7cm}{\makebox[1ex]{\ensuremath{^{\hypertarget{BI39GAMMA5}{x}}}} \ensuremath{\gamma} ray not placed in level scheme.}\\
\vspace{0.5cm}
\clearpage
\begin{figure}[h]
\begin{center}
\includegraphics{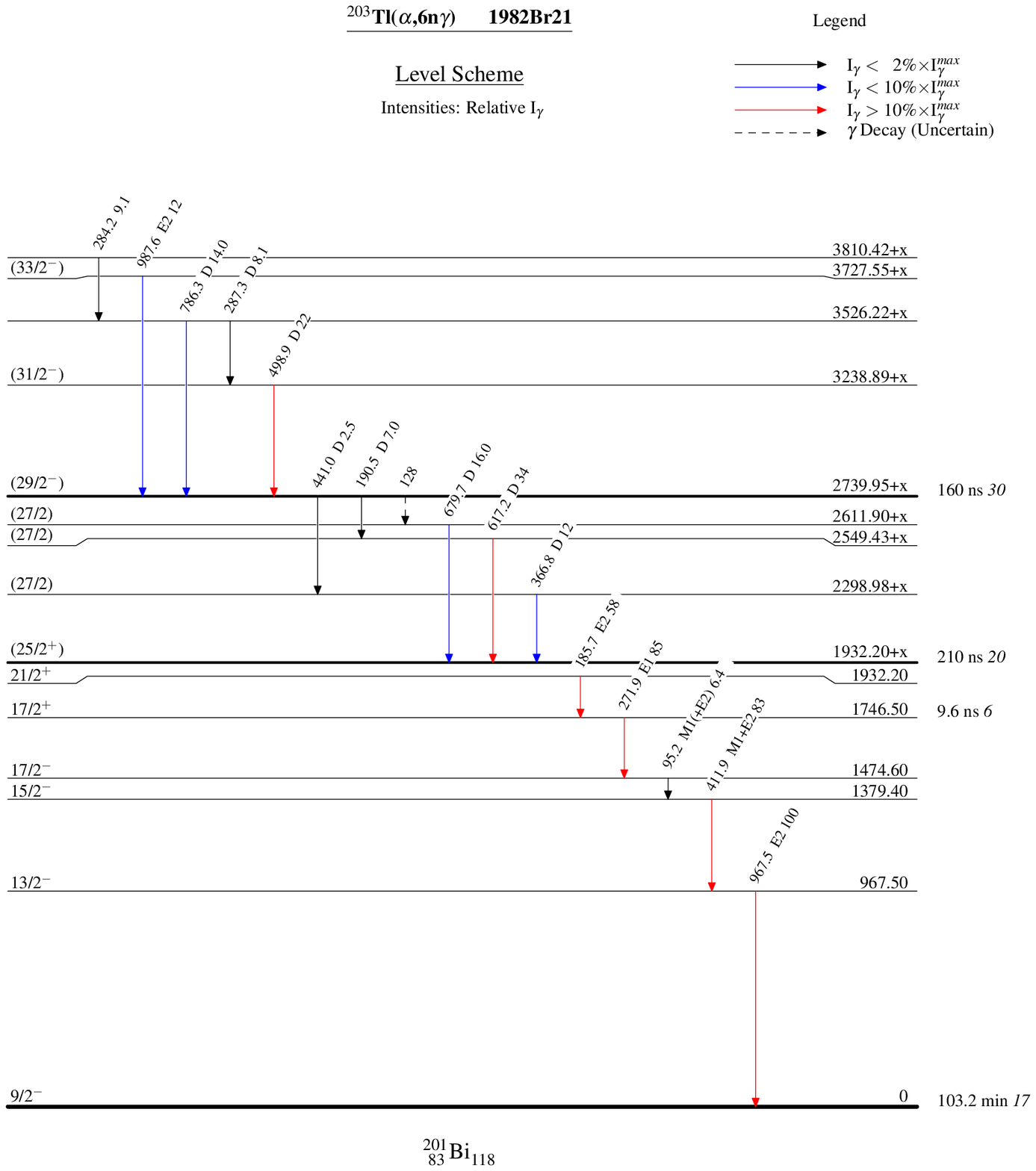}\\
\end{center}
\end{figure}
\clearpage
%196PT(10B,5NG)
\subsection[\hspace{-0.2cm}\ensuremath{^{\textnormal{196}}}Pt(\ensuremath{^{\textnormal{10}}}B,5n\ensuremath{\gamma})]{ }
\vspace{-27pt}
\vspace{0.3cm}
\hypertarget{BI40}{{\bf \small \underline{\ensuremath{^{\textnormal{196}}}Pt(\ensuremath{^{\textnormal{10}}}B,5n\ensuremath{\gamma})\hspace{0.2in}\href{https://www.nndc.bnl.gov/nsr/nsrlink.jsp?1985Pi05,B}{1985Pi05}}}}\\
\vspace{4pt}
\vspace{8pt}
\parbox[b][0.3cm]{17.7cm}{\addtolength{\parindent}{-0.2in}E(\ensuremath{^{\textnormal{10}}}B)=57-72 MeV; Target: \ensuremath{^{\textnormal{196}}}Pt, enriched \ensuremath{>}95\% and 3.6 mg/cm\ensuremath{^{\textnormal{2}}} thick; Detectors: two Ge(Li) and one planar Ge(intrinsic);}\\
\parbox[b][0.3cm]{17.7cm}{Measured: excitation functions, \ensuremath{\gamma}(\ensuremath{\theta}),DCO, \ensuremath{\gamma}\ensuremath{\gamma}, \ensuremath{\gamma}(t) {\textminus} pulsed beam with 10 ns on and 2 \ensuremath{\mu}s off periods. Deduced: level scheme,}\\
\parbox[b][0.3cm]{17.7cm}{\ensuremath{J^{\pi}}, T\ensuremath{_{\textnormal{1/2}}}.}\\
\parbox[b][0.3cm]{17.7cm}{\addtolength{\parindent}{-0.2in}Others: \href{https://www.nndc.bnl.gov/nsr/nsrlink.jsp?1974GiZX,B}{1974GiZX}, \href{https://www.nndc.bnl.gov/nsr/nsrlink.jsp?1973GiZW,B}{1973GiZW}.}\\
\vspace{12pt}
\underline{$^{201}$Bi Levels}\\

\parbox[b][0.3cm]{17.7cm}{\makebox[1ex]{\ensuremath{^{\hypertarget{BI40LEVEL0}{\dagger}}}} From a least-squares fit to E\ensuremath{\gamma}. X is expected to be less than 80 keV, otherwise a \ensuremath{\gamma}-ray transition would be observed. The}\\
\parbox[b][0.3cm]{17.7cm}{{\ }{\ }assignment is based on similarities with the \ensuremath{J^{\pi}}=7\ensuremath{^{-}} and 9\ensuremath{^{-}} states in \ensuremath{^{\textnormal{200}}}Pb, as well as with the systematics in neighboring \ensuremath{^{\textnormal{203}}}Bi}\\
\parbox[b][0.3cm]{17.7cm}{{\ }{\ }and \ensuremath{^{\textnormal{205}}}Bi isotopes.}\\
\parbox[b][0.3cm]{17.7cm}{\makebox[1ex]{\ensuremath{^{\hypertarget{BI40LEVEL1}{\ddagger}}}} From \href{https://www.nndc.bnl.gov/nsr/nsrlink.jsp?1985Pi05,B}{1985Pi05}, unless otherwise stated.}\\
\parbox[b][0.3cm]{17.7cm}{\makebox[1ex]{\ensuremath{^{\hypertarget{BI40LEVEL2}{\#}}}} An isomer with T\ensuremath{_{\textnormal{1/2}}}=14 ns \textit{3} was found above the 3526.4+X keV level in \href{https://www.nndc.bnl.gov/nsr/nsrlink.jsp?1985Pi05,B}{1985Pi05}. Note, that an isomer with T\ensuremath{_{\textnormal{1/2}}}\ensuremath{\approx}10 ns was}\\
\parbox[b][0.3cm]{17.7cm}{{\ }{\ }also reported in \ensuremath{^{\textnormal{203}}}Tl(\ensuremath{\alpha},6n\ensuremath{\gamma}) (\href{https://www.nndc.bnl.gov/nsr/nsrlink.jsp?1982Br21,B}{1982Br21}) at or above the 3810+X level.}\\
\begin{textblock}{29}(0,27.3)
Continued on next page (footnotes at end of table)
\end{textblock}
\clearpage
\vspace*{-0.5cm}
{\bf \small \underline{\ensuremath{^{\textnormal{196}}}Pt(\ensuremath{^{\textnormal{10}}}B,5n\ensuremath{\gamma})\hspace{0.2in}\href{https://www.nndc.bnl.gov/nsr/nsrlink.jsp?1985Pi05,B}{1985Pi05} (continued)}}\\
\vspace{0.3cm}
\underline{$^{201}$Bi Levels (continued)}\\
\vspace{0.3cm}
\parbox[b][0.3cm]{17.7cm}{\makebox[1ex]{\ensuremath{^{\hypertarget{BI40LEVEL3}{@}}}} Configuration=\ensuremath{\pi} h\ensuremath{_{\textnormal{9/2}}^{\textnormal{+1}}}.}\\
\parbox[b][0.3cm]{17.7cm}{\makebox[1ex]{\ensuremath{^{\hypertarget{BI40LEVEL4}{\&}}}} Configuration=\ensuremath{\pi} (h\ensuremath{_{\textnormal{9/2}}^{\textnormal{+1}}})\ensuremath{\otimes}2\ensuremath{^{\textnormal{+}}}.}\\
\parbox[b][0.3cm]{17.7cm}{\makebox[1ex]{\ensuremath{^{\hypertarget{BI40LEVEL5}{a}}}} Configuration=\ensuremath{\pi} (h\ensuremath{_{\textnormal{9/2}}^{\textnormal{+1}}})\ensuremath{\otimes}4\ensuremath{^{\textnormal{+}}}.}\\
\parbox[b][0.3cm]{17.7cm}{\makebox[1ex]{\ensuremath{^{\hypertarget{BI40LEVEL6}{b}}}} Admixture of configuration= \ensuremath{\pi} (h\ensuremath{_{\textnormal{9/2}}^{\textnormal{+1}}}) \ensuremath{\nu} (f\ensuremath{_{\textnormal{5/2}}^{\textnormal{$-$1}}},i\ensuremath{_{\textnormal{13/2}}^{\textnormal{$-$1}}})\ensuremath{_{\textnormal{5$-$}}} and configuration=\ensuremath{\pi} (h\ensuremath{_{\textnormal{9/2}}^{\textnormal{+1}}}) \ensuremath{\nu} (p\ensuremath{_{\textnormal{3/2}}^{\textnormal{$-$1}}},i\ensuremath{_{\textnormal{13/2}}^{\textnormal{$-$1}}})\ensuremath{_{\textnormal{5$-$}}}.}\\
\parbox[b][0.3cm]{17.7cm}{\makebox[1ex]{\ensuremath{^{\hypertarget{BI40LEVEL7}{c}}}} Admixture of configuration= \ensuremath{\pi} (h\ensuremath{_{\textnormal{9/2}}^{\textnormal{+1}}}) \ensuremath{\nu} (f\ensuremath{_{\textnormal{5/2}}^{\textnormal{$-$1}}},i\ensuremath{_{\textnormal{13/2}}^{\textnormal{$-$1}}})\ensuremath{_{\textnormal{7$-$}}} and configuration=\ensuremath{\pi} (h\ensuremath{_{\textnormal{9/2}}^{\textnormal{+1}}}) \ensuremath{\nu} (p\ensuremath{_{\textnormal{3/2}}^{\textnormal{$-$1}}},i\ensuremath{_{\textnormal{13/2}}^{\textnormal{$-$1}}})\ensuremath{_{\textnormal{7$-$}}}.}\\
\parbox[b][0.3cm]{17.7cm}{\makebox[1ex]{\ensuremath{^{\hypertarget{BI40LEVEL8}{d}}}} Configuration=\ensuremath{\pi} (h\ensuremath{_{\textnormal{9/2}}^{\textnormal{+1}}}) \ensuremath{\nu} (f\ensuremath{_{\textnormal{5/2}}^{\textnormal{$-$1}}},i\ensuremath{_{\textnormal{13/2}}^{\textnormal{$-$1}}})\ensuremath{_{\textnormal{9$-$}}}.}\\
\parbox[b][0.3cm]{17.7cm}{\makebox[1ex]{\ensuremath{^{\hypertarget{BI40LEVEL9}{e}}}} Configuration=\ensuremath{\pi} (h\ensuremath{_{\textnormal{9/2}}^{\textnormal{+1}}}) \ensuremath{\nu} (i\ensuremath{_{\textnormal{13/2}}^{\textnormal{$-$2}}})\ensuremath{_{\textnormal{12+}}}.}\\
\vspace{0.5cm}
\underline{$\gamma$($^{201}$Bi)}\\

\begin{textblock}{29}(0,27.3)
Continued on next page (footnotes at end of table)
\end{textblock}
\clearpage
%
\parbox[b][0.3cm]{17.7cm}{\makebox[1ex]{\ensuremath{^{\hypertarget{BI40GAMMA0}{\dagger}}}} From \href{https://www.nndc.bnl.gov/nsr/nsrlink.jsp?1985Pi05,B}{1985Pi05}.}\\
\parbox[b][0.3cm]{17.7cm}{\makebox[1ex]{\ensuremath{^{\hypertarget{BI40GAMMA1}{\ddagger}}}} From E(\ensuremath{^{\textnormal{10}}}B)=67 MeV in \href{https://www.nndc.bnl.gov/nsr/nsrlink.jsp?1985Pi05,B}{1985Pi05}.}\\
\parbox[b][0.3cm]{17.7cm}{\makebox[1ex]{\ensuremath{^{\hypertarget{BI40GAMMA2}{\#}}}} Based on \ensuremath{\gamma}(\ensuremath{\theta}) and DCO, unless otherwise stated. DCO values were obtained by gating on stretched E2 transitions.}\\
\parbox[b][0.3cm]{17.7cm}{\makebox[1ex]{\ensuremath{^{\hypertarget{BI40GAMMA3}{x}}}} \ensuremath{\gamma} ray not placed in level scheme.}\\
\vspace{0.5cm}
\clearpage
\begin{figure}[h]
\begin{center}
\includegraphics{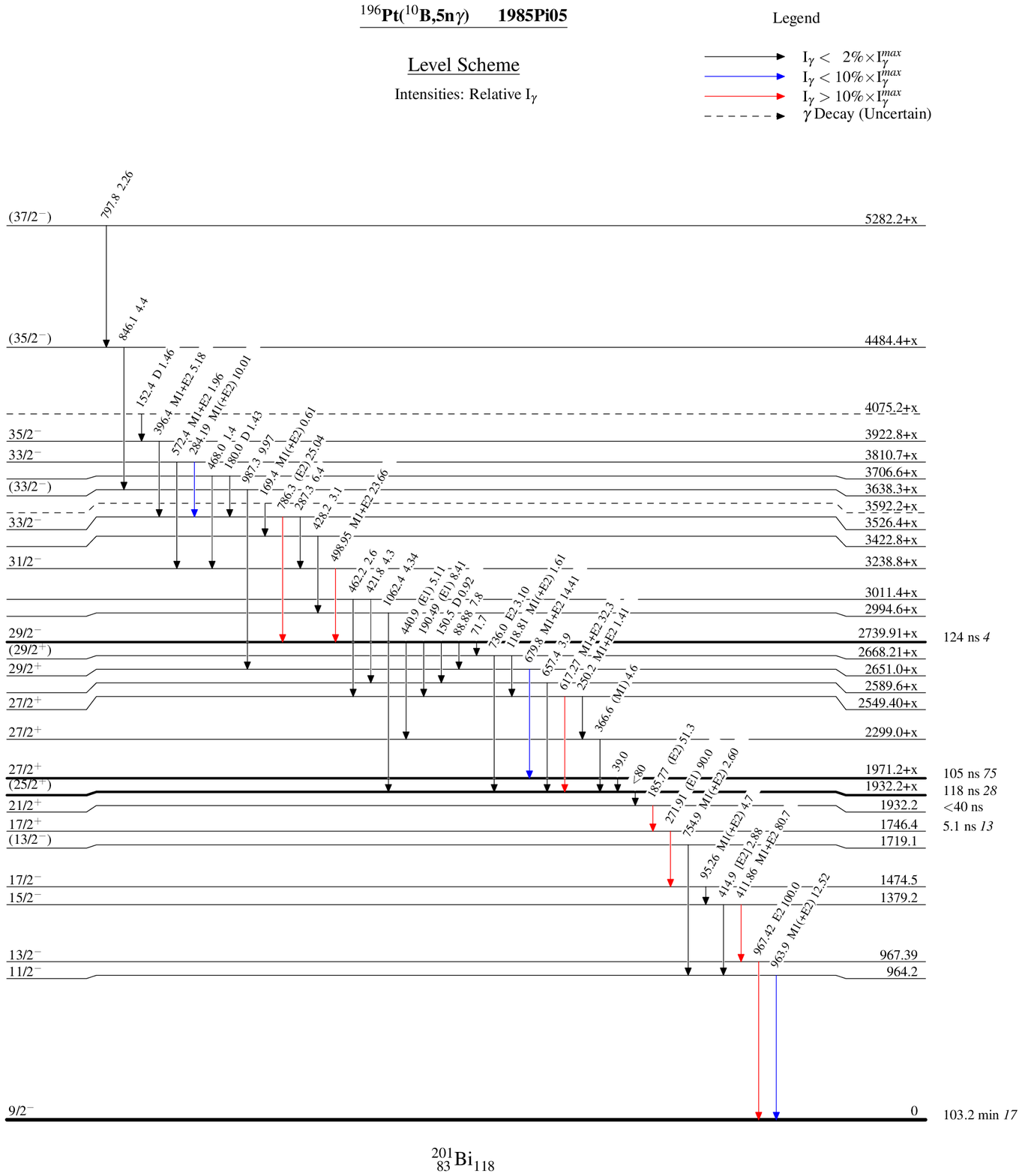}\\
\end{center}
\end{figure}
\clearpage
%ADOPTED LEVELS, GAMMAS
\section[\ensuremath{^{201}_{\ 84}}Po\ensuremath{_{117}^{~}}]{ }
\vspace{-30pt}
\setcounter{chappage}{1}
\subsection[\hspace{-0.2cm}Adopted Levels, Gammas]{ }
\vspace{-20pt}
\vspace{0.3cm}
\hypertarget{PO41}{{\bf \small \underline{Adopted \hyperlink{201PO_LEVEL}{Levels}, \hyperlink{201PO_GAMMA}{Gammas}}}}\\
\vspace{4pt}
\vspace{8pt}
\parbox[b][0.3cm]{17.7cm}{\addtolength{\parindent}{-0.2in}Q(\ensuremath{\beta^-})=$-$5732 {\it 10}; S(n)=7651 {\it 9}; S(p)=3440 {\it 23}; Q(\ensuremath{\alpha})=5799.3 {\it 17}\hspace{0.2in}\href{https://www.nndc.bnl.gov/nsr/nsrlink.jsp?2021Wa16,B}{2021Wa16}}\\

\vspace{12pt}
\hypertarget{201PO_LEVEL}{\underline{$^{201}$Po Levels}}\\

\begin{textblock}{29}(0,27.3)
Continued on next page (footnotes at end of table)
\end{textblock}
\clearpage
%
\parbox[b][0.3cm]{17.7cm}{\makebox[1ex]{\ensuremath{^{\hypertarget{PO41LEVEL0}{\dagger}}}} From a least-squares fit to E\ensuremath{\gamma}. \ensuremath{\Delta}E\ensuremath{\gamma}=0.5 keV is assumed for E\ensuremath{\gamma}$'$s without uncertainties.}\\
\parbox[b][0.3cm]{17.7cm}{\makebox[1ex]{\ensuremath{^{\hypertarget{PO41LEVEL1}{\ddagger}}}} Configuration=\ensuremath{\nu} f\ensuremath{_{\textnormal{5/2}}^{\textnormal{$-$1}}}.}\\
\parbox[b][0.3cm]{17.7cm}{\makebox[1ex]{\ensuremath{^{\hypertarget{PO41LEVEL2}{\#}}}} Configuration=\ensuremath{\nu} p\ensuremath{_{\textnormal{1/2}}^{\textnormal{$-$1}}}. The assignment is tentative.}\\
\parbox[b][0.3cm]{17.7cm}{\makebox[1ex]{\ensuremath{^{\hypertarget{PO41LEVEL3}{@}}}} Configuration=\ensuremath{\nu} p\ensuremath{_{\textnormal{3/2}}^{\textnormal{$-$1}}}.}\\
\parbox[b][0.3cm]{17.7cm}{\makebox[1ex]{\ensuremath{^{\hypertarget{PO41LEVEL4}{\&}}}} Configuration=\ensuremath{\nu} i\ensuremath{_{\textnormal{13/2}}^{\textnormal{$-$1}}}.}\\
\parbox[b][0.3cm]{17.7cm}{\makebox[1ex]{\ensuremath{^{\hypertarget{PO41LEVEL5}{a}}}} Configuration=\ensuremath{\nu} (i\ensuremath{_{\textnormal{13/2}}^{\textnormal{$-$1}}})\ensuremath{\otimes}2\ensuremath{^{\textnormal{+}}}.}\\
\parbox[b][0.3cm]{17.7cm}{\makebox[1ex]{\ensuremath{^{\hypertarget{PO41LEVEL6}{b}}}} Configuration=\ensuremath{\nu} (i\ensuremath{_{\textnormal{13/2}}^{\textnormal{$-$1}}})\ensuremath{\otimes}4\ensuremath{^{\textnormal{+}}}.}\\
\parbox[b][0.3cm]{17.7cm}{\makebox[1ex]{\ensuremath{^{\hypertarget{PO41LEVEL7}{c}}}} Possibly a mixture between configuration=\ensuremath{\nu} (i\ensuremath{_{\textnormal{13/2}}^{\textnormal{$-$1}}})\ensuremath{\otimes}6\ensuremath{^{\textnormal{+}}} and configuration=\ensuremath{\nu} (i\ensuremath{_{\textnormal{13/2}}^{\textnormal{$-$1}}}) \ensuremath{\pi} (h\ensuremath{_{\textnormal{9/2}}^{\textnormal{+2}}})\ensuremath{_{\textnormal{8+}}}.}\\
\vspace{0.5cm}
\clearpage
\vspace{0.3cm}
\begin{landscape}
\vspace*{-0.5cm}
{\bf \small \underline{Adopted \hyperlink{201PO_LEVEL}{Levels}, \hyperlink{201PO_GAMMA}{Gammas} (continued)}}\\
\vspace{0.3cm}
\hypertarget{201PO_GAMMA}{\underline{$\gamma$($^{201}$Po)}}\\

\parbox[b][0.3cm]{21.881866cm}{\makebox[1ex]{\ensuremath{^{\hypertarget{PO41GAMMA0}{\dagger}}}} From \ensuremath{^{\textnormal{201}}}At \ensuremath{\varepsilon} decay, unless otherwise stated.}\\
\parbox[b][0.3cm]{21.881866cm}{\makebox[1ex]{\ensuremath{^{\hypertarget{PO41GAMMA1}{\ddagger}}}} From \ensuremath{^{\textnormal{194}}}Pt(\ensuremath{^{\textnormal{12}}}C,5n\ensuremath{\gamma}).}\\
\parbox[b][0.3cm]{21.881866cm}{\makebox[1ex]{\ensuremath{^{\hypertarget{PO41GAMMA2}{\#}}}} From \ensuremath{\alpha}(K)exp in \ensuremath{^{\textnormal{201}}}At \ensuremath{\varepsilon} decay and \ensuremath{\gamma}(\ensuremath{\theta}) in \ensuremath{^{\textnormal{194}}}Pt(\ensuremath{^{\textnormal{12}}}C,5n\ensuremath{\gamma}), unless otherwise stated.}\\
\parbox[b][0.3cm]{21.881866cm}{\makebox[1ex]{\ensuremath{^{\hypertarget{PO41GAMMA3}{@}}}} Total theoretical internal conversion coefficients, calculated using the BrIcc code (\href{https://www.nndc.bnl.gov/nsr/nsrlink.jsp?2008Ki07,B}{2008Ki07}) with Frozen orbital approximation based on \ensuremath{\gamma}-ray energies,}\\
\parbox[b][0.3cm]{21.881866cm}{{\ }{\ }assigned multipolarities, and mixing ratios, unless otherwise specified.}\\
\vspace{0.5cm}
\end{landscape}\clearpage
\clearpage
\begin{figure}[h]
\begin{center}
\includegraphics{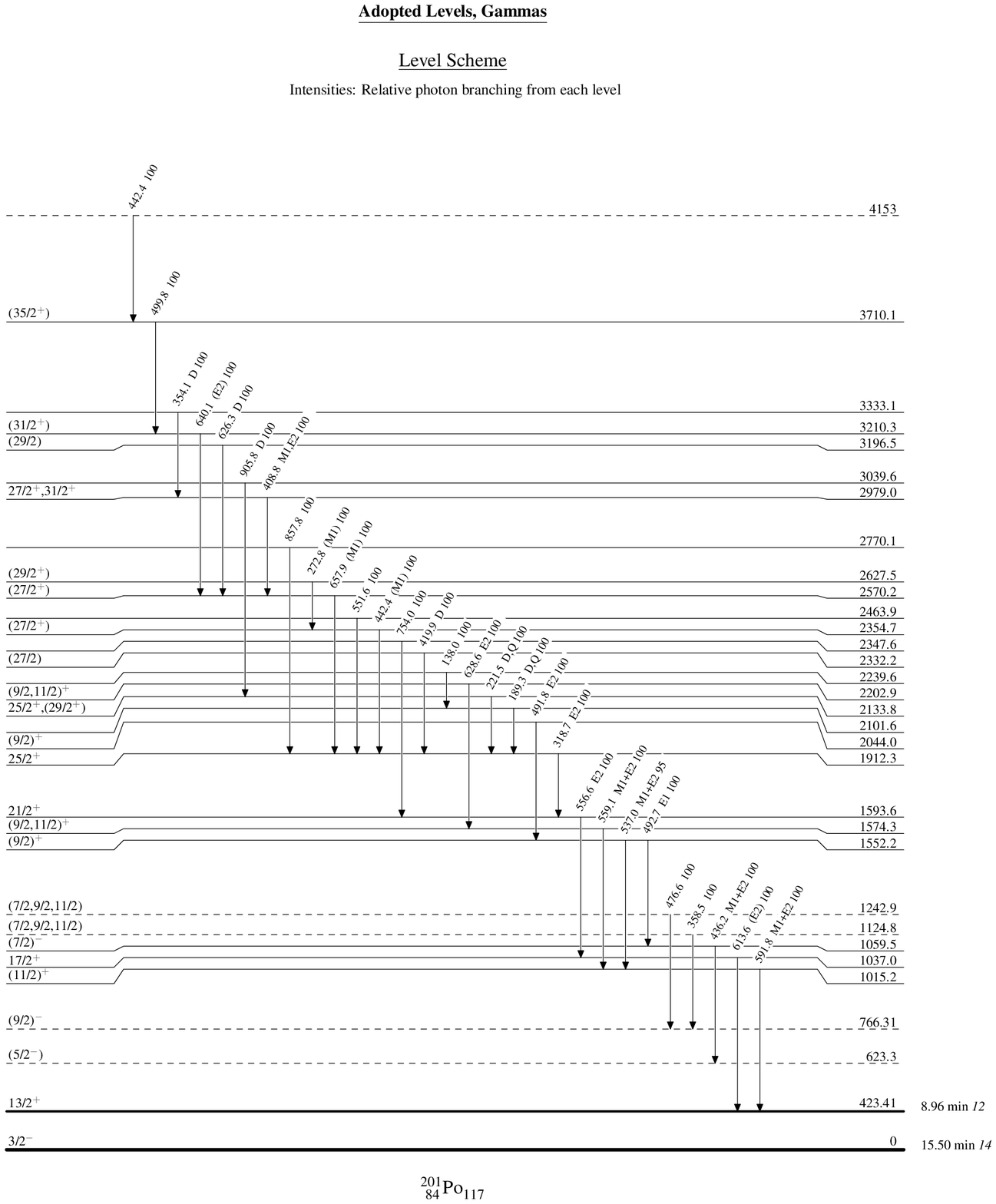}\\
\end{center}
\end{figure}
\clearpage
\begin{figure}[h]
\begin{center}
\includegraphics{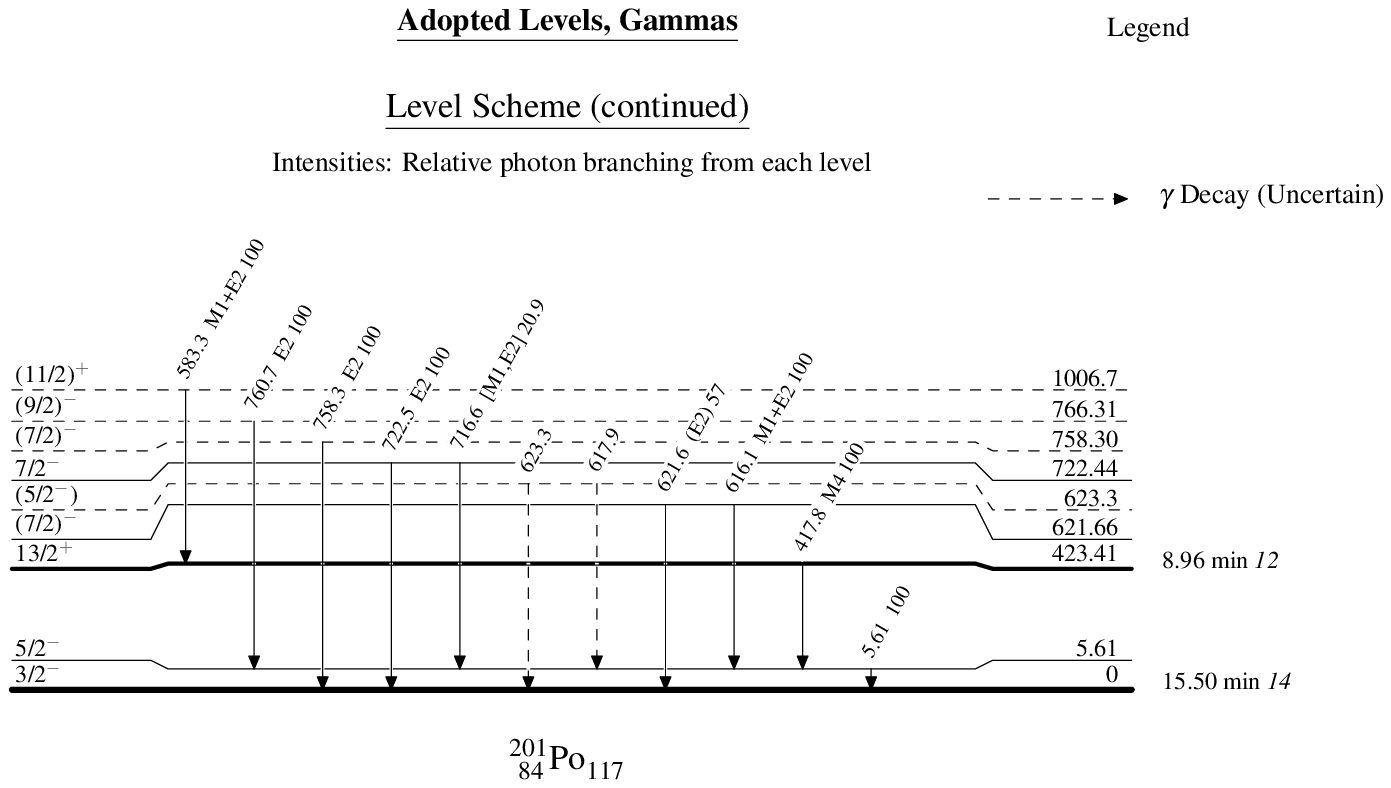}\\
\end{center}
\end{figure}
\clearpage
%201PO IT DECAY (8.96 M)
\subsection[\hspace{-0.2cm}\ensuremath{^{\textnormal{201}}}Po IT decay (8.96 min)]{ }
\vspace{-27pt}
\vspace{0.3cm}
\hypertarget{PO42}{{\bf \small \underline{\ensuremath{^{\textnormal{201}}}Po IT decay (8.96 min)}}}\\
\vspace{4pt}
\vspace{8pt}
\parbox[b][0.3cm]{17.7cm}{\addtolength{\parindent}{-0.2in}Parent: $^{201}$Po: E=423.41 {\it 22}; J$^{\pi}$=13/2\ensuremath{^{+}}; T$_{1/2}$=8.96 min {\it 12}; \%IT decay$\approx$42.6

}\\
\parbox[b][0.3cm]{17.7cm}{\addtolength{\parindent}{-0.2in}\ensuremath{^{201}}Po-\%IT decay: From Adopted Levels.}\\
\vspace{12pt}
\underline{$^{201}$Po Levels}\\
\begin{longtable}{cccccc@{\extracolsep{\fill}}c}
\multicolumn{2}{c}{E(level)$^{{\hyperlink{PO42LEVEL0}{\dagger}}}$}&J$^{\pi}$$^{{\hyperlink{PO42LEVEL0}{\dagger}}}$&\multicolumn{2}{c}{T$_{1/2}$$^{{\hyperlink{PO42LEVEL0}{\dagger}}}$}&Comments&\\[-.2cm]
\multicolumn{2}{c}{\hrulefill}&\hrulefill&\multicolumn{2}{c}{\hrulefill}&\hrulefill&
\endfirsthead
\multicolumn{1}{r@{}}{0}&\multicolumn{1}{@{}l}{\ensuremath{^{{\hyperlink{PO42LEVEL1}{\ddagger}}}}}&\multicolumn{1}{l}{3/2\ensuremath{^{-}}}&\multicolumn{1}{r@{}}{15}&\multicolumn{1}{@{.}l}{50 min {\it 14}}&&\\
\multicolumn{1}{r@{}}{5}&\multicolumn{1}{@{.}l}{61\ensuremath{^{{\hyperlink{PO42LEVEL2}{\#}}}} {\it 13}}&\multicolumn{1}{l}{5/2\ensuremath{^{-}}}&&&&\\
\multicolumn{1}{r@{}}{423}&\multicolumn{1}{@{.}l}{41\ensuremath{^{{\hyperlink{PO42LEVEL3}{@}}}} {\it 22}}&\multicolumn{1}{l}{13/2\ensuremath{^{+}}}&\multicolumn{1}{r@{}}{8}&\multicolumn{1}{@{.}l}{96 min {\it 12}}&\parbox[t][0.3cm]{12.691cm}{\raggedright \%IT\ensuremath{\approx}42.6; \%\ensuremath{\alpha}=2.4 \textit{5}; \%\ensuremath{\varepsilon}+\%\ensuremath{\beta}\ensuremath{^{\textnormal{+}}}\ensuremath{\approx}55\vspace{0.1cm}}&\\
\end{longtable}
\parbox[b][0.3cm]{17.7cm}{\makebox[1ex]{\ensuremath{^{\hypertarget{PO42LEVEL0}{\dagger}}}} From Adopted Levels.}\\
\parbox[b][0.3cm]{17.7cm}{\makebox[1ex]{\ensuremath{^{\hypertarget{PO42LEVEL1}{\ddagger}}}} Configuration=\ensuremath{\nu} p\ensuremath{_{\textnormal{3/2}}^{\textnormal{$-$1}}}.}\\
\parbox[b][0.3cm]{17.7cm}{\makebox[1ex]{\ensuremath{^{\hypertarget{PO42LEVEL2}{\#}}}} Configuration=\ensuremath{\nu} f\ensuremath{_{\textnormal{5/2}}^{\textnormal{$-$1}}}.}\\
\parbox[b][0.3cm]{17.7cm}{\makebox[1ex]{\ensuremath{^{\hypertarget{PO42LEVEL3}{@}}}} Configuration=\ensuremath{\nu} i\ensuremath{_{\textnormal{13/2}}^{\textnormal{$-$1}}}.}\\
\vspace{0.5cm}
\underline{$\gamma$($^{201}$Po)}\\

\parbox[b][0.3cm]{17.7cm}{\makebox[1ex]{\ensuremath{^{\hypertarget{PO42GAMMA0}{\dagger}}}} For absolute intensity per 100 decays, multiply by{ }\ensuremath{\approx}0.426.}\\
\parbox[b][0.3cm]{17.7cm}{\makebox[1ex]{\ensuremath{^{\hypertarget{PO42GAMMA1}{\ddagger}}}} Total theoretical internal conversion coefficients, calculated using the BrIcc code (\href{https://www.nndc.bnl.gov/nsr/nsrlink.jsp?2008Ki07,B}{2008Ki07}) with Frozen orbital approximation}\\
\parbox[b][0.3cm]{17.7cm}{{\ }{\ }based on \ensuremath{\gamma}-ray energies, assigned multipolarities, and mixing ratios, unless otherwise specified.}\\
\vspace{0.5cm}
\clearpage
\begin{figure}[h]
\begin{center}
\includegraphics{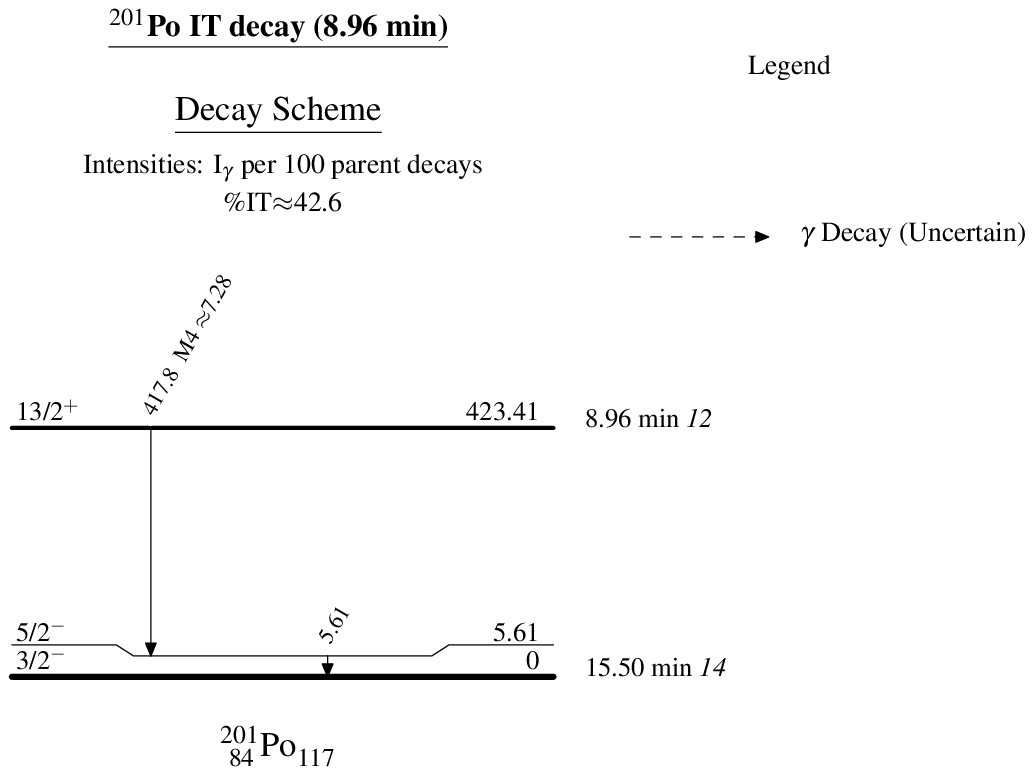}\\
\end{center}
\end{figure}
\clearpage
%201AT EC DECAY
\subsection[\hspace{-0.2cm}\ensuremath{^{\textnormal{201}}}At \ensuremath{\varepsilon} decay]{ }
\vspace{-27pt}
\vspace{0.3cm}
\hypertarget{AT43}{{\bf \small \underline{\ensuremath{^{\textnormal{201}}}At \ensuremath{\varepsilon} decay\hspace{0.2in}\href{https://www.nndc.bnl.gov/nsr/nsrlink.jsp?2010De04,B}{2010De04}}}}\\
\vspace{4pt}
\vspace{8pt}
\parbox[b][0.3cm]{17.7cm}{\addtolength{\parindent}{-0.2in}Parent: $^{201}$At: E=0; J$^{\pi}$=9/2\ensuremath{^{-}}; T$_{1/2}$=87.6 s {\it 13}; Q(\ensuremath{\varepsilon})=5732 {\it 10}; \%\ensuremath{\varepsilon}+\%\ensuremath{\beta^{+}} decay=29 {\it 7}

}\\
\parbox[b][0.3cm]{17.7cm}{\addtolength{\parindent}{-0.2in}\ensuremath{^{201}}At-Q(\ensuremath{\varepsilon}): From \href{https://www.nndc.bnl.gov/nsr/nsrlink.jsp?2021Wa16,B}{2021Wa16}.}\\
\parbox[b][0.3cm]{17.7cm}{\addtolength{\parindent}{-0.2in}\href{https://www.nndc.bnl.gov/nsr/nsrlink.jsp?2010De04,B}{2010De04}: 1.4 GeV proton beam induced spallation on a 49 mg/cm\ensuremath{^{\textnormal{2}}} UC\ensuremath{_{\textnormal{2}}}-C target at ISOLDE-CERN facility. Francium was}\\
\parbox[b][0.3cm]{17.7cm}{surface ionized, accelerated to 30 keV an a mass separated by the ISOLDE General Purpose Separator (GPS). Using tape systems,}\\
\parbox[b][0.3cm]{17.7cm}{measured E\ensuremath{\gamma}, I\ensuremath{\gamma}, \ensuremath{\gamma}\ensuremath{\gamma}, ce, \ensuremath{\gamma}(ce) coin; Detectors: two HPGe detectors located at 90\ensuremath{^\circ} and 180\ensuremath{^\circ} around Si(Li) detector placed in a}\\
\parbox[b][0.3cm]{17.7cm}{MINI-ORANGE spectrometer. \ensuremath{^{\textnormal{201}}}At source is produced from \ensuremath{\alpha} decay of \ensuremath{^{\textnormal{205}}}Fr.}\\
\parbox[b][0.3cm]{17.7cm}{\addtolength{\parindent}{-0.2in}Other: \href{https://www.nndc.bnl.gov/nsr/nsrlink.jsp?1970DaZM,B}{1970DaZM}.}\\
\vspace{12pt}
\underline{$^{201}$Po Levels}\\

\parbox[b][0.3cm]{17.7cm}{\makebox[1ex]{\ensuremath{^{\hypertarget{PO43LEVEL0}{\dagger}}}} From a least-squares fit to E\ensuremath{\gamma}$'$s, unless otherwise stated.}\\
\parbox[b][0.3cm]{17.7cm}{\makebox[1ex]{\ensuremath{^{\hypertarget{PO43LEVEL1}{\ddagger}}}} From Adopted Levels.}\\
\vspace{0.5cm}

\underline{\ensuremath{\varepsilon,\beta^+} radiations}\\
%
\parbox[b][0.3cm]{17.7cm}{\makebox[1ex]{\ensuremath{^{\hypertarget{PO43DECAY0}{\dagger}}}} From the decay scheme and the intensity balances.}\\
\parbox[b][0.3cm]{17.7cm}{\makebox[1ex]{\ensuremath{^{\hypertarget{PO43DECAY1}{\ddagger}}}} Absolute intensity per 100 decays.}\\
\vspace{0.5cm}
\clearpage
\vspace{0.3cm}
\begin{landscape}
\vspace*{-0.5cm}
{\bf \small \underline{\ensuremath{^{\textnormal{201}}}At \ensuremath{\varepsilon} decay\hspace{0.2in}\href{https://www.nndc.bnl.gov/nsr/nsrlink.jsp?2010De04,B}{2010De04} (continued)}}\\
\vspace{0.3cm}
\underline{$\gamma$($^{201}$Po)}\\
\vspace{0.34cm}
\parbox[b][0.3cm]{21.881866cm}{\addtolength{\parindent}{-0.254cm}I\ensuremath{\gamma} normalization: \ensuremath{\Sigma}I(\ensuremath{\gamma}+ce)(to g.s.)=100\% and by assuming that there is no direct feeding to the g.s. (\ensuremath{J^{\pi}}=3/2\ensuremath{^{-}}), 5.61-keV level (\ensuremath{J^{\pi}}=5/2\ensuremath{^{-}}), 423.4-keV level}\\
\parbox[b][0.3cm]{21.881866cm}{(Jp=13/2\ensuremath{^{+}}) and the 623.3-keV level (\ensuremath{J^{\pi}}=(5/2)\ensuremath{^{-}}).}\\
\vspace{0.34cm}

\parbox[b][0.3cm]{21.881866cm}{\makebox[1ex]{\ensuremath{^{\hypertarget{AT43GAMMA0}{\dagger}}}} From \href{https://www.nndc.bnl.gov/nsr/nsrlink.jsp?2010De04,B}{2010De04}, unless otherwise stated.}\\
\parbox[b][0.3cm]{21.881866cm}{\makebox[1ex]{\ensuremath{^{\hypertarget{AT43GAMMA1}{\ddagger}}}} Placement in the decay scheme is not unambiguous.}\\
\parbox[b][0.3cm]{21.881866cm}{\makebox[1ex]{\ensuremath{^{\hypertarget{AT43GAMMA2}{\#}}}} From multiple decay branches and the comparison of \ensuremath{\alpha}(K)exp (\href{https://www.nndc.bnl.gov/nsr/nsrlink.jsp?2010De04,B}{2010De04}) with theoretical values.}\\
\parbox[b][0.3cm]{21.881866cm}{\makebox[1ex]{\ensuremath{^{\hypertarget{AT43GAMMA3}{@}}}} From \ensuremath{\alpha}(K)exp and the briccmixing program.}\\
\parbox[b][0.3cm]{21.881866cm}{\makebox[1ex]{\ensuremath{^{\hypertarget{AT43GAMMA4}{\&}}}} For absolute intensity per 100 decays, multiply by 0.086 \textit{21}.}\\
\parbox[b][0.3cm]{21.881866cm}{\makebox[1ex]{\ensuremath{^{\hypertarget{AT43GAMMA5}{a}}}} Total theoretical internal conversion coefficients, calculated using the BrIcc code (\href{https://www.nndc.bnl.gov/nsr/nsrlink.jsp?2008Ki07,B}{2008Ki07}) with Frozen orbital approximation based on \ensuremath{\gamma}-ray energies,}\\
\parbox[b][0.3cm]{21.881866cm}{{\ }{\ }assigned multipolarities, and mixing ratios, unless otherwise specified.}\\
\parbox[b][0.3cm]{21.881866cm}{\makebox[1ex]{\ensuremath{^{\hypertarget{AT43GAMMA6}{b}}}} Placement of transition in the level scheme is uncertain.}\\
\parbox[b][0.3cm]{21.881866cm}{\makebox[1ex]{\ensuremath{^{\hypertarget{AT43GAMMA7}{x}}}} \ensuremath{\gamma} ray not placed in level scheme.}\\
\vspace{0.5cm}
\end{landscape}\clearpage
\clearpage
\begin{figure}[h]
\begin{center}
\includegraphics{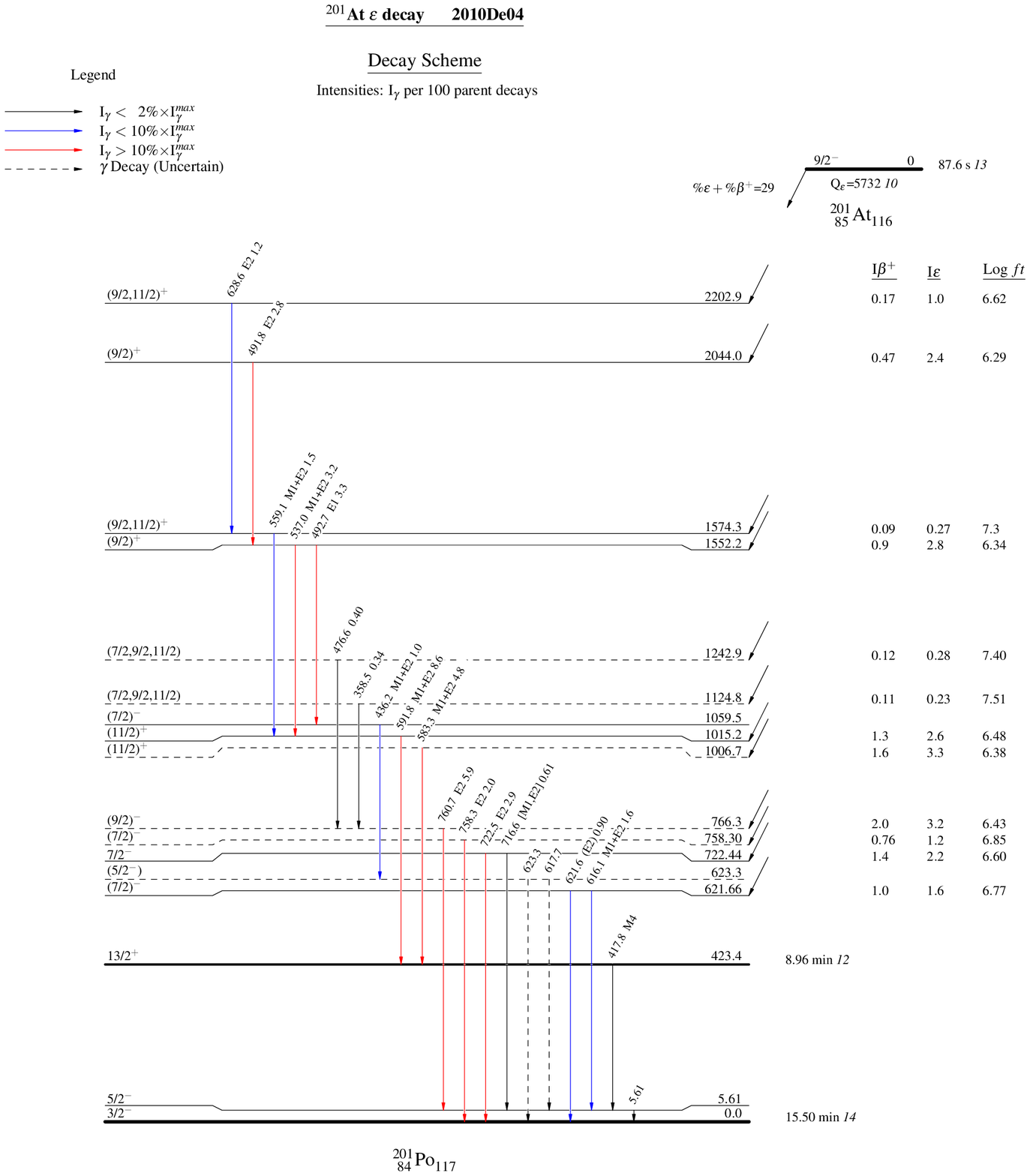}\\
\end{center}
\end{figure}
\clearpage
%205RN A DECAY
\subsection[\hspace{-0.2cm}\ensuremath{^{\textnormal{205}}}Rn \ensuremath{\alpha} decay]{ }
\vspace{-27pt}
\vspace{0.3cm}
\hypertarget{RN44}{{\bf \small \underline{\ensuremath{^{\textnormal{205}}}Rn \ensuremath{\alpha} decay}}}\\
\vspace{4pt}
\vspace{8pt}
\parbox[b][0.3cm]{17.7cm}{\addtolength{\parindent}{-0.2in}Parent: $^{205}$Rn: E=0; J$^{\pi}$=5/2\ensuremath{^{-}}; T$_{1/2}$=170 s {\it 4}; Q(\ensuremath{\alpha})=6386.5 {\it 18}; \%\ensuremath{\alpha} decay=24.6 {\it 9}

}\\
\parbox[b][0.3cm]{17.7cm}{\addtolength{\parindent}{-0.2in}\ensuremath{^{205}}Rn-\ensuremath{J^{\pi}}, T\ensuremath{_{\textnormal{1/2}}} and \%\ensuremath{\alpha} decay from \href{https://www.nndc.bnl.gov/nsr/nsrlink.jsp?2020Ko17,B}{2020Ko17}; Q(\ensuremath{\alpha}) from \href{https://www.nndc.bnl.gov/nsr/nsrlink.jsp?2021Wa16,B}{2021Wa16}.}\\
\vspace{12pt}
\underline{$^{201}$Po Levels}\\
\begin{longtable}{cccccc@{\extracolsep{\fill}}c}
\multicolumn{2}{c}{E(level)$^{{\hyperlink{PO44LEVEL0}{\dagger}}}$}&J$^{\pi}$$^{{\hyperlink{PO44LEVEL0}{\dagger}}}$&\multicolumn{2}{c}{T$_{1/2}$$^{{\hyperlink{PO44LEVEL0}{\dagger}}}$}&Comments&\\[-.2cm]
\multicolumn{2}{c}{\hrulefill}&\hrulefill&\multicolumn{2}{c}{\hrulefill}&\hrulefill&
\endfirsthead
\multicolumn{1}{r@{}}{0}&\multicolumn{1}{@{}l}{\ensuremath{^{{\hyperlink{PO44LEVEL3}{@}}}}}&\multicolumn{1}{l}{3/2\ensuremath{^{-}}}&\multicolumn{1}{r@{}}{15}&\multicolumn{1}{@{.}l}{50 min {\it 14}}&&\\
\multicolumn{1}{r@{}}{5}&\multicolumn{1}{@{.}l}{61\ensuremath{^{{\hyperlink{PO44LEVEL1}{\ddagger}}}} {\it 15}}&\multicolumn{1}{l}{5/2\ensuremath{^{-}}}&&&\parbox[t][0.3cm]{12.67816cm}{\raggedright E(level): Other: 0.3 keV \textit{25} using Q(\ensuremath{\alpha})=6386.5 \textit{18} and E(\ensuremath{\alpha}).\vspace{0.1cm}}&\\
\multicolumn{1}{r@{}}{142}&\multicolumn{1}{@{}l}{\ensuremath{^{{\hyperlink{PO44LEVEL2}{\#}}}} {\it 3}}&\multicolumn{1}{l}{(1/2\ensuremath{^{-}})}&&&\parbox[t][0.3cm]{12.67816cm}{\raggedright E(level): From \ensuremath{\alpha}-\ensuremath{\gamma} coincidences in \href{https://www.nndc.bnl.gov/nsr/nsrlink.jsp?1971Jo19,B}{1971Jo19}.\vspace{0.1cm}}&\\
\end{longtable}
\parbox[b][0.3cm]{17.7cm}{\makebox[1ex]{\ensuremath{^{\hypertarget{PO44LEVEL0}{\dagger}}}} From Adopted Levels, unless otherwise stated.}\\
\parbox[b][0.3cm]{17.7cm}{\makebox[1ex]{\ensuremath{^{\hypertarget{PO44LEVEL1}{\ddagger}}}} Configuration=\ensuremath{\nu} f\ensuremath{_{\textnormal{5/2}}^{\textnormal{$-$1}}}.}\\
\parbox[b][0.3cm]{17.7cm}{\makebox[1ex]{\ensuremath{^{\hypertarget{PO44LEVEL2}{\#}}}} Configuration=\ensuremath{\nu} p\ensuremath{_{\textnormal{1/2}}^{\textnormal{$-$1}}}. The assignment is tentative.}\\
\parbox[b][0.3cm]{17.7cm}{\makebox[1ex]{\ensuremath{^{\hypertarget{PO44LEVEL3}{@}}}} Configuration=\ensuremath{\nu} p\ensuremath{_{\textnormal{3/2}}^{\textnormal{$-$1}}}.}\\
\vspace{0.5cm}

\underline{\ensuremath{\alpha} radiations}\\

\parbox[b][0.3cm]{17.7cm}{\makebox[1ex]{\ensuremath{^{\hypertarget{PO44DECAY0}{\dagger}}}} Using r\ensuremath{_{\textnormal{0}}}(\ensuremath{^{\textnormal{201}}}Po)=1.497 \textit{2} from \href{https://www.nndc.bnl.gov/nsr/nsrlink.jsp?2020Si16,B}{2020Si16}.}\\
\parbox[b][0.3cm]{17.7cm}{\makebox[1ex]{\ensuremath{^{\hypertarget{PO44DECAY1}{\ddagger}}}} For absolute intensity per 100 decays, multiply by 0.246 \textit{9}.}\\
\vspace{0.5cm}
\clearpage
%194PT(12C,5NG)
\subsection[\hspace{-0.2cm}\ensuremath{^{\textnormal{194}}}Pt(\ensuremath{^{\textnormal{12}}}C,5n\ensuremath{\gamma})]{ }
\vspace{-27pt}
\vspace{0.3cm}
\hypertarget{PO45}{{\bf \small \underline{\ensuremath{^{\textnormal{194}}}Pt(\ensuremath{^{\textnormal{12}}}C,5n\ensuremath{\gamma})\hspace{0.2in}\href{https://www.nndc.bnl.gov/nsr/nsrlink.jsp?1985We05,B}{1985We05}}}}\\
\vspace{4pt}
\vspace{8pt}
\parbox[b][0.3cm]{17.7cm}{\addtolength{\parindent}{-0.2in}\href{https://www.nndc.bnl.gov/nsr/nsrlink.jsp?1985We05,B}{1985We05}: Produced in \ensuremath{^{\textnormal{194}}}Pt(\ensuremath{^{\textnormal{12}}}C,5n) (E(\ensuremath{^{\textnormal{12}}}C)=102 and 106 MeV) and \ensuremath{^{\textnormal{195}}}Pt(\ensuremath{^{\textnormal{12}}}C,6n) (E(\ensuremath{^{\textnormal{12}}}C)=100 MeV) reactions; Detectors:}\\
\parbox[b][0.3cm]{17.7cm}{two n-type HPGE and two liquid scin neutron detectors; Measured: E\ensuremath{\gamma}, I\ensuremath{\gamma}, \ensuremath{\gamma} singles, \ensuremath{\gamma}-\ensuremath{\gamma} coin, n-\ensuremath{\gamma} and n-\ensuremath{\gamma}-\ensuremath{\gamma} coin, \ensuremath{\gamma}(\ensuremath{\theta});}\\
\parbox[b][0.3cm]{17.7cm}{Deduced: level scheme, \ensuremath{J^{\pi}}.}\\
\parbox[b][0.3cm]{17.7cm}{\addtolength{\parindent}{-0.2in}Others: \href{https://www.nndc.bnl.gov/nsr/nsrlink.jsp?2004Ro09,B}{2004Ro09}, but no \ensuremath{^{\textnormal{201}}}Po levels and transitions were reported.}\\
\vspace{12pt}
\underline{$^{201}$Po Levels}\\

\parbox[b][0.3cm]{17.7cm}{\makebox[1ex]{\ensuremath{^{\hypertarget{PO45LEVEL0}{\dagger}}}} From a least-squares fit to E\ensuremath{\gamma} and by assuming \ensuremath{\Delta}E\ensuremath{\gamma}=0.5 keV.}\\
\parbox[b][0.3cm]{17.7cm}{\makebox[1ex]{\ensuremath{^{\hypertarget{PO45LEVEL1}{\ddagger}}}} From Adopted Levels.}\\
\parbox[b][0.3cm]{17.7cm}{\makebox[1ex]{\ensuremath{^{\hypertarget{PO45LEVEL2}{\#}}}} From \href{https://www.nndc.bnl.gov/nsr/nsrlink.jsp?1985We05,B}{1985We05}, unless otherwise stated.}\\
\parbox[b][0.3cm]{17.7cm}{\makebox[1ex]{\ensuremath{^{\hypertarget{PO45LEVEL3}{@}}}} From Adopted Levels.}\\
\parbox[b][0.3cm]{17.7cm}{\makebox[1ex]{\ensuremath{^{\hypertarget{PO45LEVEL4}{\&}}}} Configuration=\ensuremath{\nu} f\ensuremath{_{\textnormal{5/2}}^{\textnormal{$-$1}}}.}\\
\parbox[b][0.3cm]{17.7cm}{\makebox[1ex]{\ensuremath{^{\hypertarget{PO45LEVEL5}{a}}}} Configuration=\ensuremath{\nu} p\ensuremath{_{\textnormal{3/2}}^{\textnormal{$-$1}}}.}\\
\parbox[b][0.3cm]{17.7cm}{\makebox[1ex]{\ensuremath{^{\hypertarget{PO45LEVEL6}{b}}}} Configuration=\ensuremath{\nu} i\ensuremath{_{\textnormal{13/2}}^{\textnormal{$-$1}}}.}\\
\parbox[b][0.3cm]{17.7cm}{\makebox[1ex]{\ensuremath{^{\hypertarget{PO45LEVEL7}{c}}}} Configuration=\ensuremath{\nu} (i\ensuremath{_{\textnormal{13/2}}^{\textnormal{$-$1}}})\ensuremath{\otimes}2\ensuremath{^{\textnormal{+}}}.}\\
\parbox[b][0.3cm]{17.7cm}{\makebox[1ex]{\ensuremath{^{\hypertarget{PO45LEVEL8}{d}}}} Configuration=\ensuremath{\nu} (i\ensuremath{_{\textnormal{13/2}}^{\textnormal{$-$1}}})\ensuremath{\otimes}4\ensuremath{^{\textnormal{+}}}.}\\
\parbox[b][0.3cm]{17.7cm}{\makebox[1ex]{\ensuremath{^{\hypertarget{PO45LEVEL9}{e}}}} Possibly a mixture between configuration=\ensuremath{\nu} (i\ensuremath{_{\textnormal{13/2}}^{\textnormal{$-$1}}})\ensuremath{\otimes}6\ensuremath{^{\textnormal{+}}} and configuration=\ensuremath{\nu} (i\ensuremath{_{\textnormal{13/2}}^{\textnormal{$-$1}}}) \ensuremath{\pi} (h\ensuremath{_{\textnormal{9/2}}^{\textnormal{+2}}})\ensuremath{_{\textnormal{8+}}}.}\\
\parbox[b][0.3cm]{17.7cm}{\makebox[1ex]{\ensuremath{^{\hypertarget{PO45LEVEL10}{f}}}} The \href{https://www.nndc.bnl.gov/nsr/nsrlink.jsp?1985We05,B}{1985We05} authors stated that the ordering of the 318.7\ensuremath{\gamma}, 556.6\ensuremath{\gamma} and 613.6\ensuremath{\gamma}, and hence, the placement of corresponding}\\
\parbox[b][0.3cm]{17.7cm}{{\ }{\ }level energies, is based on systematics and their relative population.}\\
\vspace{0.5cm}
\underline{$\gamma$($^{201}$Po)}\\

\begin{textblock}{29}(0,27.3)
Continued on next page (footnotes at end of table)
\end{textblock}
\clearpage
%
\parbox[b][0.3cm]{17.7cm}{\makebox[1ex]{\ensuremath{^{\hypertarget{PO45GAMMA0}{\dagger}}}} From \ensuremath{^{\textnormal{194}}}Pt(\ensuremath{^{\textnormal{12}}}C,5n) reaction at 106 MeV in \href{https://www.nndc.bnl.gov/nsr/nsrlink.jsp?1985We05,B}{1985We05}, unless otherwise stated.}\\
\parbox[b][0.3cm]{17.7cm}{\makebox[1ex]{\ensuremath{^{\hypertarget{PO45GAMMA1}{\ddagger}}}} From adopted gammas.}\\
\parbox[b][0.3cm]{17.7cm}{\makebox[1ex]{\ensuremath{^{\hypertarget{PO45GAMMA2}{\#}}}} From \ensuremath{\gamma}(\ensuremath{\theta}) in \href{https://www.nndc.bnl.gov/nsr/nsrlink.jsp?1985We05,B}{1985We05}.}\\
\parbox[b][0.3cm]{17.7cm}{\makebox[1ex]{\ensuremath{^{\hypertarget{PO45GAMMA3}{@}}}} Total theoretical internal conversion coefficients, calculated using the BrIcc code (\href{https://www.nndc.bnl.gov/nsr/nsrlink.jsp?2008Ki07,B}{2008Ki07}) with Frozen orbital approximation}\\
\parbox[b][0.3cm]{17.7cm}{{\ }{\ }based on \ensuremath{\gamma}-ray energies, assigned multipolarities, and mixing ratios, unless otherwise specified.}\\
\parbox[b][0.3cm]{17.7cm}{\makebox[1ex]{\ensuremath{^{\hypertarget{PO45GAMMA4}{\&}}}} Multiply placed with undivided intensity.}\\
\vspace{0.5cm}
\clearpage
\begin{figure}[h]
\begin{center}
\includegraphics{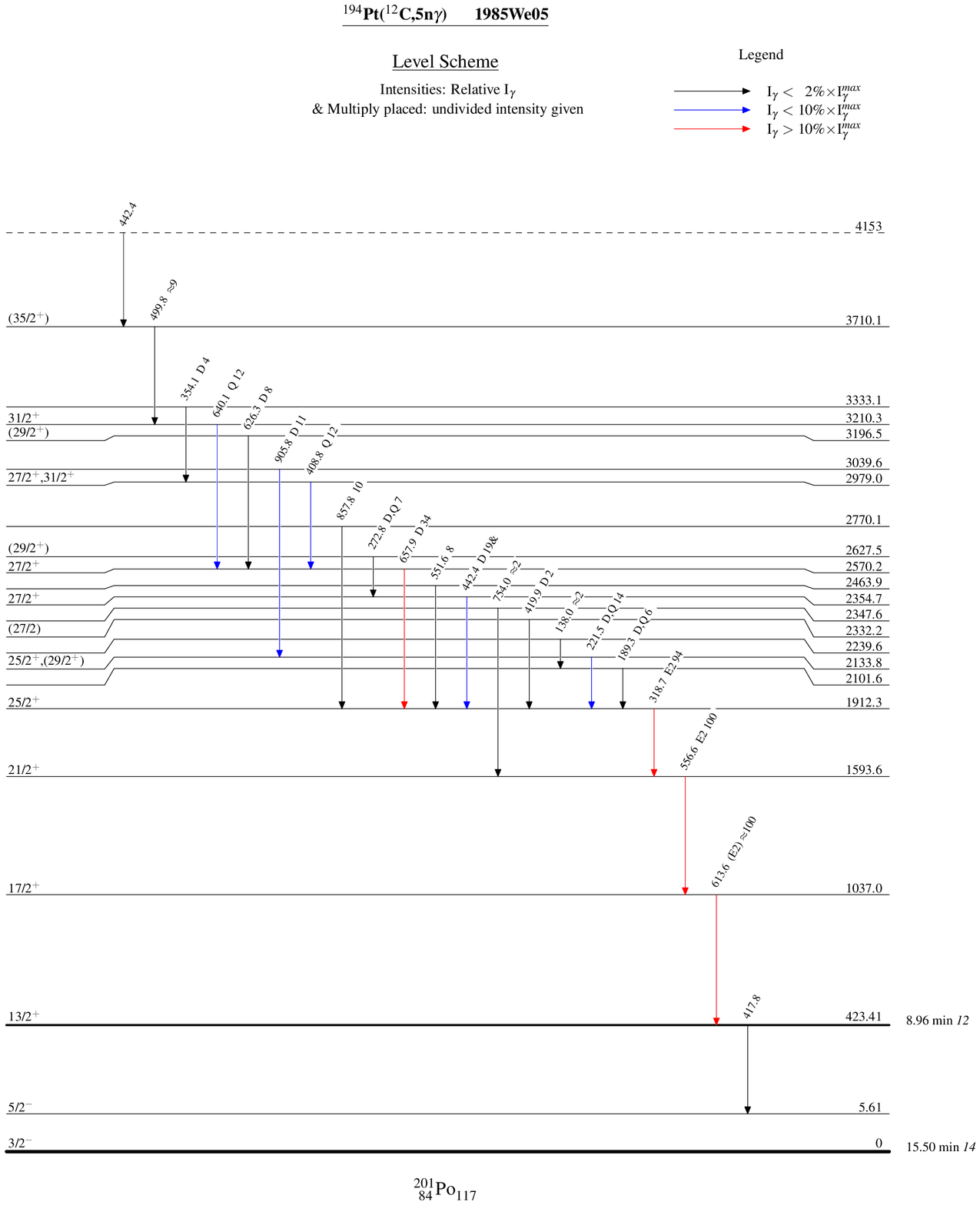}\\
\end{center}
\end{figure}
\clearpage
%ADOPTED LEVELS, GAMMAS
\section[\ensuremath{^{201}_{\ 85}}At\ensuremath{_{116}^{~}}]{ }
\vspace{-30pt}
\setcounter{chappage}{1}
\subsection[\hspace{-0.2cm}Adopted Levels, Gammas]{ }
\vspace{-20pt}
\vspace{0.3cm}
\hypertarget{AT46}{{\bf \small \underline{Adopted \hyperlink{201AT_LEVEL}{Levels}, \hyperlink{201AT_GAMMA}{Gammas}}}}\\
\vspace{4pt}
\vspace{8pt}
\parbox[b][0.3cm]{17.7cm}{\addtolength{\parindent}{-0.2in}Q(\ensuremath{\beta^-})=$-$6682 {\it 13}; S(n)=9873 {\it 26}; S(p)=1137 {\it 11}; Q(\ensuremath{\alpha})=6472.8 {\it 16}\hspace{0.2in}\href{https://www.nndc.bnl.gov/nsr/nsrlink.jsp?2021Wa16,B}{2021Wa16}}\\

\vspace{12pt}
\hypertarget{201AT_LEVEL}{\underline{$^{201}$At Levels}}\\

\begin{textblock}{29}(0,27.3)
Continued on next page (footnotes at end of table)
\end{textblock}
\clearpage
%
\begin{textblock}{29}(0,27.3)
Continued on next page (footnotes at end of table)
\end{textblock}
\clearpage
%
\parbox[b][0.3cm]{17.7cm}{\makebox[1ex]{\ensuremath{^{\hypertarget{AT46LEVEL0}{\dagger}}}} From a least-squares fit to E\ensuremath{\gamma}.}\\
\parbox[b][0.3cm]{17.7cm}{\makebox[1ex]{\ensuremath{^{\hypertarget{AT46LEVEL1}{\ddagger}}}} From deduced transition multipolarities and the observed decay pattern in \ensuremath{^{\textnormal{165}}}Ho(\ensuremath{^{\textnormal{40}}}Ar,4n\ensuremath{\gamma}) and \ensuremath{^{\textnormal{192}}}Pt(\ensuremath{^{\textnormal{14}}}N,5n\ensuremath{\gamma}), systematics in}\\
\parbox[b][0.3cm]{17.7cm}{{\ }{\ }the region and shell-model predictions, unless otherwise stated.}\\
\parbox[b][0.3cm]{17.7cm}{\makebox[1ex]{\ensuremath{^{\hypertarget{AT46LEVEL2}{\#}}}} Seq.(B): Based on \ensuremath{\pi} (d\ensuremath{_{\textnormal{3/2}}^{\textnormal{$-$1}}})\ensuremath{\otimes}\ensuremath{^{\textnormal{202}}}Rn core states (\ensuremath{J^{\pi}}=2\ensuremath{^{+}},4\ensuremath{^{+}},6\ensuremath{^{+}}).}\\
\parbox[b][0.3cm]{17.7cm}{\makebox[1ex]{\ensuremath{^{\hypertarget{AT46LEVEL3}{@}}}} Seq.(C): Based on \ensuremath{\pi} (d\ensuremath{_{\textnormal{5/2}}^{\textnormal{$-$1}}})\ensuremath{\otimes}\ensuremath{^{\textnormal{202}}}Rn core states (\ensuremath{J^{\pi}}=2\ensuremath{^{+}},4\ensuremath{^{+}}).}\\
\parbox[b][0.3cm]{17.7cm}{\makebox[1ex]{\ensuremath{^{\hypertarget{AT46LEVEL4}{\&}}}} Seq.(D): Based on \ensuremath{\pi} (h\ensuremath{_{\textnormal{9/2}}^{\textnormal{+1}}})\ensuremath{\otimes}\ensuremath{^{\textnormal{202}}}Rn core states (\ensuremath{J^{\pi}}=2\ensuremath{^{+}},4\ensuremath{^{+}},6\ensuremath{^{+}}, 8\ensuremath{^{+}}).}\\
\parbox[b][0.3cm]{17.7cm}{\makebox[1ex]{\ensuremath{^{\hypertarget{AT46LEVEL5}{a}}}} Band(A): Magnetic-dipole, shears band. Configuration=\ensuremath{\pi} (i\ensuremath{_{\textnormal{13/2}}^{\textnormal{+1}}})\ensuremath{\otimes}\ensuremath{\nu} (f\ensuremath{_{\textnormal{5/2}}^{\textnormal{$-$1}}},i\ensuremath{_{\textnormal{13/2}}^{\textnormal{$-$1}}})\ensuremath{_{\textnormal{9$-$}}} for the lower cascade and}\\
\parbox[b][0.3cm]{17.7cm}{{\ }{\ }Configuration= \ensuremath{\pi} (h\ensuremath{_{\textnormal{9/2}}^{\textnormal{+2}}},i\ensuremath{_{\textnormal{13/2}}^{\textnormal{+1}}})\ensuremath{\otimes}\ensuremath{\nu} (f\ensuremath{_{\textnormal{5/2}}^{\textnormal{$-$1}}},i\ensuremath{_{\textnormal{13/2}}^{\textnormal{$-$1}}})\ensuremath{_{\textnormal{5$-$}}} above the band crossing.}\\
\vspace{0.5cm}
\hypertarget{201AT_GAMMA}{\underline{$\gamma$($^{201}$At)}}\\

\begin{textblock}{29}(0,27.3)
Continued on next page (footnotes at end of table)
\end{textblock}
\clearpage
%
\begin{textblock}{29}(0,27.3)
Continued on next page (footnotes at end of table)
\end{textblock}
\clearpage
%
\parbox[b][0.3cm]{17.7cm}{\makebox[1ex]{\ensuremath{^{\hypertarget{AT46GAMMA0}{\dagger}}}} From \ensuremath{^{\textnormal{165}}}Ho(\ensuremath{^{\textnormal{40}}}Ar,4n\ensuremath{\gamma}), unless otherwise stated.}\\
\parbox[b][0.3cm]{17.7cm}{\makebox[1ex]{\ensuremath{^{\hypertarget{AT46GAMMA1}{\ddagger}}}} Determined by the evaluator from I(\ensuremath{\gamma}+ce) in \ensuremath{^{\textnormal{165}}}Ho(\ensuremath{^{\textnormal{40}}}Ar,4n\ensuremath{\gamma}) and \ensuremath{\alpha}.}\\
\parbox[b][0.3cm]{17.7cm}{\makebox[1ex]{\ensuremath{^{\hypertarget{AT46GAMMA2}{\#}}}} Total theoretical internal conversion coefficients, calculated using the BrIcc code (\href{https://www.nndc.bnl.gov/nsr/nsrlink.jsp?2008Ki07,B}{2008Ki07}) with Frozen orbital approximation}\\
\parbox[b][0.3cm]{17.7cm}{{\ }{\ }based on \ensuremath{\gamma}-ray energies, assigned multipolarities, and mixing ratios, unless otherwise specified.}\\
\parbox[b][0.3cm]{17.7cm}{\makebox[1ex]{\ensuremath{^{\hypertarget{AT46GAMMA3}{@}}}} Placement of transition in the level scheme is uncertain.}\\
\vspace{0.5cm}
\clearpage
\begin{figure}[h]
\begin{center}
\includegraphics{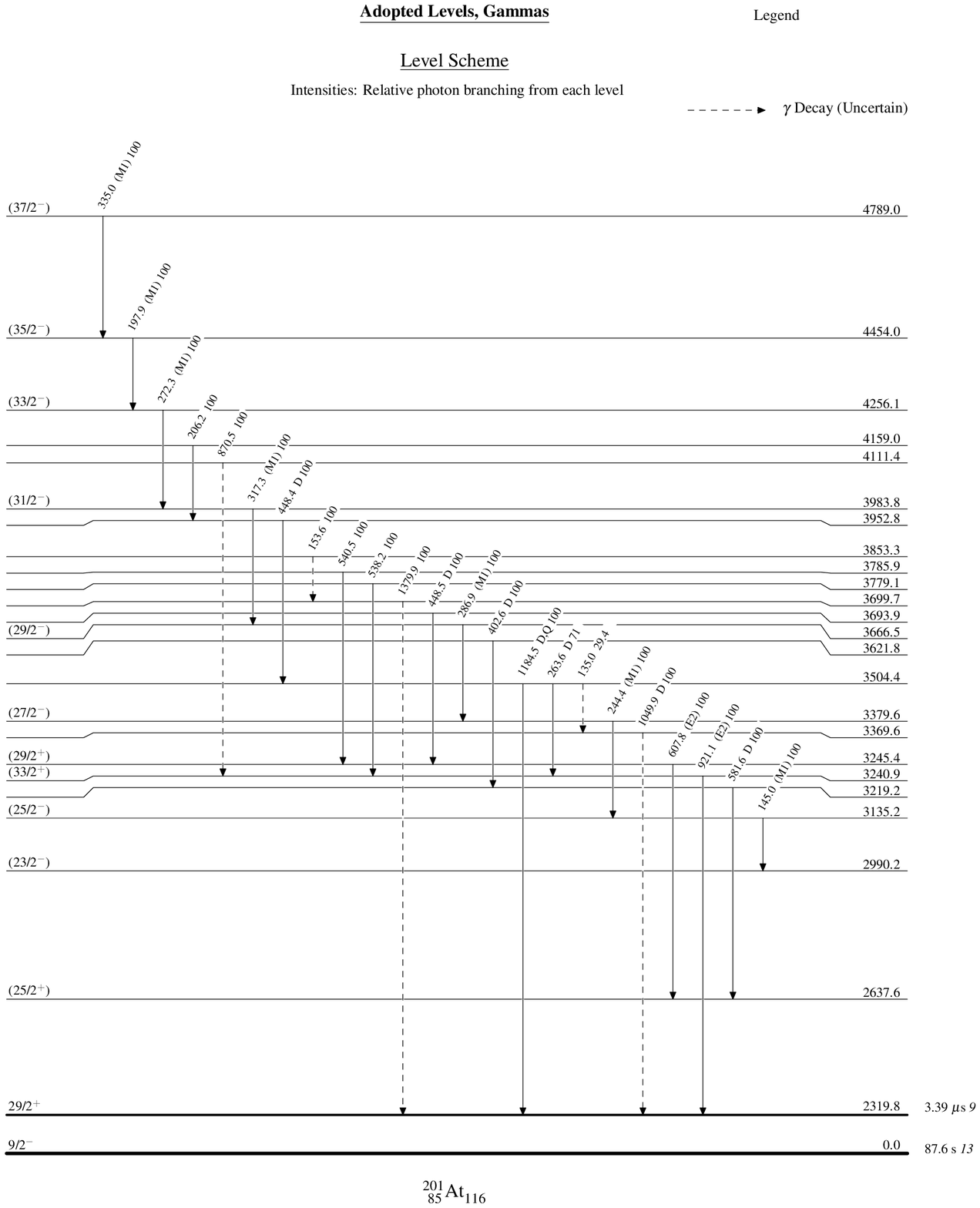}\\
\end{center}
\end{figure}
\clearpage
\begin{figure}[h]
\begin{center}
\includegraphics{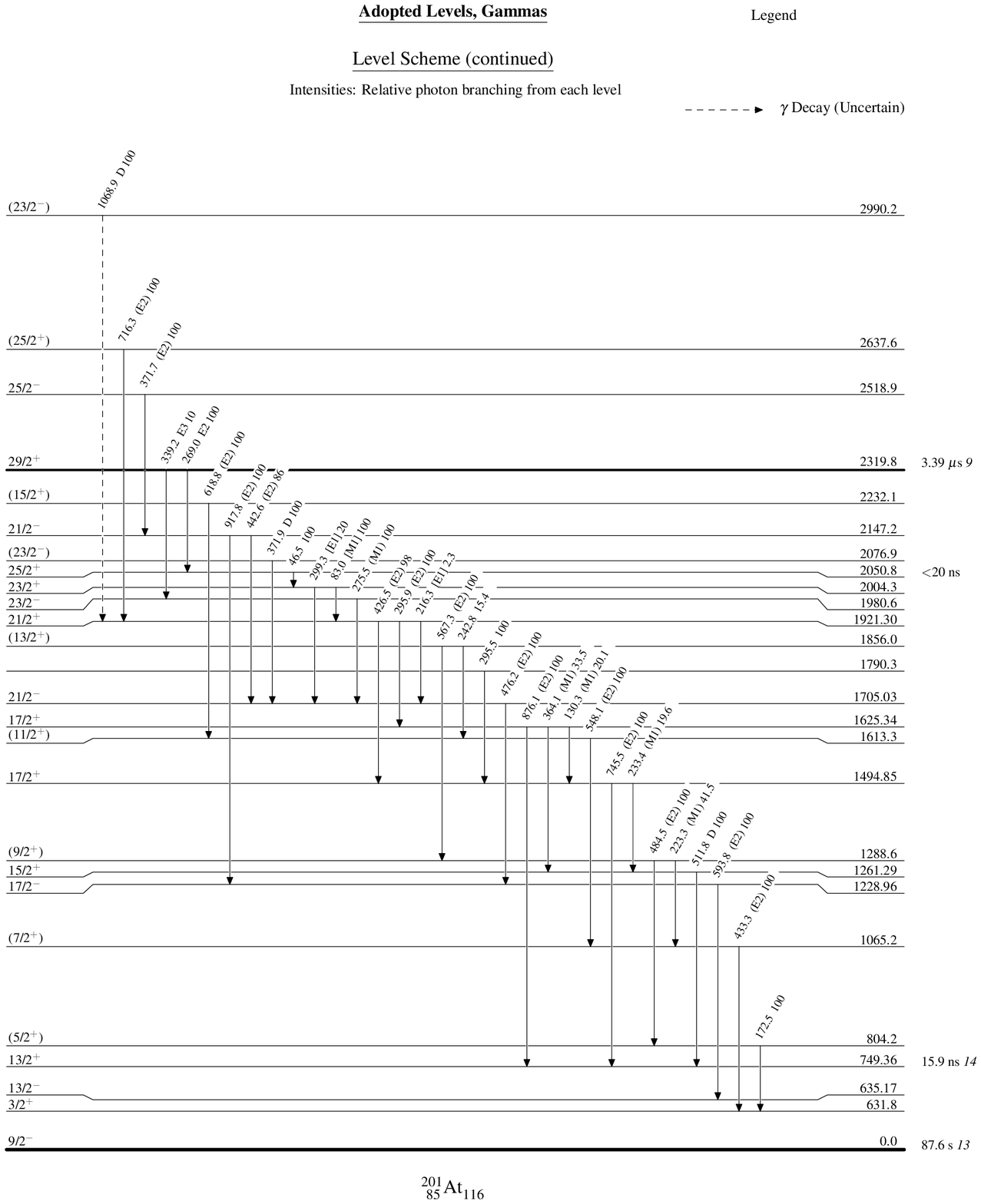}\\
\end{center}
\end{figure}
\clearpage
\begin{figure}[h]
\begin{center}
\includegraphics{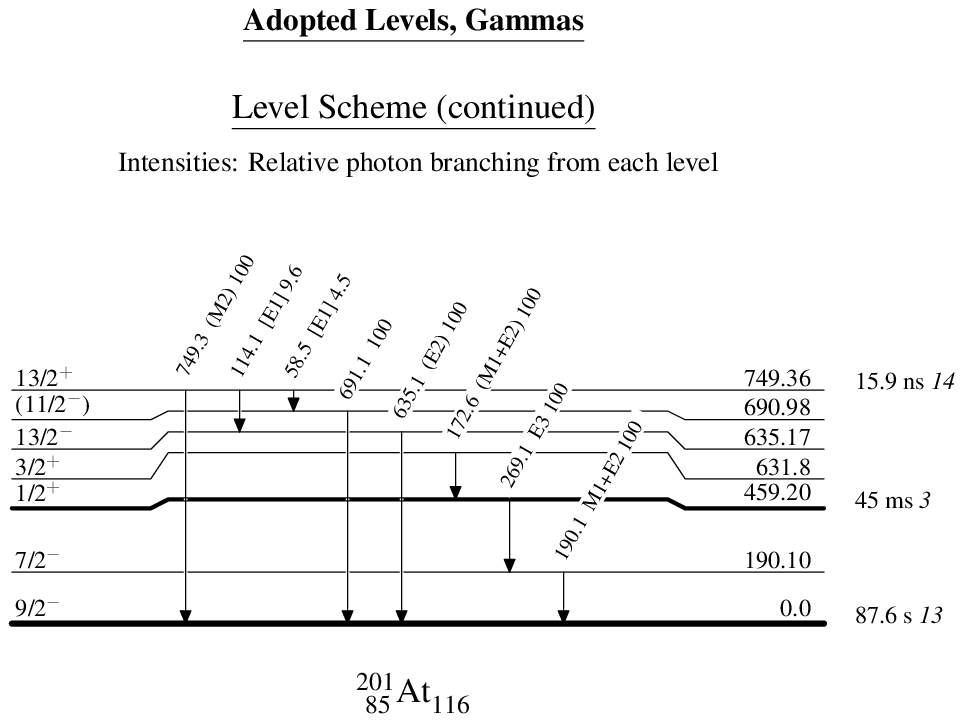}\\
\end{center}
\end{figure}
\clearpage
\clearpage
\begin{figure}[h]
\begin{center}
\includegraphics{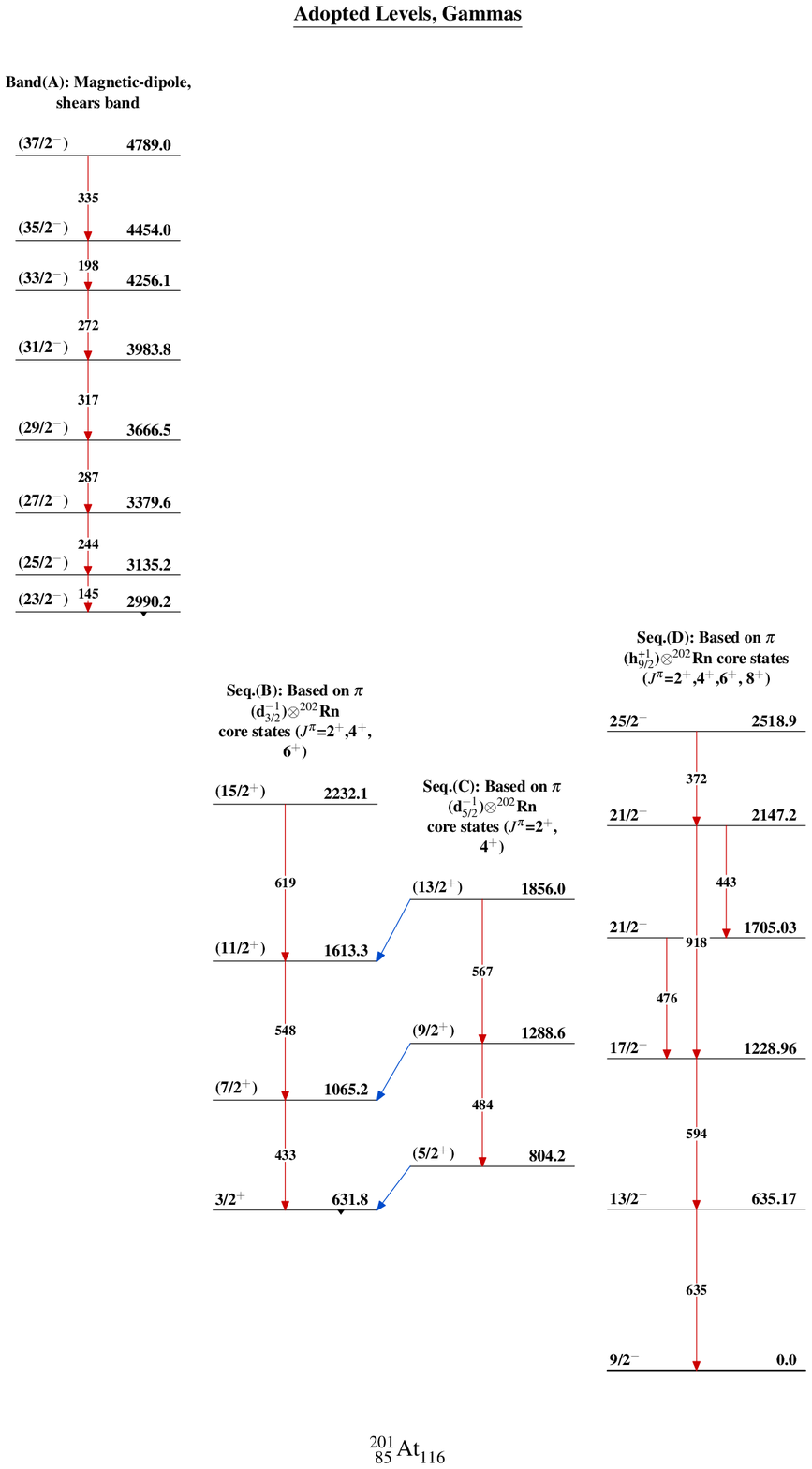}\\
\end{center}
\end{figure}
\clearpage
%205FR A DECAY
\subsection[\hspace{-0.2cm}\ensuremath{^{\textnormal{205}}}Fr \ensuremath{\alpha} decay]{ }
\vspace{-27pt}
\vspace{0.3cm}
\hypertarget{FR47}{{\bf \small \underline{\ensuremath{^{\textnormal{205}}}Fr \ensuremath{\alpha} decay\hspace{0.2in}\href{https://www.nndc.bnl.gov/nsr/nsrlink.jsp?1974Ho27,B}{1974Ho27},\href{https://www.nndc.bnl.gov/nsr/nsrlink.jsp?1967Va20,B}{1967Va20}}}}\\
\vspace{4pt}
\vspace{8pt}
\parbox[b][0.3cm]{17.7cm}{\addtolength{\parindent}{-0.2in}Parent: $^{205}$Fr: E=0; J$^{\pi}$=9/2\ensuremath{^{-}}; T$_{1/2}$=3.90 s {\it 7}; Q(\ensuremath{\alpha})=7054.7 {\it 24}; \%\ensuremath{\alpha} decay=98.5 {\it 4}

}\\
\parbox[b][0.3cm]{17.7cm}{\addtolength{\parindent}{-0.2in}\ensuremath{^{205}}Fr-J$^{\pi}$,T$_{1/2}$: From \href{https://www.nndc.bnl.gov/nsr/nsrlink.jsp?2020Ko17,B}{2020Ko17}.}\\
\parbox[b][0.3cm]{17.7cm}{\addtolength{\parindent}{-0.2in}\ensuremath{^{205}}Fr-Q(\ensuremath{\alpha}): From \href{https://www.nndc.bnl.gov/nsr/nsrlink.jsp?2021Wa16,B}{2021Wa16}.}\\
\parbox[b][0.3cm]{17.7cm}{\addtolength{\parindent}{-0.2in}\ensuremath{^{205}}Fr-\%\ensuremath{\alpha} decay: From \href{https://www.nndc.bnl.gov/nsr/nsrlink.jsp?2020Ko17,B}{2020Ko17}.}\\
\vspace{12pt}
\underline{$^{201}$At Levels}\\

\parbox[b][0.3cm]{17.7cm}{\makebox[1ex]{\ensuremath{^{\hypertarget{AT47LEVEL0}{\dagger}}}} From Adopted Levels.}\\
\vspace{0.5cm}

\underline{\ensuremath{\alpha} radiations}\\
%
\parbox[b][0.3cm]{17.7cm}{\makebox[1ex]{\ensuremath{^{\hypertarget{AT47DECAY0}{\dagger}}}} Using r\ensuremath{_{\textnormal{0}}}(\ensuremath{^{\textnormal{201}}}At)=1.516 \textit{3} from \href{https://www.nndc.bnl.gov/nsr/nsrlink.jsp?2020Si16,B}{2020Si16}.}\\
\parbox[b][0.3cm]{17.7cm}{\makebox[1ex]{\ensuremath{^{\hypertarget{AT47DECAY1}{\ddagger}}}} For absolute intensity per 100 decays, multiply by 0.985 \textit{4}.}\\
\vspace{0.5cm}
\clearpage
%192PT(14N,5NG)
\subsection[\hspace{-0.2cm}\ensuremath{^{\textnormal{192}}}Pt(\ensuremath{^{\textnormal{14}}}N,5n\ensuremath{\gamma})]{ }
\vspace{-27pt}
\vspace{0.3cm}
\hypertarget{AT48}{{\bf \small \underline{\ensuremath{^{\textnormal{192}}}Pt(\ensuremath{^{\textnormal{14}}}N,5n\ensuremath{\gamma})\hspace{0.2in}\href{https://www.nndc.bnl.gov/nsr/nsrlink.jsp?1983Dy02,B}{1983Dy02}}}}\\
\vspace{4pt}
\vspace{8pt}
\parbox[b][0.3cm]{17.7cm}{\addtolength{\parindent}{-0.2in}\ensuremath{^{\textnormal{192}}}Pt(\ensuremath{^{\textnormal{14}}}N,5n\ensuremath{\gamma}), E(\ensuremath{^{\textnormal{14}}}N)=85-100 MeV; Target: 3 mg/cm\ensuremath{^{\textnormal{2}}} thick, enriched to 57 \% in \ensuremath{^{\textnormal{192}}}Pt; Detectors: Ge(Li) with a typical}\\
\parbox[b][0.3cm]{17.7cm}{energy resolution (FWHM) of 2 keV at 1.33 MeV; Measured: excitation functions, \ensuremath{\gamma}(t), \ensuremath{\gamma}(\ensuremath{\theta}), \ensuremath{\gamma}\ensuremath{\gamma} coin (two Ge(Li) detectors);}\\
\parbox[b][0.3cm]{17.7cm}{Deduced: level scheme, \ensuremath{J^{\pi}}, T\ensuremath{_{\textnormal{1/2}}}.}\\
\vspace{12pt}
\underline{$^{201}$At Levels}\\

\parbox[b][0.3cm]{17.7cm}{\makebox[1ex]{\ensuremath{^{\hypertarget{AT48LEVEL0}{\dagger}}}} From a least-squares fit to E\ensuremath{\gamma}.}\\
\parbox[b][0.3cm]{17.7cm}{\makebox[1ex]{\ensuremath{^{\hypertarget{AT48LEVEL1}{\ddagger}}}} From deduced transition multipolarities, unless otherwise stated.}\\
\parbox[b][0.3cm]{17.7cm}{\makebox[1ex]{\ensuremath{^{\hypertarget{AT48LEVEL2}{\#}}}} Configuration=\ensuremath{\pi} h\ensuremath{_{\textnormal{9/2}}^{\textnormal{+1}}}.}\\
\parbox[b][0.3cm]{17.7cm}{\makebox[1ex]{\ensuremath{^{\hypertarget{AT48LEVEL3}{@}}}} Configuration=\ensuremath{\pi} (h\ensuremath{_{\textnormal{9/2}}^{\textnormal{+1}}})\ensuremath{\otimes}2\ensuremath{^{\textnormal{+}}}.}\\
\parbox[b][0.3cm]{17.7cm}{\makebox[1ex]{\ensuremath{^{\hypertarget{AT48LEVEL4}{\&}}}} Configuration=\ensuremath{\pi} i\ensuremath{_{\textnormal{13/2}}^{\textnormal{+1}}}.}\\
\parbox[b][0.3cm]{17.7cm}{\makebox[1ex]{\ensuremath{^{\hypertarget{AT48LEVEL5}{a}}}} Configuration=\ensuremath{\pi} (h\ensuremath{_{\textnormal{9/2}}^{\textnormal{+1}}})\ensuremath{\otimes}4\ensuremath{^{\textnormal{+}}}.}\\
\parbox[b][0.3cm]{17.7cm}{\makebox[1ex]{\ensuremath{^{\hypertarget{AT48LEVEL6}{b}}}} Configuration=\ensuremath{\pi} (i\ensuremath{_{\textnormal{13/2}}^{\textnormal{+1}}})\ensuremath{\otimes}2\ensuremath{^{\textnormal{+}}}.}\\
\parbox[b][0.3cm]{17.7cm}{\makebox[1ex]{\ensuremath{^{\hypertarget{AT48LEVEL7}{c}}}} Configuration=\ensuremath{\pi} (h\ensuremath{_{\textnormal{9/2}}^{\textnormal{+3}}})\ensuremath{_{\textnormal{21/2$-$}}}.}\\
\parbox[b][0.3cm]{17.7cm}{\makebox[1ex]{\ensuremath{^{\hypertarget{AT48LEVEL8}{d}}}} Configuration=\ensuremath{\pi} (h\ensuremath{_{\textnormal{9/2}}^{\textnormal{+2}}})\ensuremath{_{\textnormal{8$-$}}},f\ensuremath{_{\textnormal{7/2}}^{\textnormal{+1}}}).}\\
\vspace{0.5cm}
\underline{$\gamma$($^{201}$At)}\\

\parbox[b][0.3cm]{17.7cm}{\makebox[1ex]{\ensuremath{^{\hypertarget{AT48GAMMA0}{\dagger}}}} From \href{https://www.nndc.bnl.gov/nsr/nsrlink.jsp?1983Dy02,B}{1983Dy02}.}\\
\parbox[b][0.3cm]{17.7cm}{\makebox[1ex]{\ensuremath{^{\hypertarget{AT48GAMMA1}{\ddagger}}}} Estimated by the authors from the \ensuremath{\gamma}\ensuremath{\gamma} coin data.}\\
\parbox[b][0.3cm]{17.7cm}{\makebox[1ex]{\ensuremath{^{\hypertarget{AT48GAMMA2}{\#}}}} Total theoretical internal conversion coefficients, calculated using the BrIcc code (\href{https://www.nndc.bnl.gov/nsr/nsrlink.jsp?2008Ki07,B}{2008Ki07}) with Frozen orbital approximation}\\
\parbox[b][0.3cm]{17.7cm}{{\ }{\ }based on \ensuremath{\gamma}-ray energies, assigned multipolarities, and mixing ratios, unless otherwise specified.}\\
\vspace{0.5cm}
\clearpage
\begin{figure}[h]
\begin{center}
\includegraphics{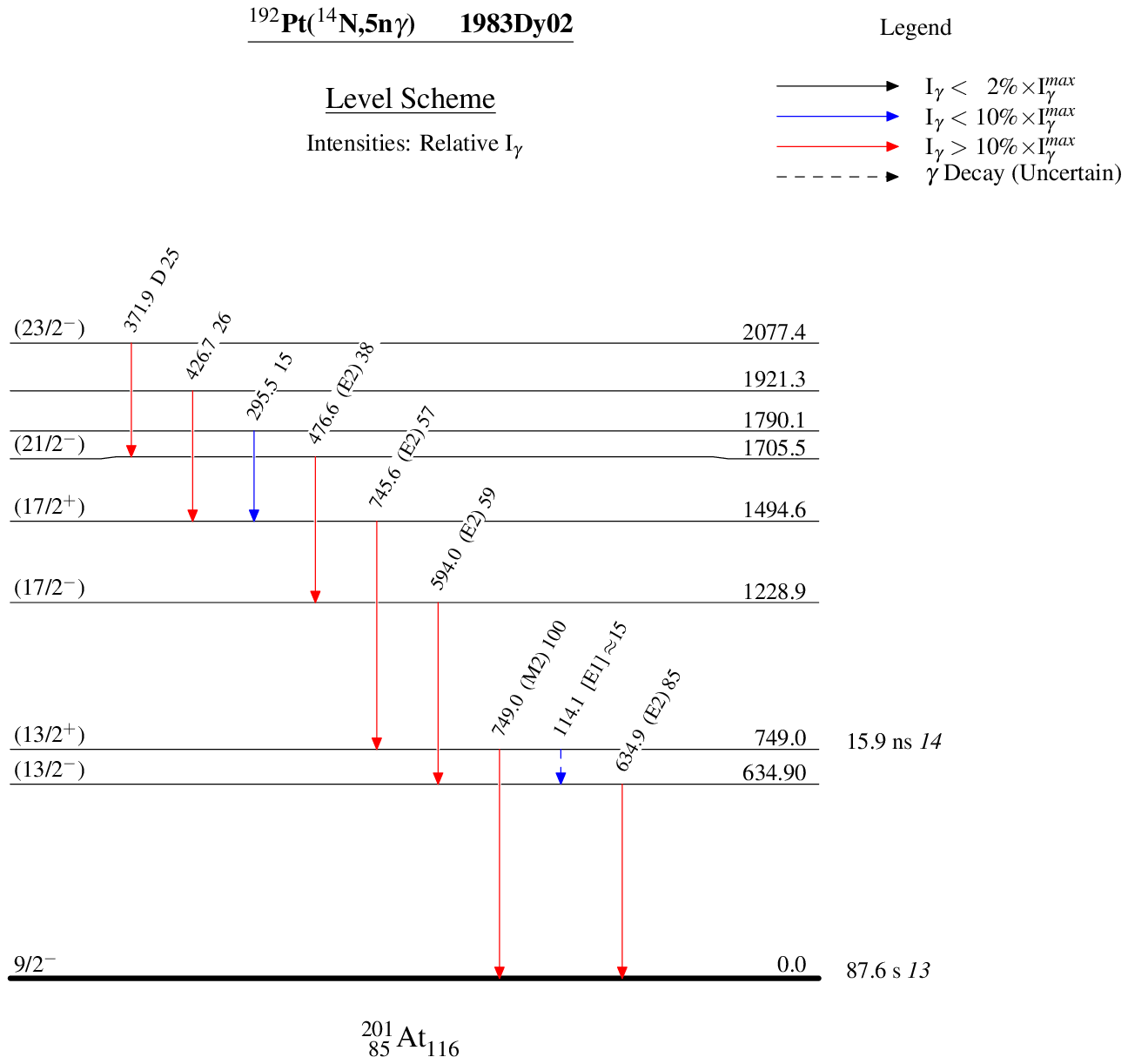}\\
\end{center}
\end{figure}
\clearpage
%165HO(40AR,4NG)
\subsection[\hspace{-0.2cm}\ensuremath{^{\textnormal{165}}}Ho(\ensuremath{^{\textnormal{40}}}Ar,4n\ensuremath{\gamma})]{ }
\vspace{-27pt}
\vspace{0.3cm}
\hypertarget{AT49}{{\bf \small \underline{\ensuremath{^{\textnormal{165}}}Ho(\ensuremath{^{\textnormal{40}}}Ar,4n\ensuremath{\gamma})\hspace{0.2in}\href{https://www.nndc.bnl.gov/nsr/nsrlink.jsp?2014Au03,B}{2014Au03},\href{https://www.nndc.bnl.gov/nsr/nsrlink.jsp?2015Au01,B}{2015Au01}}}}\\
\vspace{4pt}
\vspace{8pt}
\parbox[b][0.3cm]{17.7cm}{\addtolength{\parindent}{-0.2in}\href{https://www.nndc.bnl.gov/nsr/nsrlink.jsp?2014Au03,B}{2014Au03},\href{https://www.nndc.bnl.gov/nsr/nsrlink.jsp?2015Au01,B}{2015Au01}: \ensuremath{^{\textnormal{40}}}Ar\ensuremath{^{\textnormal{8+}}}, E=205{\textminus}MeV, I=11 pnA beam from the K-130 cyclotron at the University of Jyvaskyla Accelerator}\\
\parbox[b][0.3cm]{17.7cm}{Lab. Target: self-supporting 350{\textminus}\ensuremath{\mu}g/cm\ensuremath{^{\textnormal{2}}}\hphantom{a}thick \ensuremath{^{\textnormal{165}}}Ho. Detectors: The JUROGAM2 array consisting of Compton-suppressed 24}\\
\parbox[b][0.3cm]{17.7cm}{Clover and 15 Phase-1 and GASP type HPGe detectors. The fusion-evaporation residues (recoils) were separated from the primary}\\
\parbox[b][0.3cm]{17.7cm}{beam and other unwanted particles using the gas-filled recoil separator RITU and studied at the focal plane using the GREAT}\\
\parbox[b][0.3cm]{17.7cm}{spectrometer. The recoils were implanted onto 300 \ensuremath{\mu}m DSSD surrounded by 28 Si PIN diodes to measure the \ensuremath{\alpha} particle and ce}\\
\parbox[b][0.3cm]{17.7cm}{energies. \ensuremath{\gamma} rays at the focal plane were measured using three Clover and one planar type HPGe detector. Measured: E\ensuremath{\gamma}, I\ensuremath{\gamma}, ce,}\\
\parbox[b][0.3cm]{17.7cm}{\ensuremath{\gamma}\ensuremath{\gamma} coin, \ensuremath{\gamma}-recoil e\ensuremath{^{-}}coin, recoil e\ensuremath{^{-}} tagged prompt \ensuremath{\gamma}, recoil e\ensuremath{^{-}} tagged prompt \ensuremath{\gamma}-delayed \ensuremath{\gamma} coin, recoil e\ensuremath{^{-}} tagged prompt \ensuremath{\gamma}-prompt}\\
\parbox[b][0.3cm]{17.7cm}{\ensuremath{\gamma} coin, \ensuremath{\gamma}(\ensuremath{\theta}), and isomer half-life. Deduced levels, isomer, J, \ensuremath{\pi}, multipolarity, configuration, bands.}\\
\vspace{12pt}
\underline{$^{201}$At Levels}\\

\begin{textblock}{29}(0,27.3)
Continued on next page (footnotes at end of table)
\end{textblock}
\clearpage
%
\parbox[b][0.3cm]{17.7cm}{\makebox[1ex]{\ensuremath{^{\hypertarget{AT49LEVEL0}{\dagger}}}} From a least-squares fit to E\ensuremath{\gamma}.}\\
\parbox[b][0.3cm]{17.7cm}{\makebox[1ex]{\ensuremath{^{\hypertarget{AT49LEVEL1}{\ddagger}}}} From \href{https://www.nndc.bnl.gov/nsr/nsrlink.jsp?2014Au03,B}{2014Au03} and \href{https://www.nndc.bnl.gov/nsr/nsrlink.jsp?2015Au01,B}{2015Au01}, based on the deduced \ensuremath{\gamma}-ray transition multipolarities using the \ensuremath{\gamma}(\ensuremath{\theta}) analysis, observed decay}\\
\parbox[b][0.3cm]{17.7cm}{{\ }{\ }pattern, systematics in the region and shell-model assignments.}\\
\parbox[b][0.3cm]{17.7cm}{\makebox[1ex]{\ensuremath{^{\hypertarget{AT49LEVEL2}{\#}}}} Seq.(B): Based on \ensuremath{\pi} (d\ensuremath{_{\textnormal{3/2}}^{\textnormal{$-$1}}})\ensuremath{\otimes}\ensuremath{^{\textnormal{202}}}Rn core states (\ensuremath{J^{\pi}}=2\ensuremath{^{+}},4\ensuremath{^{+}},6\ensuremath{^{+}}).}\\
\parbox[b][0.3cm]{17.7cm}{\makebox[1ex]{\ensuremath{^{\hypertarget{AT49LEVEL3}{@}}}} Seq.(C): Based on \ensuremath{\pi} (d\ensuremath{_{\textnormal{5/2}}^{\textnormal{$-$1}}})\ensuremath{\otimes}\ensuremath{^{\textnormal{202}}}Rn core states (\ensuremath{J^{\pi}}=2\ensuremath{^{+}},4\ensuremath{^{+}}).}\\
\parbox[b][0.3cm]{17.7cm}{\makebox[1ex]{\ensuremath{^{\hypertarget{AT49LEVEL4}{\&}}}} Band(A): Magnetic-dipole, shears band. Configuration=\ensuremath{\pi} (i\ensuremath{_{\textnormal{13/2}}^{\textnormal{+1}}})\ensuremath{\otimes}\ensuremath{\nu} (f\ensuremath{_{\textnormal{5/2}}^{\textnormal{$-$1}}},i\ensuremath{_{\textnormal{13/2}}^{\textnormal{$-$1}}})\ensuremath{_{\textnormal{9$-$}}} for the lower cascade and}\\
\parbox[b][0.3cm]{17.7cm}{{\ }{\ }Configuration= \ensuremath{\pi} (h\ensuremath{_{\textnormal{9/2}}^{\textnormal{+2}}},i\ensuremath{_{\textnormal{13/2}}^{\textnormal{+1}}})\ensuremath{\otimes}\ensuremath{\nu} (f\ensuremath{_{\textnormal{5/2}}^{\textnormal{$-$1}}},i\ensuremath{_{\textnormal{13/2}}^{\textnormal{$-$1}}})\ensuremath{_{\textnormal{5$-$}}} above the band crossing.}\\
\vspace{0.5cm}
\clearpage
\vspace{0.3cm}
\begin{landscape}
\vspace*{-0.5cm}
{\bf \small \underline{\ensuremath{^{\textnormal{165}}}Ho(\ensuremath{^{\textnormal{40}}}Ar,4n\ensuremath{\gamma})\hspace{0.2in}\href{https://www.nndc.bnl.gov/nsr/nsrlink.jsp?2014Au03,B}{2014Au03},\href{https://www.nndc.bnl.gov/nsr/nsrlink.jsp?2015Au01,B}{2015Au01} (continued)}}\\
\vspace{0.3cm}
\underline{$\gamma$($^{201}$At)}\\

\parbox[b][0.3cm]{21.881866cm}{\makebox[1ex]{\ensuremath{^{\hypertarget{AT49GAMMA0}{\dagger}}}} From \href{https://www.nndc.bnl.gov/nsr/nsrlink.jsp?2015Au01,B}{2015Au01} using the JUROGAM2 data, unless otherwise stated. I\ensuremath{\gamma} normalized to I\ensuremath{\gamma}(635\ensuremath{\gamma})=100.}\\
\parbox[b][0.3cm]{21.881866cm}{\makebox[1ex]{\ensuremath{^{\hypertarget{AT49GAMMA1}{\ddagger}}}} Transition probably feeds the 2319-keV, 29/2\ensuremath{^{+}} level (\href{https://www.nndc.bnl.gov/nsr/nsrlink.jsp?2015Au01,B}{2015Au01}).}\\
\parbox[b][0.3cm]{21.881866cm}{\makebox[1ex]{\ensuremath{^{\hypertarget{AT49GAMMA2}{\#}}}} From summed spectra gated on 296\ensuremath{\gamma} and 427\ensuremath{\gamma} (\href{https://www.nndc.bnl.gov/nsr/nsrlink.jsp?2015Au01,B}{2015Au01}).}\\
\parbox[b][0.3cm]{21.881866cm}{\makebox[1ex]{\ensuremath{^{\hypertarget{AT49GAMMA3}{@}}}} From 269\ensuremath{\gamma}, 427\ensuremath{\gamma}, 635\ensuremath{\gamma}, 746\ensuremath{\gamma}, or 749\ensuremath{\gamma} delayed \ensuremath{\gamma}-ray tagged singles spectrum (\href{https://www.nndc.bnl.gov/nsr/nsrlink.jsp?2015Au01,B}{2015Au01}).}\\
\parbox[b][0.3cm]{21.881866cm}{\makebox[1ex]{\ensuremath{^{\hypertarget{AT49GAMMA4}{\&}}}} In addition to stretched dipole angular distributions, M1 character supported by high x-ray yield in coincidence with transition (\href{https://www.nndc.bnl.gov/nsr/nsrlink.jsp?2015Au01,B}{2015Au01}).}\\
\parbox[b][0.3cm]{21.881866cm}{\makebox[1ex]{\ensuremath{^{\hypertarget{AT49GAMMA5}{a}}}} From \href{https://www.nndc.bnl.gov/nsr/nsrlink.jsp?2015Au01,B}{2015Au01} using the focal-plane Clover data.}\\
\parbox[b][0.3cm]{21.881866cm}{\makebox[1ex]{\ensuremath{^{\hypertarget{AT49GAMMA6}{b}}}} From \href{https://www.nndc.bnl.gov/nsr/nsrlink.jsp?2015Au01,B}{2015Au01}. The focal plane values deduced from the focal-plane Clover data, unless otherwise stated. Normalized to I\ensuremath{\gamma}(269\ensuremath{\gamma})=100. Internal conversion}\\
\parbox[b][0.3cm]{21.881866cm}{{\ }{\ }coefficients that were used to calculate I(g\ensuremath{\pm}ce) were taken from \href{https://www.nndc.bnl.gov/nsr/nsrlink.jsp?2008Ki07,B}{2008Ki07}.}\\
\parbox[b][0.3cm]{21.881866cm}{\makebox[1ex]{\ensuremath{^{\hypertarget{AT49GAMMA7}{c}}}} From \href{https://www.nndc.bnl.gov/nsr/nsrlink.jsp?2014Au03,B}{2014Au03}. I\ensuremath{\gamma} above the \ensuremath{J^{\pi}}=1/2\ensuremath{^{+}} isomer are from recoil-ce-tagged singles \ensuremath{\gamma}-ray data.}\\
\parbox[b][0.3cm]{21.881866cm}{\makebox[1ex]{\ensuremath{^{\hypertarget{AT49GAMMA8}{d}}}} Doublet in \href{https://www.nndc.bnl.gov/nsr/nsrlink.jsp?2014Au03,B}{2014Au03}. Intensities of the two components from 173-gated, recoil-corrected \ensuremath{\gamma}\ensuremath{\gamma}-coin data.}\\
\parbox[b][0.3cm]{21.881866cm}{\makebox[1ex]{\ensuremath{^{\hypertarget{AT49GAMMA9}{e}}}} Weak transition in \href{https://www.nndc.bnl.gov/nsr/nsrlink.jsp?2014Au03,B}{2014Au03}, intensity from 173-keV gated, recoil-correlated \ensuremath{\gamma}\ensuremath{\gamma}-coin data.}\\
\parbox[b][0.3cm]{21.881866cm}{\makebox[1ex]{\ensuremath{^{\hypertarget{AT49GAMMA10}{f}}}} From \ensuremath{\gamma}(\ensuremath{\theta}) in \href{https://www.nndc.bnl.gov/nsr/nsrlink.jsp?2014Au03,B}{2014Au03} and \href{https://www.nndc.bnl.gov/nsr/nsrlink.jsp?2015Au01,B}{2015Au01}, unless otherwise stated. The reported in \href{https://www.nndc.bnl.gov/nsr/nsrlink.jsp?2014Au03,B}{2014Au03} correlation ratios, R,\hphantom{a}are defined as}\\
\parbox[b][0.3cm]{21.881866cm}{{\ }{\ }R=[I\ensuremath{\gamma}(133.6\ensuremath{^\circ})+I\ensuremath{\gamma}(157.6\ensuremath{^\circ})]/I\ensuremath{\gamma}(104.5\ensuremath{^\circ} or 75.5\ensuremath{^\circ}). Expected values are 1.30 \textit{7} for \ensuremath{\Delta}J=2, quadrupole, and 0.70 \textit{6} for \ensuremath{\Delta}J=1, dipole transitions.}\\
\parbox[b][0.3cm]{21.881866cm}{\makebox[1ex]{\ensuremath{^{\hypertarget{AT49GAMMA11}{g}}}} Total theoretical internal conversion coefficients, calculated using the BrIcc code (\href{https://www.nndc.bnl.gov/nsr/nsrlink.jsp?2008Ki07,B}{2008Ki07}) with Frozen orbital approximation based on \ensuremath{\gamma}-ray energies,}\\
\parbox[b][0.3cm]{21.881866cm}{{\ }{\ }assigned multipolarities, and mixing ratios, unless otherwise specified.}\\
\parbox[b][0.3cm]{21.881866cm}{\makebox[1ex]{\ensuremath{^{\hypertarget{AT49GAMMA12}{h}}}} Placement of transition in the level scheme is uncertain.}\\
\parbox[b][0.3cm]{21.881866cm}{\makebox[1ex]{\ensuremath{^{\hypertarget{AT49GAMMA13}{x}}}} \ensuremath{\gamma} ray not placed in level scheme.}\\
\vspace{0.5cm}
\end{landscape}\clearpage
\clearpage
\begin{figure}[h]
\begin{center}
\includegraphics{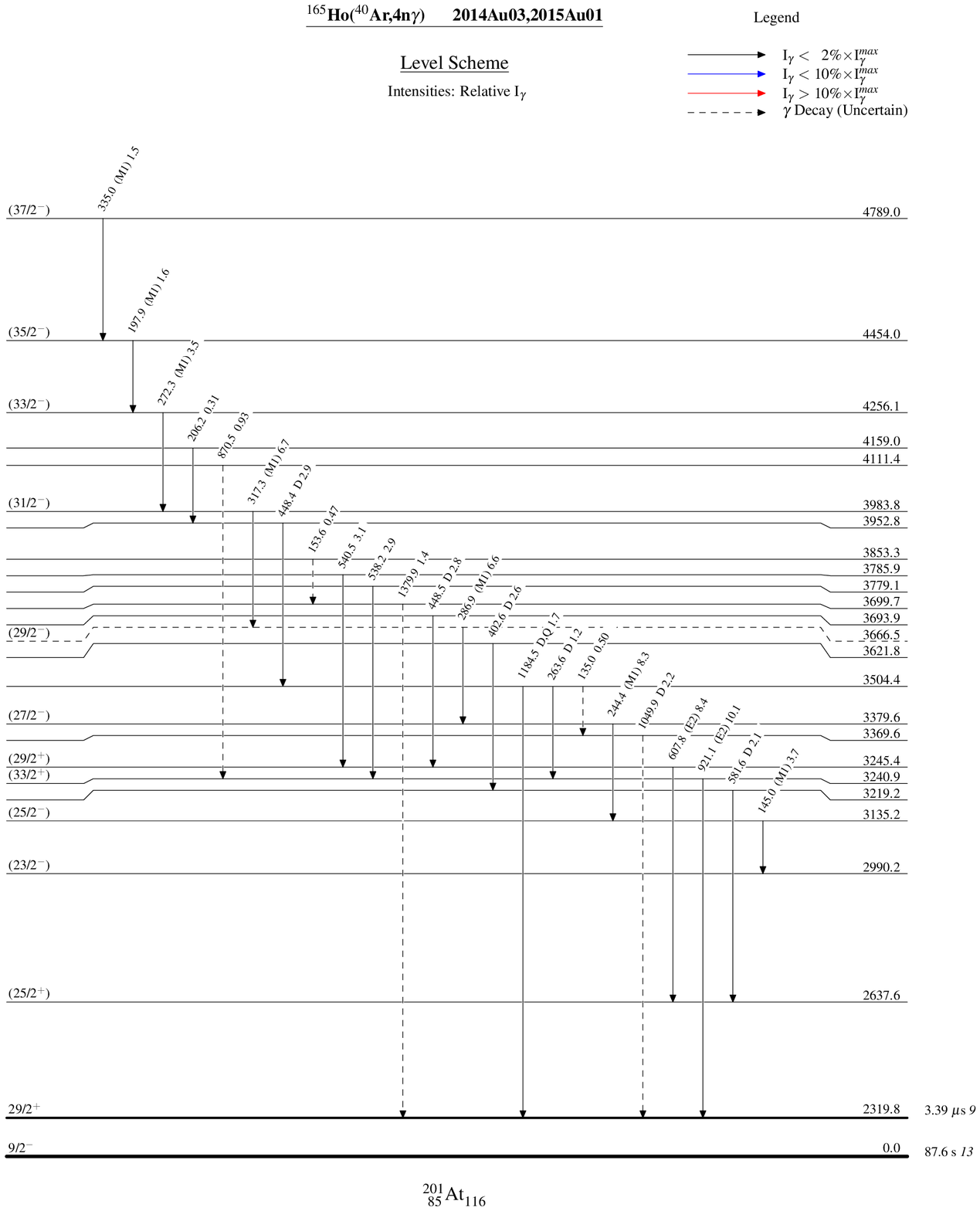}\\
\end{center}
\end{figure}
\clearpage
\begin{figure}[h]
\begin{center}
\includegraphics{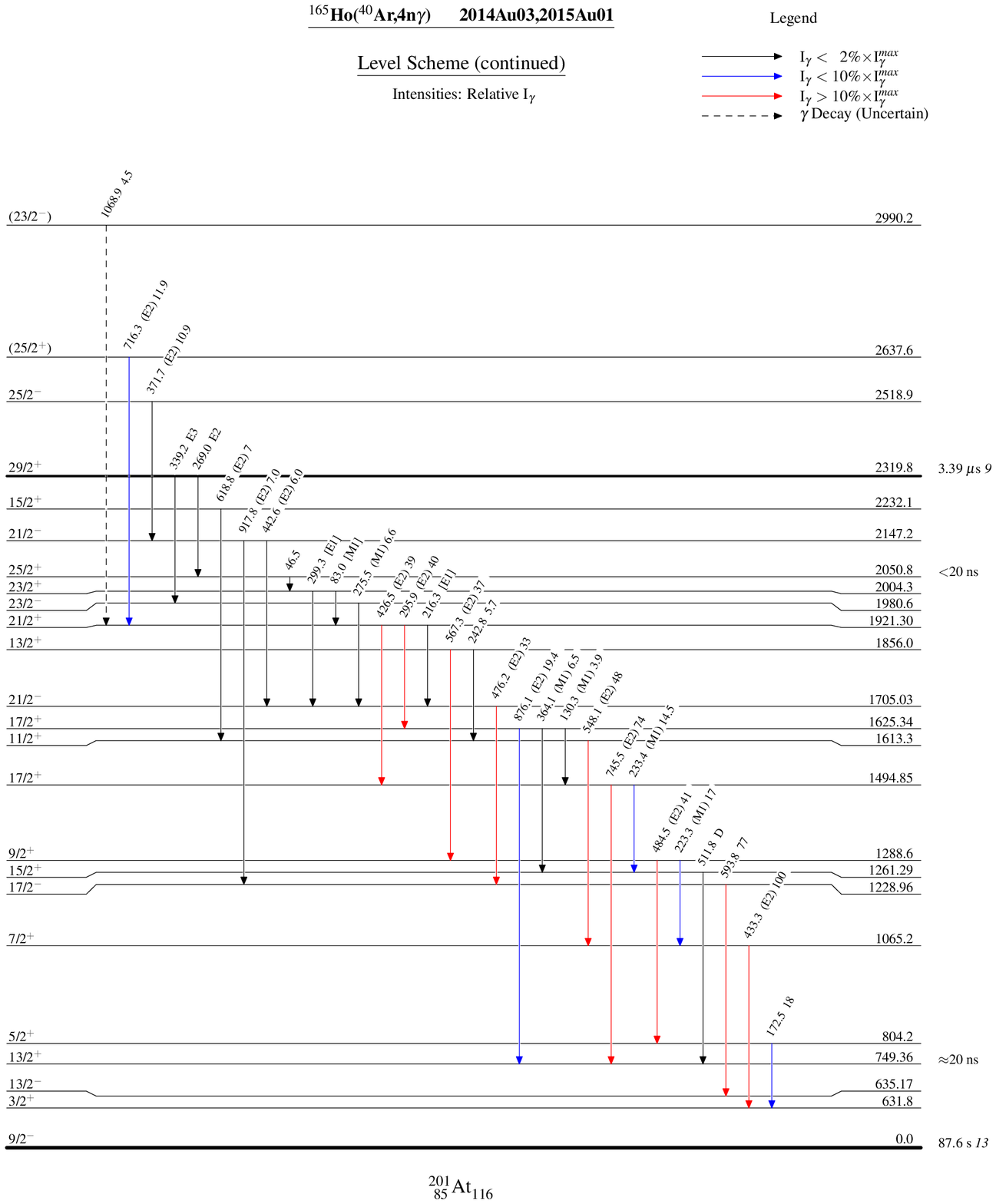}\\
\end{center}
\end{figure}
\clearpage
\begin{figure}[h]
\begin{center}
\includegraphics{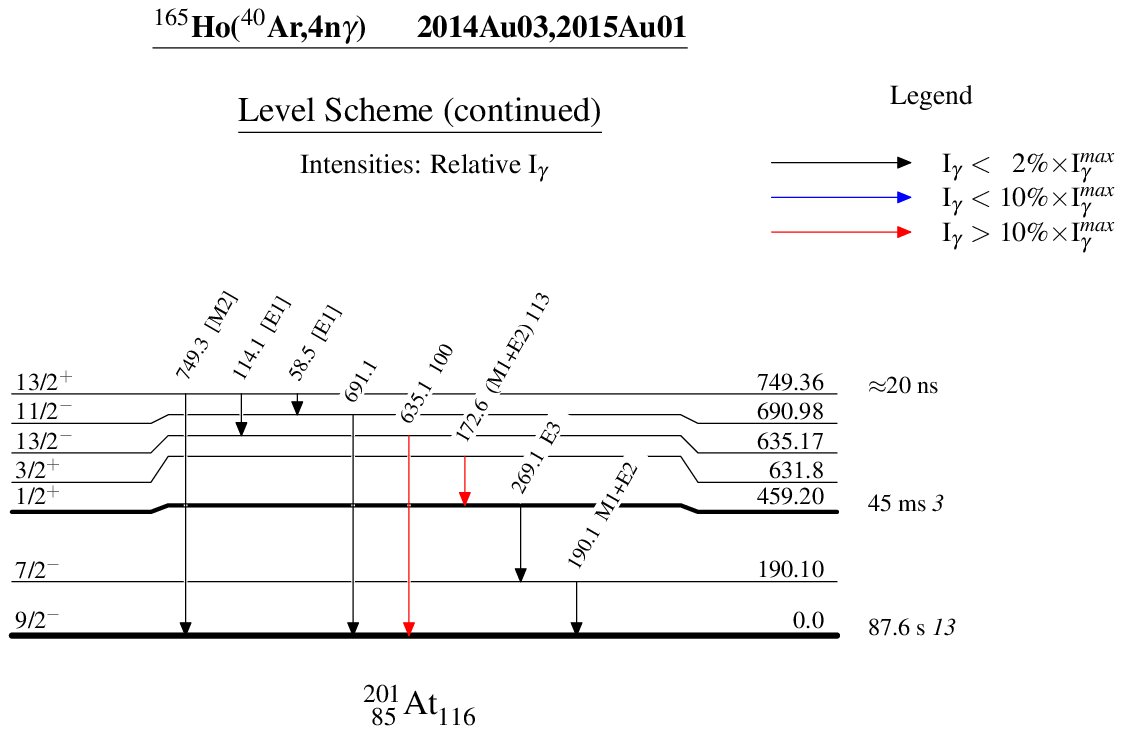}\\
\end{center}
\end{figure}
\clearpage
\clearpage
\begin{figure}[h]
\begin{center}
\includegraphics{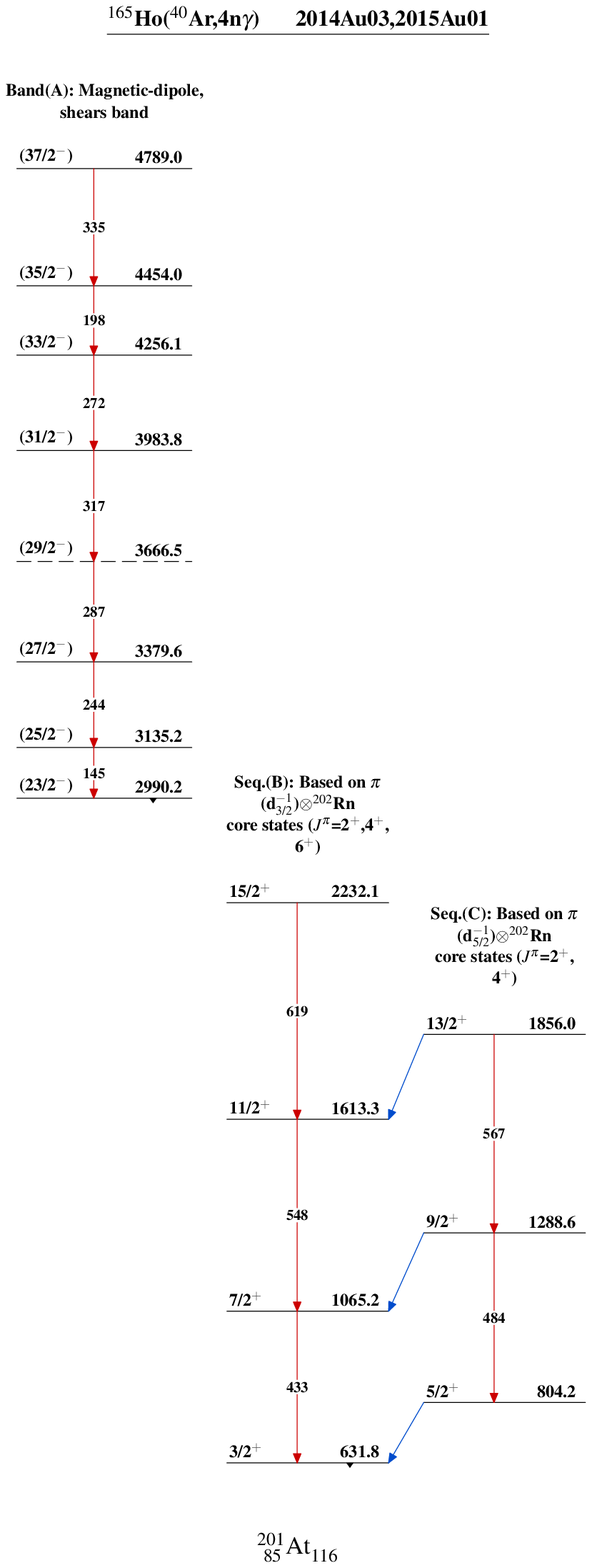}\\
\end{center}
\end{figure}
\clearpage
%ADOPTED LEVELS, GAMMAS
\section[\ensuremath{^{201}_{\ 86}}Rn\ensuremath{_{115}^{~}}]{ }
\vspace{-30pt}
\setcounter{chappage}{1}
\subsection[\hspace{-0.2cm}Adopted Levels, Gammas]{ }
\vspace{-20pt}
\vspace{0.3cm}
\hypertarget{RN50}{{\bf \small \underline{Adopted \hyperlink{201RN_LEVEL}{Levels}, \hyperlink{201RN_GAMMA}{Gammas}}}}\\
\vspace{4pt}
\vspace{8pt}
\parbox[b][0.3cm]{17.7cm}{\addtolength{\parindent}{-0.2in}Q(\ensuremath{\beta^-})=$-$7696 {\it 14}; S(n)=8178 {\it 12}; S(p)=2408 {\it 26}; Q(\ensuremath{\alpha})=6860.7 {\it 23}\hspace{0.2in}\href{https://www.nndc.bnl.gov/nsr/nsrlink.jsp?2021Wa16,B}{2021Wa16}}\\

\vspace{12pt}
\hypertarget{201RN_LEVEL}{\underline{$^{201}$Rn Levels}}\\

\parbox[b][0.3cm]{17.7cm}{\makebox[1ex]{\ensuremath{^{\hypertarget{RN50LEVEL0}{\dagger}}}} From a least-squares fit to E\ensuremath{\gamma}, unless otherwise stated.}\\
\parbox[b][0.3cm]{17.7cm}{\makebox[1ex]{\ensuremath{^{\hypertarget{RN50LEVEL1}{\ddagger}}}} From \ensuremath{^{\textnormal{122}}}Sn(\ensuremath{^{\textnormal{82}}}Kr,3n\ensuremath{\gamma}) (\href{https://www.nndc.bnl.gov/nsr/nsrlink.jsp?2008An05,B}{2008An05}), unless otherwise stated.}\\
\vspace{0.5cm}
\hypertarget{201RN_GAMMA}{\underline{$\gamma$($^{201}$Rn)}}\\

\begin{textblock}{29}(0,27.3)
Continued on next page (footnotes at end of table)
\end{textblock}
\clearpage
%
\parbox[b][0.3cm]{17.7cm}{\makebox[1ex]{\ensuremath{^{\hypertarget{RN50GAMMA0}{\dagger}}}} From \ensuremath{^{\textnormal{122}}}Sn(\ensuremath{^{\textnormal{82}}}Kr,3n\ensuremath{\gamma}) (\href{https://www.nndc.bnl.gov/nsr/nsrlink.jsp?2008An05,B}{2008An05}).}\\
\parbox[b][0.3cm]{17.7cm}{\makebox[1ex]{\ensuremath{^{\hypertarget{RN50GAMMA1}{\ddagger}}}} Placement of transition in the level scheme is uncertain.}\\
\vspace{0.5cm}
\begin{figure}[h]
\begin{center}
\includegraphics{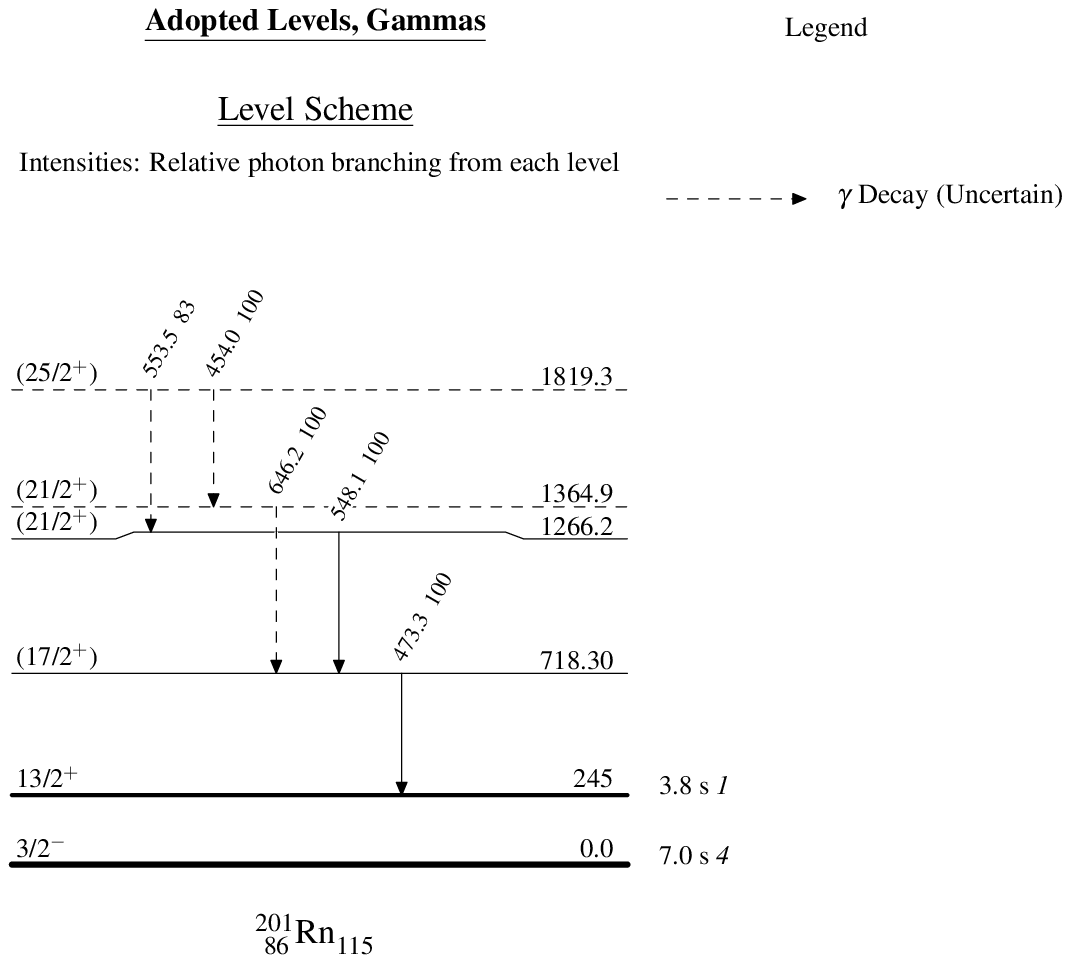}\\
\end{center}
\end{figure}
\clearpage
%205RA A DECAY (210 MS)
\subsection[\hspace{-0.2cm}\ensuremath{^{\textnormal{205}}}Ra \ensuremath{\alpha} decay (210 ms)]{ }
\vspace{-27pt}
\vspace{0.3cm}
\hypertarget{RA51}{{\bf \small \underline{\ensuremath{^{\textnormal{205}}}Ra \ensuremath{\alpha} decay (210 ms)\hspace{0.2in}\href{https://www.nndc.bnl.gov/nsr/nsrlink.jsp?1996Le09,B}{1996Le09},\href{https://www.nndc.bnl.gov/nsr/nsrlink.jsp?1987He10,B}{1987He10}}}}\\
\vspace{4pt}
\vspace{8pt}
\parbox[b][0.3cm]{17.7cm}{\addtolength{\parindent}{-0.2in}Parent: $^{205}$Ra: E=0.0; J$^{\pi}$=(3/2\ensuremath{^{-}}); T$_{1/2}$=210 ms {\it +60\textminus40}; Q(\ensuremath{\alpha})=7486 {\it 20}; \%\ensuremath{\alpha} decay$\approx$100.0

}\\
\parbox[b][0.3cm]{17.7cm}{\addtolength{\parindent}{-0.2in}\ensuremath{^{205}}Ra-J$^{\pi}$,T$_{1/2}$: From \href{https://www.nndc.bnl.gov/nsr/nsrlink.jsp?2020Ko17,B}{2020Ko17}.}\\
\parbox[b][0.3cm]{17.7cm}{\addtolength{\parindent}{-0.2in}\ensuremath{^{205}}Ra-Q(\ensuremath{\alpha}): From \href{https://www.nndc.bnl.gov/nsr/nsrlink.jsp?2021Wa16,B}{2021Wa16}.}\\
\parbox[b][0.3cm]{17.7cm}{\addtolength{\parindent}{-0.2in}\ensuremath{^{205}}Ra-\%\ensuremath{\alpha} decay: From \href{https://www.nndc.bnl.gov/nsr/nsrlink.jsp?2020Ko17,B}{2020Ko17}.}\\
\vspace{12pt}
\underline{$^{201}$Rn Levels}\\
\begin{longtable}{ccccc@{\extracolsep{\fill}}c}
\multicolumn{2}{c}{E(level)$^{}$}&J$^{\pi}$$^{}$&\multicolumn{2}{c}{T$_{1/2}$$^{}$}&\\[-.2cm]
\multicolumn{2}{c}{\hrulefill}&\hrulefill&\multicolumn{2}{c}{\hrulefill}&
\endfirsthead
\multicolumn{1}{r@{}}{0}&\multicolumn{1}{@{.}l}{0}&\multicolumn{1}{l}{3/2\ensuremath{^{-}}}&\multicolumn{1}{r@{}}{7}&\multicolumn{1}{@{.}l}{0 s {\it 4}}&\\
\end{longtable}

\underline{\ensuremath{\alpha} radiations}\\
\begin{longtable}{ccccccccc@{\extracolsep{\fill}}c}
\multicolumn{2}{c}{E$\alpha^{{}}$}&\multicolumn{2}{c}{E(level)}&\multicolumn{2}{c}{I$\alpha^{{\hyperlink{RN51DECAY1}{\ddagger}}}$}&\multicolumn{2}{c}{HF$^{{\hyperlink{RN51DECAY0}{\dagger}}}$}&Comments&\\[-.2cm]
\multicolumn{2}{c}{\hrulefill}&\multicolumn{2}{c}{\hrulefill}&\multicolumn{2}{c}{\hrulefill}&\multicolumn{2}{c}{\hrulefill}&\hrulefill&
\endfirsthead
\multicolumn{1}{r@{}}{7340}&\multicolumn{1}{@{ }l}{{\it 20}}&\multicolumn{1}{r@{}}{0}&\multicolumn{1}{@{.}l}{0}&\multicolumn{1}{r@{}}{$\approx$100}&\multicolumn{1}{@{}l}{}&\multicolumn{1}{r@{}}{$\approx$1}&\multicolumn{1}{@{.}l}{4}&\parbox[t][0.3cm]{12.855961cm}{\raggedright E$\alpha$: From \href{https://www.nndc.bnl.gov/nsr/nsrlink.jsp?1996Le09,B}{1996Le09}. Others:\hphantom{a}7350 keV \textit{25} (\href{https://www.nndc.bnl.gov/nsr/nsrlink.jsp?1995Le15,B}{1995Le15}), 7355 keV \textit{10} (\href{https://www.nndc.bnl.gov/nsr/nsrlink.jsp?1995Le04,B}{1995Le04}) and 7360\vspace{0.1cm}}&\\
&&&&&&&&\parbox[t][0.3cm]{12.855961cm}{\raggedright {\ }{\ }{\ }keV \textit{20} (\href{https://www.nndc.bnl.gov/nsr/nsrlink.jsp?1987He10,B}{1987He10}).\vspace{0.1cm}}&\\
\end{longtable}
\parbox[b][0.3cm]{17.7cm}{\makebox[1ex]{\ensuremath{^{\hypertarget{RN51DECAY0}{\dagger}}}} Using r\ensuremath{_{\textnormal{0}}}(\ensuremath{^{\textnormal{201}}}Rn)=1.527 \textit{9} from \href{https://www.nndc.bnl.gov/nsr/nsrlink.jsp?2020Si16,B}{2020Si16}.}\\
\parbox[b][0.3cm]{17.7cm}{\makebox[1ex]{\ensuremath{^{\hypertarget{RN51DECAY1}{\ddagger}}}} For absolute intensity per 100 decays, multiply by{ }\ensuremath{\approx}1.0.}\\
\vspace{0.5cm}
\clearpage
%205RA A DECAY (170 MS)
\subsection[\hspace{-0.2cm}\ensuremath{^{\textnormal{205}}}Ra \ensuremath{\alpha} decay (170 ms)]{ }
\vspace{-27pt}
\vspace{0.3cm}
\hypertarget{RA52}{{\bf \small \underline{\ensuremath{^{\textnormal{205}}}Ra \ensuremath{\alpha} decay (170 ms)\hspace{0.2in}\href{https://www.nndc.bnl.gov/nsr/nsrlink.jsp?1996Le09,B}{1996Le09},\href{https://www.nndc.bnl.gov/nsr/nsrlink.jsp?1987He10,B}{1987He10}}}}\\
\vspace{4pt}
\vspace{8pt}
\parbox[b][0.3cm]{17.7cm}{\addtolength{\parindent}{-0.2in}Parent: $^{205}$Ra: E=263 {\it 25}; J$^{\pi}$=13/2\ensuremath{^{+}}; T$_{1/2}$=170 ms {\it +60\textminus40}; Q(\ensuremath{\alpha})=7486 {\it 20}; \%\ensuremath{\alpha} decay$\approx$100.0

}\\
\parbox[b][0.3cm]{17.7cm}{\addtolength{\parindent}{-0.2in}\ensuremath{^{205}}Ra-T$_{1/2}$: From \href{https://www.nndc.bnl.gov/nsr/nsrlink.jsp?2020Ko17,B}{2020Ko17}.}\\
\parbox[b][0.3cm]{17.7cm}{\addtolength{\parindent}{-0.2in}\ensuremath{^{205}}Ra-E,J$^{\pi}$: From \href{https://www.nndc.bnl.gov/nsr/nsrlink.jsp?2021Ko07,B}{2021Ko07}.}\\
\parbox[b][0.3cm]{17.7cm}{\addtolength{\parindent}{-0.2in}\ensuremath{^{205}}Ra-Q(\ensuremath{\alpha}): From \href{https://www.nndc.bnl.gov/nsr/nsrlink.jsp?2021Wa16,B}{2021Wa16}.}\\
\parbox[b][0.3cm]{17.7cm}{\addtolength{\parindent}{-0.2in}\ensuremath{^{205}}Ra-\%\ensuremath{\alpha} decay: From \href{https://www.nndc.bnl.gov/nsr/nsrlink.jsp?2020Ko17,B}{2020Ko17}.}\\
\vspace{12pt}
\underline{$^{201}$Rn Levels}\\
\begin{longtable}{ccccc@{\extracolsep{\fill}}c}
\multicolumn{2}{c}{E(level)$^{}$}&J$^{\pi}$$^{}$&\multicolumn{2}{c}{T$_{1/2}$$^{}$}&\\[-.2cm]
\multicolumn{2}{c}{\hrulefill}&\hrulefill&\multicolumn{2}{c}{\hrulefill}&
\endfirsthead
\multicolumn{1}{r@{}}{245}&\multicolumn{1}{@{ }l}{{\it 12}}&\multicolumn{1}{l}{13/2\ensuremath{^{+}}}&\multicolumn{1}{r@{}}{3}&\multicolumn{1}{@{.}l}{8 s {\it 1}}&\\
\end{longtable}

\underline{\ensuremath{\alpha} radiations}\\
\begin{longtable}{ccccccccc@{\extracolsep{\fill}}c}
\multicolumn{2}{c}{E$\alpha^{{}}$}&\multicolumn{2}{c}{E(level)}&\multicolumn{2}{c}{I$\alpha^{{\hyperlink{RN52DECAY1}{\ddagger}}}$}&\multicolumn{2}{c}{HF$^{{\hyperlink{RN52DECAY0}{\dagger}}}$}&Comments&\\[-.2cm]
\multicolumn{2}{c}{\hrulefill}&\multicolumn{2}{c}{\hrulefill}&\multicolumn{2}{c}{\hrulefill}&\multicolumn{2}{c}{\hrulefill}&\hrulefill&
\endfirsthead
\multicolumn{1}{r@{}}{7359}&\multicolumn{1}{@{ }l}{{\it 9}}&\multicolumn{1}{r@{}}{245}&\multicolumn{1}{@{}l}{}&\multicolumn{1}{r@{}}{$\approx$100}&\multicolumn{1}{@{}l}{}&\multicolumn{1}{r@{}}{$\approx$1}&\multicolumn{1}{@{.}l}{3}&\parbox[t][0.3cm]{13.01432cm}{\raggedright E$\alpha$: From Q(\ensuremath{\alpha})=7505 keV \textit{9} in \href{https://www.nndc.bnl.gov/nsr/nsrlink.jsp?2021Hu06,B}{2021Hu06} (a least-squares adjustment of the atomic masses).\vspace{0.1cm}}&\\
&&&&&&&&\parbox[t][0.3cm]{13.01432cm}{\raggedright {\ }{\ }{\ }Individual E\ensuremath{\alpha} values are 7370 keV \textit{20} (\href{https://www.nndc.bnl.gov/nsr/nsrlink.jsp?1996Le09,B}{1996Le09}), 7355 keV \textit{10} (\href{https://www.nndc.bnl.gov/nsr/nsrlink.jsp?1995Le04,B}{1995Le04}), 7375 keV \textit{25}\vspace{0.1cm}}&\\
&&&&&&&&\parbox[t][0.3cm]{13.01432cm}{\raggedright {\ }{\ }{\ }(\href{https://www.nndc.bnl.gov/nsr/nsrlink.jsp?1995Le15,B}{1995Le15}) and 7379 keV \textit{30} (\href{https://www.nndc.bnl.gov/nsr/nsrlink.jsp?2010He25,B}{2010He25}).\vspace{0.1cm}}&\\
\end{longtable}
\parbox[b][0.3cm]{17.7cm}{\makebox[1ex]{\ensuremath{^{\hypertarget{RN52DECAY0}{\dagger}}}} Using r\ensuremath{_{\textnormal{0}}}(\ensuremath{^{\textnormal{201}}}Rn)=1.527 \textit{9} from \href{https://www.nndc.bnl.gov/nsr/nsrlink.jsp?2020Si16,B}{2020Si16}.}\\
\parbox[b][0.3cm]{17.7cm}{\makebox[1ex]{\ensuremath{^{\hypertarget{RN52DECAY1}{\ddagger}}}} For absolute intensity per 100 decays, multiply by{ }\ensuremath{\approx}1.0.}\\
\vspace{0.5cm}
\clearpage
%122SN(82KR,3NG)
\subsection[\hspace{-0.2cm}\ensuremath{^{\textnormal{122}}}Sn(\ensuremath{^{\textnormal{82}}}Kr,3n\ensuremath{\gamma})]{ }
\vspace{-27pt}
\vspace{0.3cm}
\hypertarget{RN53}{{\bf \small \underline{\ensuremath{^{\textnormal{122}}}Sn(\ensuremath{^{\textnormal{82}}}Kr,3n\ensuremath{\gamma})\hspace{0.2in}\href{https://www.nndc.bnl.gov/nsr/nsrlink.jsp?2008An05,B}{2008An05}}}}\\
\vspace{4pt}
\vspace{8pt}
\parbox[b][0.3cm]{17.7cm}{\addtolength{\parindent}{-0.2in}\ensuremath{^{\textnormal{122}}}Sn(\ensuremath{^{\textnormal{82}}}Kr,3n\ensuremath{\gamma}), E=355 MeV beam delivered at JYFL, Finland. Measured E\ensuremath{\gamma}, I\ensuremath{\gamma}, \ensuremath{\gamma}\ensuremath{\gamma}, ce, \ensuremath{\gamma}\ensuremath{\alpha} coin using recoil-decay tagging}\\
\parbox[b][0.3cm]{17.7cm}{method with the JUROGAM array of 43 EUROGAM type escape-suppressed HPGe detectors at angles of 72\ensuremath{^\circ}, 86\ensuremath{^\circ}, 94\ensuremath{^\circ}, 108\ensuremath{^\circ}, 134\ensuremath{^\circ}}\\
\parbox[b][0.3cm]{17.7cm}{and 158\ensuremath{^\circ}. Reaction products were separated with the RITU recoil separator and implanted in the double-sided silicon strip detectors}\\
\parbox[b][0.3cm]{17.7cm}{of the GREAT spectrometer. Reaction \ensuremath{^{\textnormal{152}}}Sm(\ensuremath{^{\textnormal{52}}}Cr,3n\ensuremath{\gamma}), E=231 MeV was also used.}\\
\vspace{12pt}
\underline{$^{201}$Rn Levels}\\

\parbox[b][0.3cm]{17.7cm}{\makebox[1ex]{\ensuremath{^{\hypertarget{RN53LEVEL0}{\dagger}}}} From a least-squares fit to E\ensuremath{\gamma}, unless otherwise stated.}\\
\parbox[b][0.3cm]{17.7cm}{\makebox[1ex]{\ensuremath{^{\hypertarget{RN53LEVEL1}{\ddagger}}}} From \href{https://www.nndc.bnl.gov/nsr/nsrlink.jsp?2008An05,B}{2008An05}, unless otherwise stated.}\\
\vspace{0.5cm}
\underline{$\gamma$($^{201}$Rn)}\\
%
\parbox[b][0.3cm]{17.7cm}{\makebox[1ex]{\ensuremath{^{\hypertarget{RN53GAMMA0}{\dagger}}}} \ensuremath{\gamma} ray associated with the \ensuremath{J^{\pi}}=3/2\ensuremath{^{-}} ground state. Intensities are normalized to 100 for the 397.8 \ensuremath{\gamma} ray.}\\
\parbox[b][0.3cm]{17.7cm}{\makebox[1ex]{\ensuremath{^{\hypertarget{RN53GAMMA1}{\ddagger}}}} \ensuremath{\gamma} ray associated with the \ensuremath{J^{\pi}}=13/2\ensuremath{^{+}} isomer. Intensities are normalized to 100 for the 473.3 \ensuremath{\gamma} ray.}\\
\parbox[b][0.3cm]{17.7cm}{\makebox[1ex]{\ensuremath{^{\hypertarget{RN53GAMMA2}{\#}}}} Doublet.}\\
\begin{textblock}{29}(0,27.3)
Continued on next page (footnotes at end of table)
\end{textblock}
\clearpage
\vspace*{-0.5cm}
{\bf \small \underline{\ensuremath{^{\textnormal{122}}}Sn(\ensuremath{^{\textnormal{82}}}Kr,3n\ensuremath{\gamma})\hspace{0.2in}\href{https://www.nndc.bnl.gov/nsr/nsrlink.jsp?2008An05,B}{2008An05} (continued)}}\\
\vspace{0.3cm}
\underline{$\gamma$($^{201}$Rn) (continued)}\\
\vspace{0.3cm}
\parbox[b][0.3cm]{17.7cm}{\makebox[1ex]{\ensuremath{^{\hypertarget{RN53GAMMA3}{@}}}} Placement of transition in the level scheme is uncertain.}\\
\parbox[b][0.3cm]{17.7cm}{\makebox[1ex]{\ensuremath{^{\hypertarget{RN53GAMMA4}{x}}}} \ensuremath{\gamma} ray not placed in level scheme.}\\
\vspace{0.5cm}
\begin{figure}[h]
\begin{center}
\includegraphics{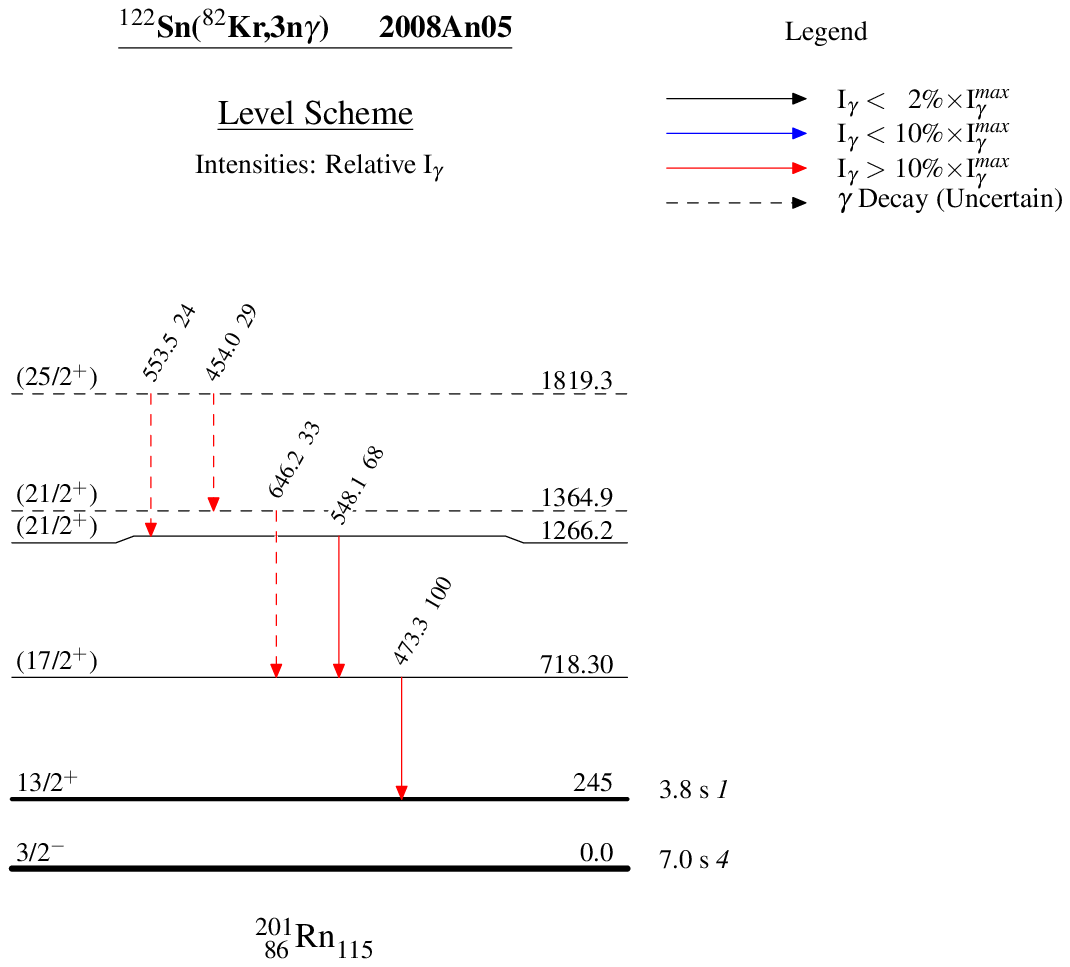}\\
\end{center}
\end{figure}
\clearpage
%ADOPTED LEVELS, GAMMAS
\section[\ensuremath{^{201}_{\ 87}}Fr\ensuremath{_{114}^{~}}]{ }
\vspace{-30pt}
\setcounter{chappage}{1}
\subsection[\hspace{-0.2cm}Adopted Levels, Gammas]{ }
\vspace{-20pt}
\vspace{0.3cm}
\hypertarget{FR54}{{\bf \small \underline{Adopted \hyperlink{201FR_LEVEL}{Levels}, \hyperlink{201FR_GAMMA}{Gammas}}}}\\
\vspace{4pt}
\vspace{8pt}
\parbox[b][0.3cm]{17.7cm}{\addtolength{\parindent}{-0.2in}Q(\ensuremath{\beta^-})=$-$8348 {\it 22}; S(n)=10620 {\it 30}; S(p)=$-$300 {\it 11}; Q(\ensuremath{\alpha})=7519 {\it 4}\hspace{0.2in}\href{https://www.nndc.bnl.gov/nsr/nsrlink.jsp?2021Wa16,B}{2021Wa16}}\\

\vspace{12pt}
\hypertarget{201FR_LEVEL}{\underline{$^{201}$Fr Levels}}\\

\parbox[b][0.3cm]{17.7cm}{\makebox[1ex]{\ensuremath{^{\hypertarget{FR54GAMMA0}{\dagger}}}} Total theoretical internal conversion coefficients, calculated using the BrIcc code (\href{https://www.nndc.bnl.gov/nsr/nsrlink.jsp?2008Ki07,B}{2008Ki07}) with Frozen orbital approximation}\\
\parbox[b][0.3cm]{17.7cm}{{\ }{\ }based on \ensuremath{\gamma}-ray energies, assigned multipolarities, and mixing ratios, unless otherwise specified.}\\
\vspace{0.5cm}
\clearpage
\begin{figure}[h]
\begin{center}
\includegraphics{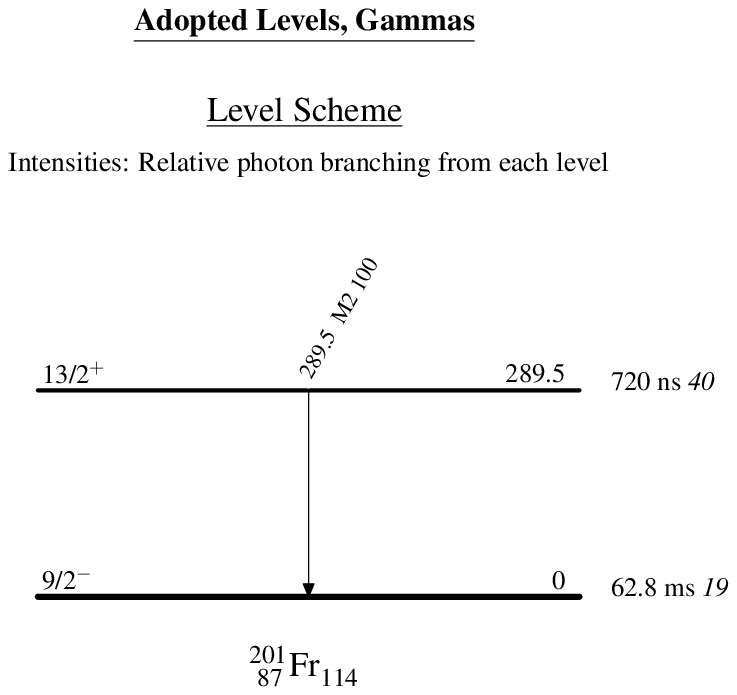}\\
\end{center}
\end{figure}
\clearpage
%205AC A DECAY
\subsection[\hspace{-0.2cm}\ensuremath{^{\textnormal{205}}}Ac \ensuremath{\alpha} decay]{ }
\vspace{-27pt}
\vspace{0.3cm}
\hypertarget{AC55}{{\bf \small \underline{\ensuremath{^{\textnormal{205}}}Ac \ensuremath{\alpha} decay\hspace{0.2in}\href{https://www.nndc.bnl.gov/nsr/nsrlink.jsp?2014Zh03,B}{2014Zh03}}}}\\
\vspace{4pt}
\vspace{8pt}
\parbox[b][0.3cm]{17.7cm}{\addtolength{\parindent}{-0.2in}Parent: $^{205}$Ac: E=0; J$^{\pi}$=9/2\ensuremath{^{-}}; T$_{1/2}$=20 ms {\it +97\textminus9}; Q(\ensuremath{\alpha})=8090 {\it 60}; \%\ensuremath{\alpha} decay$\approx$100.0

}\\
\parbox[b][0.3cm]{17.7cm}{\addtolength{\parindent}{-0.2in}\ensuremath{^{205}}Ac-E,J$^{\pi}$: From \href{https://www.nndc.bnl.gov/nsr/nsrlink.jsp?2021Ko07,B}{2021Ko07}.}\\
\parbox[b][0.3cm]{17.7cm}{\addtolength{\parindent}{-0.2in}\ensuremath{^{205}}Ac-T$_{1/2}$: From 7935\ensuremath{\alpha}(t) in \href{https://www.nndc.bnl.gov/nsr/nsrlink.jsp?2014Zh03,B}{2014Zh03}.}\\
\parbox[b][0.3cm]{17.7cm}{\addtolength{\parindent}{-0.2in}\ensuremath{^{205}}Ac-Q(\ensuremath{\alpha}): From \href{https://www.nndc.bnl.gov/nsr/nsrlink.jsp?2021Wa16,B}{2021Wa16}.}\\
\parbox[b][0.3cm]{17.7cm}{\addtolength{\parindent}{-0.2in}\ensuremath{^{205}}Ac-\%\ensuremath{\alpha} decay: From \href{https://www.nndc.bnl.gov/nsr/nsrlink.jsp?2021Ko07,B}{2021Ko07}.}\\
\parbox[b][0.3cm]{17.7cm}{\addtolength{\parindent}{-0.2in}\href{https://www.nndc.bnl.gov/nsr/nsrlink.jsp?2014Zh03,B}{2014Zh03}: \ensuremath{^{\textnormal{205}}}Ac produced in the \ensuremath{^{\textnormal{169}}}Tm(\ensuremath{^{\textnormal{40}}}Ca,4n) reaction, E(\ensuremath{^{\textnormal{40}}}Ca=196 MeV at the HIRFL facility, Lanzhou. Target: 400}\\
\parbox[b][0.3cm]{17.7cm}{\ensuremath{\mu}g/cm\ensuremath{^{\textnormal{2}}} thick covered with a 10 \ensuremath{\mu}g/cm\ensuremath{^{\textnormal{2}}}-thick carbon layer. Evaporation residues were separated in flight using SHANS recoil}\\
\parbox[b][0.3cm]{17.7cm}{separator, and implanted into position sensitive DSSD (48 vertical strips of 3 mm width). Eight non-position sensitive Si detectors}\\
\parbox[b][0.3cm]{17.7cm}{were used to detect escaping \ensuremath{\alpha} particles. Measured: recoil-\ensuremath{\alpha}\ensuremath{_{\textnormal{1}}}(t)-\ensuremath{\alpha}\ensuremath{_{\textnormal{2}}}(t)-\ensuremath{\alpha}\ensuremath{_{\textnormal{3}}}(t) correlated events. Deduced: E\ensuremath{\alpha} and half-life of \ensuremath{^{\textnormal{205}}}Ac.}\\
\vspace{12pt}
\underline{$^{201}$Fr Levels}\\
\begin{longtable}{ccccc@{\extracolsep{\fill}}c}
\multicolumn{2}{c}{E(level)$^{}$}&J$^{\pi}$$^{}$&\multicolumn{2}{c}{T$_{1/2}$$^{}$}&\\[-.2cm]
\multicolumn{2}{c}{\hrulefill}&\hrulefill&\multicolumn{2}{c}{\hrulefill}&
\endfirsthead
\multicolumn{1}{r@{}}{0}&\multicolumn{1}{@{}l}{}&\multicolumn{1}{l}{9/2\ensuremath{^{-}}}&\multicolumn{1}{r@{}}{62}&\multicolumn{1}{@{.}l}{8 ms {\it 19}}&\\
\end{longtable}

\underline{\ensuremath{\alpha} radiations}\\
\begin{longtable}{ccccccccc@{\extracolsep{\fill}}c}
\multicolumn{2}{c}{E$\alpha^{{}}$}&\multicolumn{2}{c}{E(level)}&\multicolumn{2}{c}{I$\alpha^{{\hyperlink{FR55DECAY1}{\ddagger}}}$}&\multicolumn{2}{c}{HF$^{{\hyperlink{FR55DECAY0}{\dagger}}}$}&Comments&\\[-.2cm]
\multicolumn{2}{c}{\hrulefill}&\multicolumn{2}{c}{\hrulefill}&\multicolumn{2}{c}{\hrulefill}&\multicolumn{2}{c}{\hrulefill}&\hrulefill&
\endfirsthead
\multicolumn{1}{r@{}}{7935}&\multicolumn{1}{@{ }l}{{\it 30}}&\multicolumn{1}{r@{}}{0}&\multicolumn{1}{@{}l}{}&\multicolumn{1}{r@{}}{$\approx$100}&\multicolumn{1}{@{}l}{}&\multicolumn{1}{r@{}}{$\approx$2}&\multicolumn{1}{@{.}l}{3}&\parbox[t][0.3cm]{12.855961cm}{\raggedright E$\alpha$,I$\alpha$: From \href{https://www.nndc.bnl.gov/nsr/nsrlink.jsp?2014Zh03,B}{2014Zh03}.\hphantom{a}E\ensuremath{\alpha}1=7935 keV \textit{30} correlated with E\ensuremath{\alpha}2=7406 keV \textit{30} (\ensuremath{^{\textnormal{201}}}Fr) and\vspace{0.1cm}}&\\
&&&&&&&&\parbox[t][0.3cm]{12.855961cm}{\raggedright {\ }{\ }{\ }E\ensuremath{\alpha}3=6997 keV \textit{30} (\ensuremath{^{\textnormal{197}}}At).\vspace{0.1cm}}&\\
\end{longtable}
\parbox[b][0.3cm]{17.7cm}{\makebox[1ex]{\ensuremath{^{\hypertarget{FR55DECAY0}{\dagger}}}} Using r\ensuremath{_{\textnormal{0}}}=1.498 \textit{2}, unweighted average of r\ensuremath{_{\textnormal{0}}}=1.4803 \textit{26} for \ensuremath{^{\textnormal{200}}}Po and 1.516 \textit{7} for \ensuremath{^{\textnormal{202}}}Rn.}\\
\parbox[b][0.3cm]{17.7cm}{\makebox[1ex]{\ensuremath{^{\hypertarget{FR55DECAY1}{\ddagger}}}} For absolute intensity per 100 decays, multiply by{ }\ensuremath{\approx}1.}\\
\vspace{0.5cm}
\clearpage
%(HI,XNG)
\subsection[\hspace{-0.2cm}(HI,xn\ensuremath{\gamma})]{ }
\vspace{-27pt}
\vspace{0.3cm}
\hypertarget{FR56}{{\bf \small \underline{(HI,xn\ensuremath{\gamma})\hspace{0.2in}\href{https://www.nndc.bnl.gov/nsr/nsrlink.jsp?2020Au01,B}{2020Au01},\href{https://www.nndc.bnl.gov/nsr/nsrlink.jsp?2014Ka23,B}{2014Ka23},\href{https://www.nndc.bnl.gov/nsr/nsrlink.jsp?2005Uu02,B}{2005Uu02}}}}\\
\vspace{4pt}
\vspace{8pt}
\parbox[b][0.3cm]{17.7cm}{\addtolength{\parindent}{-0.2in}\href{https://www.nndc.bnl.gov/nsr/nsrlink.jsp?2020Au01,B}{2020Au01}: \ensuremath{^{\textnormal{169}}}Tm(\ensuremath{^{\textnormal{36}}}Ar,4n\ensuremath{\gamma}) at E(\ensuremath{^{\textnormal{36}}}Ar)=178, 184, and 187 MeV. Evaporation residues separated with RITU separator and}\\
\parbox[b][0.3cm]{17.7cm}{implanted into a DSSD. Measured E\ensuremath{\gamma}, I\ensuremath{\gamma}, E\ensuremath{\alpha}, \ensuremath{\alpha}(t), ce(t) using an array of silicon PIN diodes and three Clover-type HPGe}\\
\parbox[b][0.3cm]{17.7cm}{detectors.}\\
\parbox[b][0.3cm]{17.7cm}{\addtolength{\parindent}{-0.2in}\href{https://www.nndc.bnl.gov/nsr/nsrlink.jsp?2014Ka23,B}{2014Ka23}: \ensuremath{^{\textnormal{149}}}Sm(\ensuremath{^{\textnormal{56}}}Fe,p3n) at E(\ensuremath{^{\textnormal{56}}}Fe)=275 MeV produced by the GSI accelerator facility. Target=370 \ensuremath{\mu}g/cm\ensuremath{^{\textnormal{2}}} thick enriched to}\\
\parbox[b][0.3cm]{17.7cm}{96.4\% in \ensuremath{^{\textnormal{147}}}Sm, with 40 \ensuremath{\mu}g/cm\ensuremath{^{\textnormal{2}}} thick carbon backing and covered with a 10 \ensuremath{\mu}g/cm\ensuremath{^{\textnormal{2}}} layer of carbon, and mounted on a rotating}\\
\parbox[b][0.3cm]{17.7cm}{wheel. Detectors: SHIP recoil separator, 16-strip position sensitive Si detectors (PSSD), six Si strip detectors to detect escaping \ensuremath{\alpha}}\\
\parbox[b][0.3cm]{17.7cm}{particles and one HPGe clover detector behind the PSDD. Measured: recoil-\ensuremath{\alpha}-\ensuremath{\gamma}(t) and recoil-\ensuremath{\alpha}-\ensuremath{\alpha}(t). Deduced: E\ensuremath{\alpha} and T\ensuremath{_{\textnormal{1/2}}}.}\\
\parbox[b][0.3cm]{17.7cm}{\addtolength{\parindent}{-0.2in}\href{https://www.nndc.bnl.gov/nsr/nsrlink.jsp?2005Uu02,B}{2005Uu02}: produced using \ensuremath{^{\textnormal{170}}}Yb(\ensuremath{^{\textnormal{36}}}Ar,p4n),E(\ensuremath{^{\textnormal{36}}}Ar)=180 and 185 MeV. Target: 70 \% enriched in \ensuremath{^{\textnormal{170}}}Yb. Detectors: gas filled}\\
\parbox[b][0.3cm]{17.7cm}{mass separator, position sensitive silicon detectors with a typical resolution (FWHM) of 30 keV, multi-wire proportional gas}\\
\parbox[b][0.3cm]{17.7cm}{counter. Measured: E\ensuremath{\alpha}, T\ensuremath{_{\textnormal{1/2}}}.}\\
\parbox[b][0.3cm]{17.7cm}{\addtolength{\parindent}{-0.2in}\href{https://www.nndc.bnl.gov/nsr/nsrlink.jsp?2005De01,B}{2005De01}: produced in a bombardment with a 1.4 GeV pulsed proton beam on 51 g/cm\ensuremath{^{\textnormal{2}}} thorium/graphite target. Detectors: on-line}\\
\parbox[b][0.3cm]{17.7cm}{mass separator, recoils were implanted on a carbon foil for 100 ms and subsequent \ensuremath{\alpha}-decay counted using a 400 mm\ensuremath{^{\textnormal{2}}}, 1 mm thick}\\
\parbox[b][0.3cm]{17.7cm}{silicon detector for 1100 ms; Measured: E\ensuremath{\alpha}, T\ensuremath{_{\textnormal{1/2}}}.}\\
\parbox[b][0.3cm]{17.7cm}{\addtolength{\parindent}{-0.2in}Others: \href{https://www.nndc.bnl.gov/nsr/nsrlink.jsp?1996En01,B}{1996En01}: produced using \ensuremath{^{\textnormal{170}}}Yb(\ensuremath{^{\textnormal{35}}}Cl,4n), E(\ensuremath{^{\textnormal{35}}}Cl)=205 and 213 MeV; Target: 72 \% enriched in \ensuremath{^{\textnormal{170}}}Yb; Detectors: gas}\\
\parbox[b][0.3cm]{17.7cm}{filled mass separator, position sensitive silicon detectors with a typical resolution (FWHM) of 35 keV; Measured: E\ensuremath{\alpha}, T\ensuremath{_{\textnormal{1/2}}}.}\\
\parbox[b][0.3cm]{17.7cm}{Assignment to \ensuremath{^{\textnormal{201}}}Fr is based on the observed E\ensuremath{\alpha}1-E\ensuremath{\alpha}2 correlation with the characteristic daughter \ensuremath{\alpha}-decay;\hphantom{a}\href{https://www.nndc.bnl.gov/nsr/nsrlink.jsp?1980Ew03,B}{1980Ew03}: produced}\\
\parbox[b][0.3cm]{17.7cm}{using \ensuremath{^{\textnormal{238}}}U(p,spallation); E(p)=600 MeV; Detectors: on-line mass separator, silicon charged particle detector; Measured: E\ensuremath{\alpha}, T\ensuremath{_{\textnormal{1/2}}}.}\\
\vspace{12pt}
\underline{$^{201}$Fr Levels}\\

\clearpage
\begin{figure}[h]
\begin{center}
\includegraphics{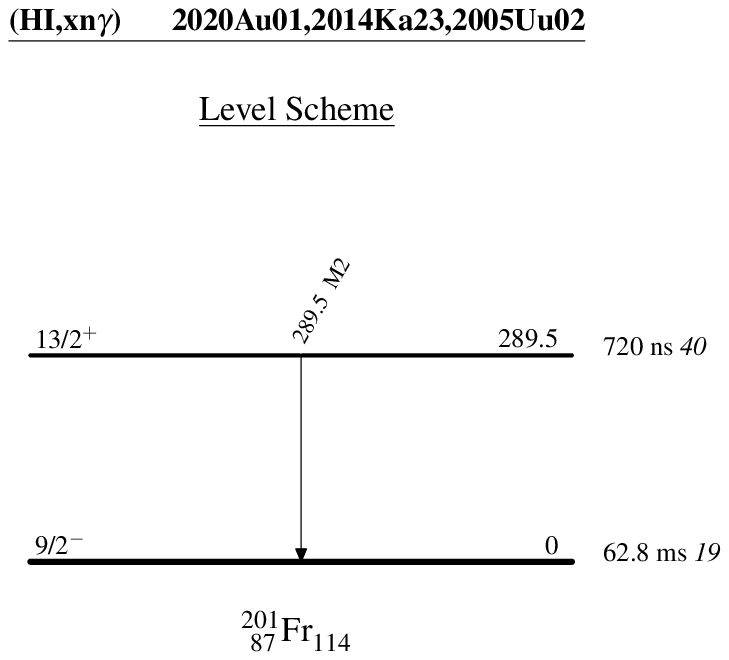}\\
\end{center}
\end{figure}
\clearpage
%ADOPTED LEVELS
\section[\ensuremath{^{201}_{\ 88}}Ra\ensuremath{_{113}^{~}}]{ }
\vspace{-30pt}
\setcounter{chappage}{1}
\subsection[\hspace{-0.2cm}Adopted Levels]{ }
\vspace{-20pt}
\vspace{0.3cm}
\hypertarget{RA57}{{\bf \small \underline{Adopted \hyperlink{201RA_LEVEL}{Levels}}}}\\
\vspace{4pt}
\vspace{8pt}
\parbox[b][0.3cm]{17.7cm}{\addtolength{\parindent}{-0.2in}S(p)=1490 {\it 40}; Q(\ensuremath{\alpha})=8002 {\it 12}\hspace{0.2in}\href{https://www.nndc.bnl.gov/nsr/nsrlink.jsp?2021Wa16,B}{2021Wa16}}\\

\parbox[b][0.3cm]{17.7cm}{\addtolength{\parindent}{-0.2in}\href{https://www.nndc.bnl.gov/nsr/nsrlink.jsp?2005Uu02,B}{2005Uu02}: \ensuremath{^{\textnormal{201}}}Ra produced using the \ensuremath{^{\textnormal{141}}}Pr(\ensuremath{^{\textnormal{63}}}Cu,3n) reaction, E(\ensuremath{^{\textnormal{63}}}Cu)=278 and 288 MeV. Detectors: gas filled mass separator,}\\
\parbox[b][0.3cm]{17.7cm}{position sensitive silicon detectors with a typical energy resolution (FWHM) of 30 keV; multi-wire proportional gas counters;}\\
\parbox[b][0.3cm]{17.7cm}{Measured: E\ensuremath{\alpha}, \ensuremath{\alpha}-\ensuremath{\alpha} correlations, T\ensuremath{_{\textnormal{1/2}}}.}\\
\parbox[b][0.3cm]{17.7cm}{\addtolength{\parindent}{-0.2in}\href{https://www.nndc.bnl.gov/nsr/nsrlink.jsp?2014Ka23,B}{2014Ka23}: \ensuremath{^{\textnormal{201}}}Ra produced using the \ensuremath{^{\textnormal{147}}}Sm(\ensuremath{^{\textnormal{56}}}Fe,2n) reactions, E(\ensuremath{^{\textnormal{56}}}Fe)=249 MeV. Target=370 \ensuremath{\mu}g/cm\ensuremath{^{\textnormal{2}}} thick enriched to 96.4\%}\\
\parbox[b][0.3cm]{17.7cm}{in \ensuremath{^{\textnormal{147}}}Sm, with 40 \ensuremath{\mu}g/cm\ensuremath{^{\textnormal{2}}} thick carbon backing and covered with a 10 \ensuremath{\mu}g/cm\ensuremath{^{\textnormal{2}}} layer of carbon, and mounted on a rotating}\\
\parbox[b][0.3cm]{17.7cm}{wheel. Detectors: SHIP recoil separator, 16-strip position sensitive Si detectors (PSSD), six Si strip detectors to detect escaping \ensuremath{\alpha}}\\
\parbox[b][0.3cm]{17.7cm}{particles and one HPGe clover detector behind the PSDD. Measured: recoil-\ensuremath{\alpha}-\ensuremath{\gamma}(t) and recoil-\ensuremath{\alpha}-\ensuremath{\alpha}(t). Deduced: E\ensuremath{\alpha} and T\ensuremath{_{\textnormal{1/2}}}.}\\
\vspace{12pt}
\hypertarget{201RA_LEVEL}{\underline{$^{201}$Ra Levels}}\\

\end{center}
\clearpage
\newpage
\pagestyle{plain}
\section[References]{ }
\vspace{-30pt}
%
\end{document}